\documentclass[12pt,a4paper,dvips]{article}
\usepackage{graphicx,epsfig}
\usepackage{hhline}

\usepackage{amsmath,amssymb}
\usepackage{times}
\usepackage[varg]{txfonts}
\DeclareMathAlphabet{\mathbold}{OML}{txr}{b}{it}

\usepackage{array,multirow,dcolumn}
\usepackage[mathlines,displaymath]{lineno}
\usepackage{rotating}
\usepackage{captcont} 

\usepackage[numbers,square,comma,sort&compress]{natbib}
\usepackage{textcomp}
\makeatletter
\g@addto@macro\bfseries{\boldmath}
\makeatother

\newcolumntype{.}{D{.}{.}{-1}}
\newcolumntype{-}{D{-}{-}{-1}}

\usepackage[dvipsnames]{color}
\definecolor{rltred}{rgb}{0.75,0,0}
\definecolor{rltgreen}{rgb}{0,0.5,0}
\definecolor{rltblue}{rgb}{0,0,0.5}

\usepackage[T1]{fontenc} 

\usepackage[hyperindex,bookmarks,bookmarksnumbered,breaklinks,a4paper,unicode]{hyperref}
\hypersetup{%
  pdftitle        = {HERAPDF2.0},
  urlcolor        = rltblue,       
  urlbordercolor  = 0 0 0.5,
  filecolor       = rltblue,       
  filebordercolor = 0 0 0.5,
  linkcolor       = rltred,        
  linkbordercolor = 0.75 0 0,
  citecolor       = rltgreen,      
  citebordercolor = 0 0.5 0,
  pagecolor       = rltgreen,      
  pagebordercolor = 0 0.5 0,
  menucolor       = rltgreen,      
  menubordercolor = 0 0.5 0,
  colorlinks    = true,
  pdfauthor     = {H1 plus ZEUS Collaborations},
  pdfsubject    = { },
  pdfkeywords   = {High-Energy Physics, Particle Physics, Proton Structure, DIS}
}

\newlength{\dinwidth}
\newlength{\dinmargin}
\newcommand{\boldfX}{\mbox{${F}_{\rm X}$}}
\newcommand{\boldfXgZ}{\mbox{${F}_{\rm X}^{\gamma Z}$}}
\newcommand{\boldfXZ}{\mbox{${F}_{\rm X}^Z$}}
\newcommand{\boldftwo}{\mbox{$\tilde{F}_2$}}
\newcommand{\boldxft}{\mbox{$x\tilde{F}_3$}}
\newcommand{\boldfl}{\mbox{$\tilde{F}_{\rm L}$}}

\newcommand{\ncred}{\mbox{$ \sigma_{r,{\rm NC}}^{\pm}$}}

\newcommand{\ncdd}{\mbox{$\frac{\textstyle {\rm d^2} \sigma^{e^{\pm}p}_{{\rm NC}}}{\textstyle {\rm d}x_{\rm Bj}{\rm d} Q^2}$}}

\newcommand{\ccdd}{\mbox{$\frac{\textstyle {\rm d^2} \sigma^{e^{\pm}p}_{{\rm CC}}}{\textstyle {\rm d}x_{\rm Bj}{\rm d} Q^2}$}}

\newcommand{\ccred}{\mbox{$ \sigma_{r,{\rm CC}}^{\pm}$}}
\newcommand{\ccredp}{\mbox{$ \sigma_{r,{\rm CC}}^{+}$}}
\newcommand{\ccredm}{\mbox{$ \sigma_{r,{\rm CC}}^{-}$}}

\newcommand{\Eq}{\mbox{Eq.}}

\newcommand{\Eqs}{\mbox{Eqs.}}

\newcommand{\asmz}{\alpha_s(M_Z^2)}
\newcommand{\msbar}{\mbox{$\overline{\rm{MS}}$}\ }
\newcommand{\bs}{\bar{s}}
\newcommand{\bc}{\bar{c}}
\newcommand{\bu}{\bar{u}}
\newcommand{\bd}{\bar{d}}

\newcommand{\bU}{\bar{U}}
\newcommand{\bD}{\bar{D}}

\begin{document}

\makeatletter \def\NAT@space{} \makeatother

\begin{titlepage}
 
\noindent
DESY-15-039 \\
June 5, 2015 \\

\vspace*{1.0cm}

\begin{center}
\begin{Large}

{\bfseries Combination of Measurements of Inclusive Deep Inelastic
  $\mathbold{e^{\pm}p}$ Scattering Cross Sections 
  and QCD Analysis of HERA Data}

\vspace*{1cm}

H1 and ZEUS Collaborations

\end{Large}
\end{center}

\vspace*{1cm}
\begin{abstract} \noindent
A combination is presented of all inclusive deep inelastic 
cross sections previously published by the H1 and ZEUS collaborations 
at HERA for neutral and charged current $e^{\pm}p$ scattering for 
zero beam polarisation.
The data were taken at proton beam 
energies of 920, 820, 575 and 460\,GeV and an electron 
beam energy of 27.5\,GeV.
The data correspond to an integrated luminosity of 
about 1\,fb$^{-1}$ and span six orders of magnitude in 
negative four-momentum-transfer squared, $Q^2$, and Bjorken $x$.
The correlations of the systematic uncertainties were evaluated and 
taken into account for the combination.
The combined cross sections were input to QCD analyses 
at leading order, next-to-leading order and at next-to-next-to-leading order, 
providing a new set of parton 
distribution functions, called HERAPDF2.0.
In addition to the experimental uncertainties,
model and parameterisation uncertainties 
were assessed for these parton distribution 
functions. 
Variants of HERAPDF2.0 with an alternative gluon parameterisation, 
HERAPDF2.0AG, and using fixed-flavour-number schemes, 
HERAPDF2.0FF, are presented.
The analysis was extended by including HERA data on charm 
and jet production, resulting in the variant HERAPDF2.0Jets.
The inclusion of jet-production cross sections 
made a simultaneous determination of these parton distributions 
and the strong coupling constant possible, resulting in 
$\asmz =0.1183 \pm 0.0009 {\rm(exp)} \pm 0.0005{\rm (model/parameterisation)} \pm 0.0012{\rm (hadronisation)} ^{+0.0037}_{-0.0030}{\rm (scale)}$.
An extraction of $xF_3^{\gamma Z}$ and results on electroweak unification
and scaling violations are also presented.

\end{abstract}

\vspace*{0.5cm}

\begin{center}
{\slshape Accepted by EPJC}
\end{center}
\vspace*{0.5cm}

\em This paper is dedicated to the memory of Professor Guido Altarelli who
sadly passed away as it went to press. The results which it presents
are founded on the principles and the formalism which he developed in
his pioneering theoretical work on Quantum Chromodynamics
in deep-inelastic lepton-nucleon scattering nearly four decades ago. 
\end{titlepage}

\newpage

{\small\raggedright
H.~Abramowicz$^{\mathrm{54},\mathrm{a1}}$,
I.~Abt$^{\mathrm{40}}$,
L.~Adamczyk$^{\mathrm{25}}$,
M.~Adamus$^{\mathrm{64}}$,
V.~Andreev$^{\mathrm{37}}$,
S.~Antonelli$^{\mathrm{8}}$,
B.~Antunovi\'{c}$^{\mathrm{4}}$,
V.~Aushev$^{\mathrm{28},\mathrm{29},\mathrm{b22}}$,
Y.~Aushev$^{\mathrm{29},\mathrm{a2},\mathrm{b22}}$,
A.~Baghdasaryan$^{\mathrm{66}}$
% ...

\newpage
\footnotesize\begin{description}\setlength{\parsep}{0em}\setlength{\itemsep}{0em}
\item[$^{1}$] 
 I. Physikalisches Institut der RWTH, Aachen, Germany
\item[$^{2}$] 
 {Institute of Physics and Technology of Ministry of Education and Science of Kazakhstan, Almaty, Kazakhstan}
\item[$^{3}$] 
 Inter-University Institute for High Energies ULB-VUB, Brussels and Universiteit Antwerpen, Antwerpen, Belgium$^{\mathrm{b1}}$
\item[$^{4}$] 
 Univerzitet u Banjoj Luci, Arhitektonsko-gra\dj{}ko-geodetski fakultet, Banja Luka, Bosnia-Herzegovina
\item[$^{5}$] 
 {Universit\"at Bielefeld, Bielefeld, Germany}
\item[$^{6}$] 
 School of Physics and Astronomy, University of Birmingham, Birmingham, UK$^{\mathrm{b2}}$
\item[$^{7}$] 
 {INFN Bologna, Bologna, Italy{}}$^{\mathrm{b3}}$
\item[$^{8}$] 
 {University and INFN Bologna, Bologna, Italy{}}$^{\mathrm{b3}}$
\item[$^{9}$] 
 {Physikalisches Institut der Universit\"at Bonn, Bonn, Germany{}}$^{\mathrm{b4}}$
\item[$^{10}$] 
 National Institute for Physics and Nuclear Engineering (NIPNE) , Bucharest, Romania$^{\mathrm{b5}}$
\item[$^{11}$] 
 {Panjab University, Department of Physics, Chandigarh, India}
\item[$^{12}$] 
 {Calabria University, Physics Department and INFN, Cosenza, Italy{}}$^{\mathrm{b3}}$
\item[$^{13}$] 
 STFC, Rutherford Appleton Laboratory, Didcot, Oxfordshire, UK$^{\mathrm{b2}}$
\item[$^{14}$] 
 Institut f\"ur Physik, TU Dortmund, Dortmund, Germany$^{\mathrm{b6}}$
\item[$^{15}$] 
 Joint Institute for Nuclear Research, Dubna, Russia
\item[$^{16}$] 
 CERN, Geneva, Switzerland
\item[$^{17}$] 
 CEA, DSM/Irfu, CE-Saclay, Gif-sur-Yvette, France
\item[$^{18}$] 
 {School of Physics and Astronomy, University of Glasgow, Glasgow, United Kingdom{}}$^{\mathrm{b2}}$
\item[$^{19}$] 
 II. Physikalisches Institut, Universit\"at G\"ottingen, G\"ottingen, Germany
\item[$^{20}$] 
 Institut f\"ur Experimentalphysik, Universit\"at Hamburg, Hamburg, Germany$^{\mathrm{b6},\mathrm{b7}}$
\item[$^{21}$] 
 {Deutsches Elektronen-Synchrotron DESY, Hamburg, Germany}
\item[$^{22}$] 
 Physikalisches Institut, Universit\"at Heidelberg, Heidelberg, Germany$^{\mathrm{b6}}$
\item[$^{23}$] 
 Institute of Experimental Physics, Slovak Academy of Sciences, Ko\v{s}ice, Slovak Republic$^{\mathrm{b8}}$
\item[$^{24}$] 
 {The Henryk Niewodniczanski Institute of Nuclear Physics, Polish Academy of \ Sciences, Krakow, Poland{}}$^{\mathrm{b9},\mathrm{b16}}$
\item[$^{25}$] 
 {AGH-University of Science and Technology, Faculty of Physics and Applied Computer Science, Krakow, Poland{}}$^{\mathrm{b9}}$
\item[$^{26}$] 
 {Department of Physics, Jagellonian University, Krakow, Poland}
\item[$^{27}$] 
 {National Centre for Particle Physics, Universiti Malaya, 50603 Kuala Lumpur, Malaysia{}}$^{\mathrm{b10}}$
\item[$^{28}$] 
 {Institute for Nuclear Research, National Academy of Sciences, Kyiv, Ukraine}
\item[$^{29}$] 
 {Department of Nuclear Physics, National Taras Shevchenko University of Kyiv, Kyiv, Ukraine}
\item[$^{30}$] 
 Department of Physics, University of Lancaster, Lancaster, UK$^{\mathrm{b2}}$
\item[$^{31}$] 
 Department of Physics, University of Liverpool, Liverpool, UK$^{\mathrm{b2}}$
\item[$^{32}$] 
 School of Physics and Astronomy, Queen Mary, University of London, London, UK$^{\mathrm{b2}}$
\item[$^{33}$] 
 {Physics and Astronomy Department, University College London, London, United Kingdom{}}$^{\mathrm{b2}}$
\item[$^{34}$] 
 Aix Marseille Universit\'{e}, CNRS/IN2P3, CPPM UMR 7346, 13288 Marseille, France
\item[$^{35}$] 
 Departamento de Fisica Aplicada, CINVESTAV, M\'erida, Yucat\'an, M\'exico$^{\mathrm{b11}}$
\item[$^{36}$] 
 Institute for Theoretical and Experimental Physics, Moscow, Russia$^{\mathrm{b12}}$
\item[$^{37}$] 
 Lebedev Physical Institute, Moscow, Russia
\item[$^{38}$] 
 Institute for Information Transmission Problems RAS, Moscow, Russia$^{\mathrm{b13}}$
\item[$^{39}$] 
 {Lomonosov Moscow State University, Skobeltsyn Institute of Nuclear Physics, Moscow, Russia{}}$^{\mathrm{b14}}$
\item[$^{40}$] 
 {Max-Planck-Institut f\"ur Physik, M\"unchen, Germany}
\item[$^{41}$] 
 {Department of Physics, York University, Ontario, Canada M3J 1P3{}}$^{\mathrm{b15}}$
\item[$^{42}$] 
 LAL, Universit\'e Paris-Sud, CNRS/IN2P3, Orsay, France
\item[$^{43}$] 
 {Department of Physics, University of Oxford, Oxford, United Kingdom{}}$^{\mathrm{b2}}$
\item[$^{44}$] 
 {INFN Padova, Padova, Italy{}}$^{\mathrm{b3}}$
\item[$^{45}$] 
 {Dipartimento di Fisica e Astronomia dell' Universit\`a and INFN, Padova, Italy{}}$^{\mathrm{b3}}$
\item[$^{46}$] 
 LLR, Ecole Polytechnique, CNRS/IN2P3, Palaiseau, France
\item[$^{47}$] 
 Faculty of Science, University of Montenegro, Podgorica, Montenegro$^{\mathrm{b17}}$
\item[$^{48}$] 
 Institute of Physics, Academy of Sciences of the Czech Republic, Praha, Czech Republic$^{\mathrm{b18}}$
\item[$^{49}$] 
 Faculty of Mathematics and Physics, Charles University, Praha, Czech Republic$^{\mathrm{b18}}$
\item[$^{50}$] 
 {Department of Physics, McGill University, Montr\'eal, Qu\'ebec, Canada H3A 2T8{}}$^{\mathrm{b15}}$
\item[$^{51}$] 
 {Department of Particle Physics and Astrophysics, Weizmann Institute, Rehovot, Israel}
\item[$^{52}$] 
 Dipartimento di Fisica Universit\`a di Roma Tre and INFN Roma~3, Roma, Italy
\item[$^{53}$] 
 Institute for Nuclear Research and Nuclear Energy, Sofia, Bulgaria
\item[$^{54}$] 
 {Raymond and Beverly Sackler Faculty of Exact Sciences, School of Physics, \ Tel Aviv University, Tel Aviv, Israel{}}$^{\mathrm{b19}}$
\item[$^{55}$] 
 {Polytechnic University, Tokyo, Japan{}}$^{\mathrm{b20}}$
\item[$^{56}$] 
 {Department of Physics, Tokyo Institute of Technology, Tokyo, Japan{}}$^{\mathrm{b20}}$
\item[$^{57}$] 
 {Universit\`a di Torino and INFN, Torino, Italy{}}$^{\mathrm{b3}}$
\item[$^{58}$] 
 {Universit\`a del Piemonte Orientale, Novara, and INFN, Torino, Italy{}}$^{\mathrm{b3}}$
\item[$^{59}$] 
 {Institute of Particle and Nuclear Studies, KEK, Tsukuba, Japan{}}$^{\mathrm{b20}}$
\item[$^{60}$] 
 Institute of Physics and Technology of the Mongolian Academy of Sciences, Ulaanbaatar, Mongolia
\item[$^{61}$] 
 IPNL, Universit\'e Claude Bernard Lyon 1, CNRS/IN2P3, Villeurbanne, France
\item[$^{62}$] 
 Paul Scherrer Institut, Villigen, Switzerland
\item[$^{63}$] 
 {Faculty of Physics, University of Warsaw, Warsaw, Poland}
\item[$^{64}$] 
 {National Centre for Nuclear Research, Warsaw, Poland}
\item[$^{65}$] 
 Fachbereich C, Universit\"at Wuppertal, Wuppertal, Germany
\item[$^{66}$] 
 Yerevan Physics Institute, Yerevan, Armenia
\item[$^{67}$] 
 {Meiji Gakuin University, Faculty of General Education, Yokohama, Japan{}}$^{\mathrm{b20}}$
\item[$^{68}$] 
 {Deutsches Elektronen-Synchrotron DESY, Zeuthen, Germany}
\item[$^{69}$] 
 Institut f\"ur Teilchenphysik, ETH, Z\"urich, Switzerland$^{\mathrm{b21}}$
\item[$^{70}$] 
 Physik-Institut der Universit\"at Z\"urich, Z\"urich, Switzerland$^{\mathrm{b21}}$
\item[$^{\dagger}$] 
 {Deceased}
\end{description}
\medskip
\begin{description}\setlength{\parsep}{0em}\setlength{\itemsep}{0em}\item[$^{\mathrm{a1}}$] 
 Also at Max Planck Institute for Physics, Munich, Germany, External Scientific Member
\item[$^{\mathrm{a2}}$] 
 Member of National Technical University of Ukraine, Kyiv Polytechnic Institute, Kyiv, Ukraine
\item[$^{\mathrm{a3}}$] 
 Now at Department of Physics, Saint-Petersburg State University, Saint-Petersburg, Russia
\item[$^{\mathrm{a4}}$] 
 Also at Moscow Institute of Physics and Technology, Moscow, Russia
\item[$^{\mathrm{a5}}$] 
 Also at \L\'{o}d\'{z} University, Poland
\item[$^{\mathrm{a6}}$] 
 Now at Rockefeller University, New York, NY 10065, USA
\item[$^{\mathrm{a7}}$] 
 Also at Rechenzentrum, Universit\"at Wuppertal, Wuppertal, Germany
\item[$^{\mathrm{a8}}$] 
 Alexander von Humboldt Professor; also at DESY and University of Oxford
\item[$^{\mathrm{a9}}$] 
 Now at School of Physics and Astronomy, University of Birmingham, UK
\item[$^{\mathrm{a10}}$] 
 Now at European X-ray Free-Electron Laser facility GmbH, Hamburg, Germany
\item[$^{\mathrm{a11}}$] 
 Now at DESY
\item[$^{\mathrm{a12}}$] 
 Now at LNF, Frascati, Italy
\item[$^{\mathrm{a13}}$] 
 Also at Agensi Nuklear Malaysia, 43000 Kajang, Bangi, Malaysia
\item[$^{\mathrm{a14}}$] 
 Now at Humboldt-Universit\"at zu Berlin, Berlin, Germany
\item[$^{\mathrm{a15}}$] 
 Member of \L\'{o}d\'{z} University, Poland
\item[$^{\mathrm{a16}}$] 
 Also at Department of Physics, University of Toronto, Toronto, Ontario, Canada M5S 1A7
\item[$^{\mathrm{a17}}$] 
 Also at LAPP, Universit\'e de Savoie, CNRS/IN2P3, Annecy-le-Vieux, France
\item[$^{\mathrm{a18}}$] 
 Now at BNL, USA
\item[$^{\mathrm{a19}}$] 
 Member of National University of Kyiv - Mohyla Academy, Kyiv, Ukraine
\item[$^{\mathrm{a20}}$] 
 Also at University of Tokyo, Japan
\item[$^{\mathrm{a21}}$] 
 Now at Polish Air Force Academy in Deblin
\item[$^{\mathrm{a22}}$] 
 Also at Ulaanbaatar University, Ulaanbaatar, Mongolia
\item[$^{\mathrm{a23}}$] 
 Also at Universit\"{a}t Hamburg and supported by DESY and the Alexander von Humboldt Foundation
\item[$^{\mathrm{a24}}$] 
 Now at Kobe University, Japan
\end{description}
\medskip
\begin{description}\setlength{\parsep}{0em}\setlength{\itemsep}{0em}\item[$^{\mathrm{b1}}$] 
 Supported by FNRS-FWO-Vlaanderen, IISN-IIKW and IWT and by InteruniversityAttraction Poles Programme, Belgian Science Policy
\item[$^{\mathrm{b2}}$] 
 Supported by the UK Science and Technology Facilities Council, and formerly by the UK Particle Physics and Astronomy Research Council
\item[$^{\mathrm{b3}}$] 
 Supported by the Italian National Institute for Nuclear Physics (INFN)
\item[$^{\mathrm{b4}}$] 
 Supported by the German Federal Ministry for Education and Research (BMBF), under contract No. 05 H09PDF
\item[$^{\mathrm{b5}}$] 
 Supported by the Romanian National Authority for Scientific Research under the contract PN 09370101
\item[$^{\mathrm{b6}}$] 
 Supported by the Bundesministerium f\"ur Bildung und Forschung, FRG, under contract numbers 05H09GUF, 05H09VHC, 05H09VHF, 05H16PEA
\item[$^{\mathrm{b7}}$] 
 Supported by the SFB 676 of the Deutsche Forschungsgemeinschaft (DFG)
\item[$^{\mathrm{b8}}$] 
 Supported by VEGA SR grant no. 2/7062/ 27
\item[$^{\mathrm{b9}}$] 
 Supported by the National Science Centre under contract No. DEC-2012/06/M/ST2/00428
\item[$^{\mathrm{b10}}$] 
 Supported by HIR grant UM.C/625/1/HIR/149 and UMRG grants RU006-2013, RP012A-13AFR and RP012B-13AFR from Universiti Malaya, and ERGS grant ER004-2012A from the Ministry of Education, Malaysia
\item[$^{\mathrm{b11}}$] 
 Supported by CONACYT, M\'exico, grant 48778-F
\item[$^{\mathrm{b12}}$] 
 Russian Foundation for Basic Research (RFBR), grant no 1329.2008.2 and Rosatom
\item[$^{\mathrm{b13}}$] 
 Russian Foundation for Sciences, project no 14-50-00150
\item[$^{\mathrm{b14}}$] 
 Supported by RF Presidential grant N 3042.2014.2 for the Leading Scientific Schools and by the Russian Ministry of Education and Science through its grant for Scientific Research on High Energy Physics
\item[$^{\mathrm{b15}}$] 
 Supported by the Natural Sciences and Engineering Research Council of Canada (NSERC)
\item[$^{\mathrm{b16}}$] 
 Partially Supported by Polish Ministry of Science and Higher Education, grant DPN/N168/DESY/2009
\item[$^{\mathrm{b17}}$] 
 Partially Supported by Ministry of Science of Montenegro, no. 05-1/3-3352
\item[$^{\mathrm{b18}}$] 
 Supported by the Ministry of Education of the Czech Republic under the project INGO-LG14033
\item[$^{\mathrm{b19}}$] 
 Supported by the Israel Science Foundation
\item[$^{\mathrm{b20}}$] 
 Supported by the Japanese Ministry of Education, Culture, Sports, Science and Technology (MEXT) and its grants for Scientific Research
\item[$^{\mathrm{b21}}$] 
 Supported by the Swiss National Science Foundation
\item[$^{\mathrm{b22}}$] 
 Supported by DESY, Germany
\item[$^{\mathrm{b23}}$] 
 Also funded by Max Planck Institute for Physics, Munich, Germany, now at Sant Longowal Institute of Engineering and Technology, Longowal, Punjab, India
\item[$^{\mathrm{b24}}$] 
 Supported by the Alexander von Humboldt Foundation
\item[$^{\mathrm{b25}}$] 
 Also funded by Max Planck Institute for Physics, Munich, Germany, now at Sri Guru Granth Sahib World University, Fatehgarh Sahib, India
\item[$^{\mathrm{b26}}$] 
 Partially supported by the Polish National Science Centre projects DEC-2011/01/B/ST2/03643 and DEC-2011/03/B/ST2/00220
\end{description}}

\newpage


\section{Introduction \label{sec:int}}
Deep inelastic scattering (DIS) of electrons\footnote{In this paper, 
the word ``electron" refers to both electrons and positrons, 
unless otherwise stated.} 
on protons at centre-of-mass energies of up to $\sqrt{s} \simeq 320\,$GeV
at HERA has been central to the exploration
of proton structure and quark--gluon dynamics as
described by perturbative Quantum Chromo Dynamics (pQCD)~\cite{saturation}. 
The two collaborations, H1 and ZEUS, have explored a large
phase space in Bjorken $x$, $x_{\rm Bj}$, 
and negative four-momentum-transfer squared, $Q^2$.
Cross sections for neutral current (NC) interactions have been
published for
$0.045 \leq Q^2 \leq 50000 $\,GeV$^2$
and  $6 \cdot 10^{-7} \leq x_{\rm Bj} \leq 0.65$ 
at values of the inelasticity, $y = Q^2/(sx_{\rm Bj})$, 
between $0.005$ and $0.95$.
Cross sections for
charged current (CC) interactions 
have been published
for 
$200 \leq Q^2 \leq 50000 $\,GeV$^2$ and
$1.3 \cdot 10^{-2} \leq x_{\rm Bj} \leq 0.40$  
at values of $y$ between $0.037$ and $0.76$.

HERA was operated in two phases: HERA\,I, from 1992 to 2000, and HERA\,II, 
from 2002 to 2007. From 1994 onwards, and for all data used here, HERA operated with an electron beam energy of
$E_e \simeq 27.5$\,GeV.
For most of HERA\,I and~II, the proton beam energy 
was $E_p = 920$\,GeV, resulting in the highest centre-of-mass energy of 
$\sqrt{s} \simeq 320\,$GeV.
During the HERA\,I period, 
each experiment collected about 100\,pb$^{-1}$ of $e^+p$ and 
15\,pb$^{-1}$ of $e^-p$ data. 
These HERA\,I data were the
basis of a combination and pQCD analysis published
previously~\cite{HERAIcombi}. 
During the HERA\,II period, 
each experiment added about 150\,pb$^{-1}$ of $e^+p$ and 
235\,pb$^{-1}$ of $e^-p$ data.  
As a result,
the H1 and ZEUS collaborations
collected total integrated luminosities of 
approximately 500\,pb$^{-1}$ each, 
divided about equally between $e^+p$ and $e^-p$ scattering. 
The paper presented here is based on the
combination of all published 
H1~\cite{Collaboration:2009bp,Collaboration:2009kv,Adloff:1999ah,Adloff:2000qj,Adloff:2003uh,H1allhQ2,H1FL1,H1FL2} 
and ZEUS~\cite{Breitweg:1997hz,Breitweg:2000yn,Breitweg:1998dz,Chekanov:2001qu,zeuscc97,Chekanov:2002ej,Chekanov:2002zs,Chekanov:2003yv,Chekanov:2003vw,ZEUS2NCe,ZEUS2CCe,ZEUS2NCp,ZEUS2CCp,ZEUSFL} 
measurements
from both HERA\,I and~II on inclusive DIS in NC and CC reactions.
This includes data taken 
with proton beam energies of
$E_p = 920$, 820,  575 and 460\,GeV
corresponding to 
 $\sqrt{s}\simeq$\,320, 300, 251 and 225\,GeV.
During the HERA\,II period, the electron beam was longitudinally polarised.
The data considered in this paper are cross sections
corrected to zero beam polarisation as
published by the collaborations.

The combination of the data and the pQCD analysis were performed using the 
packages  HERAverager~\cite{HERAveragerweb} and
HERAFitter~\cite{HERAfitter,HERAfitterweb}. 
The method~\cite{glazov,Collaboration:2009bp} also  
allowed a model-independent demonstration of the consistency of the data. 
The correlated systematic uncertainties
and global normalisations were treated 
such that one coherent data set was obtained. 
Since H1 and
ZEUS employed different experimental techniques, 
using different detectors and methods
of kinematic reconstruction, the combination 
also led to a significantly reduced systematic uncertainty.

Within the framework of pQCD, the proton is described
in terms of parton density functions, $f(x)$, which 
provide the probability to find a parton, either gluon or quark, 
with a fraction $x$ of the proton's momentum. 
This probability is predicted to depend on the scale
at which the proton is probed, called the factorisation scale, 
$\mu_{\rm f}^2$, which for inclusive DIS is usually taken as $Q^2$.
These functions are usually presented as parton momentum distributions, 
$xf(x)$, and are called parton distribution functions (PDFs).
The PDFs are convoluted with the fundamental point-like 
scattering cross sections for partons to calculate cross sections.
Perturbative QCD provides
the framework to evolve the PDFs to other scales once they are provided
at a starting scale. 
However, pQCD does not predict the PDFs at the starting scale. 
They must be determined by fits to data using ad hoc parameterisations.
 
The name HERAPDF stands for a pQCD analysis within the
DGLAP~\cite{Gribov:1972ri,Gribov:1972rt,Lipatov:1974qm,Dokshitzer:1977sg,Altarelli:1977zs}
formalism. The $x_{\rm Bj}$ and $Q^2$ dependences of the NC and CC DIS
cross sections from both the H1 and ZEUS collaborations
are used to determine sets of quark and gluon 
momentum distributions in the proton.
The set of PDFs denoted as
HERAPDF1.0~\cite{HERAIcombi} was based on the combination of all 
inclusive DIS scattering cross sections obtained   
from HERA\,I data. A preliminary set of PDFs, 
HERAPDF1.5~\cite{HERAPDF15}, was obtained
using HERA\,I and selected HERA\,II data, some of which were still preliminary.
In this paper, a new set of PDFs, HERAPDF2.0, is presented, based on
combined inclusive DIS cross sections from all of HERA\,I and HERA\,II.

Several groups, JR~\cite{Jimenez-Delgado:2014twa}, MSTW/MMHT~\cite{Martin:2009iq,MMHT2014}, 
CTEQ/CT~\cite{CTEQ6L,CT10NLO}, ABM~\cite{ABM1,ABM2,ABM3}
and NNPDF~\cite{Ball:2008by,NNPDF3.0}, 
provide PDF sets using HERA, fixed-target and 
hadron-collider data. 
The strength of the HERAPDF approach is that 
a single coherent high-precision data set containing NC and CC cross sections
is used as input.
The new combined data used for the HERAPDF2.0 analysis span
four orders of magnitude in $Q^2$ and $x_{\rm Bj}$.
The availability of precision NC and CC cross sections 
over this large phase space allows HERAPDF to use only
$ep$ scattering data and thus makes HERAPDF independent 
of any heavy nuclear (or deuterium) corrections.
The difference between the NC $e^+p$ and $e^-p$ cross sections 
at high $Q^2$, together with the high-$Q^2$ CC data,
constrain the valence-quark distributions.
The CC $e^+p$ data especially constrain
the valence down-quark distribution in the proton
without assuming strong isospin symmetry
as done in the analysis of deuterium data.     
The lower-$Q^2$ NC data 
constrain the low-$x$ sea-quark distributions and through
their precisely measured $Q^2$ variations 
they also constrain the gluon distribution.
A  further constraint on the gluon distribution 
comes from the inclusion of NC data at different beam energies
such that the longitudinal structure function is probed 
through the $y$ dependence of the cross sections~\cite{CooperSarkar:1988}. 

The consistency of the input data allowed the determination of the 
experimental uncertainties of the HERAPDF2.0 parton distributions  
using rigorous statistical methods.  
The uncertainties resulting from model assumptions
and from the choice of the parameterisation of the PDFs 
were considered separately. 

Both H1 and ZEUS also published charm production cross sections, some of
which were combined and analysed previously~\cite{HERAccombi}, 
and jet production cross 
sections~\cite{zeus9697jets,zeusdijets,h1lowq2jets,h1highq2oldjets,h1highq2newjets}.
These data were included to obtain the variant HERAPDF2.0Jets.
The inclusion of jet cross sections allowed for a simultaneous
determination of the PDFs and the strong coupling constant.

The paper is structured as follows. Section~\ref{xsecns} gives
an introduction to the connection between cross sections and
the partonic structure of the proton. Section~\ref{sec:meas} introduces
the data used in the analyses presented here. Section~\ref{sec:comb}
describes the combination of data while Section~\ref{subsec:comb:results}
presents the results of the combination. Section~\ref{sec:qcdan}
describes the 
pQCD
analysis to extract PDFs from the combined inclusive
cross sections. The PDF set HERAPDF2.0 and its variants are presented in 
Section~\ref{sec:pdf20}.
In Section~\ref{sec:legplots}, results on
electroweak unification as well as scaling violations 
and the extraction of $xF_3^{\gamma Z}$  are presented.
The paper closes with a summary.

 \section{Cross sections and parton distributions}
\label{xsecns}
The reduced NC deep inelastic $e^{\pm}p$ scattering cross sections 
are given by a linear combination of generalised structure functions. 
For unpolarised $e^{\pm}p$ scattering,
reduced cross sections after
correction for QED radiative effects 
may be expressed in terms of structure functions as
\begin{eqnarray} \label{ncsi}     
 \ncred =\ncdd \cdot \frac{Q^4 x_{\rm Bj}}{2\pi \alpha^2 Y_+}                                                     
  =            \tilde{F_2} \mp \frac{Y_-}{Y_+} x\tilde{F_3} -\frac{y^2}{Y_+} \tilde{F_{\rm L}}~,
\end{eqnarray}                                                                  
where 
the fine-structure constant, $\alpha$, which is defined 
at zero momentum transfer,
the photon propagator and a helicity factor are absorbed 
in the definitions of \ncred~and $Y_{\pm}=1 \pm (1-y)^2$.
The overall structure functions, $\boldftwo$, $\boldfl$  and $\boldxft$,
are sums of structure functions, $\boldfX$, $\boldfXgZ$ and $\boldfXZ$,
relating to photon exchange, photon--$Z$ interference and $Z$ exchange, 
respectively, and depend on the electroweak parameters as~\cite{PDG12}
\begin{eqnarray} \label{strf}                                                   
 \boldftwo &=& F_2 - \kappa_Z v_e  \cdot F_2^{\gamma Z} +                      
  \kappa_Z^2 (v_e^2 + a_e^2 ) \cdot F_2^Z~, \nonumber \\   
 \boldfl &=& F_{\rm L} - \kappa_Z v_e  \cdot F_{\rm L}^{\gamma Z} +                      
  \kappa_Z^2 (v_e^2 + a_e^2 ) \cdot F_{\rm L}^Z~, \nonumber \\                     
 \boldxft &=&  - \kappa_Z a_e  \cdot xF_3^{\gamma Z} +                    
  \kappa_Z^2 \cdot 2 v_e a_e  \cdot xF_3^Z~,                                   
\end{eqnarray} 
where $v_e$ and $a_e$ are the vector and axial-vector weak couplings of 
the electron to the $Z$ boson, 
and $\kappa_Z(Q^2) =   Q^2 /[(Q^2+M_Z^2)(4\sin^2 \theta_W \cos^2 \theta_W)]$. 
In the analysis presented here, electroweak effects were treated
at leading order. 
The values of $\sin^2 \theta_W=0.23127$ and $M_Z=91.1876$\,GeV were used
for the electroweak mixing angle and the $Z$-boson mass~\cite{PDG12}.

At low $Q^2$, i.e. $Q^2 \ll M_Z^2$, 
the contribution of $Z$ exchange is negligible and 
\begin{equation} \label{eq:fl}
 \ncred = F_2  - \frac{y^2}{Y_+} F_{\rm L}~.
\end{equation}
The contribution of the term containing the 
longitudinal structure function $\tilde{F_{\rm L}}$ is 
only significant for values of $y$ larger than approximately 0.5.

In the analysis presented in this paper, the full formulae of pQCD 
at the relevant order in 
the strong coupling, $\alpha_s$, are used. 
However, to demonstrate the sensitivity
of the data, it is useful to discuss the simplified equations 
of the  Quark Parton Model (QPM), where gluons are not present and   
$\boldfl=0$~\cite{PhysRevLett.22.156}. In the QPM, 
the kinematic variable $x_{\rm Bj}$ is  equal to
the fractional momentum of the struck quark, $x$. The 
structure functions in \Eq\,\ref{strf} become
\begin{eqnarray} \label{ncfu}                                                   
  (F_2, F_2^{\gamma Z}, F_2^Z) & \approx &  [(e_u^2, 2e_uv_u, v_u^2+a_u^2)(xU+ x\bar{U})
  +  (e_d^2, 2e_dv_d, v_d^2+a_d^2)(xD+ x\bar{D})]~,            
                                 \nonumber \\                                   
  (xF_3^{\gamma Z}, xF_3^Z) & \approx & 2  [(e_ua_u, v_ua_u) (xU-x\bar{U})
  +  (e_da_d, v_da_d) (xD-x\bar{D})]~,                        
\end{eqnarray} 
where  $e_u$ and $e_d$ denote the electric charge of up- and
down-type quarks, while $v_{u,d}$ and $a_{u,d}$ are 
the vector and axial-vector weak couplings of the up- and
down-type quarks to the $Z$ boson.
The terms  $xU$, $xD$, $x\bU$ and $x\bD$ denote
the sums of parton distributions for up-type and down-type quarks 
and anti-quarks, respectively. 
Below the $b$-quark mass threshold,
these sums are related to the quark distributions as follows
\begin{equation}  \label{ud}
  xU  = xu + xc\,,    ~~~~~~~~
 x\bU = x\bu + x\bc\,, ~~~~~~~~
  xD  = xd + xs\,,    ~~~~~~~~
 x\bD = x\bd + x\bs\,, 
\end{equation}
where $xs$ and $xc$ are the strange- and charm-quark distributions.
Assuming symmetry between the quarks and anti-quarks in the sea, 
the valence-quark distributions can be expressed as
\begin{equation} \label{valq}
xu_v = xU -x\bU\,, ~~~~~~~~~~~~~ xd_v = xD -x\bD\,.
\end{equation}


It follows from \Eq~\ref{ncsi} 
that the structure function
$x\tilde{F_3}$ can be determined from the difference between
the $e^+p$ and $e^-p$ reduced cross sections:
\begin{equation}
 x\tilde{F_3}
  =  \frac {Y_{+}} {2Y_{-}}( \sigma^{-}_{r,{\rm NC}} - \sigma^{+}_{r,{\rm NC}}).
\label{eqn:txf3}
\end{equation} 

Equations~\ref{strf}, \ref{ncfu} and~\ref{valq} demonstrate 
that in the QPM, $x{\tilde{F_3}}$ is directly related
to the valence-quark distributions. 
In the HERA kinematic range, its dominant contribution is
from the photon--$Z$ exchange interference and
the simple relation
\begin{equation}
 xF_3^{\gamma Z} \approx \frac{x}{3}(2u_{v} + d_{v})
\label{eqn:xf3gz_simple}
\end{equation} 
emerges. 
The measurement of $xF_3^{\gamma Z}$ therefore provides access
to the lower-$x$ behaviour of the valence-quark distribution,
under the assumption that sea-quark and anti-quark distributions 
are the same.


The
reduced cross sections for 
inclusive unpolarised CC $e^{\pm} p$ 
scattering are defined as
\begin{equation}
 \label{Rnc}
 \ccred =  
  \frac{2 \pi  x_{\rm Bj}}{G_F^2}
 \left[ \frac {M_W^2+Q^2} {M_W^2} \right]^2
          \ccdd ~. 
\end{equation}
In HERAFitter, the values of $G_F=1.16638\times 10^{-5} $~GeV$^{-2}$
and $M_W=80.385$~GeV \cite{PDG12}  
were used for the Fermi constant and $W$-boson mass.
In analogy to \Eq~\ref{ncsi}, CC structure functions are defined such that
\begin{eqnarray}
 \label{ccsi}
 \sigma_{r,{\rm CC}}^{\pm}=
  \frac{Y_+}{2}W_2^\pm   \mp \frac{Y_-}{2} xW_3^\pm - \frac{y^2}{2} W_{\rm L}^\pm \,~.
\end{eqnarray}

In the QPM, $W_{\rm L}^\pm = 0$ and
$W_2^\pm$, $xW_3^\pm$  represent sums and differences
of quark and anti-quark distributions, depending on the 
charge of the lepton beam:
\begin{eqnarray}
 \label{ccstf}
    W_2^{+}  \approx  x\bU+xD\,,\hspace{0.05cm} ~~~~~~~
  xW_3^{+}  \approx   xD-x\bU\,,\hspace{0.05cm}  ~~~~~~~ 
    W_2^{-}  \approx  xU+x\bD\,,\hspace{0.05cm} ~~~~~~~
 xW_3^{-}  \approx  xU-x\bD\,.
\end{eqnarray}
From these equations, it follows that
\begin{equation}
\label{ccupdo}
 \ccredp \approx  (x\bU+ (1-y)^2xD)\,, ~~~~~~~
 \ccredm \approx  (xU +(1-y)^2 x\bD)\,. 
\end{equation}
The combination 
of NC and CC measurements makes it possible
to determine both the combined sea-quark distributions, 
$x\bU$ and $x\bD$,
and the valence-quark distributions, $xu_v$ and $xd_v$. 

The relations within the QPM illustrate in a simple way
which data contribute which information.
However, the parton distributions are determined 
by a fit to the $x_{\rm Bj}$  and $Q^2$
dependence of the new combined data using the linear 
DGLAP equations~\cite{Gribov:1972ri,Gribov:1972rt,Lipatov:1974qm,Dokshitzer:1977sg,Altarelli:1977zs} at leading order (LO),
next-to-leading order (NLO) and next-to-next-to-leading order (NNLO) in pQCD. 
These are convoluted with coefficient functions
(matrix elements) at the appropriate order~\cite{vanNeerven:10,Moch:06}. 
Already at LO, the gluon PDF enters the equations giving rise to 
logarithmic scaling violations which make the
parton distributions depend on the scale of the process. 
This factorisation scale, $\mu_{\rm f}^2$, is taken as $Q^2$ and the 
experimentally measured scaling violations determine 
the gluon distribution\footnote{The definition of what is 
meant by LO can differ; it can be taken to mean $\mathcal O(1)$ in $\alpha_s$, 
or it can be taken to mean the first non-zero order. 
For example, the longitudinal structure function $F_{\rm L}$ is zero at  
$\mathcal O(1)$ such that its first non-zero order is $\mathcal O(\alpha_s$). 
This is what is meant by LO here unless otherwise stated.
Higher orders follow suit such that at NLO, $F_2$ has coefficient functions 
calculated up to $\mathcal O(\alpha_s)$, whereas $F_{\rm L}$ has coefficient 
functions calculated up to $\mathcal O(\alpha_s^2)$.}.


%


\section{Measurements \label{sec:meas}}

\subsection{Detectors} 
\label{sec:detectors}
The H1~\cite{h1det,h1det2,spacalc}
and ZEUS~\cite{ZEUSDETECTOR,ZEUSCAL,ZEUSCTD,ZEUSMVD} 
detectors 
were both multi-purpose detectors with
an almost $4\pi$ hermetic coverage\footnote{Both experiments 
used a right-handed Cartesian coordinate system, 
with the $Z$ axis pointing in the proton
beam direction, referred to as the ``forward direction", 
and the $X$ axis pointing 
towards the centre of HERA. The
coordinate origins were at the nominal interaction points. 
The polar angle, $\theta$, was measured with respect to 
the proton beam direction.}.
They were built following similar physics considerations but 
the collaborations opted for different technical 
solutions resulting in slightly different capabilities~\cite{Max0805.3334}.
The discussion here focuses on general ideas;
details of the construction and performance are not discussed.

In both detectors, the calorimeters had an inner part to
measure electromagnetic energy and identify
electrons and an outer, less-segmented, part to measure hadronic
energy and determine missing energy. 
Both main calorimeters were divided into
barrel and forward sections. The H1 collaboration
chose a liquid-argon calorimeter while the ZEUS collaboration
opted for a uranium--scintillator device. These choices are
somewhat complementary. The liquid-argon technology allowed
a finer segmentation and thus the identification of electrons
down to lower energies. The uranium--scintillator calorimeter
was intrinsically ``compensating'' making jet studies easier.
In the backward region, ZEUS also opted for
a uranium--scintillator device.
The H1 collaboration chose a lead--scintillating fibre or so-called ``spaghetti"
calorimeter. The backward region is particularly important to identify
electrons in events with $Q^2 < 100$\,GeV$^2$.

Both detectors were operated with a solenoidal magnetic field.
The field strength
was  $1.16$\,T and  $1.43$\,T within the tracking volumes of 
the H1 and ZEUS detectors, respectively. 
The main tracking devices were in both cases cylindrical
drift chambers. The H1 device consisted of two concentric
drift chambers while ZEUS featured one large chamber.
Both tracking systems were
augmented with special devices in the forward and backward region.
Over time, both collaborations
upgraded their tracking systems
by installing  silicon microvertex detectors to enhance the
capability to identify events with heavy-quark production.
In the backward direction, the vertex detectors were also important
to identify the electrons in low-$Q^2$ events.

During the HERA\,I running period, 
special devices to measure very backward electrons
were operated and events
with very low $Q^2$ were reconstructed.
This became impossible after the luminosity
upgrade for HERA\,II due to the
placement of  final-focus magnets 
further inside the detectors. This also required some
significant changes in both main detectors.
Detector elements had to be retracted, and as a result the 
acceptance for low-$Q^2$ events in the main detectors was reduced.


Both experiments measured the luminosity using the Bethe--Heitler reaction 
$ ep \rightarrow e\gamma p$.
In HERA\,I, H1 and ZEUS both had photon taggers 
positioned about 100\,m down the electron beam line.
For the  higher luminosity of the HERA\,II period, 
both H1~\cite{H1lumi1,H1lumi2,H1allhQ2}  and
ZEUS~\cite{Zlumi1,Zlumi2,Zlumi3} 
had to upgrade their luminosity detectors 
and analysis methods.
The uncertainties on the integrated luminosities were
typically about 2\,\%.




\subsection{Reconstruction of kinematics}
\label{diskine}
The usage of different reconstruction techniques,
due to differences in the strengths of the detector components
of the two experiments,
contributes to the reduction of systematic uncertainties
when combining data sets.
The choice of the most appropriate kinematic reconstruction method
for a given phase-space region and experiment is 
based on resolution, possible biases of the measurements  
and effects due to initial- or final-state radiation.
The different methods are described in the following.

The deep inelastic $ep$ scattering cross sections
of the inclusive neutral and charged current reactions
depend on the centre-of-mass energy, $\sqrt{s}$, and 
on the two kinematic variables
$Q^2$ and $x_{\rm Bj}$. 
The variable $x_{\rm Bj}$ is related to $y$,  $Q^2$ and $s$ through the relationship
$x_{\rm Bj}=Q^2/(sy)$. 
The  HERA collider experiments were able
to determine the NC event kinematics from the scattered
electron, $e$, 
or from the hadronic final state, $h$, 
or from a combination of the
two.

The ``electron method'' was applied to NC scattering events for which
the quantities $y$ and $Q^2$
were calculated using only the variables measured 
for the scattered electron:

\begin{equation}
 y_e = 1-\frac{\Sigma_e}{2 E_e}\,~,~~~~~~~~~~~~ 
Q^2_e = \frac{P_{T,e}^2} {1 - y_e}\,~,~~~~~~~~~~~~x_e = \frac{Q^2_e}{s y_e}\,~,
 \label{eq:emeth}
\end{equation}
where $\Sigma_e = E'_e(1-\cos\theta_e)$, 
$E'_e$ is the energy of the scattered electron, 
$\theta_e$ is its angle with respect to
the proton beam, and 
$P_{T,e}$ is its transverse momentum.

The ``hadron method'' was applied to CC scattering events.
The reconstruction of the hadronic final state $h$
allowed the usage of similar relations~\cite{yjb}:  
\begin{equation}
 y_h  = \frac{\Sigma_h}{2 E_e}\,~,~~~~~~~~~~~~ 
Q^2_h = \frac{P_{T,h}^2} {1 - y_h}\,~,~~~~~~~~~~~~x_h = \frac{Q^2_h}{s y_h}\,~, 
 \label{yjb}
\end{equation}
where
$\Sigma_h = (E-P_{\rm Z})_h=\sum_i{(E_i-p_{{\rm Z},i})}$
is the hadronic $E-P_{\rm Z}$ variable with the sum extending over
the energies, $E_i$, and the longitudinal components of the momentum, $p_{{\rm Z},i}$ 
of the reconstructed hadronic final-state particles, $i$.
The quantity $P_{T,h} = \left| \sum_i \mathbold{p}_{T,i} \right|$
is the total transverse momentum of the hadronic final state
with $\mathbold{p}_{T,i}$ 
being the transverse-momentum vector of the particle $i$.
A hadronic scattering angle, $\theta_h$, was defined as
\begin{equation}\label{eq:thh}
\tan \frac{\theta_h}{2} = \frac{\Sigma_h}{P_{T,h}} ~.
\end{equation}
In the framework of the QPM, 
$\theta_h$ corresponds to the direction of the struck quark.

In the ``sigma method''~\cite{ysigma}, the total $E-P_{\rm Z}$ variable,
\begin{equation} \label{eq:sigma}
 E-P_{\rm Z} = E'_e (1-\cos{\theta_e}) + 
 \sum_i \left(E_i - p_{{\rm Z},i}\right) =  \Sigma_e + \Sigma_h ~,
\end{equation}
was introduced. For events without initial- or final-state radiation,
the relation $E-P_{\rm Z} = 2E_e$ holds. Thus,
\Eqs~\ref{eq:emeth} and \ref{yjb} become
\begin{equation}\label{eq:yh}
 y_{\Sigma}  = \frac{\Sigma_h} {E-P_{\rm Z}}\,~, ~~~~~~~~~~~~~ Q^2_{\Sigma}=\frac{P^2_{T,e}}{1-y_{\Sigma}}\,~,~~~~~~~~~~~~ x_{\Sigma} = \frac{Q^2_{\Sigma}}{s y_{\Sigma}}\,~.
\end{equation} 
An extension of the  
sigma method~\cite{Collaboration:2009bp,Collaboration:2009kv} 
introduced the variables 
\begin{equation} \label{eq:sigma2}
 y_{\Sigma'}  = y_{\Sigma}~, ~~~~~~~~~~~~~ Q^2_{\Sigma'}=Q^2_{\Sigma}\,~,~~~~~~~~~~~~ 
x_{\Sigma'} = \frac{Q^2_{\Sigma}} {2 E_p (E-P_{\rm Z}) y_{\Sigma}} = \frac{Q_{\Sigma}^2} {2 E_p \Sigma_h}~.
\end{equation} 
This method
allowed  
radiation at the lepton vertex to be taken into account by replacing
the electron beam energy in the calculation of $x_{\Sigma'}$ in a way similar 
to its replacement in the calculation of $y_{\Sigma}$. 

In the hybrid ``e--sigma method''~\cite{ysigma,Adloff:1999ah,Breitweg:2000yn}, 
$Q^2_e$ and $x_\Sigma$ are used to reconstruct the event kinematics as
\begin{equation} \label{eq:esigma}
  y_{e\Sigma} =  \frac{Q^2_e}{s x_{\Sigma}} = \frac{2E_e}{E-P_{\rm Z}}\,y_{\Sigma}\,~,~~~~~~~~~~~~~ Q^2_{e\Sigma} = Q^2_e\,~,~~~~~~~~~~~~ x_{e\Sigma} = x_{\Sigma}\,~.
\end{equation}

The  ``double-angle method'' \cite{standa,hoegerda} is used  to  
reconstruct  $Q^2$ and $x_{\rm Bj}$ from the electron and hadronic scattering angles as
\begin{equation}\label{qxda}
y_{DA} = \frac{\tan{(\theta_h/2)}}{\tan{(\theta_e/2)} + \tan{(\theta_h/2)}}\,,
~~~~~~Q^2_{DA}= 4 E_e^{~2} \cdot  
\frac{\cot{(\theta_e/2)}}{\tan{(\theta_e/2)} + \tan{(\theta_h/2)}}\,,~~~~~~~ x_{DA} = \frac{Q^2_{DA}}{s y_{DA}}\,.
\end{equation}
This method is largely insensitive to hadronisation effects. 
To first order, it is also 
independent of the detector energy scales. However, 
the hadronic angle is experimentally not as well determined
as the electron angle due to particle loss in the beampipe.

In the ``PT method'' of reconstruction~\cite{Derrick:1996hn},  
the well-measured 
electron variables are used to obtain a good
event-by-event estimate of the loss of hadronic energy 
by employing $\delta_{PT}=P_{T,h}/P_{T,e}$. This  improves
both the resolution and uncertainties on the reconstructed $y$ and $Q^2$.
The PT method uses all measured variables to optimise the 
resolution over the entire kinematic range measured. A variable
$\theta_{PT}$ is introduced as
\begin{equation} \label{eq:ptmeth}
 \tan{\frac{\theta_{PT}}{2}} = \frac{\Sigma_{PT}}{P_{T,e}}~,{\rm~~~where}~~~~~~
 \Sigma_{PT} = 2E_e\frac{{C(\theta_h,P_{T,h},\delta_{PT})}\cdot\Sigma_h}
                       {\Sigma_e+{C(\theta_h,P_{T,h},\delta_{PT})}\cdot\Sigma_h}~.
\end{equation}
The variable $\theta_{PT}$ is then substituted for $\theta_h$ in the formulae
for the double-angle method to determine $x_{\rm Bj}$, $y$ and $Q^2$. The 
detector-specific function, $C$, is calculated  
using Monte Carlo simulations as $\Sigma_{{\rm true},h}/\Sigma_{h}$, 
depending on $\theta_h$, $P_{T,h}$ and $\delta_{PT}$.

\subsection{Inclusive data samples}

A summary of the 41 data sets used 
in the combination is presented in Table~\ref{tab:data}.
From 1994 onwards, HERA was operated 
with an electron beam energy of $E_e \simeq 27.5$\,GeV.
In the first years, until 1997,
the proton beam energy, $E_p$, was set to 820\,GeV. 
In 1998, it was increased to 920\,GeV. 
In 2007, it was lowered to 575\,GeV and 460\,GeV.
The values for the centre-of-mass energies given in
Table~\ref{tab:data} are those for which the cross sections are
quoted in the individual publications. The two collaborations
did not always choose the same reference values
for $\sqrt{s}$ for the same $E_p$. 
The methods of reconstruction used
by H1 and ZEUS for the individual data sets are also given
in the table. 
The integrated luminosities for a given period as provided
by the collaborations can be different. One reason is the
fact that H1 quotes luminosities
for the data within the $Z$-vertex acceptance and ZEUS luminosities
are given without any acceptance cut.

The very low-$Q^2$ region is covered by data 
from both experiments taken during the HERA\,I period.
The lowest, $Q^2 \ge 0.045$~GeV$^2$, data 
come from measurements with the ZEUS detector using special tagging
devices. They are named ZEUS BPT in Table~\ref{tab:data}.
During the course of this analysis, it was discovered that 
in the HERA\,I analysis~\cite{HERAIcombi}, 
values given for $F_2$ were erroneously
treated as reduced cross sections. This was corrected for the
analysis presented in this paper.
All other individual data sets from HERA\,I were used in the new combination
exactly as in the previously 
published combination~\cite{HERAIcombi}.


The $Q^2$ range from $0.2$\,GeV$^2$ to $1.5$\,GeV$^2$
was covered using special HERA\,I runs, in which the interaction vertex
position was shifted forward, bringing backward scattered electrons
with small scattering angles into the acceptance of the 
detectors~\cite{Adloff:1997mf,Breitweg:1998dz,Collaboration:2009bp}.
The lowest-$Q^2$ values for these shifted-vertex data 
were reached using events in which the  electron
energy was reduced by initial-state radiation~\cite{Collaboration:2009bp}. 

The $Q^2 \ge 1.5$\,GeV$^2$ range was covered by 
HERA\,I and HERA\,II data
in various configurations. 
The high-statistics HERA\,II data sets increase the accuracy at high $Q^2$,
particularly for $e^-p$ scattering, for which
the integrated luminosity for HERA\,I was very limited.

The 2007 running periods with lowered 
proton energies~\cite{H1FL1,H1FL2,ZEUSFL} were
included in the combination and 
provide data with reduced $\sqrt{s}$ and $Q^2$ up to 800\,GeV$^2$.
These data were originally taken to measure $F_{\rm L}$.

\subsection{Data on charm, beauty and jet production}
\label{sec:adddata}

The QCD analyses presented in Section~\ref{sec:qcdan}
also used selected results on heavy-quark and
jet production.

The charm production cross sections were taken
from a publication~\cite{HERAccombi} in 
which data from nine data sets published by H1 and ZEUS, 
covering both the HERA~I and~II periods, were combined.
The beauty production cross sections were taken 
from two publications, 
one from ZEUS~\cite{zeusf2b} and one from H1~\cite{h1f2b}.
The heavy-quark events form small subsets 
of the inclusive data. Correlations between the charm and the
inclusive data are small and were not taken
into account.

The data on jet production cross sections were taken from
selected publications:
ZEUS inclusive-jet production data 
from HERA\,I~\cite{zeus9697jets}, 
ZEUS dijet production data from HERA\,II~\cite{zeusdijets}, 
H1 inclusive-jet production data at low $Q^2$~\cite{h1lowq2jets} and 
high $Q^2$ from HERA\,I~\cite{h1highq2oldjets} 
and HERA\,II~\cite{h1highq2newjets}.  
The HERA\,II H1 publication provides 
inclusive-jet, dijet and trijet 
cross sections normalised to the inclusive NC DIS cross sections
in the respective $Q^2$ range. This largely reduces  
the correlations with the H1 inclusive DIS reduced cross sections.
The HERA\,I H1 high-$Q^2$ jet data are similarly normalised.
The  other ZEUS and H1 jet data sets are small subsamples of the respective
inclusive sample; correlations are small and are thus ignored.

For the heavy-quark and jet data sets used, the statistical, uncorrelated systematic
and correlated systematic uncertainties were used as published.

\section{Combination of the inclusive cross sections \label{sec:comb}}

In order to combine the published cross sections
from the 41 data sets listed in Table~\ref{tab:data}, 
they were translated onto common grids and averaged.

\subsection{Common $\boldsymbol{\sqrt{s}}$ values, 
            common $\boldsymbol{(x_{\rm Bj},Q^2)}$ grids
            and translation of data}
\label{subsec:extrapol}
%
%
The data were taken with several $E_p$ values and
the double-differential cross sections were published 
by the two experiments
for  different reference $\sqrt{s}$  
and $(x_{\rm Bj},Q^2)$ grids.
In order to average a set of data points, the points had to be translated
to common  $\sqrt{s}_{{\rm com}}$ values and 
common $(x_{\rm Bj,grid},Q^2_{\rm grid})$ grids.
The following choices were made.

Three common centre-of-mass values, 
$\sqrt{s}_{{\rm com},i}$, were chosen to combine data onto two common grids: 
\begin{align}
\nonumber
  E_p=920 \,{\rm GeV} \rightarrow  \sqrt{s}_{{\rm com},1}=318 \,{\rm GeV}
 \rightarrow  {\rm grid}~1 ~~,\\  
\nonumber
  E_p=820 \,{\rm GeV} \rightarrow  \sqrt{s}_{{\rm com},1}=318 \,{\rm GeV} 
 \rightarrow  {\rm grid}~1 ~~,\\  
\nonumber
  E_p=575 \,{\rm GeV} \rightarrow  \sqrt{s}_{{\rm com},2}=251 \,{\rm GeV} 
 \rightarrow  {\rm grid}~2 ~~,\\  
\nonumber
  E_p=460 \,{\rm GeV} \rightarrow  \sqrt{s}_{{\rm com},3}=225 \,{\rm GeV} 
 \rightarrow  {\rm grid}~2 ~~.  
\end{align}
Exceptions were made for data with $E_p=820$\,GeV 
and $y \ge 0.35$. 
These cross sections were not translated 
to $\sqrt{s}_{{\rm com},1}$, but 
were kept separately in grid~1
in order to retain their $y$ dependence.

\begin{figure}
\centerline{\epsfig{file=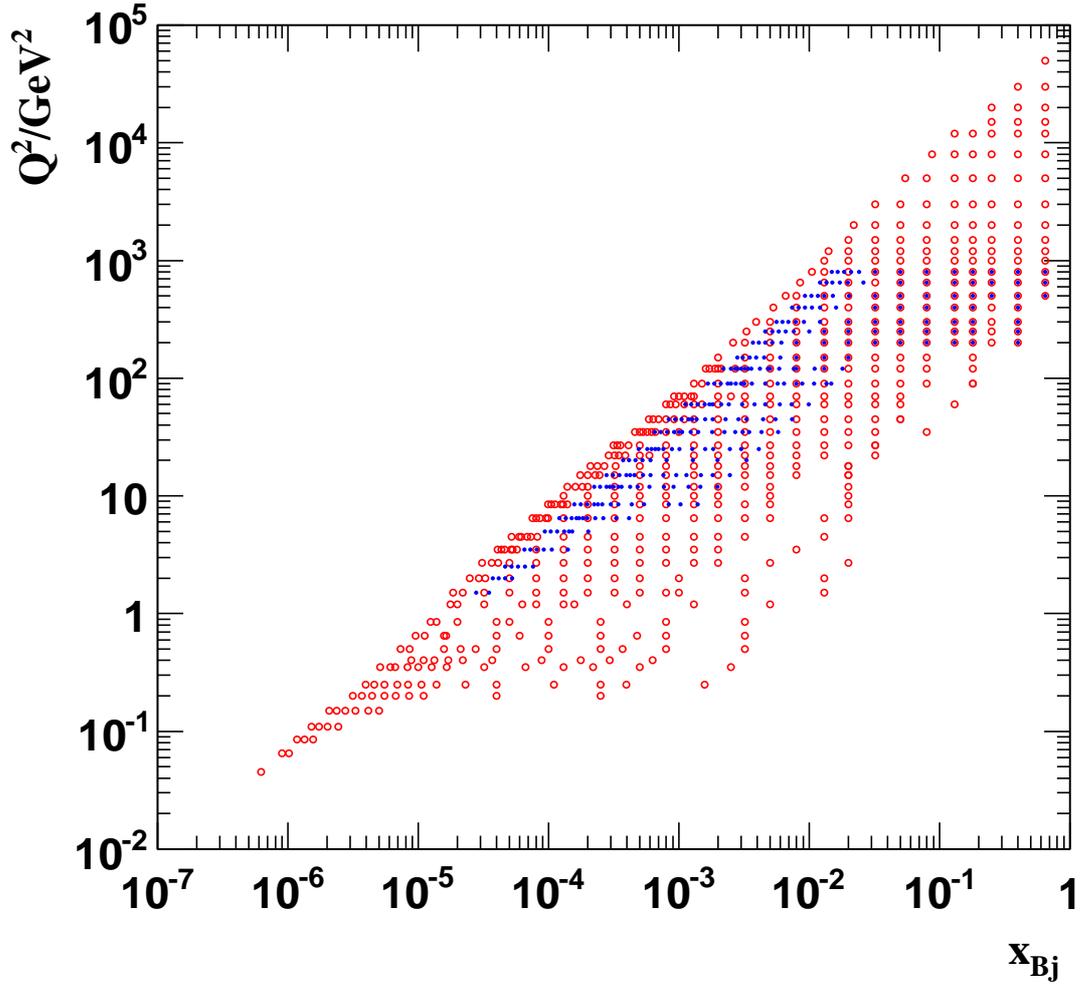  ,width=\linewidth}}
\caption{The points of the two grids used for the combination.
Grid\,1 (open circles) was used for data with 
$\sqrt{s}_{{\rm com},1}=318$\,GeV. 
Grid\,2 (dots) was used for data with 
$\sqrt{s}_{{\rm com},2}=251$\,GeV
or
$\sqrt{s}_{{\rm com},3}=225$\,GeV.
The latter grid 
has a finer binning in $x_{\rm Bj}$ in accordance with its
special structure in $y$.}
\label{fig:grid}
\end{figure}

The two grids have a different structure in $y$ such that the 
corrections due to translation were minimised.
The grids are depicted in
Figure~\ref{fig:grid}.
For a given data point with $\sqrt{s}_{{\rm com},1}$,
the grid point was in general chosen to be closest in $Q^2$ and 
then in $x_{\rm Bj}$. 
However, for some data points, the grid point closest in $y$ was chosen.
This occurs for data sets  marked with $^{*y}$ or $^{*y.5}$ in 
Table~\ref{tab:data}. 
The markers indicate that it happens for all $y$ or $y>0.5$, 
respectively.  
For a given data point at $\sqrt{s}_{{\rm com},2}$ 
or $\sqrt{s}_{{\rm com},3}$,
the grid point closest in $Q^2$ and then closest in $y$ was always chosen.

In most of the phase space, separate measurements
from the same data set were not translated to the same grid  
point. Only 9 
out of 1307 grid points 
accumulated two and in one case three points from
the same data set. 
Up to 10 data sets were available for a given process. 
The vast majority of grid points 
accumulated data from both H1 and ZEUS measurements; 
the typical case is six measurements from six different data sets.
However, 22\,\% of all grid points have only one measurement,
predominantly at low $Q^2$. For $Q^2$ above 3.5\,GeV$^2$, only 13\,\%
of the grid points have only one measurement.

For the translation of the cross-section values, predictions for the ratios 
of the double-differential cross section at the $(x_{\rm Bj},Q^2)$ and $\sqrt{s}$
where the measurements took place, and the $(x_{\rm Bj,grid},Q^2_{\rm grid})$ 
to which they were translated, were needed.
These predictions, $T_{\rm grid}$, were obtained from the data themselves
by performing fits to the data using the
HERAFitter~\cite{HERAfitter,HERAfitterweb} tool.
For $Q^2 \ge 3$\,GeV$^2$, a next-to-leading-order QCD fit 
using the DGLAP formalism was 
performed\footnote{As a cross check, predictions 
using HERAPDF1.0 were used instead.
The induced changes were negligible.}.
In addition, a fit using the fractal 
model\footnote{The {\it ansatz} of the
fractal model is based on the self-similar properties in $x_{\rm Bj}$ and $Q^2$
of the proton structure function
at low $x_{\rm Bj}$. They are represented by 
two continuous, variable and correlated fractal dimensions.}~\cite{Collaboration:2009bp,Lastovicka:2002hw}
was performed for $Q^2 \le 4.9$\,GeV$^2$.
For $Q^2 < 3$\,GeV$^2$, 
the fit to the
fractal model was used\footnote{A cross check was performed using the 
colour dipole model~\cite{GolecBiernat:1998js}
as implemented in HERAFitter. The results did not change significantly.}
to obtain factors $T_{\rm grid,FM}$.
For $Q^2 > 4.9$\,GeV$^2$, the QCD fit was used to provide $T_{\rm grid,QCD}$.  
For $3 \le Q^2 \le 4.9$\,GeV$^2$, the factors were averaged as 
$T_{\rm grid} = T_{\rm grid,FM} (1-(Q^2-3)/1.9)  + T_{\rm grid,QCD} (Q^2-3)/1.9 $ 
where $Q^2$ is in GeV$^2$. 
The upper edge of the application of the
fractal fit was varied between 3\,GeV$^2$ and 5\,GeV$^2$;
the effect was negligible.



\subsection{Averaging cross sections \label{sec:comb:averaging}}
The original double-differential
cross-section measurements were published with their statistical and systematic
uncertainties. 
The systematic uncertainties were classified as either 
point-to-point correlated or point-to-point uncorrelated.
For each data set, all uncorrelated systematic uncertainties were added in 
quadrature before averaging. 
Correlated systematic uncertainties were kept separately.
Some of the systematic uncertainties
were originally reported as asymmetric.
They were symmetrised by the collaborations before entering the 
averaging procedure. 

The averaging of the data points
was performed using the  
HERAverager~\cite{HERAveragerweb} 
tool which is based on a
$\chi^2$ minimisation method~\cite{Collaboration:2009bp}.
This method imposes that there is one and only one correct value for
the cross section of each process at each point of the phase space. 
These values are estimated by optimising a vector, $\boldsymbol{m}$,
which is the result of the averaging for the cross sections.
The $\chi^2$ definition used takes into account the correlated 
and uncorrelated systematic 
uncertainties
of the H1 and ZEUS cross-section measurements and allows for shifts of the data
to accommodate the correlated uncertainties.
For a single data set, $ds$, the $\chi^2$ is defined as
\begin{equation}
 \chi^2_{{\rm exp},ds}\left(\boldsymbol{m},\boldsymbol{b}\right) = 
  \sum_{i}^{ds} + \sum_{j}^{b} =
 \sum_i
 \frac{\left[m^i
- \sum_j \gamma^{i,ds}_j m^i b_j  - \mu^{i,ds} \right]^2}
{ \textstyle \delta^2_{i,ds,{\rm stat}}\,{\mu^{i,ds}}  \left(m^i -  \sum_j \gamma^{i,ds}_j m^i b_j\right)+
\left(\delta_{i,ds,{\rm uncor}}\,  m^i\right)^2}
 + \sum_j b^2_j\, ,
\label{eq:ave1}
\end{equation}
where  ${\mu^{i,ds}}$ is the  measured  value  at the point $i$ and
$\gamma^{i,ds}_j $, 
$\delta_{i,ds,{\rm stat}} $ and 
$\delta_{i,ds,{\rm uncor}}$ are the relative
correlated systematic, relative statistical and 
relative uncorrelated systematic uncertainties,
respectively.
For the reduced cross-section  measurements,  ${\mu^{i,ds}} = \sigma_r^{i,ds}$,
$i$ runs over all points on the $(x_{\rm Bj,{\rm grid}},Q^2_{\rm grid})$ 
plane for which a measurement
exists in $ds$. 
The components $b_j$ of the vector $\boldsymbol{b}$ represent 
correlated shifts of the cross sections
in units of sigma of the respective correlated systematic
uncertainties; the summations over $j$ extend 
over all correlated systematic uncertainties.


The leading systematic uncertainties
on the cross-section measurements
used for the combination 
arose from the uncertainties on the
acceptance corrections and luminosity determinations. 
Thus, both the correlated and uncorrelated systematic 
uncertainties are multiplicative in nature, 
i.e.\ they increase proportionally to the central values. 
In \Eq~\ref{eq:ave1}, the multiplicative nature of 
these uncertainties is taken 
into account by multiplying the
relative errors $\gamma^{i,ds}_j$ and $\delta_{i,ds,{\rm uncor}}$  
by the estimate $m^i$. 
The denominator in the first right-hand-side term in 
\Eq~\ref{eq:ave1} contains
an estimate of the squared statistical uncertainty of the cross-section 
measurement, $\delta_{i,ds,stat}^2 \mu^{i,ds} 
(m^{i} - \sum_j \gamma^{i,ds}_j m^i b_j)$,
which is 
assumed\footnote{For the DIS cross-section measurements,
the background contributions were small and thus it is justified
to take the square root of the number of events as the statistical uncertainty.}
to scale with the expected number of events in bin $i$, 
as calculated from $m^{i}$. Corrections due to the shifts to
accommodate the correlated systematic uncertainties 
are introduced through the term $\sum_{j} \gamma^{i,ds}_j m^i b_j$.

For 
several data sets, 
a total $\chi^2$ function is defined as
\begin{equation}
\chi^2_{\rm tot} = \sum_{ds} \sum_{i}^{ds} + \sum_{j}^{b} , \label{eq:tot}
\end{equation}
with  $\sum_{i}^{ds}$ and  $\sum_{j}^{b}$ as introduced 
for a single measurement in Eq.~\ref{eq:ave1}.
The total $\chi^2$ function in Eq.~\ref{eq:tot} can be approximated by
\begin{equation}
\chi^2_{\rm tot} \approx
 \chi^2_{\rm min} +
 \sum_{i=1,N_M}
 \frac{\left[m^i
- \sum_j \gamma^{i}_j m^i b'_j  - {\mu^{i}} \right]^2}
{ \textstyle \delta^2_{i,{\rm stat}}\, \mu^{i}\left(m^i -  \sum_j
\gamma^{i}_j m^i b'_j\right)+
\left(\delta_{i,{\rm uncor}}\,  m^i\right)^2}
 + \sum_j (b'_j)^2,
\label{eq:ave1tot}
\end{equation}
where
$\chi^2_{\rm min}$ is the minimum of $\chi^2_{\rm tot}$,
$N_M$ is the number of combined measurements,
$\mu^{i}$ is the average value at point $i$,  and
$\gamma^{i}_j $,
$\delta_{i,{\rm stat}} $ and
$\delta_{i,{\rm uncor}}$ are its relative
correlated systematic, relative statistical and relative uncorrelated
systematic uncertainties,
respectively.
To determine the average of the data as defined in Eq.~\ref{eq:ave1tot},
an iterative procedure is used.
For the first iteration, for all terms in Eqs.~\ref{eq:ave1} and~\ref{eq:ave1tot} 
related to uncertainties or correlated shifts of the data,
the expectation values $m^i$  are replaced by $\mu^{i,ds}$
and the term $\sum_j \gamma^{i}_j m^i b'_j$ is set to zero for the calculation
of the statistical uncertainty\footnote {For the first iteration, terms are modified as  
$\gamma^{i,ds}_j m^i \to  \gamma^{i,ds}_j \mu^{i,ds},~~
\delta_{i,ds,{\rm uncor}}\,  m^i \to \delta_{i,ds,{\rm uncor}}\, \mu^{i,ds}$  and
$\delta^2_{i,ds,{\rm stat}}\, \mu^{i,ds}\left(m^i -  \sum_j \gamma^{i,ds}_j m^i
b_j\right) \to (\delta_{i,ds,{\rm stat}}\, \mu^{i,ds})^2$, respectively.}.
The average values $\mu^i$ and systematic shifts $b_j$ are determined
analytically from a system of linear
equations $\partial \chi^2_{\rm tot} / \partial m^i = 0$ and 
$\partial \chi^2_{\rm tot} / \partial b_j = 0$. 
For the next iterations,
the average values $\mu^i$ from the previous 
iteration are used\footnote {For subsequent iterations, terms are modified as  
$\gamma^{i,ds}_j m^i \to  \gamma^{i,ds}_j \mu^{i},~~
\delta_{i,ds,{\rm uncor}}\,  m^i \to \delta_{i,ds,{\rm uncor}}\, \mu^{i}$  and
$\delta^2_{i,ds,{\rm stat}}\, \mu^{i,ds}\left(m^i -  \sum_j \gamma^{i,ds}_j m^i
b_j\right) \to 
\delta^2_{i,ds,{\rm stat}}\, \mu^{i,ds}\left(\mu^i -  
\sum_j \gamma^{i,ds}_j \mu^i b_j\right)$, respectively.}.
The procedure converges after two iterations.
The shifts $b'_j$, also called nuisance parameters, are related to
the original shifts $b_j$ through an orthogonal
transformation
which is also used to determine $\gamma^{i}_j$~\cite{HERAIcombi}.

The ratio of 
$\chi^2_{\rm min}$
and the number of degrees of freedom, 
$\chi^2_{\rm min}/{\rm d.o.f.}$, is a measure 
of the consistency of the data sets.
The number ${\rm d.o.f.}$ is the difference between the total 
number of measurements and the number of averaged points $N_M$. 



Some systematic uncertainties $\gamma_j^i$,  which were treated as 
having point-to-point correlations, may be common for several data sets.
A full table of
the correlations of the systematic uncertainties across the data sets
can be found elsewhere~\cite{fullcorr}.
The systematic uncertainties were in general 
treated as independent between H1 and ZEUS. 
However, an overall normalisation uncertainty of $0.5\%$,
due to uncertainties on higher-order corrections 
to the Bethe--Heitler cross-section calculations,  
was assumed for all data sets which
were normalised with data from the luminosity monitors. 

All the NC and CC cross-section data from H1 and ZEUS are combined in 
one simultaneous minimisation. 
Therefore, the resulting shifts of 
the correlated systematic uncertainties propagate 
coherently to both NC and CC data. 
Even in cases where there are data only from a single data set,
the procedure can still produce shifts with respect to the 
original measurement due to the
correlation of systematic uncertainties.

\subsection{Combination procedure}

The combination procedure is iterative.
Each iteration has two steps:

\begin{enumerate}
\item
    the data are translated to the common 
    $\sqrt{s}_{\rm com}$ values and $(x_{{\rm Bj,grid}},Q^2_{\rm grid})$ grids
    as described in Section~\ref{subsec:extrapol};  
\item
    the data are averaged
    as described in Section~\ref{sec:comb:averaging}.
\end{enumerate}

In the first iteration, the fits to provide the predictions needed
for the translation were performed on the uncombined data.
Starting with the second iteration, 
the fits were performed on combined data.
The process was stopped after the third iteration.
An investigation showed that further
iterations did not induce significant changes in the 
resulting averaged cross sections.

\subsection{Consistency of the data}
The 2927 published cross sections were combined to become 1307 
combined cross-section measurements.
For the resulting 1620 degrees of freedom, a $\chi^2_{\rm min} = 1687$
was obtained. 
For data points $k$ contributing  
to point $i$ on the $(x_{\rm Bj, grid},Q^2_{\rm grid})$, 
pulls ${\rm p}^{i,k}$ were defined as
\begin{equation}
{\rm p}^{i,k} = \frac{\mu^{i,k}  - \mu^{i}\left(1- \sum_j \gamma^{i,k}_j b'_{j}\right)}{\sqrt{\Delta_{i,k}^2 - \Delta_{i}^2}}~,
\end{equation}
where $\Delta_{i,k}$ and $\Delta_{i}$ are 
the statistical and uncorrelated systematic uncertainties added in quadrature
for the point $k$ and the average, respectively. 
The pull distribution 
for the different data sets is shown Fig.~\ref{fig:pulls}. The RMS values of these
distributions are close to unity, indicating
good consistency of all data.

\subsection{Procedural uncertainties \label{subsubsec:proc_errors}}
Procedural uncertainties are introduced by the choices made for
the combination. Three kinds of such uncertainties were considered.

\subsubsection{Multiplicative versus additive treatment of systematic uncertainties}
\label{subsec:procerr1}

The $\chi^2$ definition from \Eq~\ref{eq:ave1} treats all 
systematic uncertainties as multiplicative, i.e.\ their size is expected to be
proportional to the ``true'' values $\boldsymbol{m}$.
While this is a good assumption for normalisation uncertainties,
this might not be the case for other uncertainties.
Therefore an alternative combination was performed, 
in which only the normalisation uncertainties were taken as
multiplicative, while all other uncertainties were treated as additive. 
The differences between this alternative combination and the 
nominal combination were defined as correlated procedural 
uncertainties $\delta_{\rm rel}$. 
This is a conservative approach  but still yields quite small
uncertainties.
The typical values
of $\delta_{\rm rel}$ for the
$\sqrt{s}_{\rm com,1}=318$\,GeV ($\sqrt{s}_{\rm com,2/3}$)
combination were below 0.5\,\% (1\,\%) 
for medium-$Q^2$ data, increasing to a few percent for 
low- and high-$Q^2$ data.

\subsubsection{Correlations between systematic uncertainties on different data sets}
\label{subsec:procerr2}

Similar methods were often used to calibrate different data sets
obtained by one or by both collaborations. In addition,
the same Monte Carlo simulation packages were used 
to analyse different data sets. 
These similar approaches might have  
led to correlations between data sets 
from one or both collaborations.
This was investigated in depth for the combination of
HERA\,I data~\cite{HERAIcombi}. 
The important correlations for this period were found to be related to 
the background from photoproduction and the hadronic energy scales.
The correlations for the HERA\,I period were taken into account as 
before~\cite{HERAIcombi}.

The correlations between the experiments for the HERA\,II period were
considered much less important, because both experiments developed 
different methods to address calibration and normalisation.
In the case of H1, some potential correlations between the data from the 
HERA\,I and HERA\,II periods were identified.
In the case of ZEUS, no such correlations were found; this is due to 
significant changes in the detector and in the data processing.

The differences between the nominal combination and the combinations, 
in which 
systematic sources for the photoproduction background and 
hadronic energy scale were taken as correlated across data sets,
were defined as additional signed procedural uncertainties  
$\delta_{\gamma p}$ and $\delta_{\rm had}$.
Typical values of $\delta_{\gamma p}$ and $\delta_{\rm had}$ 
are below $1\%$ (0.5\%) for NC (CC) scattering.
For the  data at low $Q^2$, they can reach a few percent.
 

\subsubsection{Pull distribution of correlated systematic uncertainties}

\begin{figure}
\centerline{\epsfig{file=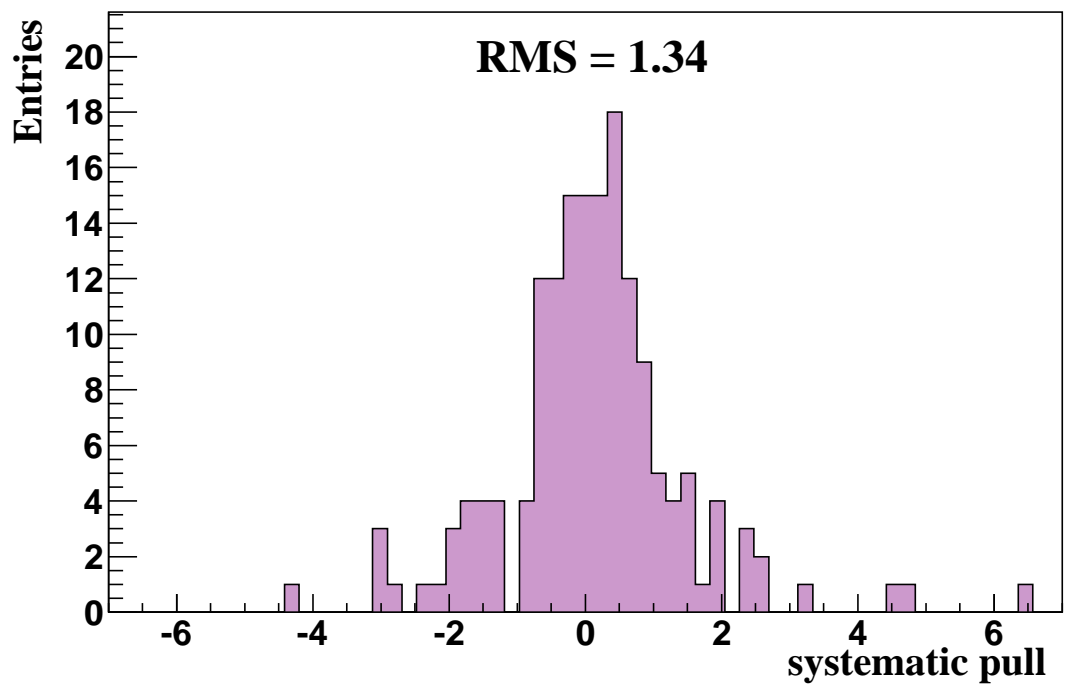  ,width=\linewidth}}
\caption{Distribution of pulls ${\rm p}_{j}$ for the correlated
systematic uncertainties including global normalisations. 
There are no entries outside the histogram range. 
The root mean square, RMS, of the distribution is given.
\label{fig:syspulls}}
\end{figure}

There are in total 162 sources of correlated 
systematic uncertainty including global
normalisations characterising 
the separate data sets. 
In the procedure applied, all these sources were assumed to be 
fully point-to-point correlated. 
None of these sources was shifted
by more than $2.4\,\sigma$ from its nominal 
value in the combination procedure. 
The pull on any such source $j$ is defined as 
$
{\rm p}_j = b'_{j}/(1 - \Delta^2_{b'_{j}})^{1/2}~,
$
where $\Delta_{b'_{j}}$ is the uncertainty on the source $j$
after the averaging.
The distribution of ${\rm p}_{j}$ is shown in
Fig.~\ref{fig:syspulls}.
Some large values for $|{\rm p}_{j}|$ are observed. 
They are connected to small relative uncertainties, below 1\,\%,
for which there is only a small reduction
in the uncertainty. 
Such cases are, for example, expected if the point-to-point correlation within 
a data set is not 100\,\% as was assumed.

The distribution of pulls shown in Fig.~\ref{fig:syspulls}
is not Gaussian; it has a root-mean-square value of 1.34.
Out of the 162 point-to-point correlated uncertainties, 
40 were identified with $ {\rm p}_j > 1.3$. 
This might indicate that these uncertainties were either
underestimated or do not fulfil the implicit assumptions
of the linear procedure applied.
Scaling these 40 uncertainties by a factor of two would reduce the
root-mean-square value to 1.03 and the 
$\chi^2_{\rm min}$ of the combination would be reduced from 1687 to 1614 
for the 1620 degrees of freedom. 

Each of these 40 uncertainties could give rise to an individual procedural
uncertainty if scaled. 
However, an extensive study revealed cross correlations between them. 
These cross correlations were used to form
four groups related to 
\begin{enumerate}
\item very low-$Q^2$ data from HERA\,I 
      (14 uncertainties); 
\item  low-$Q^2$ data from HERA\,II with lowered proton beam energies 
      (10 uncertainties);
\item medium- and high-$Q^2$ data from HERA\,I and~II (11 uncertainties);
\item normalisation issues from HERA\,I and~II  (5 uncertainties).
\end{enumerate}

The normalisation related uncertainties were also found to be correlated to some
of the uncertainties in the other groups but they were kept separate.
Signed procedural uncertainties $\delta_{(1,2,3,4)}$ were assigned to 
the four groups by scaling the
uncertainties within each group by a factor of two and
taking the difference between the result of this  
combination and that of the nominal combination as
the uncertainty. 
Such cross correlations 
as observed here between different systematic uncertainties
are not unexpected, even though different methods were used for different
regions of phase space by two different experiments.
Both experiments contribute about equally to the 40 sources discussed.


Since H1 and ZEUS used, as described for example in Section~\ref{diskine}, 
different reconstruction methods, similar systematic sources influence 
the measured cross section differently as a function of $x_{\rm Bj}$ and $Q^2$. 
Therefore, requiring the cross sections to agree at all $x_{\rm Bj}$ and $Q^2$ 
constrains the systematics efficiently. 
In addition, for certain regions of the phase space, one of the two 
experiments has superior precision compared to the other.
For these regions, 
the less precise measurement is fitted to the more precise measurement, 
with a simultaneous reduction of the correlated systematic uncertainty.
This reduction propagates to the other points, 
including those which are based solely on the measurement 
from the less precise experiment.
However, over most of the phase space, the precision of the H1 and
ZEUS measurements is very similar and the systematic 
uncertainties are reduced uniformly.

\section{Combined inclusive  $\mathbold{e^{\pm}p}$ cross sections\label{subsec:comb:results}}

The combined reduced cross sections for NC and CC $ep$ scattering 
together with their statistical, uncorrelated and total correlated 
systematic uncertainties, 
as well as procedural uncertainties as defined in Section~\ref{sec:comb},
are listed in 
Appendix~\ref{appendix:C}\,%
\footnote{The 
full information about correlations
between cross-section measurements is available elsewhere~\cite{fullcorr}.}.
The new values supersede those published previously~\cite{HERAIcombi}. 

The total uncertainties are below 1.5\,\% over the $Q^2$ range
of $3 \le  Q^2 \le 500$\,GeV$^2$ and below 3\,\% up to  $Q^2 = 3000$\,GeV$^2$.
Cross sections are provided for values of 
$Q^2$ between $Q^2=0.045$\,GeV$^2$
and $Q^2=50000$\,GeV$^2$ and values of $x_{\rm Bj}$ 
between $x_{\rm Bj}=6\times10^{-7}$ and $x_{\rm Bj}=0.65$. 
The events have a minimum invariant mass of the hadronic system, $W$, 
of $15$\,GeV. 

In Fig.~\ref{fig:quality:NCepp}, 
the individual and the combined reduced cross sections
for NC $e^+p$  DIS scattering are shown as a function of
$Q^2$ for selected values of $x_{\rm Bj}$.
The improvement due to combination is clearly visible.
In Fig.~\ref{fig:Hera1:NCepp},
a comparison between the new combination and the combination
of HERA\,I data alone is shown. 
The improvement is especially significant at high $Q^2$.
The results for NC $e^-p$ scattering are depicted in
Figs.~\ref{fig:quality:NCemp} and~\ref{fig:Hera1:NCemp}.
As the integrated luminosity for $e^-p$ scattering was very
limited for the HERA\,I period, the improvements due to the
new combination are even more substantial than for $e^+p$ scattering.

The results
of the combination of the data with lower proton beam energies
are shown in Figs.~\ref{fig:quality:575} and \ref{fig:quality:460} as
a function of $x_{\rm Bj}$ in selected bins of $Q^2$. 
These data augment the data with standard proton energy to provide
increased sensitivity to the gluon density in the proton.

The combined NC $e^+p$ data for
very low $Q^2$ with proton beam energies of 920 and 820\,GeV are shown in
Figs.~\ref{fig:NCvlQ2-920} and \ref{fig:NCvlQ2-820}.  
These data were taken during the HERA\,I period, but due to the
systematic shifts introduced by the combination with HERA\,II data, 
the numbers are not always the same as in the old HERA\,I combination.

The combined CC cross sections 
are shown in Figs.~\ref{fig:quality:CCepp}--\ref{fig:Hera1:CCemp},  
together with the
input data from H1 and ZEUS and the comparison 
to the HERA\,I combination 
results for $e^+p$ and $e^-p$ scattering. As for the NC data,
the power of combination and the improved precision due to the
high statistics data from HERA\,II are demonstrated.

The high-precision DIS cross sections provided here 
form a coherent set spanning six orders of magnitude, 
both in $Q^2$ and $x_{\rm Bj}$.
They are a major legacy of HERA.

\section{QCD analysis}
\label{sec:qcdan}
In this section, the pQCD analysis
of the combined data resulting in the PDF set HERAPDF2.0 
and its released variants is presented.
The framework established for HERAPDF1.0~\cite{HERAIcombi} was followed 
in this analysis.
A breakdown of pQCD is expected for 
$Q^2$ approaching 1\,GeV$^2$.
To safely remain in the kinematic region where
pQCD is expected to be applicable,
only cross sections for $Q^2$ starting from
$Q^2_{\rm min} = 3.5$\,GeV$^2$ were used in the analysis. 
In this kinematic region,
target-mass corrections are expected to be negligible. 
Since the centre-of-mass energy at the $\gamma p$ vertex $W$
is above 15\,GeV
for all the data, 
large-$x_{\rm Bj}$ higher-twist corrections are also expected to be negligible. 
The $Q^2$ range of the cross sections entering the fit 
is $3.5 \le Q^2 \le 50000\,$GeV$^2$.
The corresponding $x_{\rm Bj}$ range 
is $0.651 \times 10^{-4} \le x_{\rm Bj} \le 0.65 $.
 
In addition to experimental uncertainties, 
model and parameterisation uncertainties were also considered. 
The latter were evaluated by 
variations of the values 
of various input settings at the starting scale and the 
form of the parameterisation.

\subsection{Theoretical formalism and settings}
\label{sec:choices}
Predictions from pQCD are fitted to data.
These predictions
were obtained by solving the DGLAP evolution 
equations~\cite{Gribov:1972ri,Gribov:1972rt,Lipatov:1974qm,Dokshitzer:1977sg,Altarelli:1977zs}
at LO, NLO and NNLO in the \msbar scheme~\cite{MSbar}.
This was done using the programme 
QCDNUM~\cite{QCDNUM} within the HERAFitter 
framework~\cite{HERAfitter,HERAfitterweb} and an independent
programme, which was already used to analyse the combined
HERA\,I data~\cite{HERAIcombi}. 
The results obtained by the two programmes 
were in excellent agreement, well within fit uncertainties.
The numbers on fit quality and resulting parameters given in this paper 
were obtained using HERAFitter. 

The DGLAP equations yield the PDFs
at all scales $\mu_{\rm f}^2$ and $x$, if they are provided
as functions of $x$ at some starting scale, $\mu^2_{\rm f_{0}}$. 
In variable-flavour schemes, this scale has to be below the
charm-quark mass parameter, $M_c$, squared.
It was chosen to be $\mu^2_{\rm f_{0}} = 1.9\,$GeV$^2$ 
as for HERAPDF1.0. 
The renormalisation and factorisation scales were 
chosen to be $\mu^2_{\rm r} = \mu^2_{\rm f} = Q^2$.
The predictions for the structure functions~\cite{saturation} 
which appear in the
calculation of the cross sections, see Eq.~\ref{ncsi},
were obtained by convoluting the parton distribution functions 
with coefficient functions appropriate to the order of the calculation. 
The light-quark coefficient functions were calculated using QCDNUM. 
The heavy-quark coefficient functions were calculated 
in the general-mass variable-flavour-number scheme 
called RTOPT~\cite{Thorne:1997ga,Thorne:2006qt,Thorne:RTopt} 
for the NC structure functions. 
For the CC structure functions, the zero-mass approximation was used, 
since all HERA CC data have $Q^2 \gg M_b^2$, where $M_b$ is
the beauty-quark mass parameter in the calculation.

\begin{figure}[tbp]
\centerline{
\epsfig{file=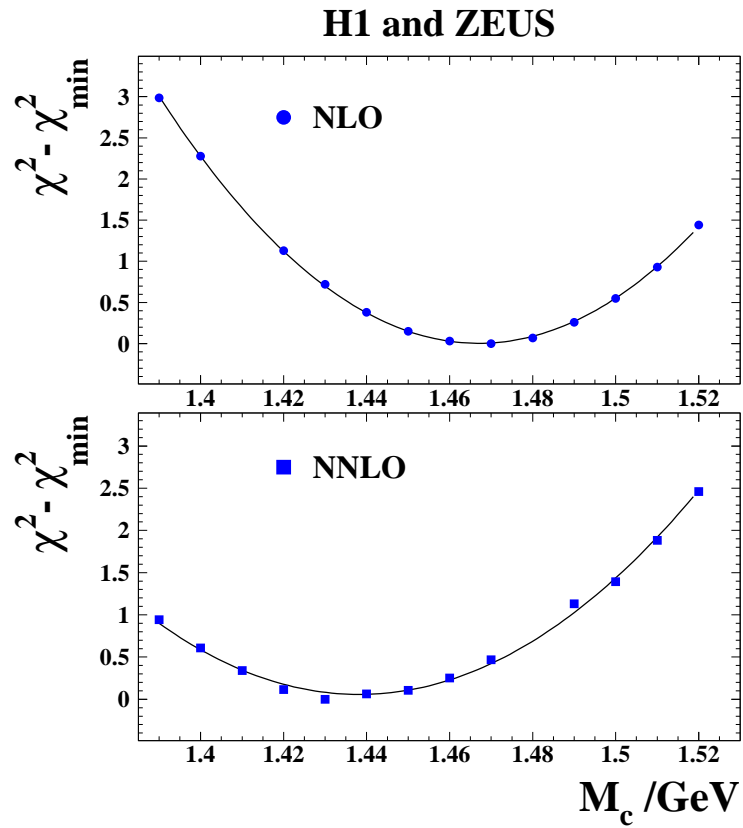  ,width=0.7\textwidth}}
\caption { 
The $\Delta \chi^2 = \chi^2 - \chi^2_{\rm min}$ 
versus the charm mass parameter $M_c$ for 
NLO and NNLO fits based on the combined data on charm production 
in addition to the combined inclusive data. 
}
\label{fig:charmscan}
\end{figure}

The value of $M_c$ 
was chosen after performing $\chi^2$ scans of NLO and NNLO pQCD fits 
to the combined inclusive data from the analysis presented here 
and the HERA combined charm data~\cite{HERAccombi}.
The procedure is described in detail in 
the context of the combination of the reduced charm cross-section
measurements~\cite{HERAccombi}. 
All correlations of the inclusive  
and of the charm data were considered in the fits.
Figure~\ref{fig:charmscan} 
shows the $\Delta\chi^2 =  \chi^2 -\chi^2_{\rm min}$, where $\chi^2_{\rm min}$
is the minimum $ \chi^2$ obtained,
of these fits versus $M_c$ at NLO and NNLO.
As a result, the value of $M_c$ was chosen 
as $M_c=1.47\,$GeV at NLO  and $M_c=1.43\,$GeV at NNLO.
The settings for LO were chosen as for NLO unless otherwise stated.

\begin{figure}[tbp]
\centerline{
\epsfig{file=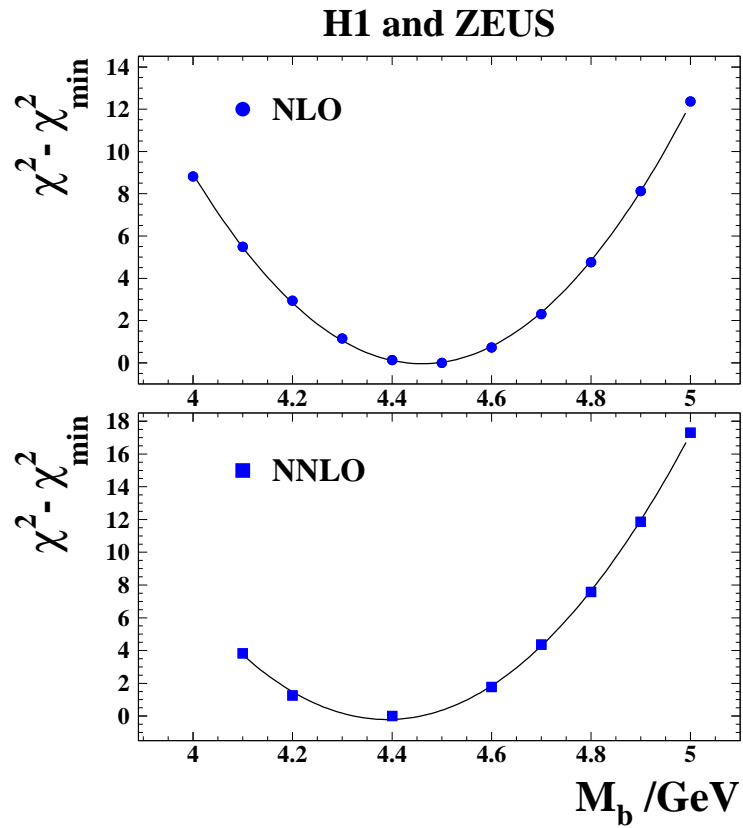  ,width=0.7\textwidth}}
\caption { 
The $\Delta \chi^2= \chi^2 - \chi2_{\rm min}$  
versus the beauty mass parameter $M_b$ 
for NLO and NNLO fits based on H1 and ZEUS data on beauty production 
in addition to the combined inclusive data. 
}
\label{fig:beautyscan}
\end{figure}

The value of the beauty-quark mass parameter $M_b$ was chosen
after performing $\chi^2$ scans of NLO and NNLO pQCD fits 
using the combined inclusive data and
data on beauty production from ZEUS~\cite{zeusf2b} and H1~\cite{h1f2b}.
The $\chi^2$ scans are shown in Fig.~\ref{fig:beautyscan}.
The value of $M_b$ was chosen to be $M_b=4.5$\,GeV 
at LO, NLO and NNLO. 
The value of the top-quark mass parameter was chosen 
to be 173\,GeV~\cite{PDG12} at all orders.

The value of the strong coupling constant was chosen to be 
$\asmz = 0.118$~\cite{PDG12} at both NLO and NNLO and
$\asmz = 0.130$~\cite{CTEQ6L} for the LO fit.

\subsection{Parameterisation}
\label{sec:param}

In the appoach of HERAPDF, 
the PDFs of the proton, $xf$, 
are generically parameterised at the starting scale 
$\mu^2_{\rm f_{0}}$ as
\begin{equation}
 xf(x) = A x^{B} (1-x)^{C} (1 + D x + E x^2)~~,
\label{eqn:pdf}
\end{equation}
where $x$ is the fraction of the proton's momentum taken by the struck parton 
in the infinite momentum frame. 
The PDFs parameterised  are the gluon distribution, $xg$, 
the valence-quark distributions, $xu_v$, $xd_v$, and 
the $u$-type and $d$-type anti-quark distributions,
$x\bar{U}$, $x\bar{D}$. The relations $x\bar{U} = x\bar{u}$ and 
$x\bar{D} = x\bar{d} +x\bar{s}$ are assumed at the starting scale $\mu^2_{\rm f_{0}}$.
  
The central parameterisation is
\begin{eqnarray}
\label{eq:xgpar}
xg(x) &=   & A_g x^{B_g} (1-x)^{C_g} - A_g' x^{B_g'} (1-x)^{C_g'}  ,  \\
\label{eq:xuvpar}
xu_v(x) &=  & A_{u_v} x^{B_{u_v}}  (1-x)^{C_{u_v}}\left(1+E_{u_v}x^2 \right) , \\
\label{eq:xdvpar}
xd_v(x) &=  & A_{d_v} x^{B_{d_v}}  (1-x)^{C_{d_v}} , \\
\label{eq:xubarpar}
x\bar{U}(x) &=  & A_{\bar{U}} x^{B_{\bar{U}}} (1-x)^{C_{\bar{U}}}\left(1+D_{\bar{U}}x\right) , \\
\label{eq:xdbarpar}
x\bar{D}(x) &= & A_{\bar{D}} x^{B_{\bar{D}}} (1-x)^{C_{\bar{D}}} .
\end{eqnarray}

The gluon distribution, $xg$, is an exception from Eq.~\ref{eqn:pdf}, 
for which an additional term of the form 
$A_g'x^{B_g'}(1-x)^{C_g'}$is subtracted\footnote{In the analysis presented here, 
$C_g'$ is fixed to $C_g' = 25$~\cite{Martin:2009iq}.
The fits are not sensitive to the exact value of $C_g'$ once   
$C_g' \gg C_g$, such that the term does not contribute at large $x$.}.
This additional term was added to make the parameterisation more flexible 
at low $x$, such that it is not controlled by the single 
power $B_g$ as $x$ approaches zero~\cite{Martin:2009iq}. 
This requires that the powers $B_g$  and $B_g'$ are different. 
Therefore a restriction was placed on $B_g'$,
such that $B_g'$ values in the range $ 0.95 < B_g'/B_g < 1.05 $ were excluded 
for all PDFs released.
The term $A_g'x^{B_g'}(1-x)^{C_g'}$ was subtracted at NLO and NNLO,
but not at LO, since such a term could lead to $xg(x)$ becoming negative
which is not physical at LO,
because the  LO gluon distribution function
at low $x$ is directly related to the observable longitudinal 
structure function $\tilde{F_{\rm L}}$~\cite{CooperSarkar:1988}.

The normalisation parameters, 
$A_{u_v}, A_{d_v}, A_g$, are constrained 
by the quark-number sum rules and the momentum sum rule. 
The $B$ parameters  $B_{\bar{U}}$ and $B_{\bar{D}}$ were set as equal,
$B_{\bar{U}}=B_{\bar{D}}$, 
such that 
there is a single $B$ parameter for the sea distributions. 
The strange-quark distribution is expressed 
as an $x$-independent fraction, $f_s$, of the $d$-type sea, 
$x\bar{s}= f_s x\bar{D}$ at $\mu^2_{\rm f_{0}}$.  
The value $f_s=0.4$ 
was chosen as a compromise between the determination of a suppressed 
strange sea from neutrino-induced di-muon 
production~\cite{Martin:2009iq,Nadolsky:2008zw} and a recent 
determination of an unsuppressed strange sea, published by the ATLAS 
collaboration~\cite{atlasstrange}. 
A further constraint was applied by setting
$A_{\bar{U}}=A_{\bar{D}} (1-f_s)$.
This,
together with the requirement $B_{\bar{U}}=B_{\bar{D}}$, ensures that 
$x\bar{u} \rightarrow x\bar{d}$ as $x \rightarrow 0$.

The parameters appearing in Eqs.~\ref{eq:xgpar}--\ref{eq:xdbarpar} 
were selected by first fitting with
all $D$ and $E$ parameters and $A_g'$ set to zero. 
This left 10 free parameters. The other parameters were then included
in the fit one at a time.  The improvement of the  $\chi^2$ of the fits
was monitored and the procedure was ended when no further 
improvement in  $\chi^2$
was observed.
This led to a $15$-parameter fit at NLO and a $14$-parameter fit 
at NNLO.
A common parameterisation with $14$-parameters was chosen as ``central'', 
both at NLO and at NNLO,
such that any differences between these fits reflect only the change in order. 
The central fits satisfy the criterion that all the PDFs are positive in the 
measured region.
The $15$-parameter NLO fit was used as a parameterisation variation, 
see Section~\ref{sec:assumpt}.

\subsection{Definition of $\boldsymbol{\chi^2}$} 
\label{sec:defc}
The pQCD predictions were 
fit to the data using a $\chi^2$ method 
similar to that described in
Section~\ref{sec:comb:averaging}. 
The definition of $\chi^2$ is
\begin{equation}
 \chi^2_{\rm exp}\left(\boldsymbol{m},\boldsymbol{s}\right) = 
 \sum_i
 \frac{\left[m^i
- \sum_j \gamma^i_j m^i s_j  - {\mu^i} \right]^2}
{ \textstyle \delta^2_{i,{\rm stat}}\,{\mu^i} m^i +
\delta^2_{i,{\rm uncor}}\,  (m^i)^2}
 + \sum_j s^2_j\, + \sum_i \ln \frac{ \delta^2_{i,{\rm stat}} \mu^i m^i + (\delta_{i,{\rm uncor}} m^i)^2}{ (\delta^2_{i,{\rm stat}} +\delta^2_{i,{\rm uncor}})(\mu^i)^2} ~,
\label{eq:avefit}
\end{equation} 
where the notation is equivalent to that in \Eq~\ref{eq:ave1};
here $\boldsymbol{s}$ is used to denote systematic shifts. 
The additional logarithmic term in 
\Eq~\ref{eq:avefit} compared to \Eq~\ref{eq:ave1} was introduced to
minimise biases~\cite{H1allhQ2}.

Correlated
systematic uncertainties were treated as for the combination of data,
see Section~\ref{sec:comb:averaging}. 
For the combined inclusive data, the correlated systematic uncertainties
are smaller or comparable to the  statistical and uncorrelated uncertainties. 
Nevertheless, the remaining correlations are significant and thus the 
$162$ systematic uncertainties 
present for the 
H1 and ZEUS data sets plus the seven sources of 
procedural uncertainty which resulted from the 
combination procedure, see Section~\ref{subsubsec:proc_errors}, were all 
individually treated as correlated uncertainties.

\subsection{Experimental uncertainties}
\label{sec:exp:unc}

Experimental uncertainties were determined using the Hessian method
with the criterion $\Delta\chi^2=1$.
The use of a consistent input data set with common correlations justifies this approach.

A cross check was performed using the Monte Carlo method~\cite{MC1,MC2}.
It is based on analysing a large number of pseudo
data sets called replicas. For this cross check, 
1000 replicas were created by taking the combined data and fluctuating
the values of the reduced cross sections 
randomly within their given statistical
and systematic uncertainties taking into account correlations. 
All uncertainties were assumed to follow Gaussian distributions. 
The PDF central values and uncertainties 
were estimated using the mean and RMS values
over the replicas.

\begin{figure}[tbp]
\vspace{-0.3cm} 
\centerline{
\epsfig{file=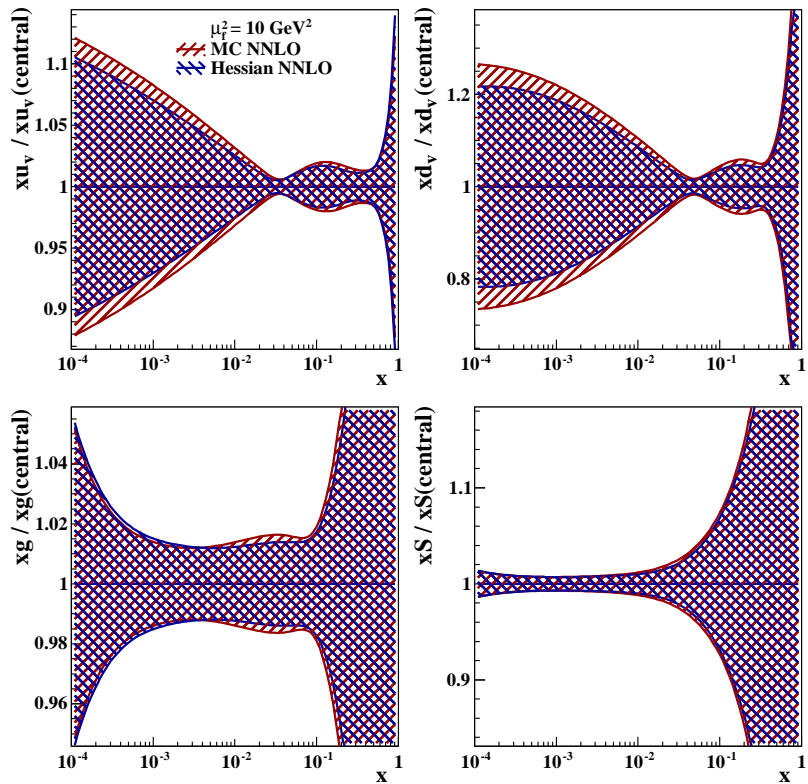 ,width=0.7\textwidth}}
\caption {
          Comparison of the PDF uncertainties as 
          determined by the Hessian and Monte Carlo 
          (MC) methods at NNLO 
          for the valence 
          distributions $xu_v$ and $xd_v$, the
          gluon distribution $xg$ and the sea distribution, 
          $xS=2x(\bar{U}+\bar{D})$, 
          at the scale 
          $\mu_{\rm f}^{2}$ = 10\,GeV$^{2}$.
}
\label{fig:MCcomp:NLO+NNLO}
\end{figure}

The uncertainties obtained by the Monte Carlo method 
and the Hessian method were
consistent within the kinematic reach of HERA.
This is demonstrated in Fig.~\ref{fig:MCcomp:NLO+NNLO} where
experimental uncertainties obtained for HERAPDF2.0 NNLO by the Hessian
and Monte Carlo methods are compared for the valence, the gluon and the
total sea-quark distributions.
The RMS values taken as Monte Carlo uncertainties tend 
to be slightly larger than the standard deviations obtained 
in the Hessian approach.

\subsection{Model and parameterisation uncertainties}
\label{sec:assumpt}


For the NLO and NNLO PDFs, the uncertainties on HERAPDF2.0 due to  
the choice of model settings and the form of the 
parameterisation 
were evaluated
by  varying the assumptions. 
A summary of the variations
on model parameters is given in Table~\ref{tab:model}. 
The variations of $M_c$ and $M_b$ were chosen in accordance
with the $\chi^2$ scans related to the heavy-quark mass parameters 
as shown in 
Figs.~\ref{fig:charmscan} and~\ref{fig:beautyscan}.
The data on heavy-quark production from HERA\,II
led to a considerably reduced uncertainty on the 
heavy-quark mass parameters compared to 
the HERAPDF1.0 and HERAPDF1.5 analyses, see Appendix~\ref{appendix:A}.

The variation of $f_s$ was chosen to span the ranges between a suppressed 
strange sea~\cite{Martin:2009iq,Nadolsky:2008zw} and an 
unsuppressed strange sea~\cite{atlasstrange}. 
In addition to this, two more 
variations of the assumptions about the strange sea were made. 
Instead of assuming 
that the strange contribution is a fixed fraction of the 
$d$-type sea, an $x$-dependent shape,  
$x\bar{s}= f_s' \, 0.5 \tanh(-20(x-0.07))\, x\bar{D}$, 
was used in which high-$x$ strangeness is highly suppressed. 
This was suggested by
measurements published by the HERMES collaboration~\cite{HERMES1,HERMES2}. 
The normalisation of  $f_s'$ 
was also varied between $f_s'=0.3$ 
and $f_s'=0.5$. 
 
In addition to these model variations, $Q^2_{\rm min}$ 
was varied 
as for the HERAPDF1.0 and HERAPDF1.5 analyses, see Appendix~\ref{appendix:A}.
%
The differences between the central fit and the fits corresponding  
to the variations of 
$Q^2_{\rm min}$,
$f_s$, 
$M_c$ and 
$M_b$ 
are
added in quadrature, separately for positive and negative deviations, and 
represent the model uncertainty of the HERAPDF2.0 sets.

Two kinds of parameterisation uncertainties were considered,
the variation in $\mu^2_{\rm f_{0}}$ and the addition of parameters
$D$ and $E$, see Eq.~\ref{eqn:pdf}.
The variation in $\mu^2_{\rm f_{0}}$ mostly increased the PDF uncertainties of 
the sea and gluon  at small $x$. 
The parameters $D$ and $E$ were added separately for each PDF.
The only significant difference from the $14$-parameter central fit came 
from the $15$-parameter fit, 
for which $D_{u_v}$ was non zero. 
This affected the shape of the $U$-type sea as well as the shape of $u_v$.
The final parameterisation uncertainty for a given quantity is taken as the largest
of the uncertainties. 
This uncertainty is valid in the $x$-range covered by the QCD fits
to HERA data.

%

\subsection{Total uncertainties}

The total PDF uncertainty is obtained by adding in quadrature 
the experimental, 
the model and the parameterisation uncertainties 
described in Sections~\ref{sec:exp:unc} and~\ref{sec:assumpt}.
Differences arising from using 
alternative values of $\asmz$, alternative forms of parameterisations, 
different heavy-flavour schemes or 
a very different $Q^2_{\rm min}$
are not included in these uncertainties.
Such changes result in the different variants of the 
PDFs to be discussed in the subsequent sections.

\subsection{Alternative values of $\boldsymbol{\asmz}$}  
\label{sec:altasmz}
The HERAPDF2.0 NLO and NNLO standard fits 
were additionally made for a series 
of $\asmz$ values from $\asmz=0.110$ to $\asmz=0.130$ in steps of $0.001$. 
These variants are also released.
They can be used to assess the uncertainty on any predicted cross section 
due to the choice of $\asmz$ and for $\asmz$ determinations 
using independent data. 

\subsection{Alternative forms of parameterisation}  
\label{sec:altparam}

An ``alternative gluon parameterisation'', AG, was considered
at NNLO and NLO.
The value of $A_g'$ in Eq.~\ref{eq:xgpar} was set to zero 
and a polynominal term for $xg(x)$ as 
in Eq.~\ref{eqn:pdf} was substituted. 
This potentially resulted in a different $14$-parameter fit.
However, in practice a $13$-parameter fit with a non-zero $D_g$ 
was sufficient for the AG parameterisation, since
there was no improvement in $\chi^2$ for a non-zero $E_g$.
Note that AG was the only parameterisation considered at LO.

The standard parameterisation fits the HERA data better;
however, especially at NNLO, it
produces a negative gluon distribution
for very low $x$, i.e. $x < 10^{-4}$. 
This is outside the kinematic region of the fit, 
but may cause problems if  
the PDFs are used at very low $x$ within the conventional formalism.
Therefore, a variant HERAPDF2.0AG 
using the alternative gluon parameterisation 
is provided for predictions
of cross sections at very low $x$, 
such as very high-energy neutrino cross sections. 

HERAPDF has a certain {\it ansatz} for the parameterisation of the PDFs,
see Section~\ref{sec:param}.
Different ways of using the polynomial form, such as parameterising
$xg$, $xu_v$, $xd_v$, $x\bar{d}+x\bar{u}$ and $x\bar{d}-x\bar{u}$ or $xg$, 
$xU$, $xD$, $x\bar{U}$ and $x\bar{D}$ were investigated.  
The resulting PDFs agreed with the standard PDFs within uncertainties and 
no improvement of fit quality resulted.
Therefore, these alternative parameterisations were not
pursued further.

\subsection{Alternative heavy-flavour schemes}
\label{sec:althfs}
 
The standard choice of heavy-flavour scheme for HERAPDF2.0 is the 
variable-flavour-number scheme RTOPT~\cite{Thorne:RTopt}. 
Investigations using other heavy-flavour schemes were also carried out.

Two other variable-flavour-number schemes,
FONLL~\cite{Cacciari:1998it,Forte:2010ta} and
ACOT~\cite{ACOT}, 
were considered, as implemented in HERAFitter at the time of the analysis. 
The FONLL scheme is implemented via an interface to the 
APFEL program~\cite{Bertone:2013vaa} and was used at NLO and NNLO.
The ACOT scheme is implemented using $k$-factors for the NLO corrections.
The three heavy-flavour schemes differ in the order at 
which $F_{\rm L}$ is evaluated. 
At NLO, the massless contribution to $F_{\rm L}$ is evaluated to 
${\cal O} (\alpha_s^2)$ for RTOPT and to ${\cal O}(\alpha_s)$ for 
FONNL-B and ACOT. 
At NNLO, the massless contribution to $F_{\rm L}$ is evaluated to 
${\cal O} (\alpha_s^3)$ for RTOPT and to ${\cal O}(\alpha_s^2)$ for 
FONNL-C. 
Fixed-flavour-number schemes were also investigated.
In such schemes, 
the number of (massless) light flavours in the 
PDFs remains fixed across ``flavour thresholds'' and (massive) heavy 
flavours only occur in the matrix elements.

For some calculations, e.g.\,charm production at HERA, the availability
of fixed-flavour variants of the PDFs is useful or even mandatory.
Many PDF groups provide either fixed-flavour fits only, or 
variable-flavour fits only, with a fixed-flavour variant calculated from the 
variable-flavour parton distributions at the starting scale using theory.
For HERAPDF2.0, fixed-flavour variants are provided which
were actually fitted to the data.

Two schemes with three active flavours in the PDFs, 
FF3A and FF3B, were considered:
\begin{itemize}
\item scheme FF3A:
 \begin{itemize}
 \item Three-flavour running of $\alpha_s$;
 \item $F_{\rm L}$ calculated to 
       ${\cal O}(\alpha_s^2)$;
 \item pole masses for charm, $m_c^{\rm pole}$, and beauty, $m_b^{\rm pole}$;
 \end{itemize}
\item scheme FF3B:
 \begin{itemize}
 \item Variable-flavour running of $\alpha_s$ \cite{Gluck:1994}. 
       This is sometimes called the 
       ``mixed scheme'' \cite{QCDNUM};
 \item massless (light flavour) part of the $F_{\rm L}$ contribution 
       calculated to ${\cal O}(\alpha_s)$; 
 \item $\overline{\rm MS}$~\cite{MSbar} 
       running masses for charm, $m_c(m_c)$, 
       and beauty $m_b(m_b)$.
 \end{itemize}
\end{itemize}
The input parameters to the fits are given in Table~\ref{tab:FF}.

The fits providing the variants HERAPDF2.0FF3A and HERAPDF2.0FF3B
were obtained using the OPENQCDRAD \cite{OPENQCDRAD} package as 
implemented in HERAFitter, partially interfaced to QCDNUM. 
This was proven to be consistent with the standalone version of 
OPENQCDRAD and, in the case of the A variant, with the FFNS definition  used by 
the ABM \cite{ABM1,ABM2,ABM3} fitting group. 
The HERAFitter implementation allows an external steering of the order 
of $\alpha_s$ in $F_{\rm L}$, as listed in Table~\ref{tab:FF}.

\subsection{Adding data on charm production to the HERAPDF2.0 fit}
\label{sec:addcharm}

The data on charm production described in
Section~\ref{sec:adddata} were used to find the optimal value of
$M_c$ for the HERAPDF2.0 fits as described
in Section~\ref{sec:choices}.

The impact of adding charm data to inclusive 
data as input to NLO QCD fits 
has been extensively discussed in a previous publication \cite{HERAccombi}.
This previous analysis was based on the 
HERA\,I combined inclusive data
and combined charm data. 
It was established that the main 
impact of the charm data on the PDF fits is 
a reduction of  the uncertainty on $M_c$.
It was also established that the optimal value of $M_c$ 
can differ according to the particular general-mass variable-flavour-number 
scheme chosen for the fit.
The fits for all schemes considered were of similar quality.

For the HERAPDF2.0 analysis, 
a total of 47 data points on charm production \cite{HERAccombi} 
with $Q^2$ larger than $Q^2_{\rm min} = 3.5\,$GeV$^2$ 
were added as input to the NLO fits.
The 42 sources of correlated systematic uncertainty 
from the H1 and ZEUS data 
sets on charm production and  two additional sources due to the 
combination procedure were taken into account.
The correlations between the normalisation of the inclusive data 
and the normalisation of the charm data was not taken into account in the 
PDF fits but it was verified that this has a negligible effect. 

The inclusion of the charm data 
had little influence on the result of the fit.
This was not unexpected, since the main effect of the charm data,
i.e. to constrain $M_c$, was already used for the fit to the
inclusive data.
The charm data were proven to be consistent with the inclusive data,
but only a marginal reduction 
in the uncertainty on the low-$x$ gluon PDF was obtained. 
The situation is similar at NNLO. Therefore no HERAPDF2.0 variants 
with only the addition of data on charm production are released.

\subsection{Adding data on jet production to the HERAPDF2.0 fit}
\label{sec:addjfit}

In pQCD fits to inclusive DIS data only, 
the gluon PDF is determined via the DGLAP equations using
the observed
scaling violations.
This results in a strong 
correlation between the shape of the gluon distribution 
and the value of $\asmz$. 
In most PDF fits, the value of $\asmz$ is not fitted but taken from external 
information~\cite{PDG12}. 
The uncertainty on the gluon distribution is reduced
for fits with fixed $\asmz$ compared to fits with free 
$\asmz$.
Data on jet production cross sections
provide an independent measurement of the 
gluon distribution.
They  are sensitive to $\asmz$ and 
already at LO to the gluon distribution 
at lower $Q^2$ and  
to the valence-quark distribution at higher $Q^2$.
Therefore the inclusion of jet  data  
not only reduces the uncertainty on the high-$x$ gluon 
distribution in fits with 
fixed $\asmz$  but also allows 
the accurate simultaneous determination of $\asmz$ 
and the gluon distribution.

%
The jet data were included in the fits at NLO  by calculating
predictions for the jet cross sections
with NLOjet++~\cite{nlojet1,nlojet2}, which was interfaced to
FastNLO~\cite{fastnlo1,fastnlo2,fastnlo3} 
in order to achieve the speed necessary
for iterative PDF fits.
The predictions were multiplied by  
corrections for hadronisation and 
$Z^0$ exchange 
before they were used to fit the 
data~\cite{zeusdijets,zeus9697jets,h1lowq2jets,h1highq2oldjets,h1highq2newjets}.
A running electro-magnetic $\alpha$ as implemented in the 2012 version of 
the programme EPRC~\cite{Spiesberger:95} was used for the treatment
of jet cross sections when they were included in the PDF fits.
The factorisation scale was chosen as 
$\mu_{\rm f}^2 = Q^2$,
while the renormalisation scale was linked to the transverse
momenta, $p_T$, of the jets by $\mu_{\rm r}^2 = (Q^2 + p_{T}^2)/2$.
Jet data  could not be included at NNLO 
for the analysis presented here, because the matrix elements
were not available at the time of writing.


The normalisations of the ZEUS jet data~\cite{zeusdijets,zeus9697jets} 
and the H1 low-$Q^2$ jet data~\cite{h1lowq2jets}
are correlated with the inclusive 
samples but because of the combination procedure these correlations 
cannot be recovered. 
Thus they are treated conservatively as uncorrelated.
However, cross checks performed by using the uncombined
H1 and ZEUS inclusive 
data  have shown that this does not have a significant 
impact on the result. 
In the case of 
the H1 high-$Q^2$ jet data~\cite{h1highq2oldjets,h1highq2newjets},
the correlations due to the uncertainty 
on the integrated luminosity 
are  accounted for by the normalisation of the jet cross sections
to the inclusive cross sections. 
The statistical correlations present between the jet data and the inclusive data
were neglected, with no significant impact on the result.

Fits including these jet data and including the combined charm data 
were performed with $\asmz=0.118$ fixed and with $\asmz$  as a free 
parameter in the fit. 
The resulting HERAPDF variant with free $\asmz$ 
is called HERAPDF2.0Jets.
A full uncertainty analysis was performed for the HERAPDF2.0Jets variant, 
including model and parameterisation uncertainties and additional hadronisation 
uncertainties on the jet data 
as evaluated for the original publications~\cite{zeusdijets,zeus9697jets,h1lowq2jets,h1highq2oldjets,h1highq2newjets}.

\subsection{The {$\chi^2$}  values of the HERAPDF2.0 fits and alternative {$Q^2_{\rm min}$}}
\label{sec:cuts}

\begin{figure}[tbp]
\centerline{
\epsfig{file=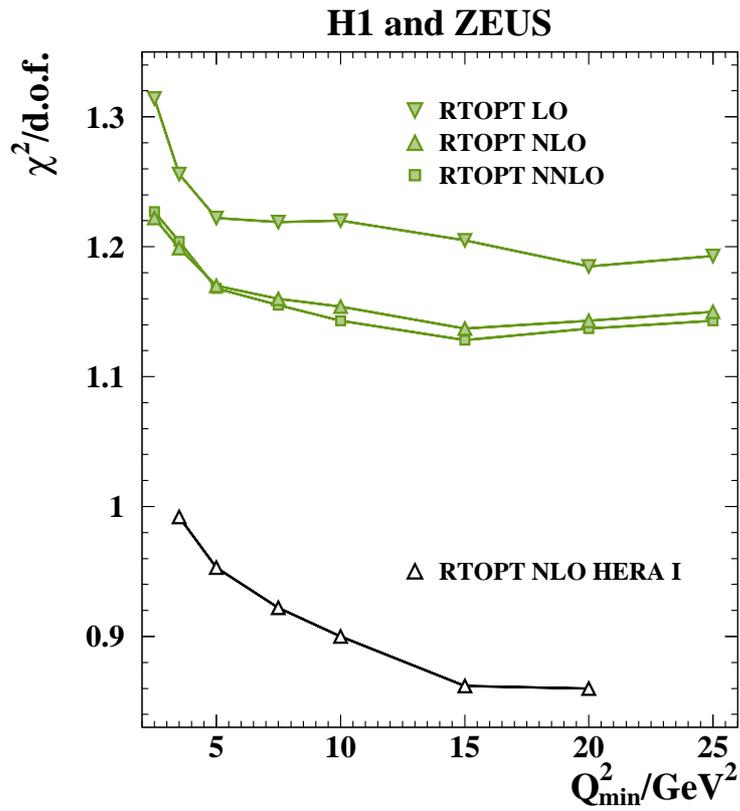 ,width=0.7\textwidth}}
\caption { 
The dependence of $\chi^2/{\rm d.o.f.}$ on $Q^2_{\rm min}$ 
of the LO, NLO and NNLO fits to the HERA combined inclusive data.
Also shown are values for an NLO fit to the 
combined HERA\,I data~\cite{HERAIcombi}.
All fits were performed using the RTOPT heavy-flavour scheme.
}
\label{fig:chiscan}
\end{figure}

The $\chi^2/{\rm d.o.f.}$ of the fits for HERAPDF2.0
and its variants are listed in Table~\ref{tab:chi2}.
These values are somewhat large, typically around 1.2. 
The dependence of 
$\chi^2$ on $Q^2_{\rm min}$ was investigated in detail. 
Figure~\ref{fig:chiscan} shows the
$\chi^2/{\rm d.o.f.}$ values for the LO, NLO and NNLO 
fits versus $Q^2_{\rm min}$. 
The  $\chi^2/{\rm d.o.f.}$ drop steadily until  
$Q^2_{\rm min} \approx 10\,$GeV$^2$.
Also shown are  $\chi^2$ values obtained for an NLO fit
to HERA\,I data only. 
These values are substantially closer to one, but they show the
same trend as seen for HERAPDF2.0.

The  $\chi^2/{\rm d.o.f.}$ values rise again for 
$Q^2_{\rm min} > 15\,$GeV$^2$.
If only data with $Q^2$ between $Q^2=15\,$GeV$^2$ and $Q^2=150\,$GeV$^2$ were used,
the  $\chi^2/{\rm d.o.f.}$ became close to unity.
The addition of either data with lower or higher $Q^2$ increased
the $\chi^2/{\rm d.o.f.}$.
The lower- and
middle-$Q^2$ data clearly show tension.
The higher-$Q^2$ data generally cannot be fitted very well.
It was not possible to attribute this to any 
particular region in $x_{\rm Bj}$ or a particular NC or CC process. 
For the standard value $Q^2_{\rm min} = 3.5\,$GeV$^2$, the data between 
$Q^2=3.5\,$GeV$^2$ and $Q^2=15\,$GeV$^2$ create about one third of the
excess  $\chi^2/{\rm d.o.f.}$ while two thirds originate from the data with 
$Q^2>150\,$GeV$^2$.

\begin{figure}[tbp]
  \centering
  \setlength{\unitlength}{0.1\textwidth}
  \begin{picture} (9,12)
  \put(0,0){\includegraphics[width=0.9\textwidth]{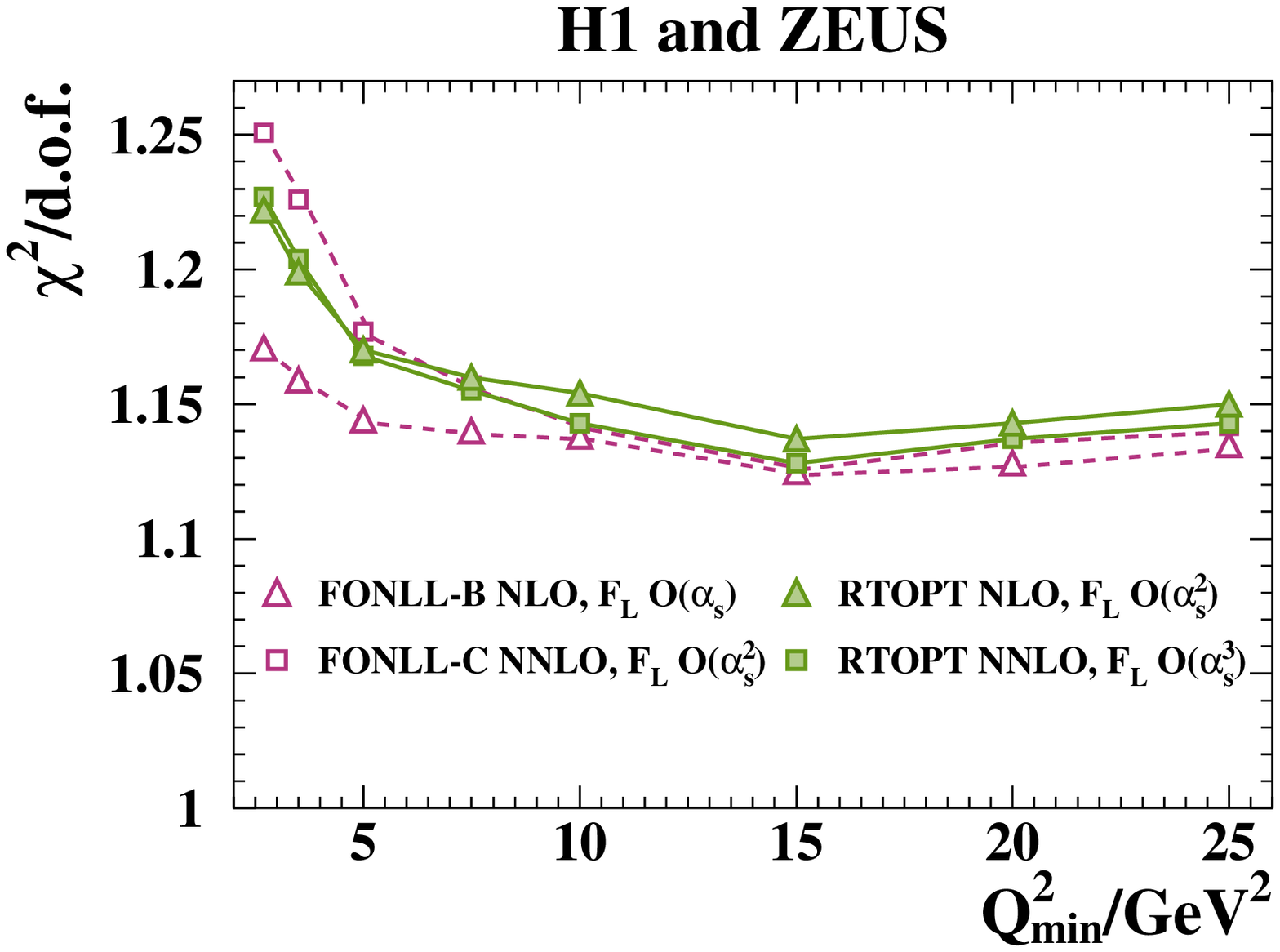}}
  \put(0,5.8){\includegraphics[width=0.9\textwidth]{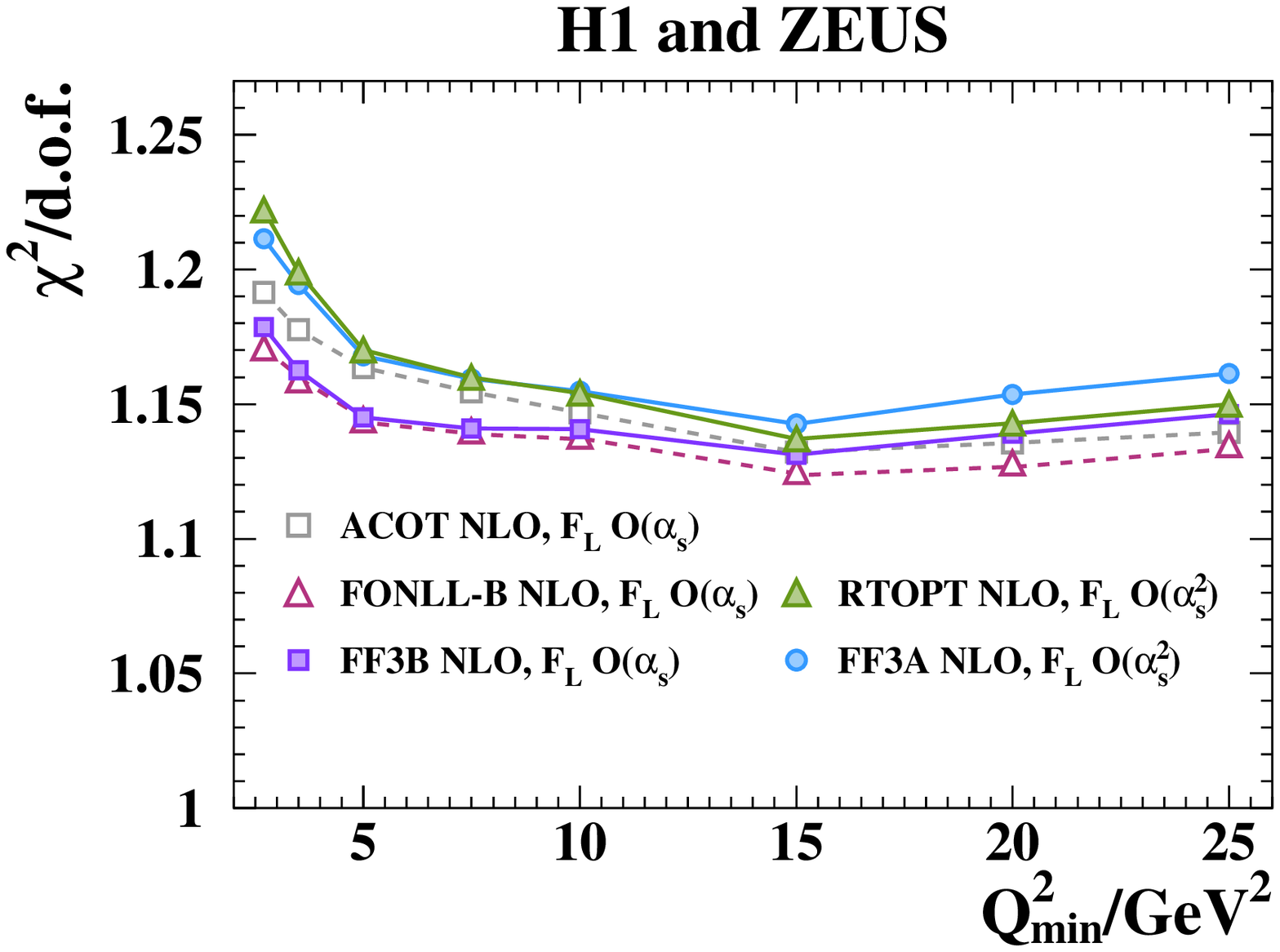}}
  \put (0.3,6.6) {a)}
  \put (0.3,0.8) {b)}
  \end{picture}
%
\caption{
     The dependence of $\chi^2/{\rm d.o.f.}$  on $Q^2_{\rm min}$  
     for HERAPDF2.0 fits using a) 
     the RTOPT~\cite{Thorne:RTopt}, 
     FONNL-B~\cite{Cacciari:1998it}, 
     ACOT~\cite{Kramer:2000hn} 
     and fixed-flavour (FF) 
     schemes at NLO and 
     b) the RTOPT 
     and FONNL-B/C~\cite{Forte:2010ta}
     schemes at NLO and NNLO. 
     The $F_{\rm L}$ contributions are calculated 
     using matrix elements of the 
     order of $\alpha_s$ indicated in the legend.
     The number of degrees of freedom drops from 
     1148 for $Q^2_{\rm min}=2.7$\,GeV$^2$ to
     1131 for the nominal $Q^2_{\rm min}=3.5$\,GeV$^2$ and to
     868 for $Q^2_{\rm min}=25$\,GeV$^2$.
     }
\label{fig:altscan}
\end{figure}

The influence of the choice of heavy-flavour scheme,
and the order at which the massless contribution to $F_{\rm L}$ is evaluated,
on the $\chi^2/{\rm d.o.f.}$ behaviour was also investigated.
Scans at NLO and NNLO 
of the $\chi^2/{\rm d.o.f.}$ versus $Q^2_{\rm min}$ for 
fits done with the heavy-flavour schemes described in 
Section~\ref{sec:althfs}
are illustrated in Fig.~\ref{fig:altscan}. 
The decrease of the $\chi^2/{\rm d.o.f.}$ with increasing $Q^2_{\rm min}$  
is observed for every scheme.
At NLO and low $Q^2_{\rm min}$, 
all fits using schemes for which the $F_{\rm L}$  
contributions are calculated using matrix elements of the 
order of $\alpha_s$ result in slightly 
lower $\chi^2/{\rm d.o.f.}$ than fits for schemes using matrix elements  
of the order of $\alpha_s^2$.
The increase of  $\chi^2/{\rm d.o.f.}$ for lower $Q^2_{\rm min}$
is also less pronounced for fits using the ``${\cal O}(\alpha_s)$-schemes''.
However, at NNLO, the trend reverses and RTOPT,
 which uses matrix elements of
order $\alpha_s^3$ 
in the calculation of $F_{\rm L}$, 
results in lower $\chi^2/{\rm d.o.f.}$ than the
FONNL scheme, for which matrix elements of order $\alpha_s^2$ are used.
The $\chi^2/{\rm d.o.f.}$ values for fits with the RTOPT scheme are
quite similar at NLO and NNLO.

The two fixed-flavour-number schemes considered, see Section~\ref{sec:althfs},
also differ in using light-flavour 
matrix elements of order $\alpha_s$ (FF3B)  
and $\alpha_s^2$ (FF3A).
The FF3A fit variant results in $\chi^2/{\rm d.o.f.}$ values very similar
to the values from the standard fit using RTOPT while the 
values for the FF3B variant closely follow the results for fits using the
FONNL scheme.
This suggests that
the determining factor for the $\chi^2$ of the fits 
is the order of $\alpha_s$ 
of the matrix elements
used to calculate the massless $F_{\rm L}$ contribution.
Other differences between FF3A and FF3B as well as
differences~\cite{GMVFNSdiff} between different 
variable-flavour-number schemes,  
and differences between fixed-flavour-number and 
variable-flavour-number schemes, seem to have less influence on  $\chi^2$.

At HERA, the low-$Q^2$ data are also dominantly at low $x_{\rm Bj}$. 
Some of the poor
$\chi^2$ values in this kinematic region could be 
due to low-$x_{\rm Bj}$ physics not accounted for in
the current framework~\cite{saturation,Caola}. 
This could mean that the inclusion of low-$x_{\rm Bj}$, low-$Q^2$ data 
into the fits introduces bias. 
To study this, NLO and NNLO fits 
with $Q^2_{\rm min} = 10\,$GeV$^2$ were also fully evaluated.
This variant is called HERAPDF2.0HiQ2. 
As part of the evaluation, the settings were reexamined. No significant 
changes for the optimal parameterisation or for the optimal value 
of $M_c$ or $M_b$ were observed. 
Model and parameterisation variations were also performed 
in order to better assess possible bias.
For the NLO fits, the  $\chi^2/{\rm d.o.f.}$  of 
$1156/1002$ for the $Q^2_{\rm min} = 10\,$GeV$^2$ fit
can be compared to the
$1357/1131$ for the $Q^2_{\rm min} = 3.5\,$GeV$^2$ fit. 
This is a significant improvement, but still larger than observed
for HERAPDF1.0. 
The values are similar at NNLO, see Table~\ref{tab:chi2}.
In particular, the NNLO fit does not fit 
the lower-$Q^2$ data better than the NLO fit, see Fig.~\ref{fig:chiscan}, 
just as, at NLO, the higher-order evaluation of $F_{\rm L}$ does not fit 
these data better, see Fig.~\ref{fig:altscan}.

Fits were also performed with the alternative gluon
parameterisation and  $Q^2_{\rm min} = 10\,$GeV$^2$.
The $\chi^2/{\rm d.o.f.}$ was always worse than for 
the standard parameterisation, see Table~\ref{tab:chi2}.
 
The $\chi^2/{\rm d.o.f.}$ values obtained for HERAPDF2.0Jets,
both for fixed and for free $\asmz$ are better than the value
for the standard HERAPDF2.0 NLO fit, see Table~\ref{tab:chi2}.
The partial $\chi^2$ for the jet data is 161 for 162 data points,
while it is 41 for 47 data points for the charm data.
The partial $\chi^2$ for the inclusive data remains practically the
same as for HERAPDF2.0 NLO.
This demonstrates 
the compatibility of the data on charm and jet 
production 
with the inclusive data.

\section{HERAPDF2.0}
\label{sec:pdf20}

The analysis described  in Section~\ref{sec:qcdan}
resulted in  
a set of PDFs called HERAPDF2.0.
The HERAPDF2.0 analysis has the following notable features:
\begin{itemize}
\item the data include four different processes,
      NC and CC for $e^+p$ and $e^-p$ scattering, such that there is sufficient 
      information to extract the $xd_v$, $xu_v$, $x\bar{U}$  and $x\bar{D}$ PDFs, 
      and the gluon PDF from the scaling violations; 
\item the NC $e^+p$ data include data at 
      centre-of-mass energies sufficiently different to access
      different values of $y$  at the same $x_{\rm Bj}$ and $Q^2$; 
      this makes the data sensitive to $F_{\rm L}$ and thus gives further 
      information on the low-$x$ gluon distribution;
\item it is based on a consistent data set with small correlated 
      systematic uncertainties; 
\item the experimental uncertainties are Hessian uncertainties;
\item the uncertainties introduced both 
      by model assumptions 
      and by assumptions about the form of the parameterisation 
      are provided;
\item no heavy-target corrections were needed as all data are
      on $ep$ scattering; 
      the assumption of $u_{\rm neutron}=d_{\rm proton}$ was not needed.    
\end{itemize}

An overview about HERAPDF2.0 and its variants as released
is given in Appendix~\ref{appendix:B}.

\subsection{HERAPDF2.0 NLO, NNLO and 2.0AG} 
\label{sec:pdfs:orders}


 
\begin{figure}[tbp]
\vspace{-0.5cm} 
\centerline{
\epsfig{file=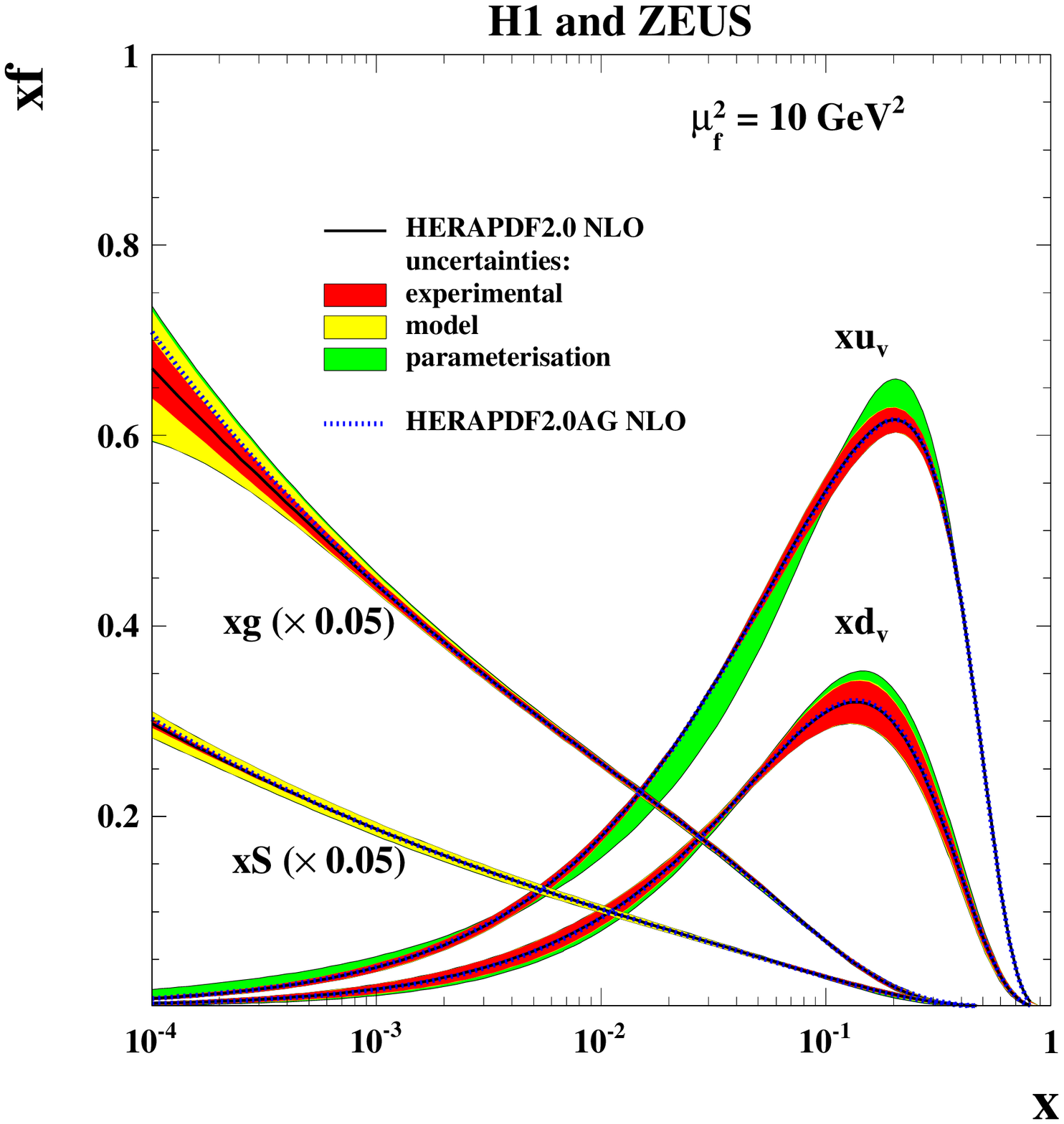 ,width=0.9\textwidth}}
\caption { 
The parton distribution functions 
$xu_v$, $xd_v$, $xS=2x(\bar{U}+\bar{D})$ and $xg$ of 
HERAPDF2.0 NLO
at $\mu_{\rm f}^{2} = 10\,$GeV$^{2}$.
The gluon and sea distributions are scaled down 
by a factor of $20$.
The experimental, model and parameterisation 
uncertainties are shown. 
The dotted lines represent HERAPDF2.0AG NLO with the alternative
gluon parameterisation, see Section~\ref{sec:altparam}.
}
\label{fig:summarynlo}
\end{figure}

A summary of HERAPDF2.0 NLO is shown in Fig.~\ref{fig:summarynlo}
at the scale   
$\mu^2_{\rm f}=10$\,GeV$^2$.
The experimental, model and parameterisation uncertainties, 
see Sections~\ref{sec:exp:unc} and~\ref{sec:assumpt},
are shown separately.
The model and parameterisation uncertainties are asymmetric.  
The uncertainties arising from the 
variation of 
$\mu^2_{\rm f_{0}}=1.9$\,GeV$^2$ and $Q^2_{\rm min}=3.5$\,GeV$^2$ 
affect predominantly the low-$x$ region
of the sea and gluon distributions.
The parameterisation uncertainty from adding the $D_{u_{v}}$ parameter is important 
for the valence distributions
for all $x$.

The gluon distribution of HERAPDF2.0 NLO does not become negative
within the fitted kinematic region. 
The distributions of
HERAPDF2.0AG with the alternative gluon parameterisation as described in 
Section~\ref{sec:param}
and discussed in Section~\ref{sec:altparam}
are shown superimposed on the standard PDFs. 

\begin{figure}[tbp]
\vspace{-0.3cm} 
\centerline{
\epsfig{file=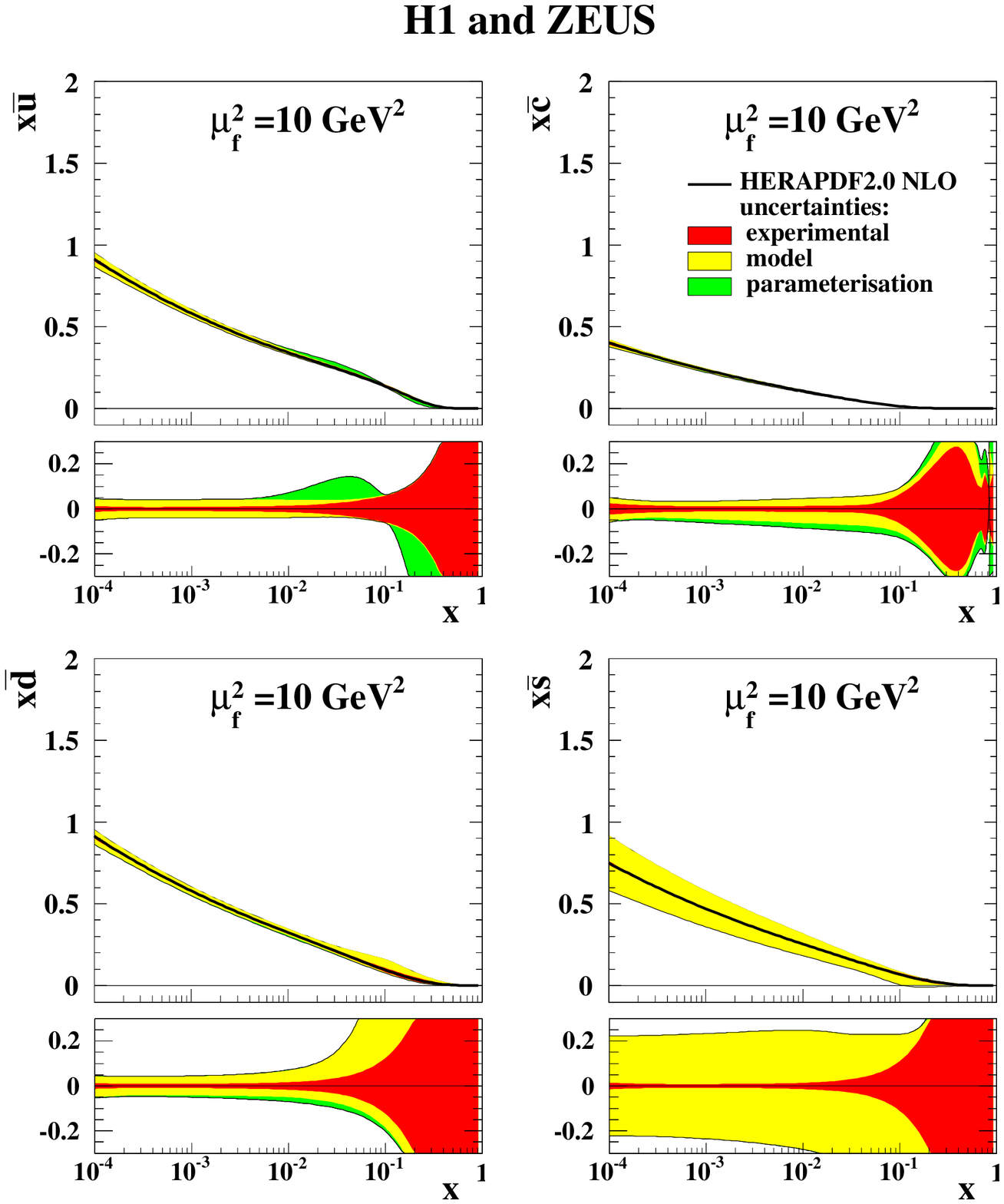  ,width=0.9\textwidth}}
\vspace{0.5cm}
\caption {The flavour breakdown of the sea distribution of HERAPDF2.0 NLO
          at $\mu_{\rm f}^{2}$ = 10\,GeV$^{2}$.
          Shown are the distributions 
          $x\bar{u}$, $x\bar{d}$, $x\bar{c}$ and $x\bar{s}$ 
          together with their experimental, 
          model and parameterisation uncertainties.
          The fractional uncertainties are also shown.
}
\label{fig:flavour1}
\end{figure}

The flavour breakdown of the sea
into $x\bar{u}$, $x\bar{d}$, $x\bar{c}$ and $x\bar{s}$
for HERAPDF2.0 NLO at the scale $\mu_{\rm f}^2=10\,$GeV$^2$ 
is shown in Fig.~\ref{fig:flavour1}. 
The fractional uncertainties are also shown.
The model uncertainties from 
the variation of $f_s$ and $M_c$ affect the  $x\bar{s}$
and  $x\bar{c}$ distributions. 
The $x\bar{c}$ uncertainties also derive from the uncertainty 
on the gluon distribution, since charm is generated 
from $g \to c \bar{c}$ splitting. 
The variation of $M_c$ also affects 
the $x\bar{u}$ distribution since the  
suppression (enhancement) of $x\bar{c}$ 
results in an enhancement (suppression) 
of $x\bar{u}$ in the $u$-type sea. 
Similarly the strangeness variations also affect 
$x\bar{d}$, since the suppressed strangeness must be 
compensated by enhanced $x\bar{d}$ in the $d$-type sea. 
However, since $x\bar{d}$ is fixed to $x\bar{u}$ at low $x$, 
this mostly affects the high-$x$ uncertainty on $x\bar{d}$. 
The central fit gives $x\bar{d} - x\bar{u}$ negative 
at $x \approx 0.1$. 
However, the uncertainty is very large 
because HERA data are not very sensitive to this difference. 
The uncertainty on $x\bar{u}$ has a significant contribution from
the parameterisation uncertainties.
The values of the parameters of HERAPDF2.0 NLO
are given in Table~\ref{tab:param}.

\begin{figure}[tbp]
\vspace{-0.5cm} 
\centerline{
\epsfig{file=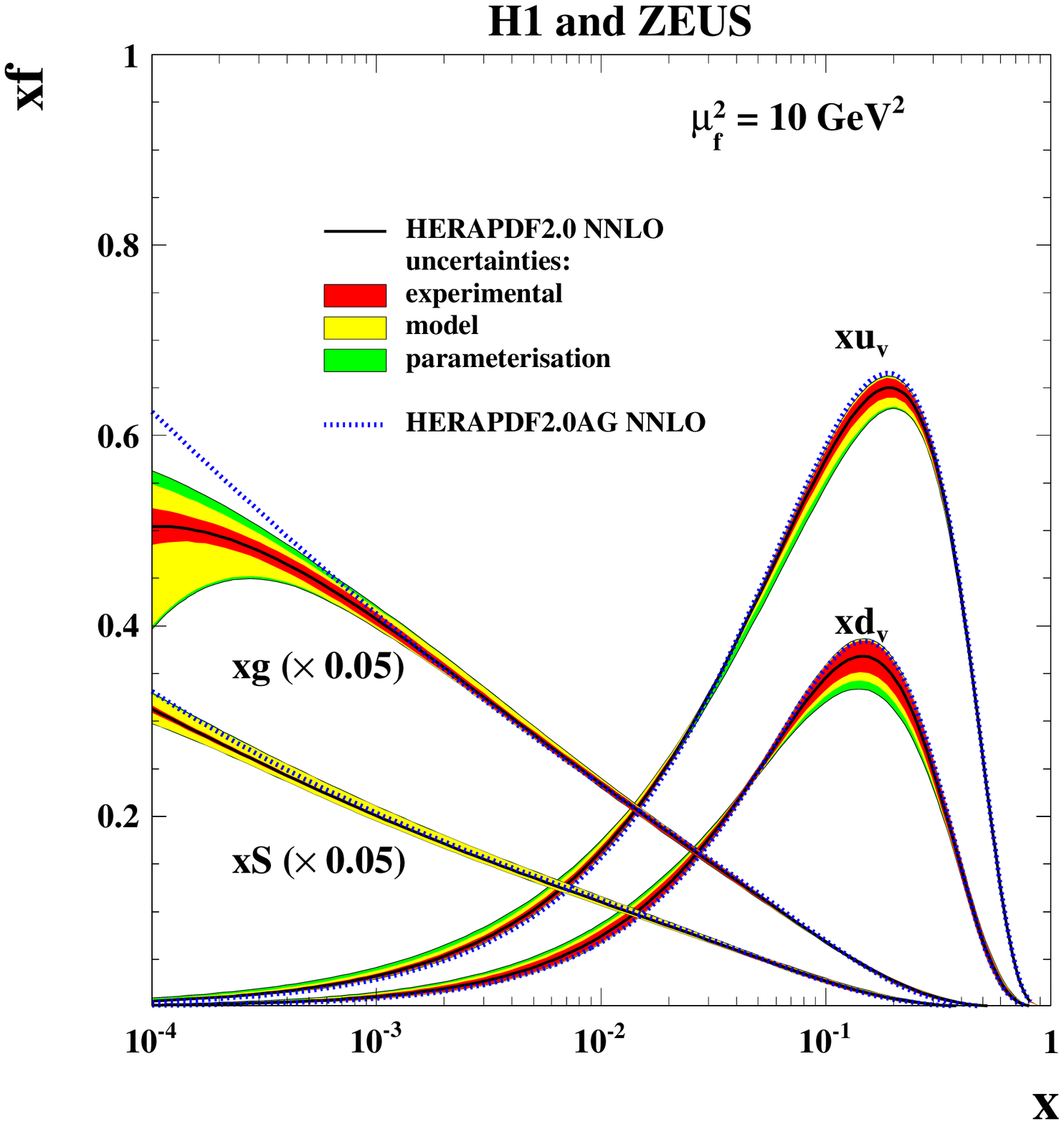 ,width=0.9\textwidth}}
\vspace*{-0.6cm}
\caption { 
The parton distribution functions 
$xu_v$, $xd_v$, $xS=2x(\bar{U}+\bar{D})$ and $xg$ of 
HERAPDF2.0 NNLO
at $\mu_{\rm f}^{2} = 10\,$GeV$^{2}$.
The gluon and sea distributions are scaled down by a factor $20$.
The experimental, model and parameterisation 
uncertainties are shown. 
The dotted lines represent HERAPDF2.0AG NNLO with the alternative 
gluon parameterisation, see Section~\ref{sec:altparam}.
}
\label{fig:summarynnlo}
\end{figure}

A summary of HERAPDF2.0 NNLO is shown in Fig.~\ref{fig:summarynnlo}
at the scale $\mu^2_{\rm f}=10$\,GeV$^2$.
At NNLO, the gluon distribution of HERAPDF2.0 ceases to rise
at low $x$. Consequently, $xg$ from HERAPDF2.0AG NNLO deviates significantly.
As at NLO, the uncertainties arising from the 
variation of 
$\mu^2_{\rm f_{0}}$ and $Q^2_{\rm min}$ 
affect predominantly the low-$x$ region
of the sea and gluon distributions. 
The parameterisation uncertainty from adding 
the $D_{u_v}$ parameter is not important for the NNLO fit, 
since there was no significant improvement in $\chi^2$ 
from the addition of the 15th parameter.
The parameters of the NNLO fit are listed in Table~\ref{tab:nnloparam}.

\begin{figure}[tbp]
\vspace{-0.3cm} 
\centerline{
\epsfig{file=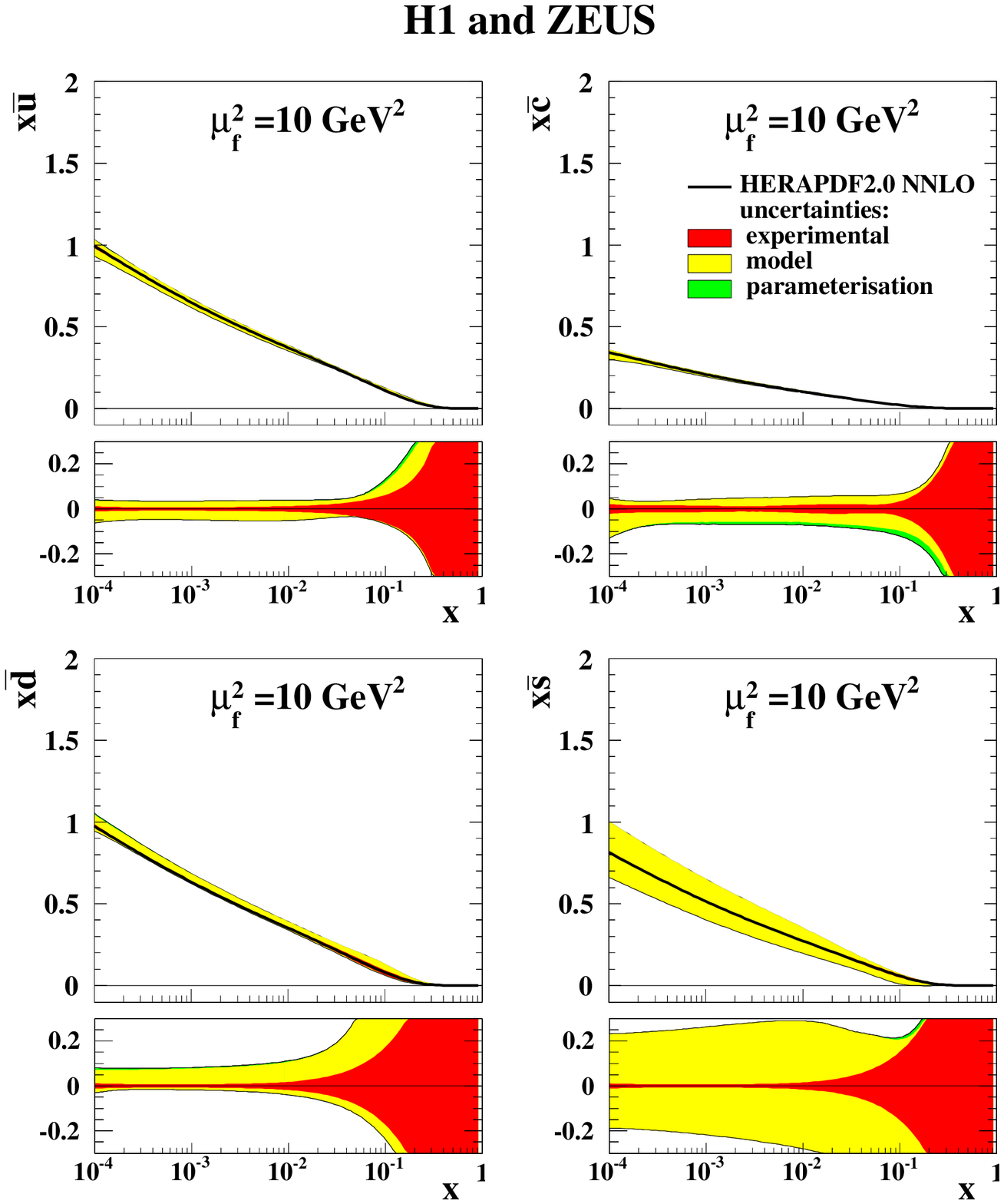 ,width=0.9\textwidth}}
\vspace{0.5cm}
\caption {The flavour breakdown of the sea distribution of HERAPDF2.0 NNLO
          at $\mu_{\rm f}^{2}$ = 10\,GeV$^{2}$.
          Shown are the distributions 
          $x\bar{u}$, $x\bar{d}$, $x\bar{c}$ and $x\bar{s}$ 
          together with their experimental, 
          model and parameterisation uncertainties.
          The fractional uncertainties are also shown.
}
\label{fig:flavour2}
\end{figure}

The flavour breakdown of the sea
into $x\bar{u}$, $x\bar{d}$, $x\bar{c}$ and $x\bar{s}$
for HERAPDF2.0 NNLO is shown 
in Fig.~\ref{fig:flavour2}.
The uncertainties are also shown as
fractional uncertainties. 
They are dominated by model uncertainties 
and derive from the same sources as already described at NLO. 
The parameterisation uncertainties are less 
important at NNLO than at NLO.

\begin{figure}[tbp]
\vspace{-0.5cm} 
\centerline{
\epsfig{file=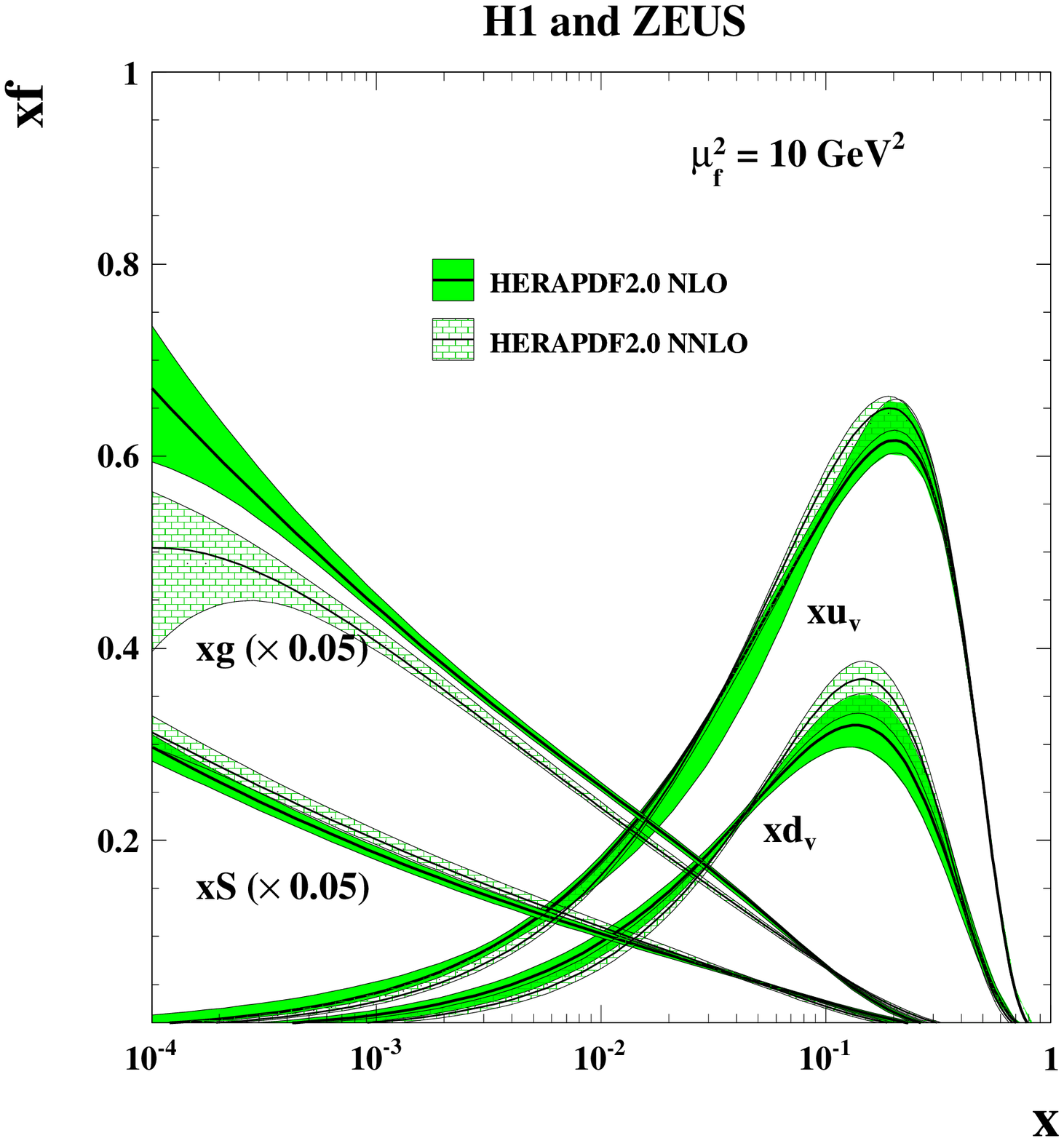,width=0.65\textwidth}}
\centerline{
\epsfig{file=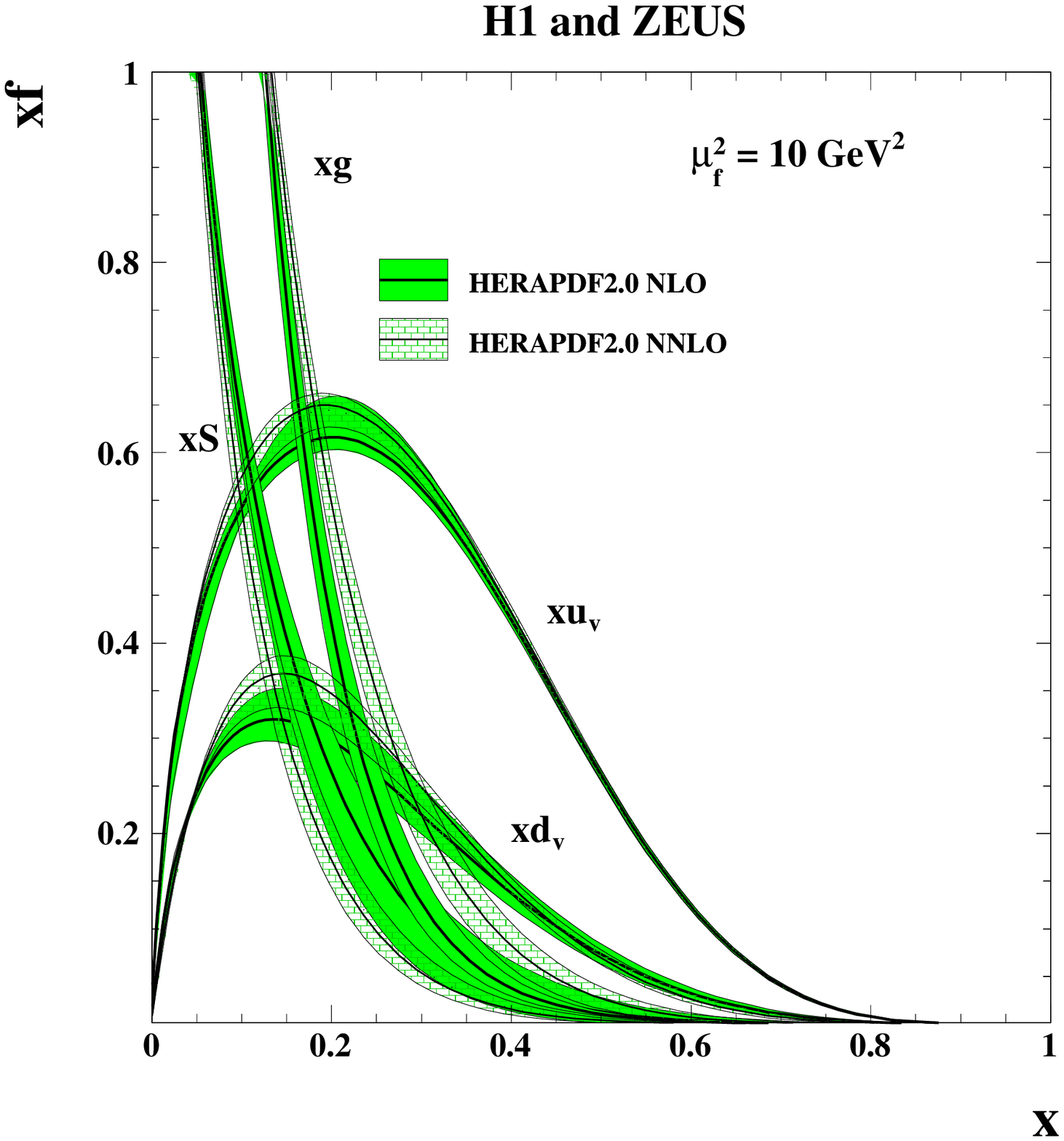,width=0.65\textwidth}}
\caption { 
The parton distribution functions 
$xu_v$, $xd_v$, $xS=2x(\bar{U}+\bar{D})$ and $xg$ of 
HERAPDF2.0 NLO
at $\mu_{\rm f}^{2} = 10\,$GeV$^{2}$
compared to those of HERAPDF2.0 NNLO on logarithmic (top)
and linear (bottom) scales. 
The bands represent the total uncertainties.
}
\label{fig:nlovsnnlo}
\end{figure}

A comparison between 
HERAPDF2.0  NNLO and NLO 
is shown in Fig.~\ref{fig:nlovsnnlo}
with total uncertainties, using both linear and logarithmic $x$ scales.
The main difference is the different shapes of the gluon 
distributions as expected from the 
differing evolution at NLO and NNLO.

\begin{figure}[tbp]
\vspace{-0.5cm} 
\centerline{
\epsfig{file=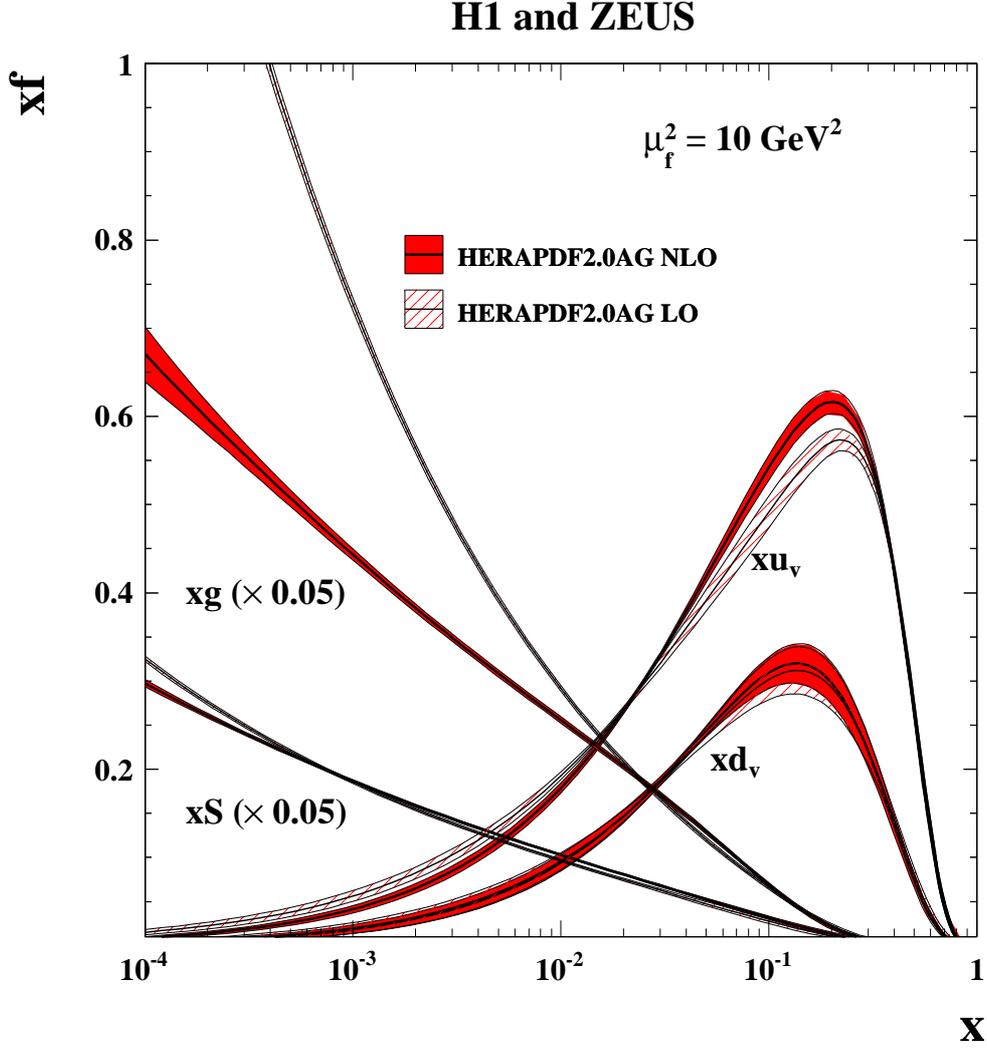 ,width=0.9\textwidth}}
\vspace*{-0.6cm}
\caption { 
The parton distribution functions 
$xu_v$, $xd_v$, $xS=2x(\bar{U}+\bar{D})$ and $xg$ of 
HERAPDF2.0AG LO
at $\mu_{\rm f}^{2} = 10\,$GeV$^{2}$ 
compared to those of HERAPDF2.0AG NLO.
The bands represent the experimental uncertainties only.
}
\label{fig:lovsnlo}
\end{figure}

At leading order, HERAPDF2.0 is only available as HERAPDF2.0AG LO
with the alternative gluon
parameterisation. It has thus to be compared to HERAPDF2.0AG NLO.
HERAPDF2.0AG LO was determined 
with experimental uncertainties only, because 
its main purpose is to be used 
in LO Monte Carlo programmes. 
A comparison between the 
distributions of HERAPDF2.0AG LO and 
HERAPDF2.0AG NLO is shown in Fig.~\ref{fig:lovsnlo}.  
The gluon distribution at LO rises much faster 
than at NLO, as expected from the different evolution. 
The $xu_v$ distribution
is softer at LO than at NLO.

\subsubsection{Comparisons to inclusive HERA data}
\label{sec:comp:fit:data} 

The data with the proton beam energy 
of 920\,GeV ($\sqrt{s}=318\,$GeV) 
are the most precise data
due to the large integrated luminosity, see Table~\ref{tab:data}.
HERAPDF2.0 predictions are compared at NNLO, NLO and LO
to these high-precision data. 

The predictions of HERAPDF2.0 NNLO, NLO and AG LO
are compared to
the high-$Q^2$ NC $e^+p$ data in
Figs.~\ref{fig:nnloQ23pt5ncepc},~\ref{fig:nloQ23pt5ncepc} 
and~\ref{fig:loQ23pt5ncepc}. 
The data are well described by the predictions at all orders.
Figure~\ref{fig:5mod} shows the cross sections
already shown in Fig.~\ref{fig:Hera1:NCepp} together with the 
predictions of HERAPDF2.0 NNLO and NLO.
The predictions at NNLO and NLO are very similar.

The predictions of HERAPDF2.0 NNLO, NLO and AG LO
are compared to the NC $e^-p$ data
in Figs.~\ref{fig:nnloQ23pt5ncem},~\ref{fig:nloQ23pt5ncem} 
and~\ref{fig:loQ23pt5ncem}. 
The description of the $e^-p$ data is as good as for the $e^+p$ data.

For $e^+p$ scattering, data at low  $Q^2$ are available.
Figures~\ref{fig:nnloQ23pt5ncepb}, \ref{fig:nloQ23pt5ncepb}, 
and~\ref{fig:loQ23pt5ncepb}
show comparisons between the predictions of 
HERAPDF2.0 NNLO, NLO and AG LO  
and these low-$Q^2$ data. 
The description of the data is generally good and for the
predictions at NNLO and NLO, it remains so
even for $Q^2$ below the fitted kinematic region.
However, at low $x_{\rm Bj}$ and low $Q^2$, the turnover 
in the cross sections related to $F_{\rm L}$ is not well described,
and HERAPDF2.0 NNLO does not describe these data 
better than HERAPDF2.0 NLO.  
The HERAPDF2.0AG\,LO predictions show a clear turnover, but the prediction
is significantly too high at all $x_{\rm Bj}$ for the lowest $Q^2$.

The predictions of the NNLO and NLO fits 
are compared to the  CC $e^+p$ data with $\sqrt{s} = 318\,$GeV
in Figs.~\ref{fig:nnloQ23pt5ccep} and~\ref{fig:nloQ23pt5ccep} and to
CC $e^-p$ data
in Figs.~\ref{fig:nnloQ23pt5ccem} and~\ref{fig:nloQ23pt5ccem}. 
The precise predictions describe the CC cross sections well.
The CC data are in general less precise than the NC data.

The predictions of HERAPDF2.0 NLO
compared to low-$Q^2$ and high-$Q^2$  NC $e^+p$ data for 
$\sqrt{s} = 300\,$GeV are shown in
Figs.~\ref{fig:nloQ23pt5ncepb820}
and~\ref{fig:nloQ23pt5ncepc820}.
Equivalent comparisons for 
$\sqrt{s} = 251\,$GeV and $\sqrt{s} = 225\,$GeV
are shown in
Figs.~\ref{fig:nloQ23pt5ncepb575} and~\ref{fig:nloQ23pt5ncepc575},
and 
Figs.~\ref{fig:nloQ23pt5ncepb460} and~\ref{fig:nloQ23pt5ncepc460},
respectively.
The data with reduced proton beam energy are also
reasonably well described. 

\subsubsection{Comparisons to HERAPDF1.0 and 1.5}

\begin{figure}[tbp]
\vspace{-0.5cm} 
\centerline{
\epsfig{file=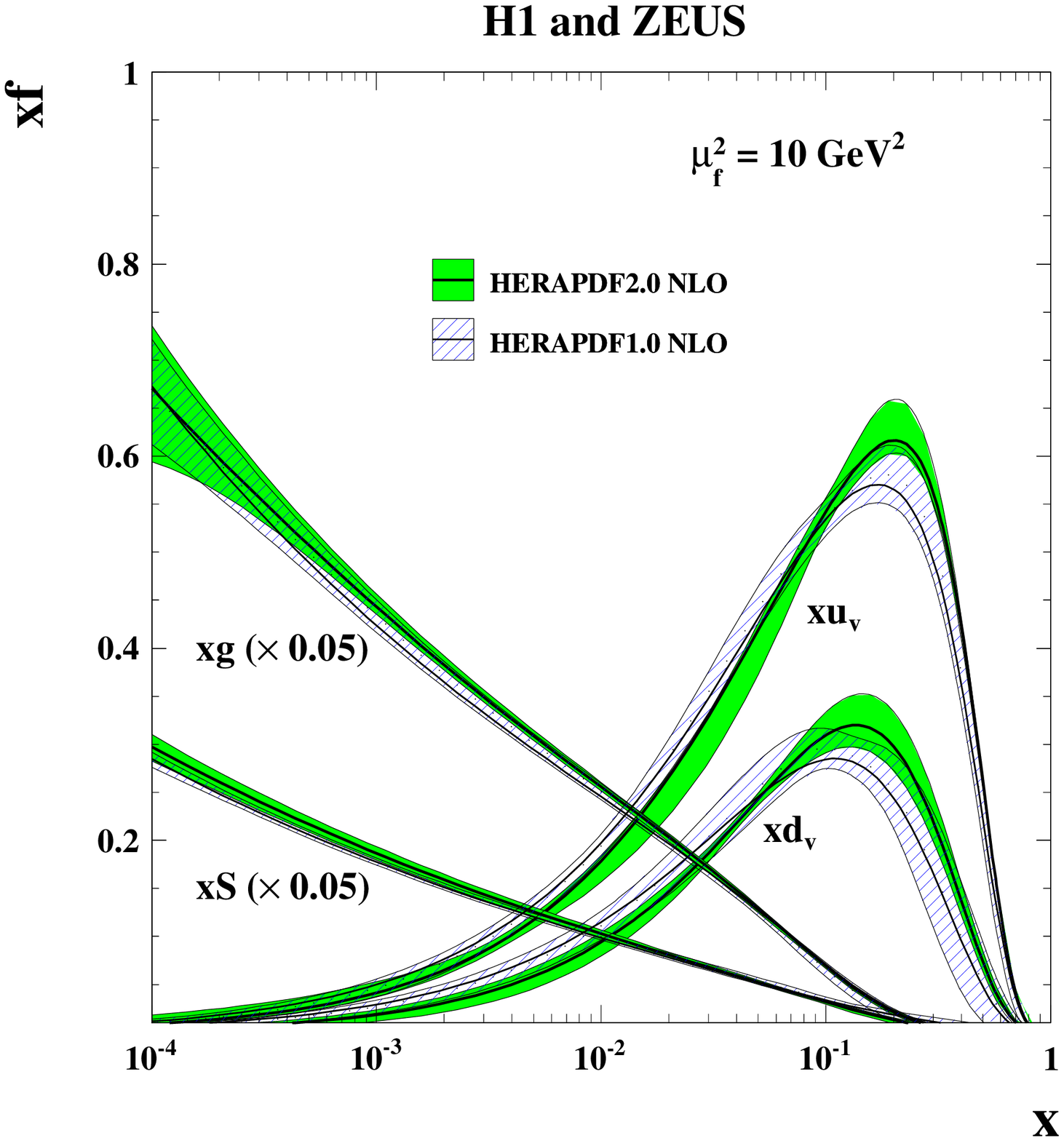,width=0.65\textwidth}}
\centerline{
\epsfig{file=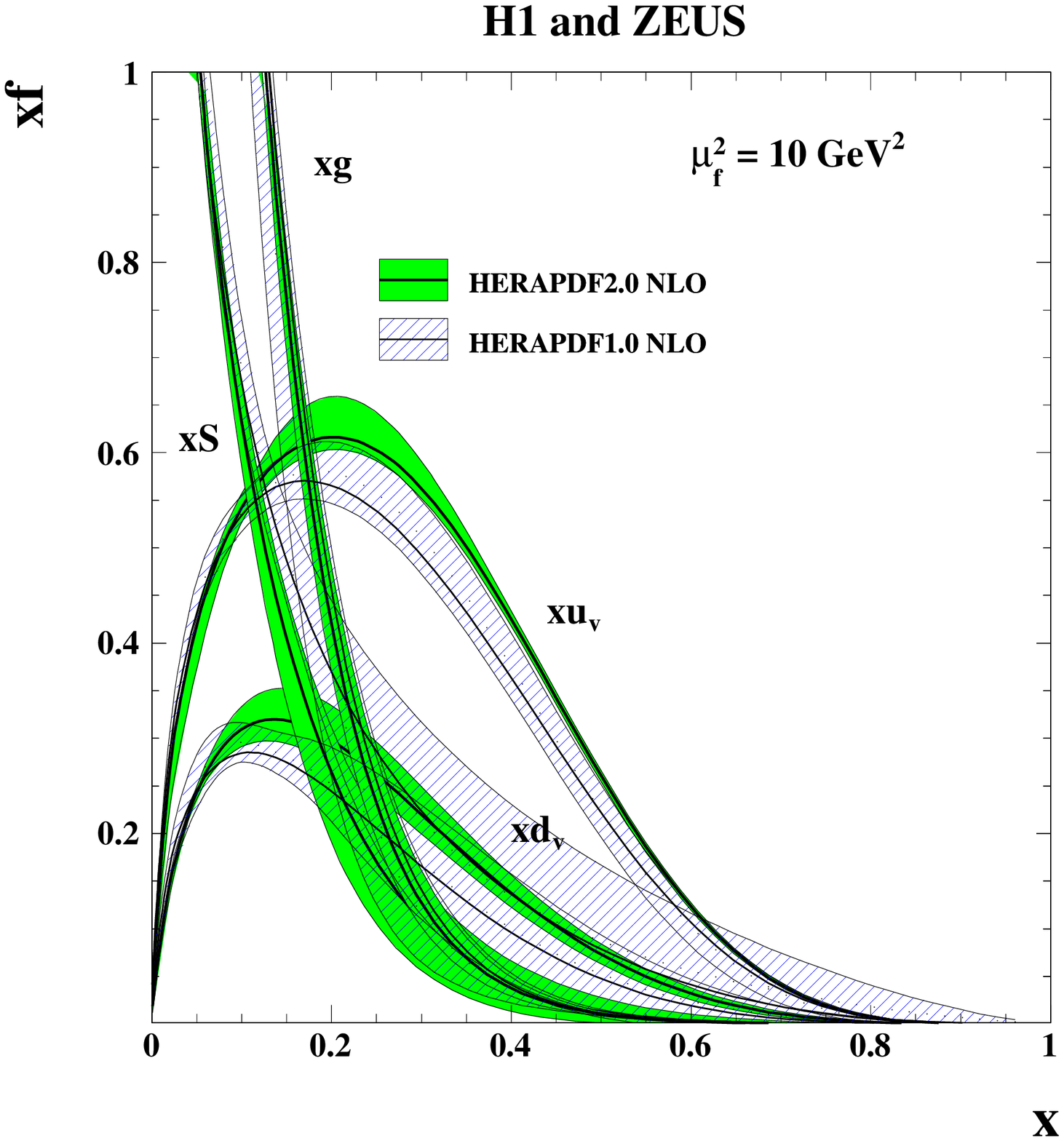,width=0.65\textwidth}}
\caption { 
The parton distribution functions 
$xu_v$, $xd_v$, $xS=2x(\bar{U}+\bar{D})$ and $xg$ of
HERAPDF2.0 NLO 
at $\mu_{\rm f}^{2}=10\,$GeV$^{2}$ 
compared to those of HERAPDF1.0 on logarithmic (top)
and linear (bottom) scales. 
The bands represent the total uncertainties.
}
\label{fig:vsherapdf10}
\end{figure}

\begin{figure}[tbp]
\vspace{-0.5cm} 
\centerline{
\epsfig{file=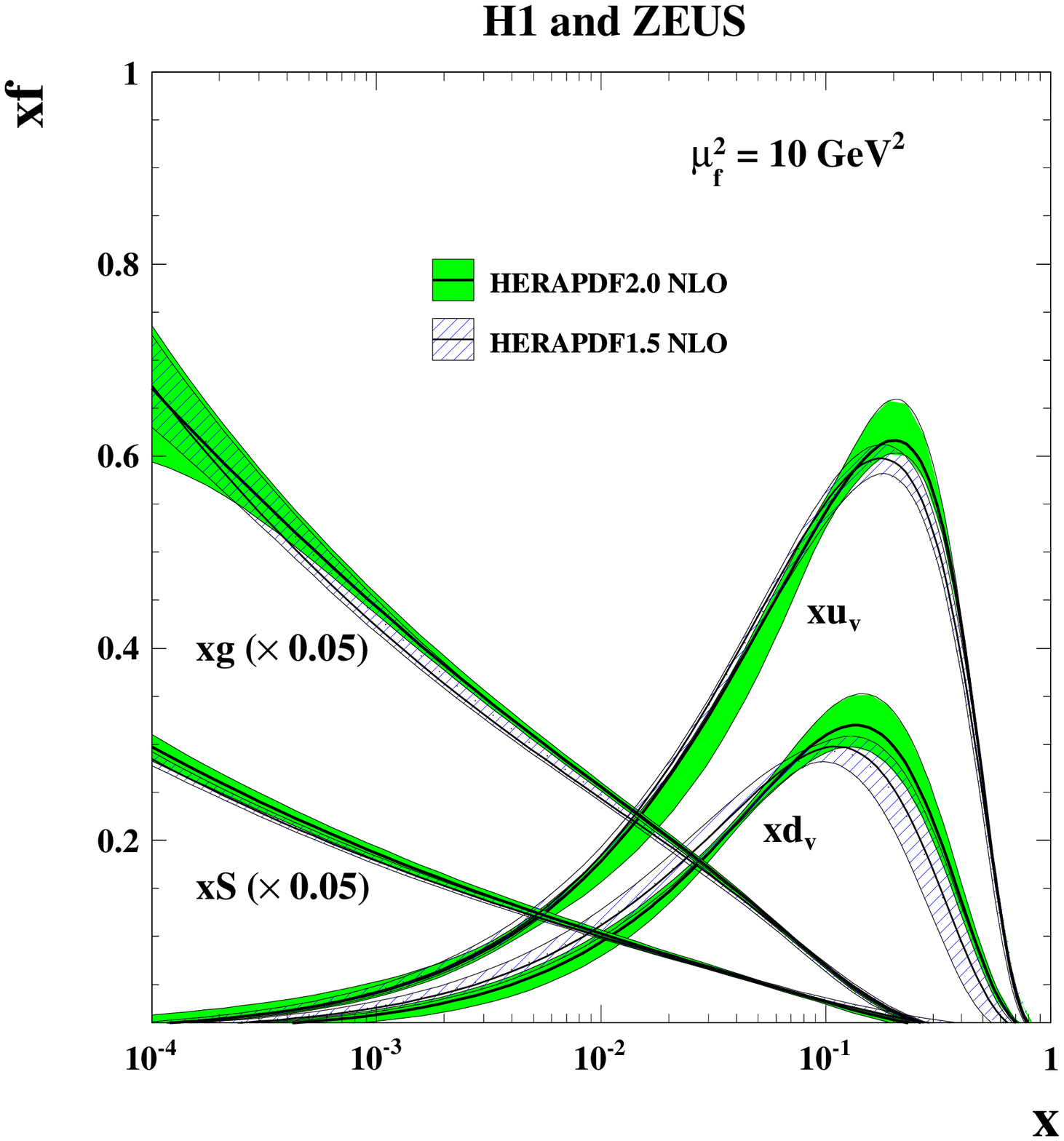,width=0.65\textwidth}}
\centerline{
\epsfig{file=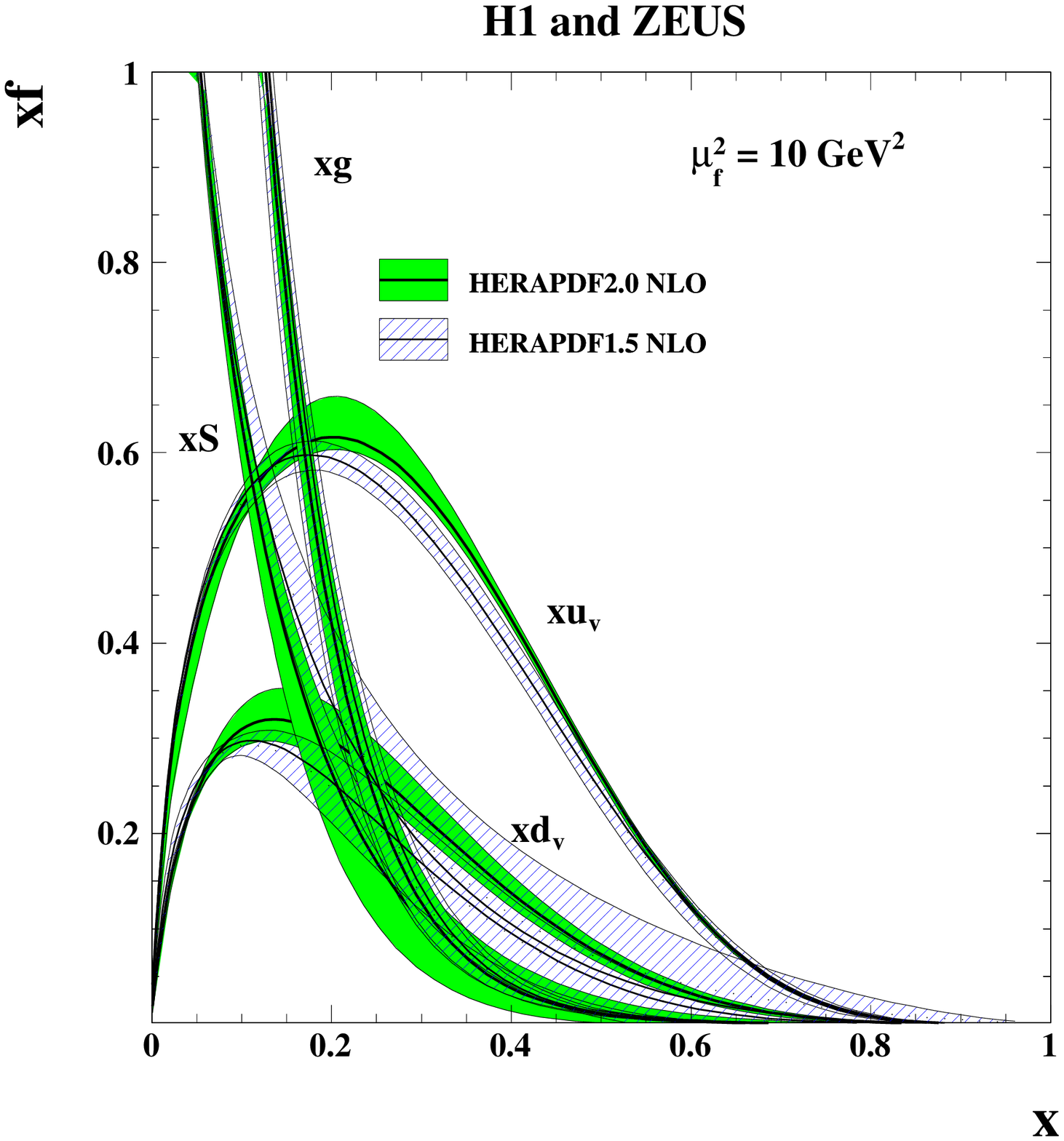,width=0.65\textwidth}}
\caption { 
The parton distribution functions 
$xu_v$, $xd_v$, $xS=2x(\bar{U}+\bar{D})$ and $xg$ of
HERAPDF2.0 NLO
at $\mu_{\rm f}^{2} = 10\,$GeV$^{2}$
compared to those of HERAPDF1.5 on logarithmic (top)
and linear (bottom) scales. 
The bands represent the total uncertainties.
}
\label{fig:vsherapdf15}
\end{figure}

Comparisons of HERAPDF2.0 NLO to
HERAPDF1.0 NLO and HERAPDF1.5 NLO
are shown in
Figs.~\ref{fig:vsherapdf10} and~\ref{fig:vsherapdf15}, 
respectively.  
Whereas HERAPDF1.5 already had a somewhat smaller uncertainty
than HERAPDF1.0, the use of all HERA\,II data for HERAPDF2.0 has led to
a much larger reduction of the uncertainties on all PDFs. 
The shapes of 
the PDFs have also changed somewhat. 
The shape of the valence distributions have become a little harder.
This was caused by the additional  data with high $x_{\rm Bj}$ which
were not yet available for HERAPDF1.5. 
The  HERAPDF2.0 high-$x$ 
gluon distribution is also slightly harder than HERAPDF1.5 while the
sea distribution of HERAPDF2.0 at high $x$ is considerably softer.

\begin{figure}[tbp]
\vspace{-0.5cm} 
\centerline{
\epsfig{file=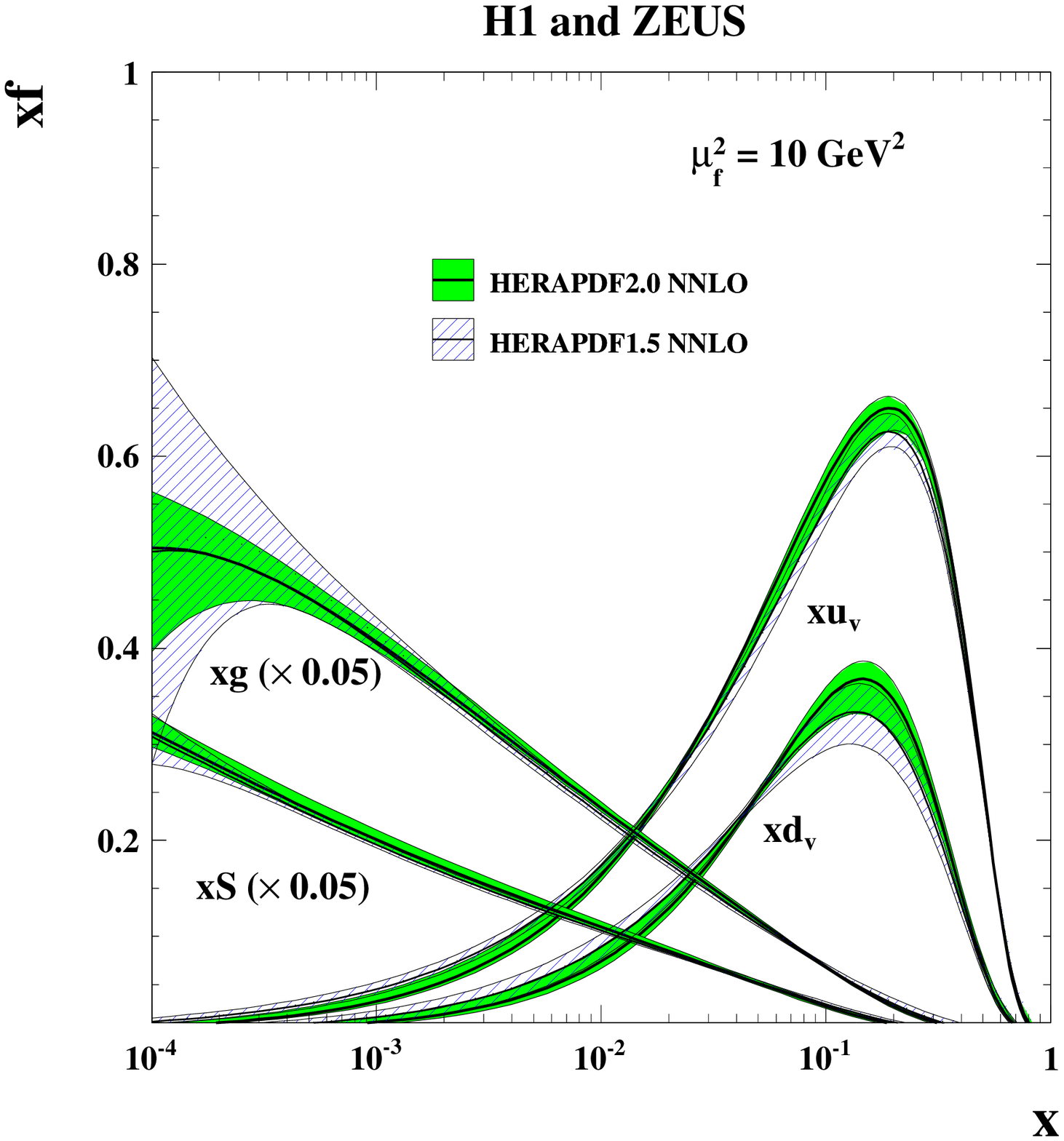,width=0.65\textwidth}}
\centerline{
\epsfig{file=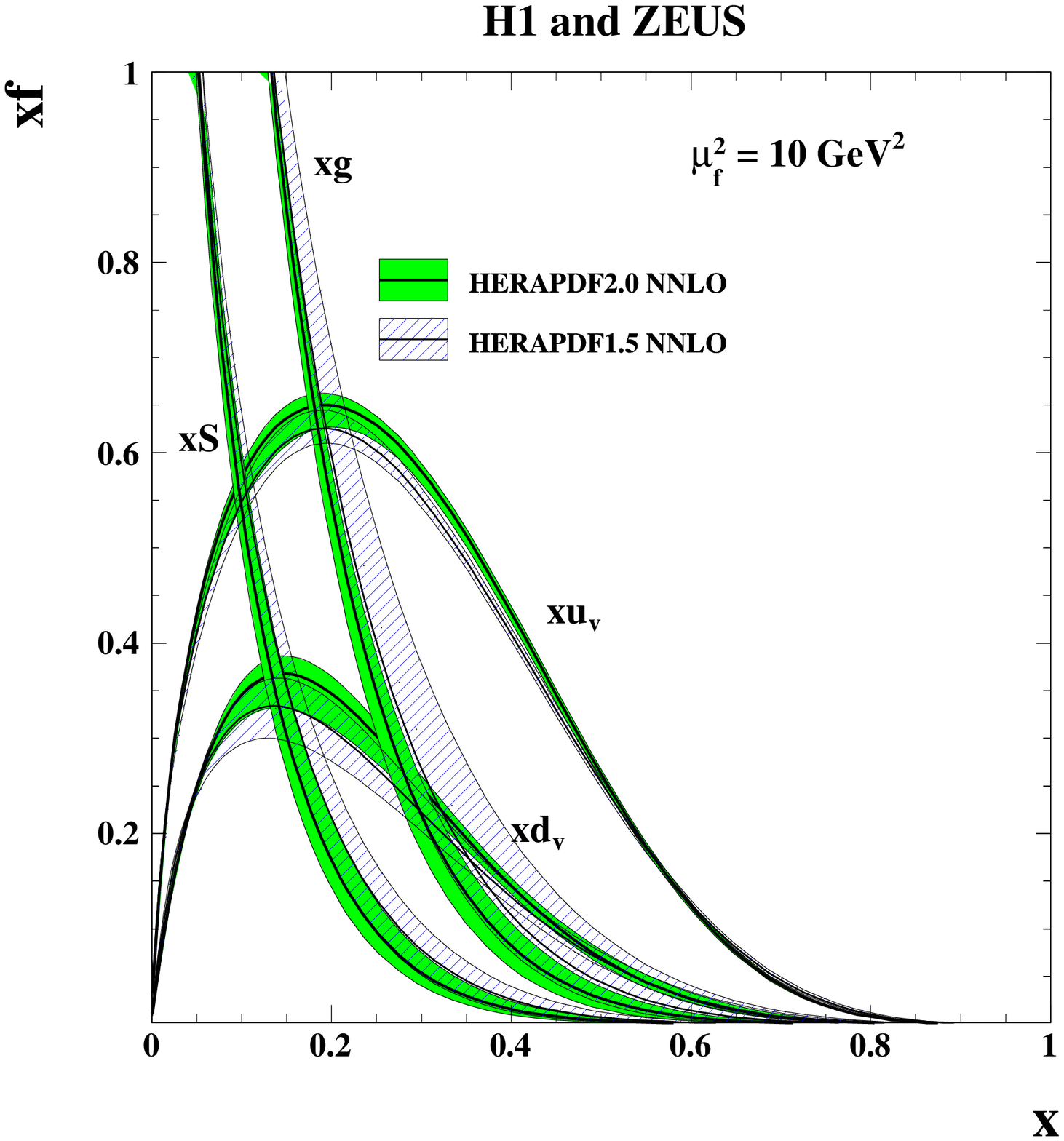,width=0.65\textwidth}}
\caption { 
The parton distribution functions 
$xu_v$, $xd_v$, $xS=2x(\bar{U}+\bar{D})$ and $xg$ of
HERAPDF2.0 NNLO
at $\mu_{\rm f}^{2} = 10\,$GeV$^{2}$ 
compared to the ones of  HERAPDF1.5 on logarithmic (top)
and linear (bottom) scales. 
The bands represent the total uncertainties.
}
\label{fig:vsherapdf15nnlo}
\end{figure}

A comparison between HERAPDF2.0 NNLO
and HERAPDF1.5 NNLO is provided 
in Fig.~\ref{fig:vsherapdf15nnlo}. 
As in the case of the NLO PDFs, a reduction of the uncertainty 
at high $x$ has been achieved by
including further high-$x_{\rm Bj}$ data.
There is also a reduction of uncertainties at low $x$.
This is mostly due to the 
better stability of the fit under the variation of 
$Q^2_{\rm min}$, which is part of the model uncertainties.
The shapes of the HERAPDF1.5  and HERAPDF2.0 at NNLO are rather similar, 
but the gluon distribution at high $x$  has moved 
to the lower end of its previous uncertainty band.

\subsubsection{Comparisons to other sets of PDFs}
\begin{figure}[tbp]
\vspace{-0.5cm} 
\centerline{
\epsfig{file=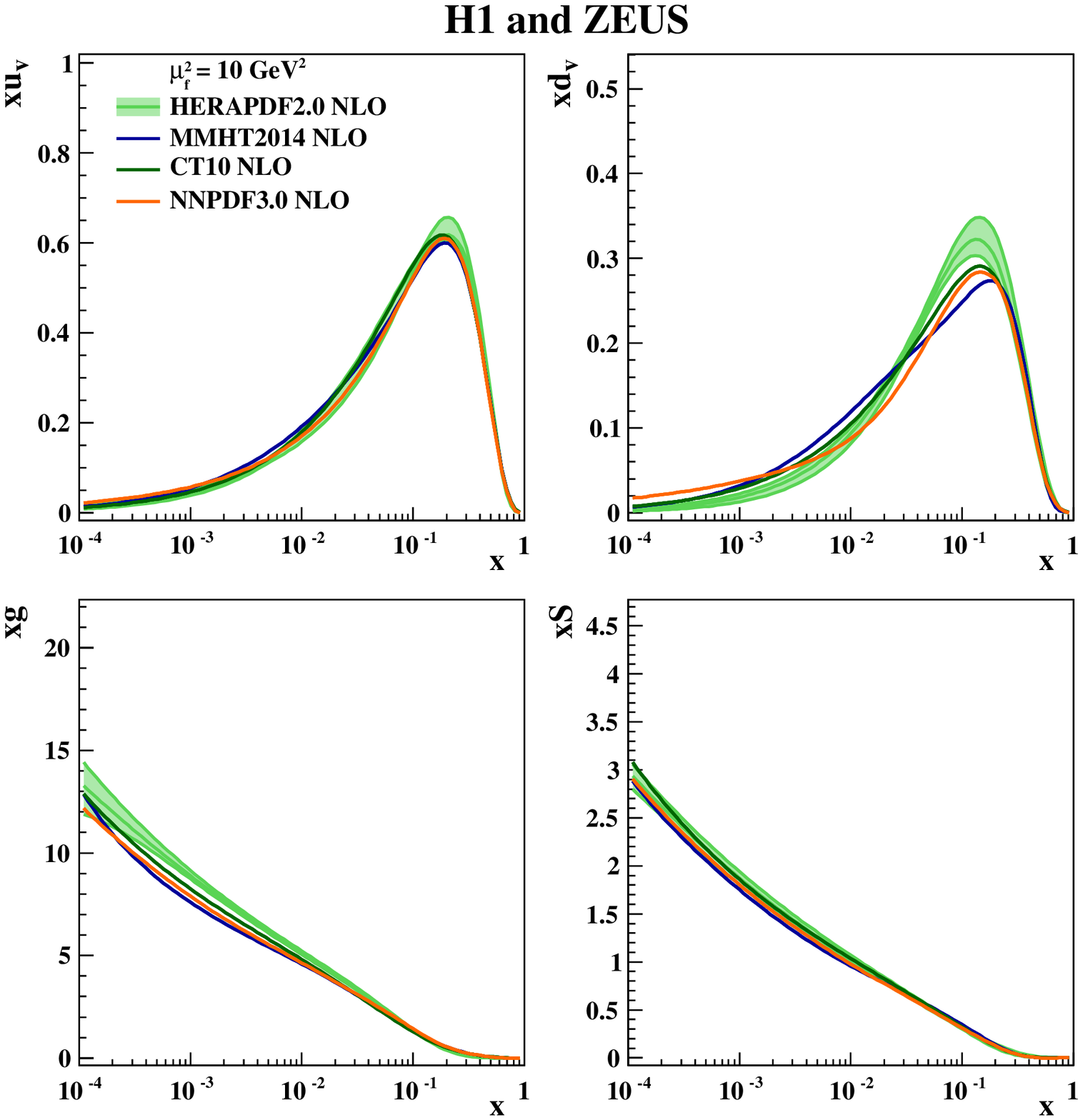,width=0.65\textwidth}}
\centerline{
\epsfig{file=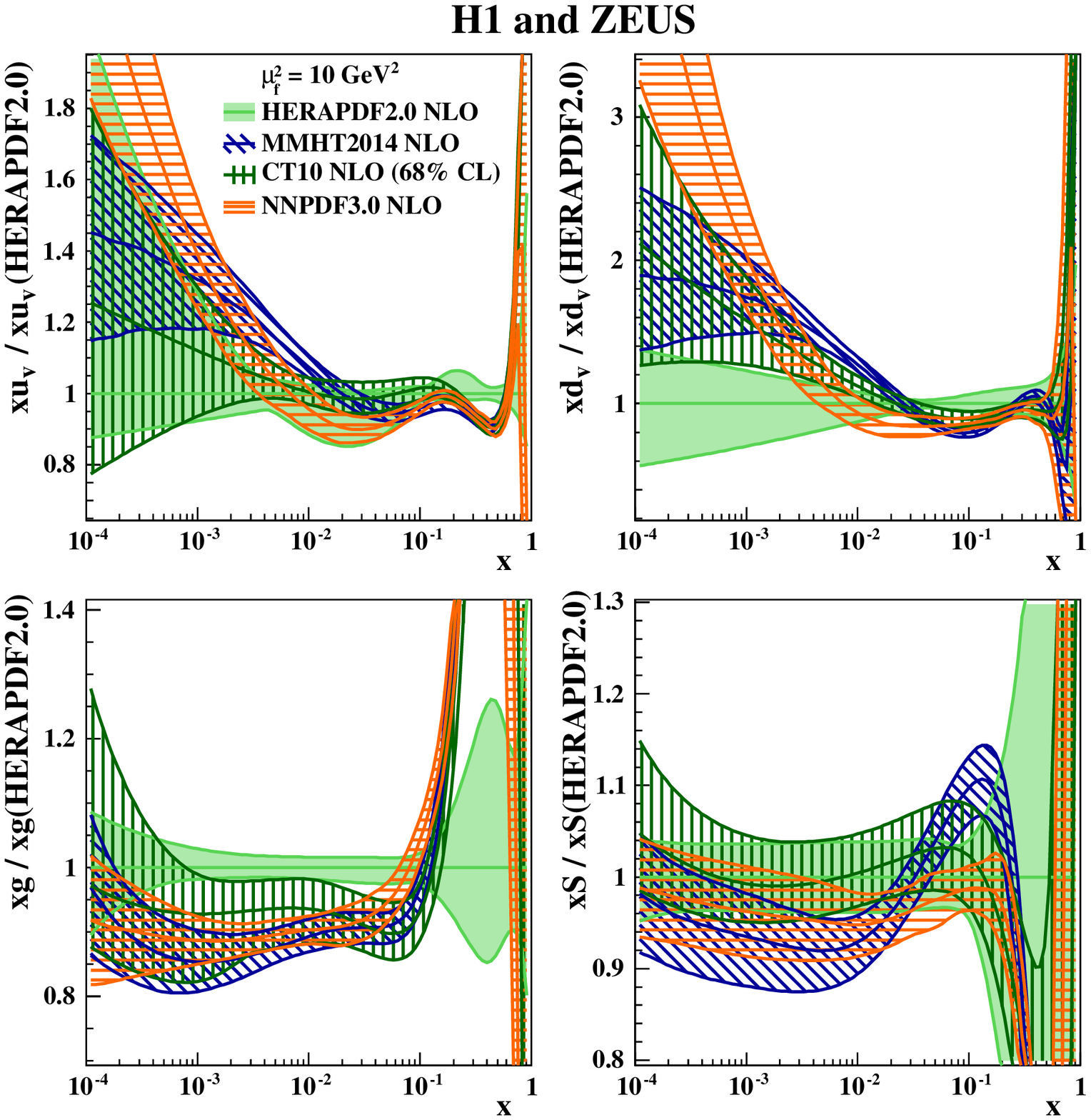,width=0.65\textwidth}}
\caption { 
The parton distribution functions 
$xu_v$, $xd_v$, $xg$ and $xS=2x(\bar{U}+\bar{D})$ of
HERAPDF2.0 NLO
at $\mu_{\rm f}^{2} = 10\,$GeV$^{2}$
compared to those  of MMHT2014~\cite{MMHT2014}, CT10~\cite{CT10NLO} and 
NNPDF3.0~\cite{NNPDF3.0}. 
The top panel
shows the distribution with uncertainties only for HERAPDF2.0. 
The bottom panel shows the PDFs normalised to HERAPDF2.0 and
with uncertainties for all PDFs.
}
\label{fig:20NLO-others}
\end{figure}

\begin{figure}[tbp]
\vspace{-0.5cm} 
\centerline{
\epsfig{file=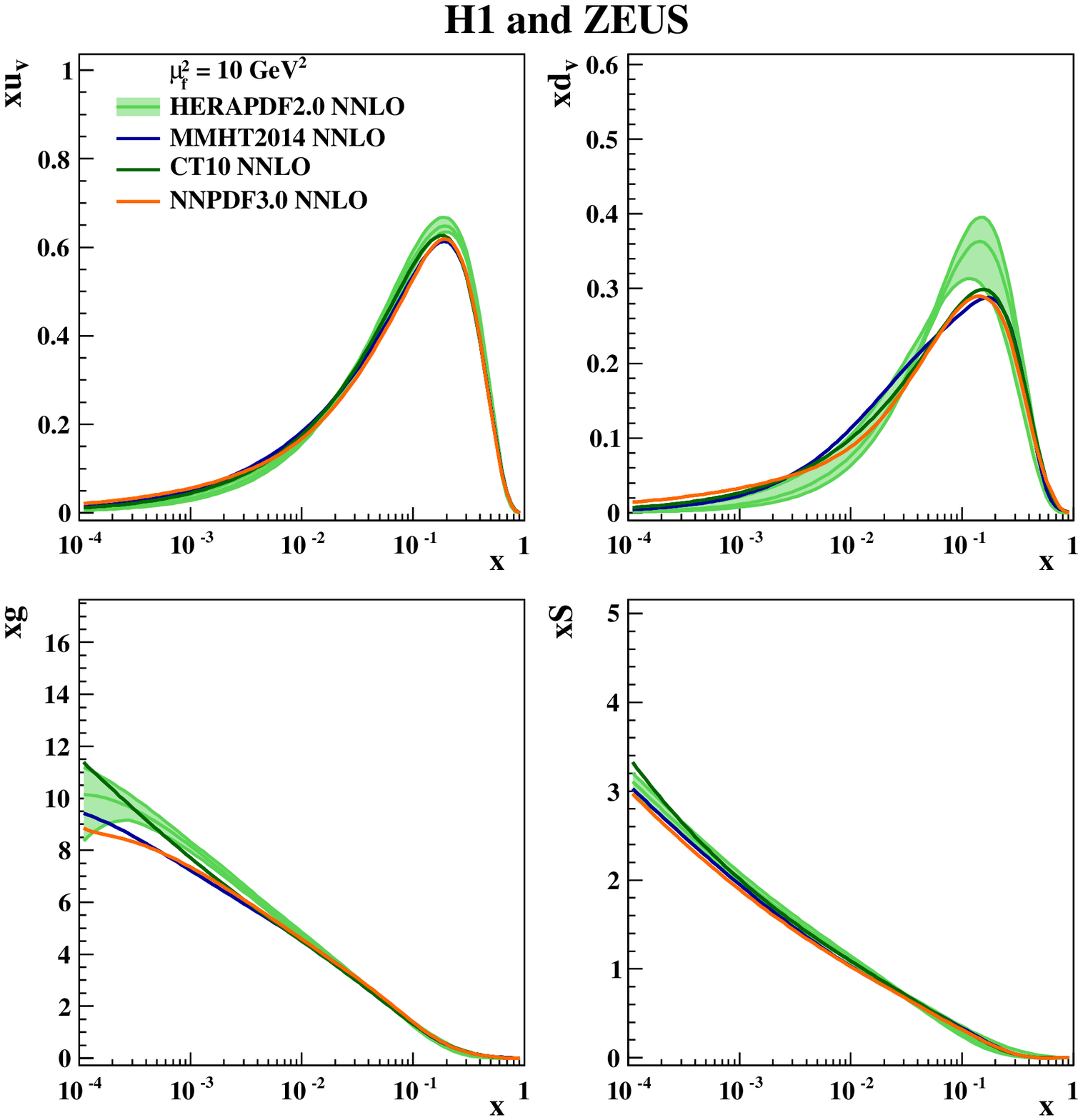,width=0.65\textwidth}}
\vspace*{0.6cm}
\centerline{
\epsfig{file=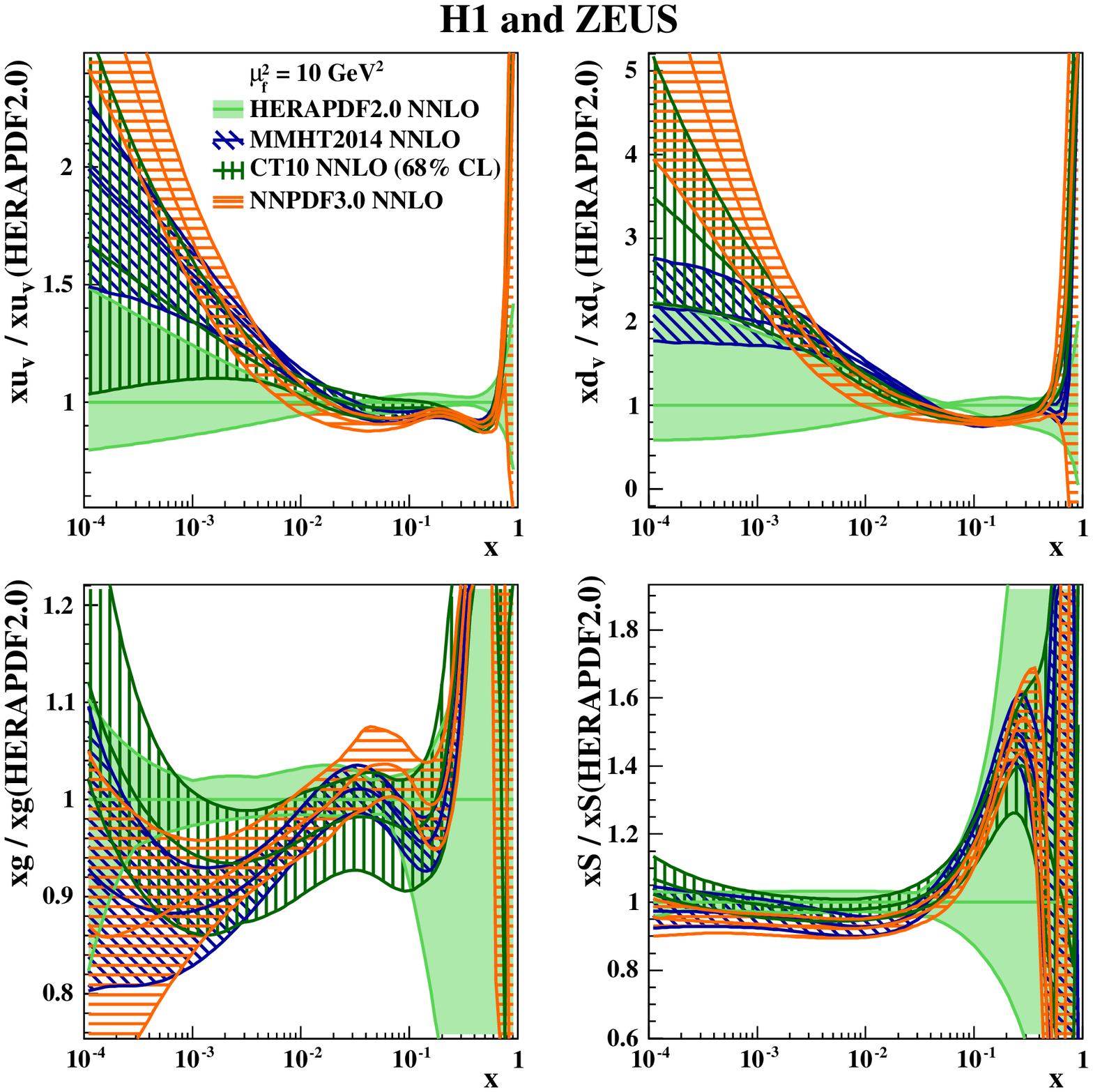,width=0.65\textwidth}}
\caption { 
The parton distribution functions 
$xu_v$, $xd_v$, $xg$ and $xS=2x(\bar{U}+\bar{D})$ of
HERAPDF2.0 NNLO
at $\mu_{\rm f}^{2} = 10\,$GeV$^{2}$
compared to those of MMHT2014~\cite{MMHT2014}, CT10~\cite{CT10NNLO} and 
NNPDF3.0~\cite{NNPDF3.0}. 
The top panel
shows the distribution with uncertainties only for HERAPDF2.0. 
The bottom panel shows the PDFs normalised to HERAPDF2.0 and
with uncertainties for all PDFs.
}
\label{fig:20NNLO-others}
\end{figure}

The PDFs of HERAPDF2.0 NLO and NNLO  can be directly compared to
the PDFs of 
MMHT 2014~\cite{MMHT2014}, for which the same heavy-flavour scheme,
i.e. RTOPT, was used.
Comparisons are also made to the PDFs of 
CT10~\cite{CT10NLO,CT10NNLO}, for which a heavy-flavour-scheme based on 
ACOT was used,  
and NNPDF3.0~\cite{NNPDF3.0}, for which the FONLL scheme was used.
The results are shown in 
Figs.~\ref{fig:20NLO-others} and~\ref{fig:20NNLO-others} for NLO and NNLO, respectively.
For the PDFs themselves, the uncertainties are only
shown for HERAPDF2.0.
All uncertainties  are shown when
the ratios of the other PDFs with respect to HERAPDF2.0 are illustrated.
Taking the full uncertainties into account, all PDFs are compatible. 
The largest relative discrepancy ($\approx 2.5\sigma$) 
is found in the shape of the $xu_v$ distribution at $x\approx0.4$ 
for both NLO and NNLO PDFs.
In addition, at NLO, the gluon distribution of HERAPDF2.0 at high $x$ is 
softer than that of the other PDFs, 
whereas at NNLO it is close to their $68\%$ uncertainty bands.

\subsection{HERAPDF2.0HiQ2}

\begin{figure}[tbp]
\vspace{-0.5cm} 
\centerline{
\epsfig{file=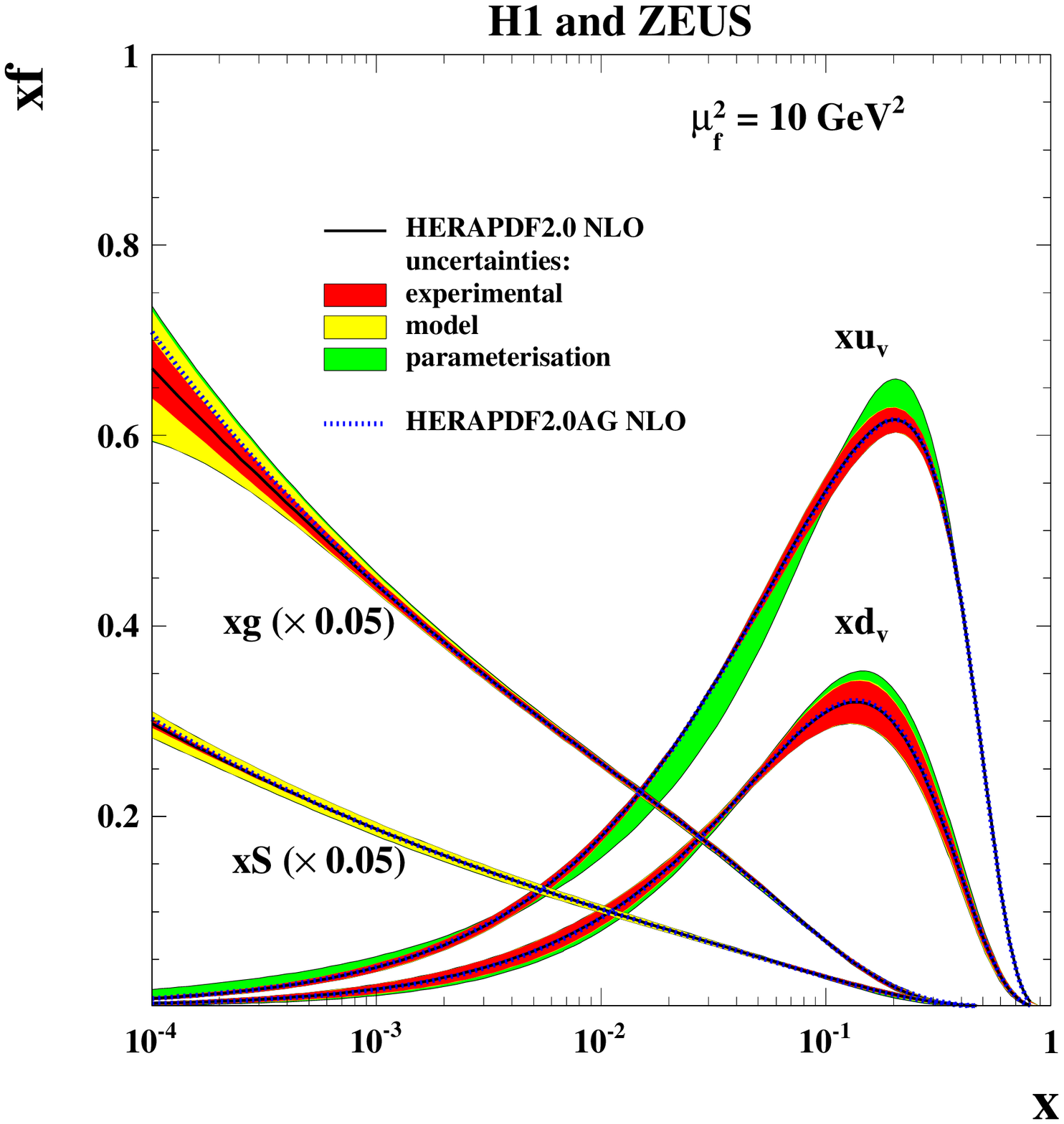,width=0.65\textwidth}}
\centerline{
\epsfig{file=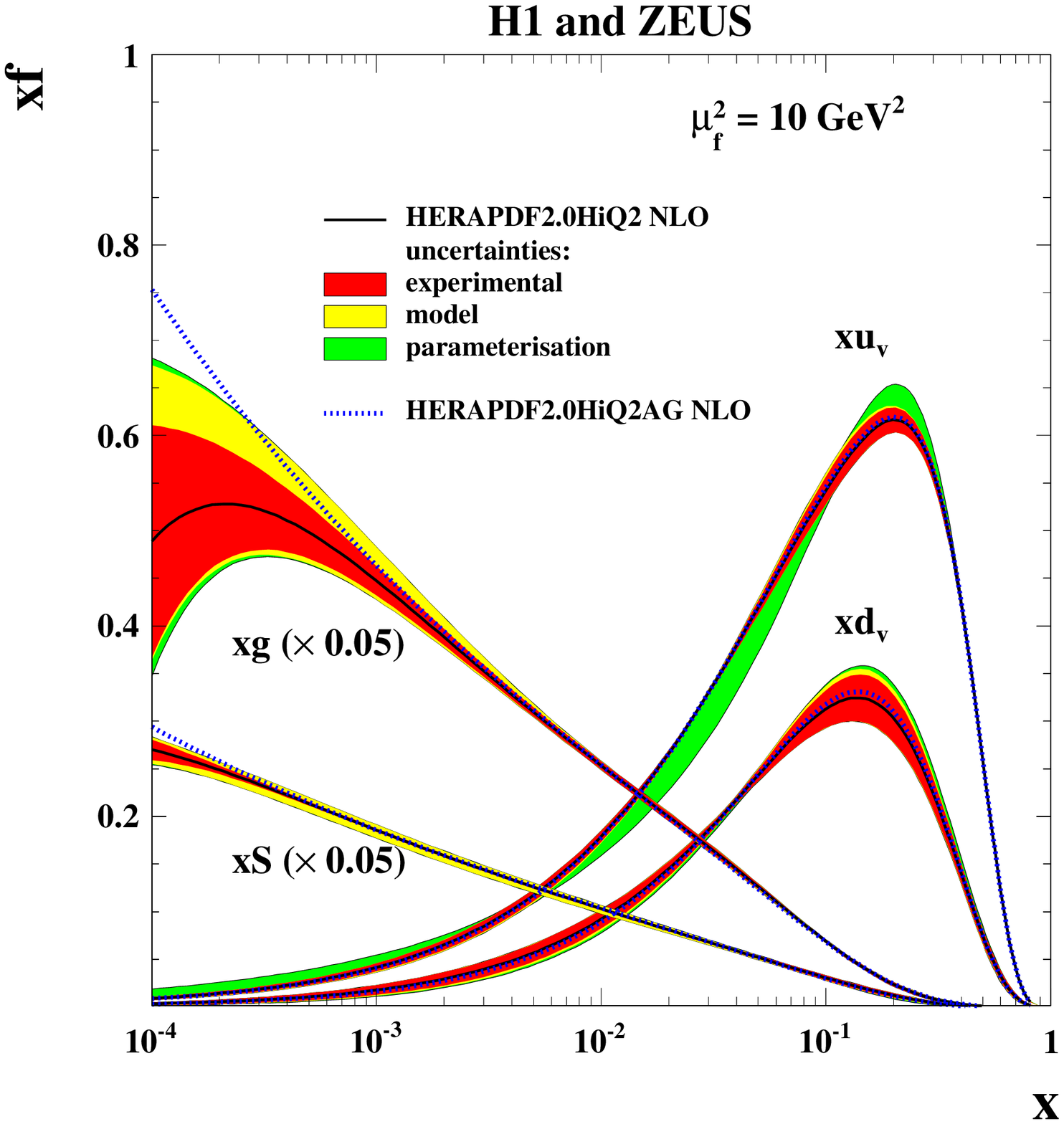,width=0.65\textwidth}}
\caption { 
The parton distribution functions 
$xu_v$, $xd_v$, $xS=2x(\bar{U}+\bar{D})$ and $xg$ of 
HERAPDF2.0 NLO
at $\mu_{\rm f}^{2} = 10\,$GeV$^{2}$ 
with $Q^{2}_{\rm min} = 3.5$\,GeV$^{2}$ (top) 
and of HERAPDF2.0HiQ2 NLO with $Q^2_{\rm min} = 10$\,GeV$^2$ (bottom).
The gluon and sea distributions are scaled 
down by a factor of $20$.
The experimental, model and parameterisation 
uncertainties are shown. 
The dotted lines represent HERAPDF2.0AG NLO and HERAPDF2.0AG HiQ2 NLO.
}
\label{fig:hiQ2nlo}
\end{figure}

\begin{figure}[tbp]
\vspace{-0.5cm} 
\centerline{
\epsfig{file=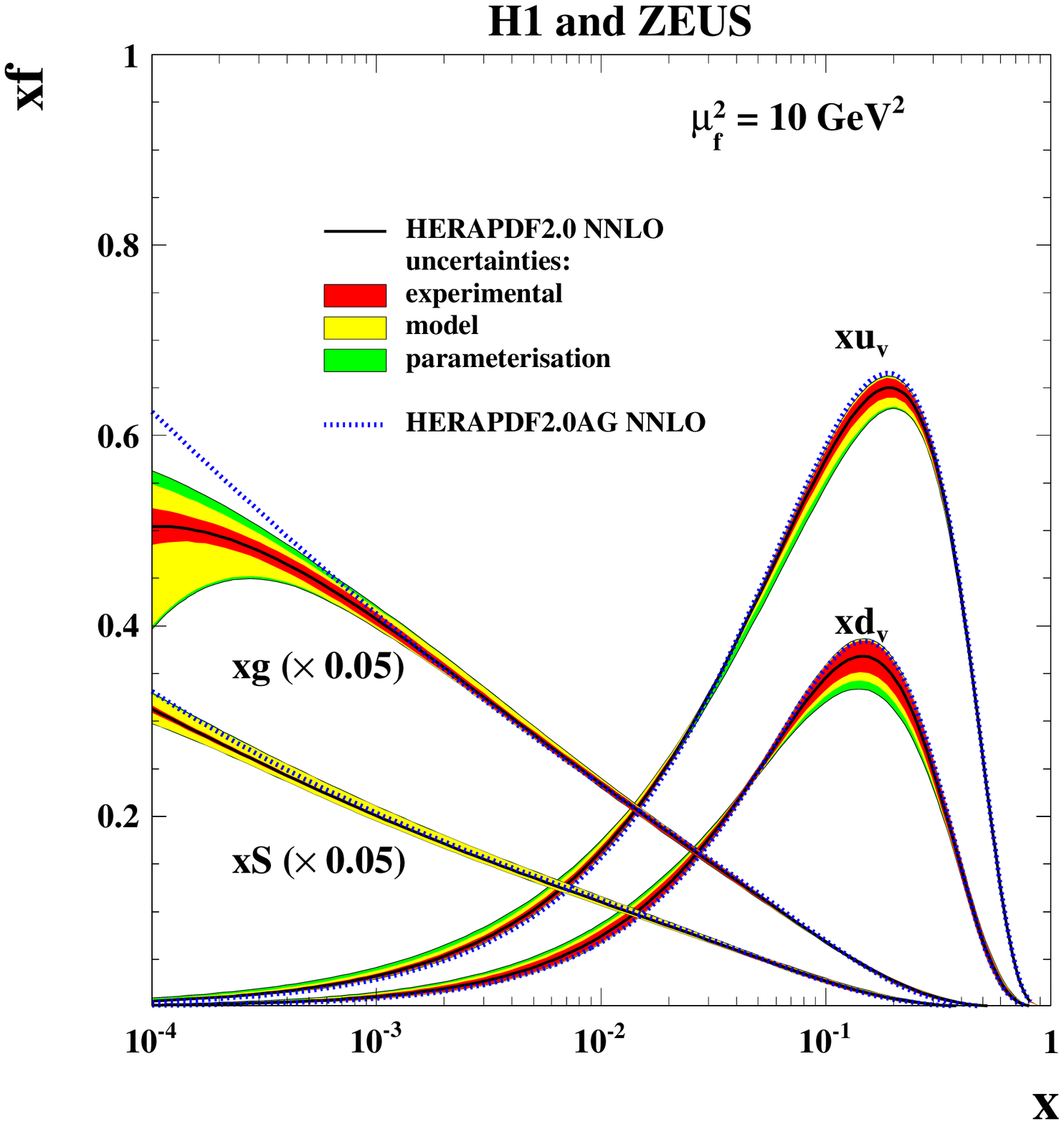,width=0.65\textwidth}}
\centerline{
\epsfig{file=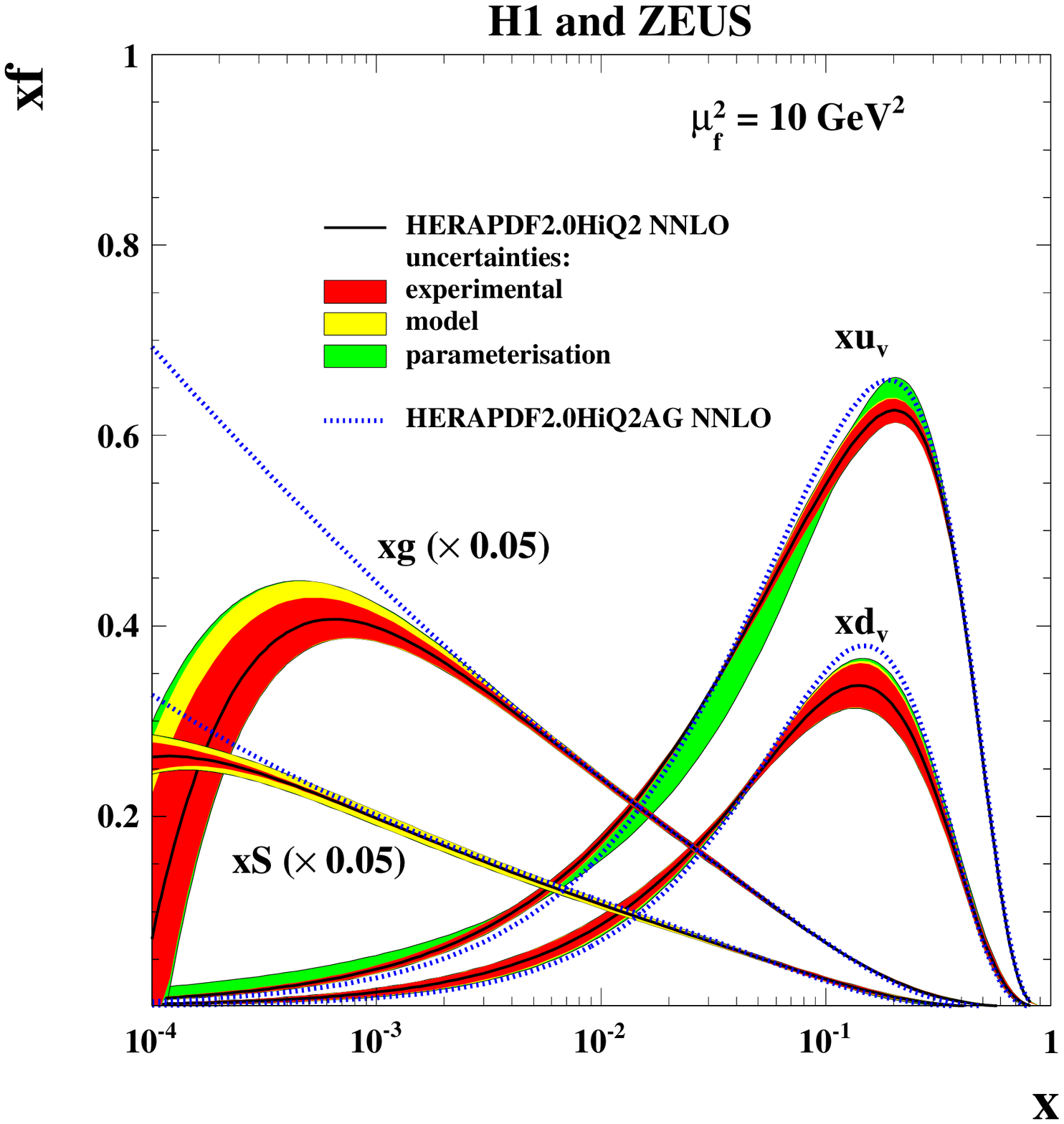,width=0.65\textwidth}}
\caption { 
The parton distribution functions 
$xu_v$, $xd_v$, $xS=2x(\bar{U}+\bar{D})$ and $xg$ of  
HERAPDF2.0 NNLO
at $\mu_{\rm f}^{2} = 10\,$GeV$^{2}$ 
with $Q^{2}_{\rm min} = 3.5$\,GeV$^{2}$ (top) 
and of HERAPDF2.0HiQ2 NNLO with $Q^2_{\rm min} = 10$\,GeV$^2$ (bottom).
The gluon and sea distributions are scaled down by a factor $20$.
The experimental, model and parameterisation 
uncertainties are shown. 
The dotted lines represent HERAPDF2.0AG NNLO and HERAPDF2.0AG HiQ2 NNLO.
}
\label{fig:hiQ2nnlo}
\end{figure}

Figures~\ref{fig:hiQ2nlo} and~\ref{fig:hiQ2nnlo} show summaries
for HERAPDF2.0 NLO and NNLO as already shown in 
Figs.~\ref{fig:summarynlo} and~\ref{fig:summarynnlo} 
together with the equivalent plots for HERAPDF2.0HiQ2.
The only difference is that HERAPDF2.0 has
$Q^2_{\rm min} = 3.5\,$GeV$^2$ while HERAPDF2.0HiQ2 has
$Q^2_{\rm min} = 10\,$GeV$^2$.
At NLO, the gluon distributions of 
HERAPDF2.0 and HERAPDF2.0HiQ2 are compatible within uncertainties.
At NNLO, the two gluon distributions differ significantly.
Using the higher $Q^2_{\rm min}$ at NNLO causes the gluon distribution 
to  turn over significantly at low $x$. 
The distributions of HERAPDF2.0AG are also shown in Figs.~\ref{fig:hiQ2nlo} 
and~\ref{fig:hiQ2nnlo}. They
are not very different for the two $Q^2_{\rm min}$ values. At NNLO, this causes the 
gluon distribution of HERAPDF2.0AG
to be completely different than that of the standard parameterisation for
$x < 10^{-3}$.

\subsubsection{Comparison of HERADPF2.0HiQ2 to HERAPDF2.0}

\begin{figure}[tbp]
\vspace{-0.5cm} 
\centerline{
\epsfig{file=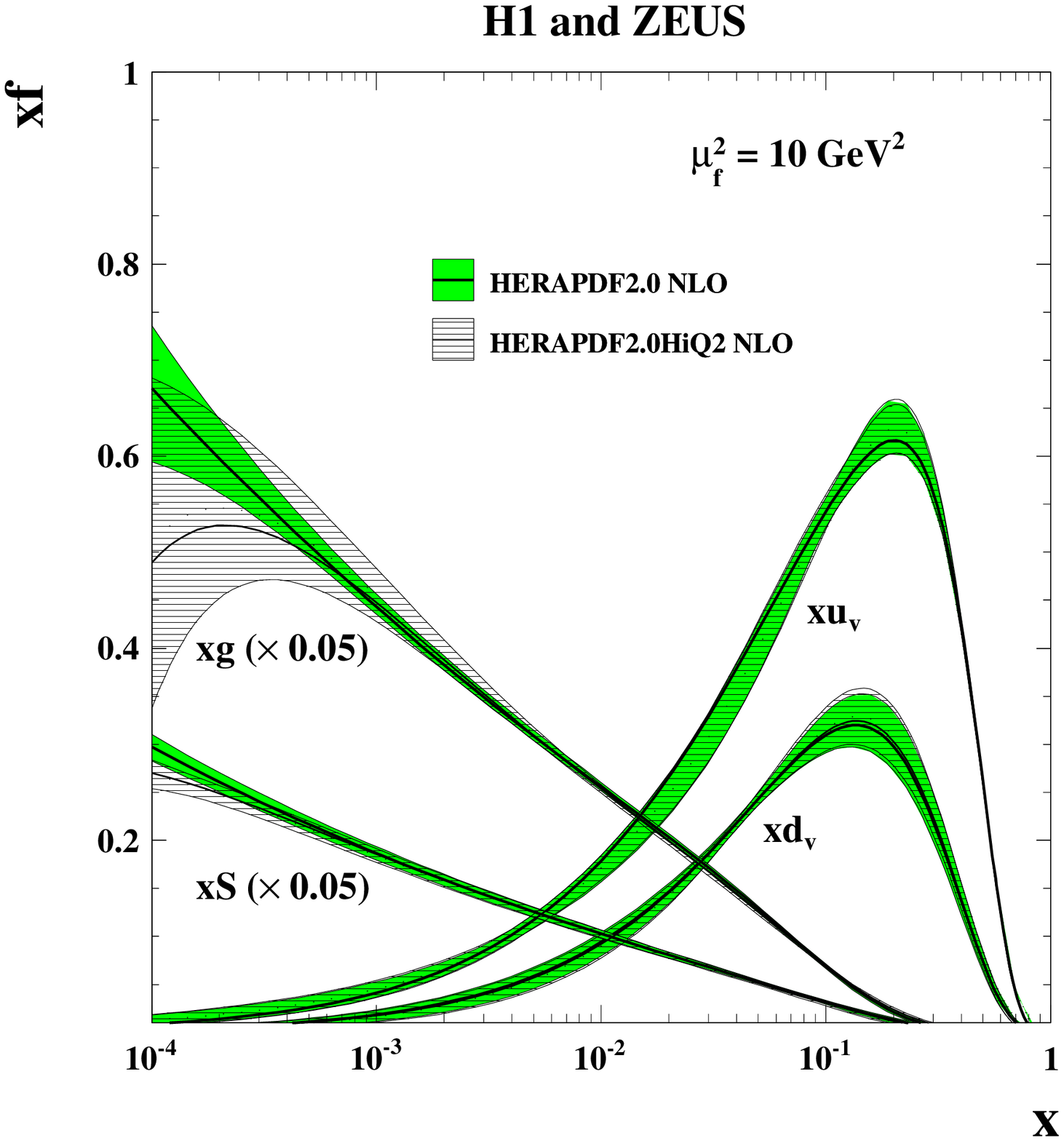,width=0.65\textwidth}}
\centerline{
\epsfig{file=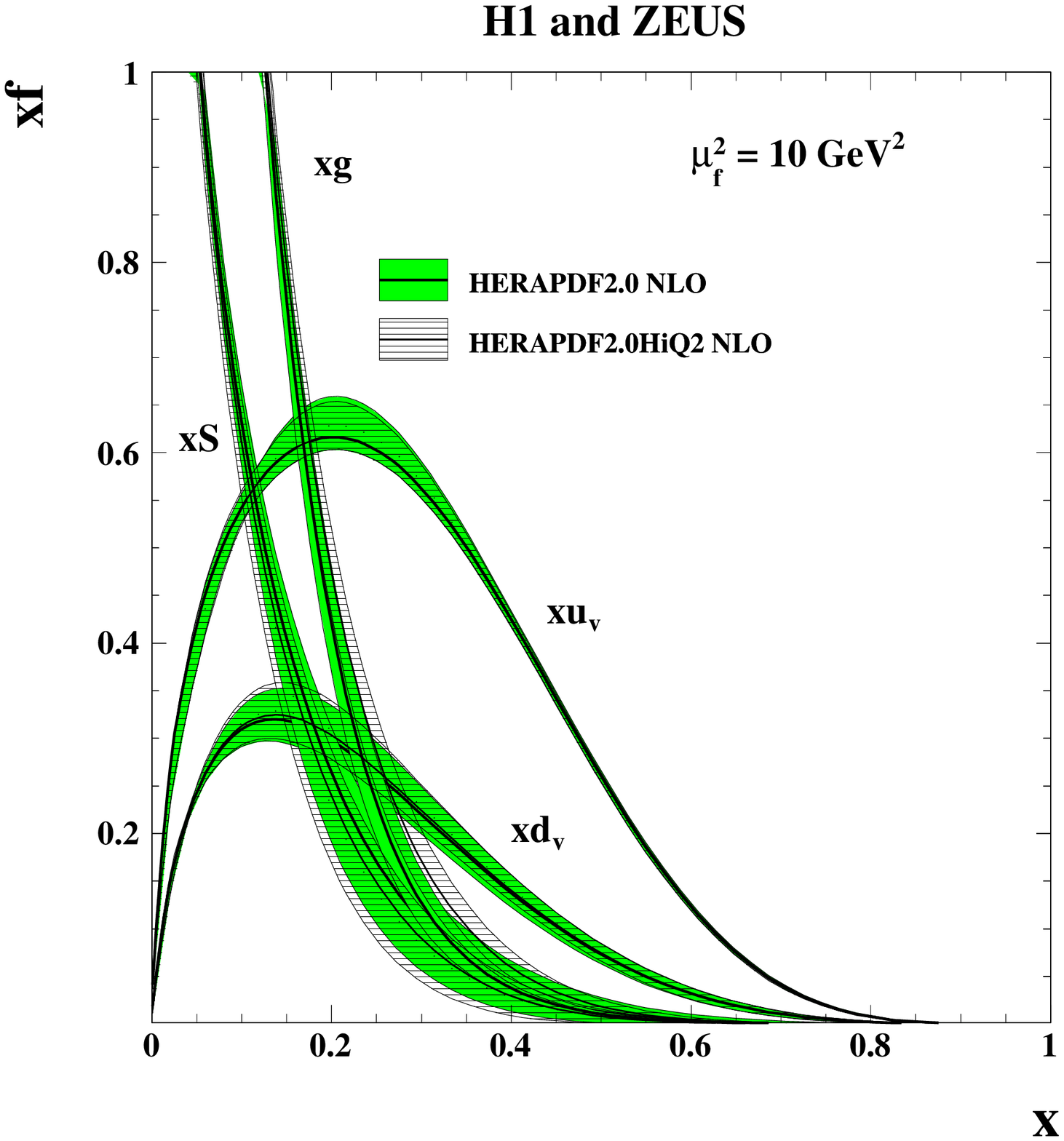,width=0.65\textwidth}}
\caption { 
The parton distribution functions 
$xu_v$, $xd_v$, $xS=2x(\bar{U}+\bar{D})$ and $xg$ of  
HERAPDF2.0 NLO
at $\mu_{\rm f}^{2} = 10\,$GeV$^{2}$ 
compared to those of HERAPDF2.0HiQ2 NLO 
on logarithmic (top) and linear (bottom) scales. 
The bands represent the total uncertainties.
}
\label{fig:nlo10vs3pt5}
\end{figure}

\begin{figure}[tbp]
\vspace{-0.5cm} 
\centerline{
\epsfig{file=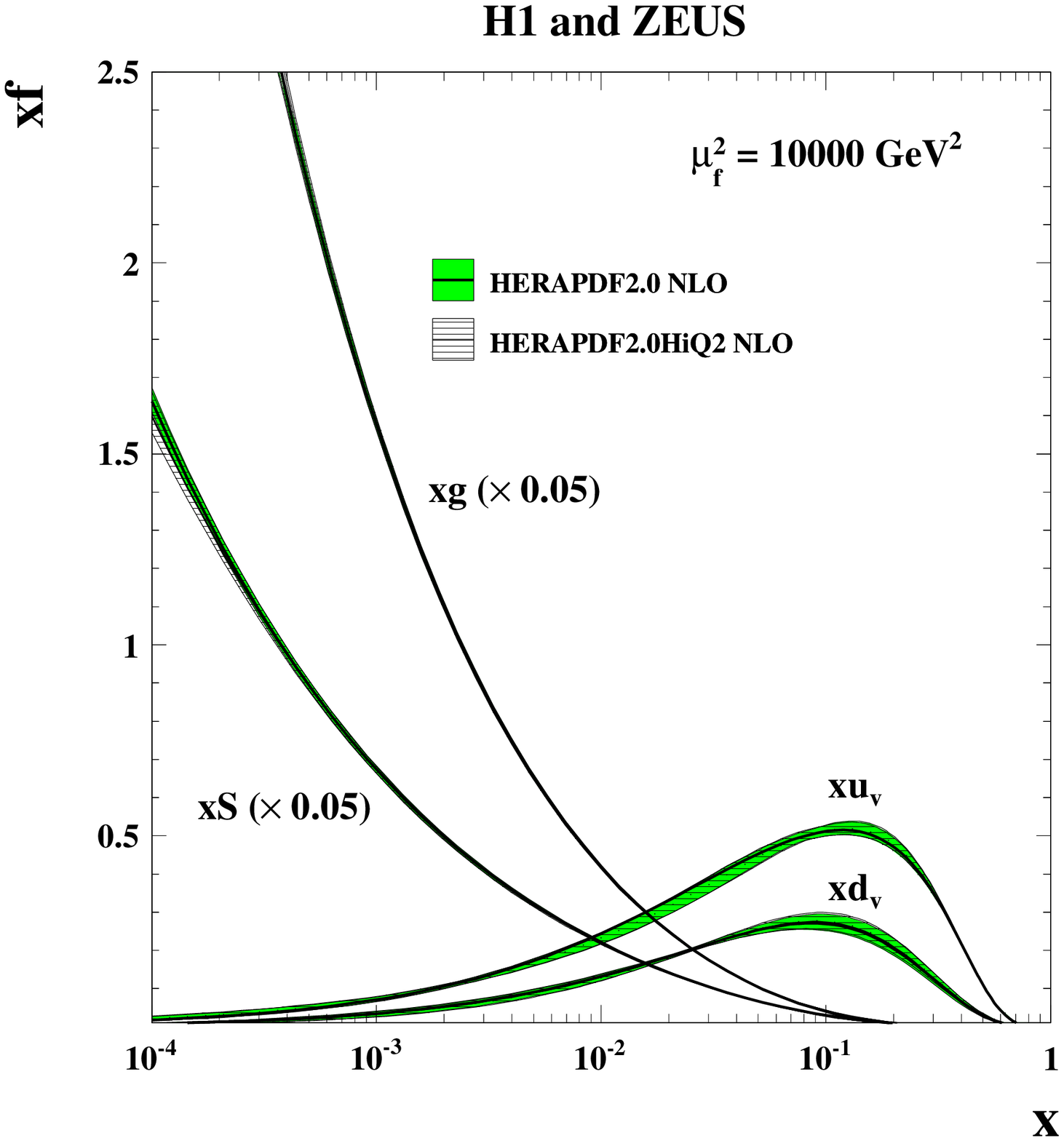,width=0.65\textwidth}}
\centerline{
\epsfig{file=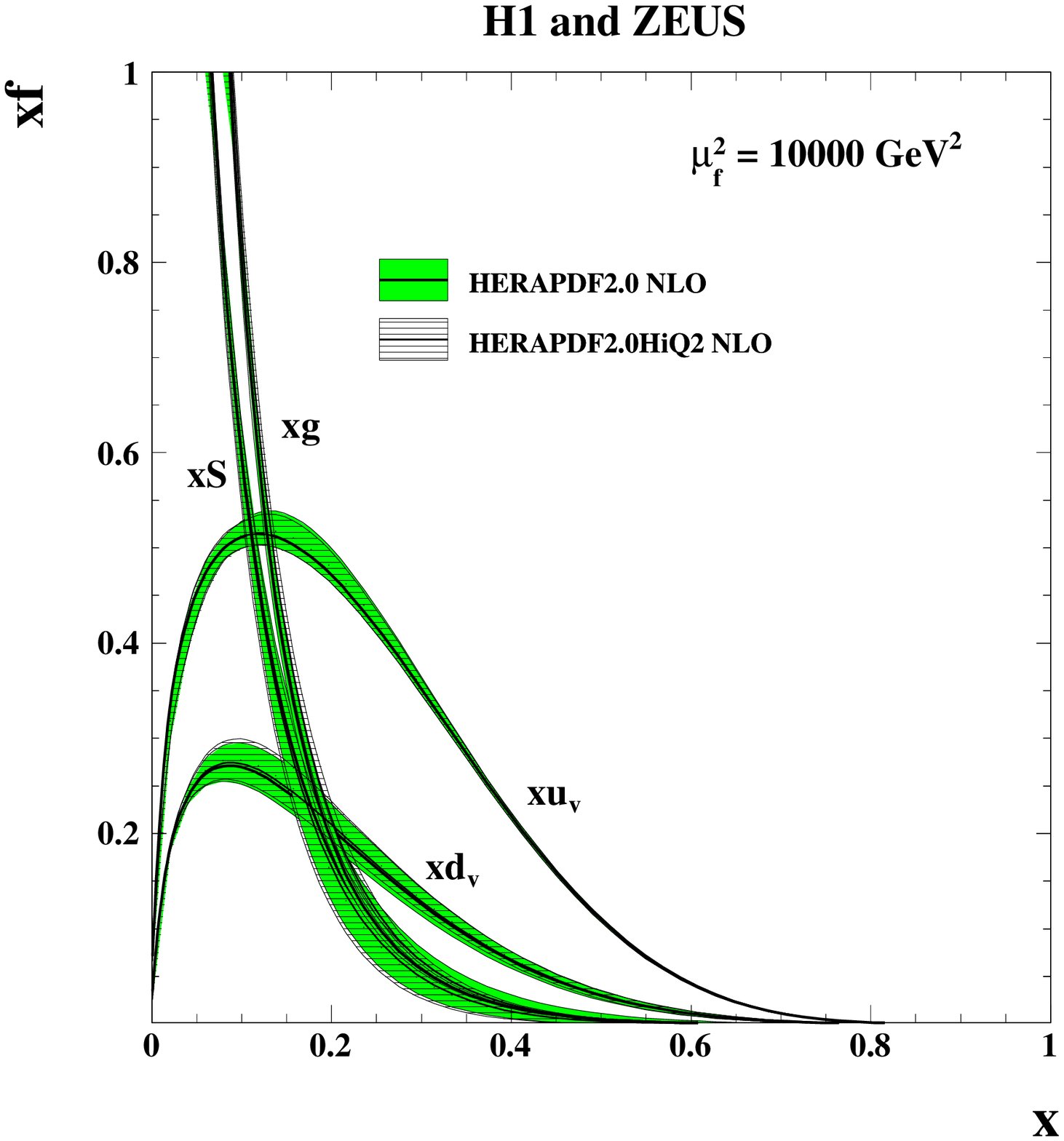,width=0.65\textwidth}}
\caption { 
The parton distribution functions 
$xu_v$, $xd_v$, $xS=2x(\bar{U}+\bar{D})$ and $xg$ of  
HERAPDF2.0 NLO
at $\mu_{\rm f}^{2} = 10000\,$GeV$^{2}$ 
compared to those of HERAPDF2.0HiQ2 NLO 
on logarithmic (top)
and linear (bottom) scales. The bands represent the total uncertainties.
}
\label{fig:highscale}
\end{figure}

A comparison of the NLO PDFs 
of HERAPDF2.0 to
HERAPDF2.0HiQ2 
at the  scale $\mu_{\rm f}^2=10\,$GeV$^2$
is shown in Fig.~\ref{fig:nlo10vs3pt5}.
The different shapes
of the gluon distribution at low $x$ are compatible within uncertainties. 
In Section~\ref{sec:cuts}, the question arose whether including data
from the 
kinematic region 
of low $x_{\rm Bj}$ and low $Q^2$, 
i.e.\ below 10\,GeV$^2$, 
in the PDF fits would introduce
a bias on
predictions for high $x_{\rm Bj}$ and high $Q^2$.
Figure~\ref{fig:highscale}   
demonstrates that at the high scale of $\mu_{\rm f}^2=10000\,$GeV$^2$, 
the PDFs resulting from the two fits are very similar. 
This confirms that the value of $Q^2_{\rm min} = 3.5\,$GeV$^2$ is a safe value
for pQCD fits to HERA data and no bias is introduced for applications
at higher scales like cross-section predictions for LHC.

\begin{figure}[tbp]
\vspace{-0.5cm} 
\centerline{
\epsfig{file=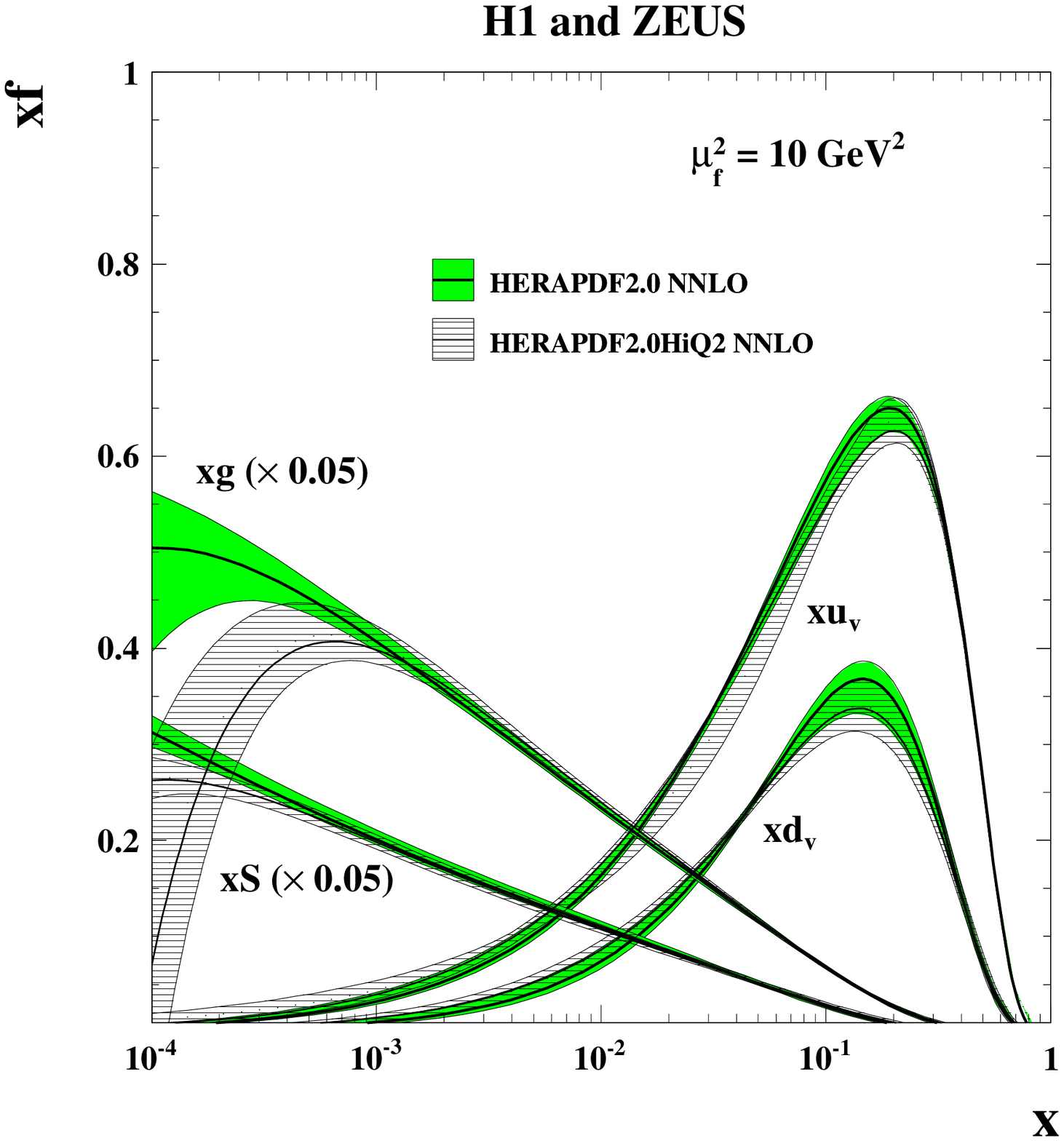,width=0.65\textwidth}}
\centerline{
\epsfig{file=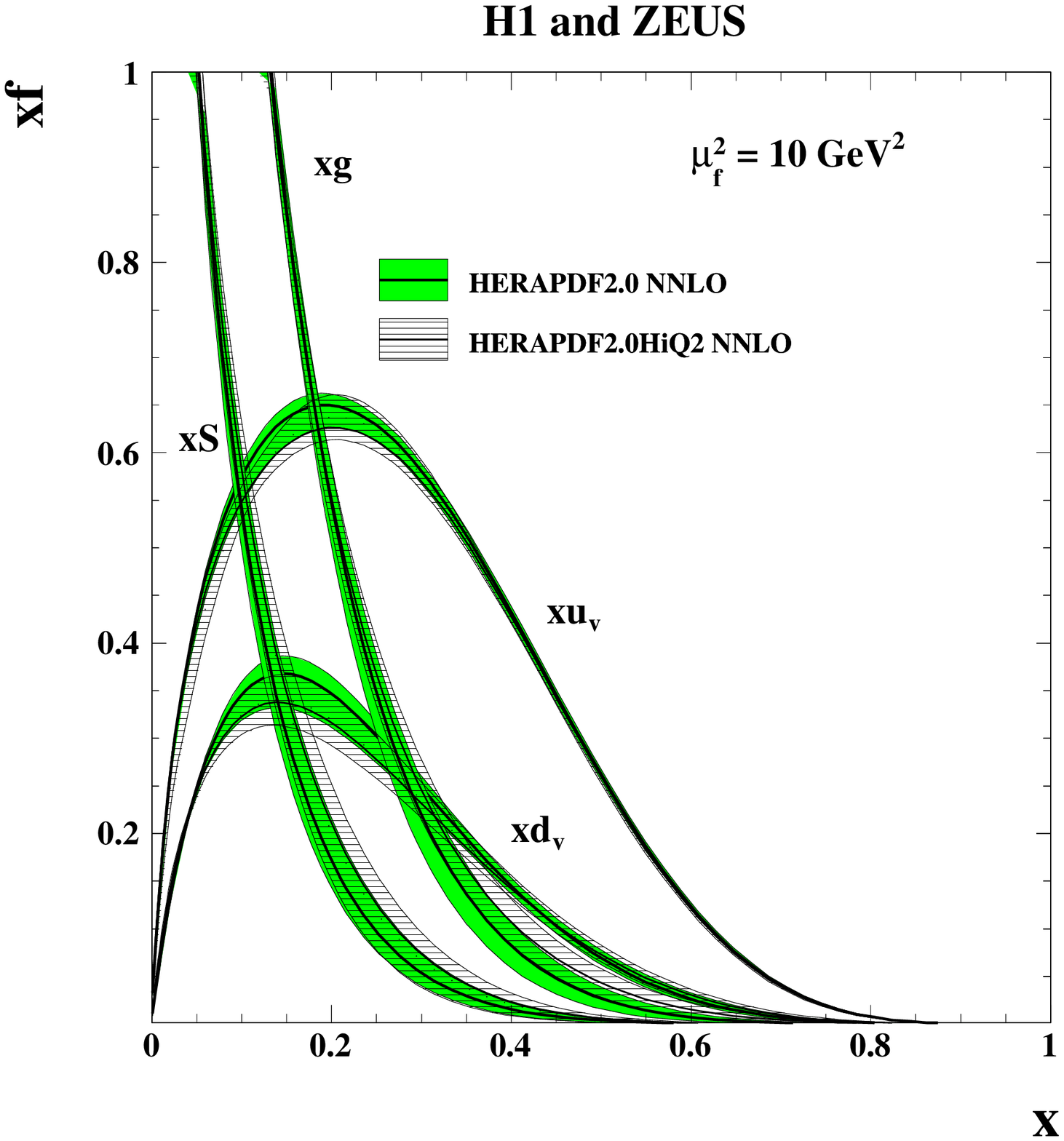,width=0.65\textwidth}}
\caption { 
The parton distribution functions 
$xu_v$, $xd_v$, $xS=2x(\bar{U}+\bar{D})$ and $xg$ of  
HERAPDF2.0 NLO
at $\mu_{\rm f}^{2} = 10\,$GeV$^{2}$ 
compared to those of HERAPDF2.0HiQ2 NLO 
on logarithmic (top) and linear (bottom) scales. 
The bands represent the total uncertainties.
}
\label{fig:nnlo10vs3pt5}
\end{figure}

\begin{figure}[tbp]
\vspace{-0.5cm} 
\centerline{
\epsfig{file=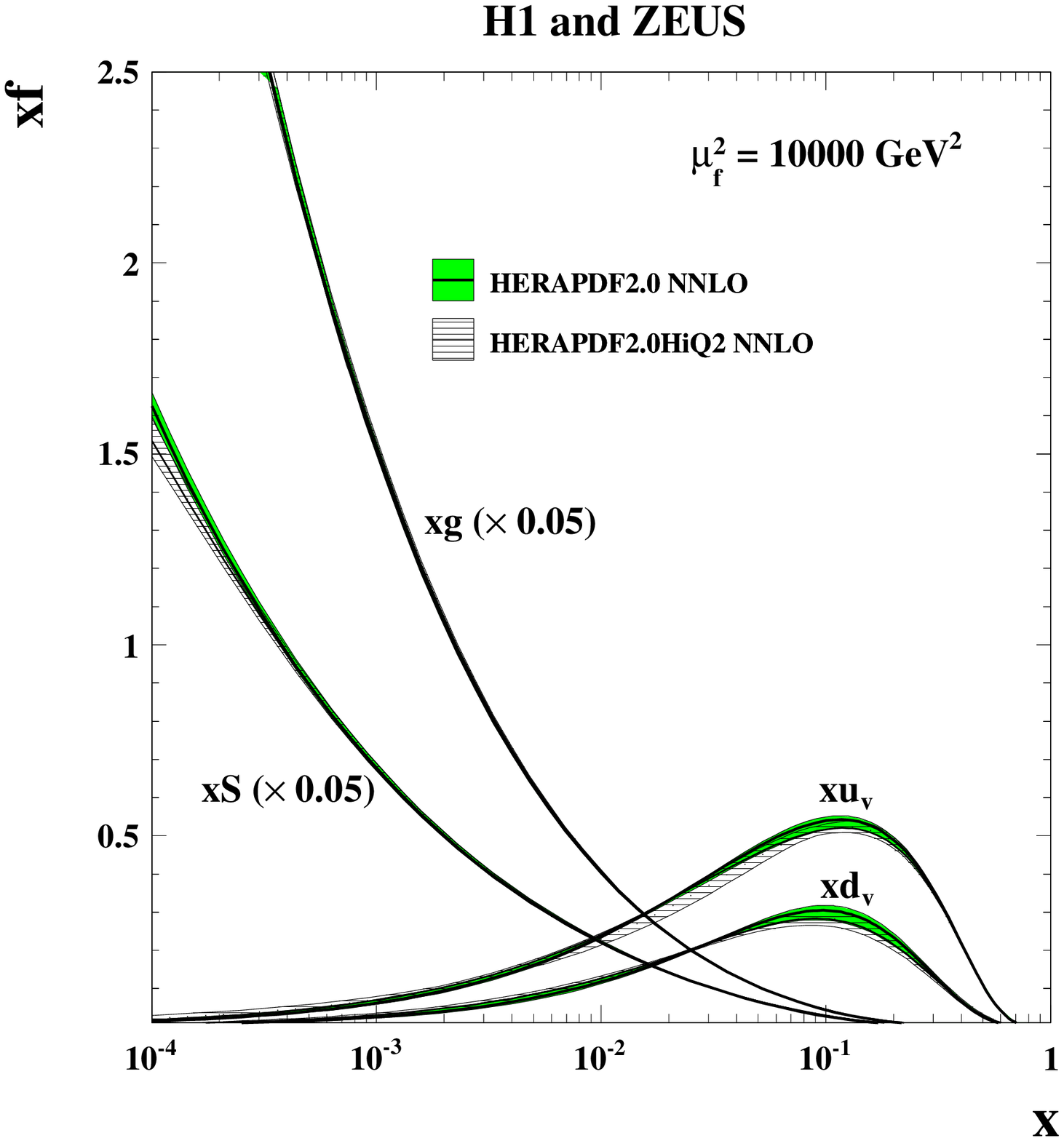,width=0.65\textwidth}}
\centerline{
\epsfig{file=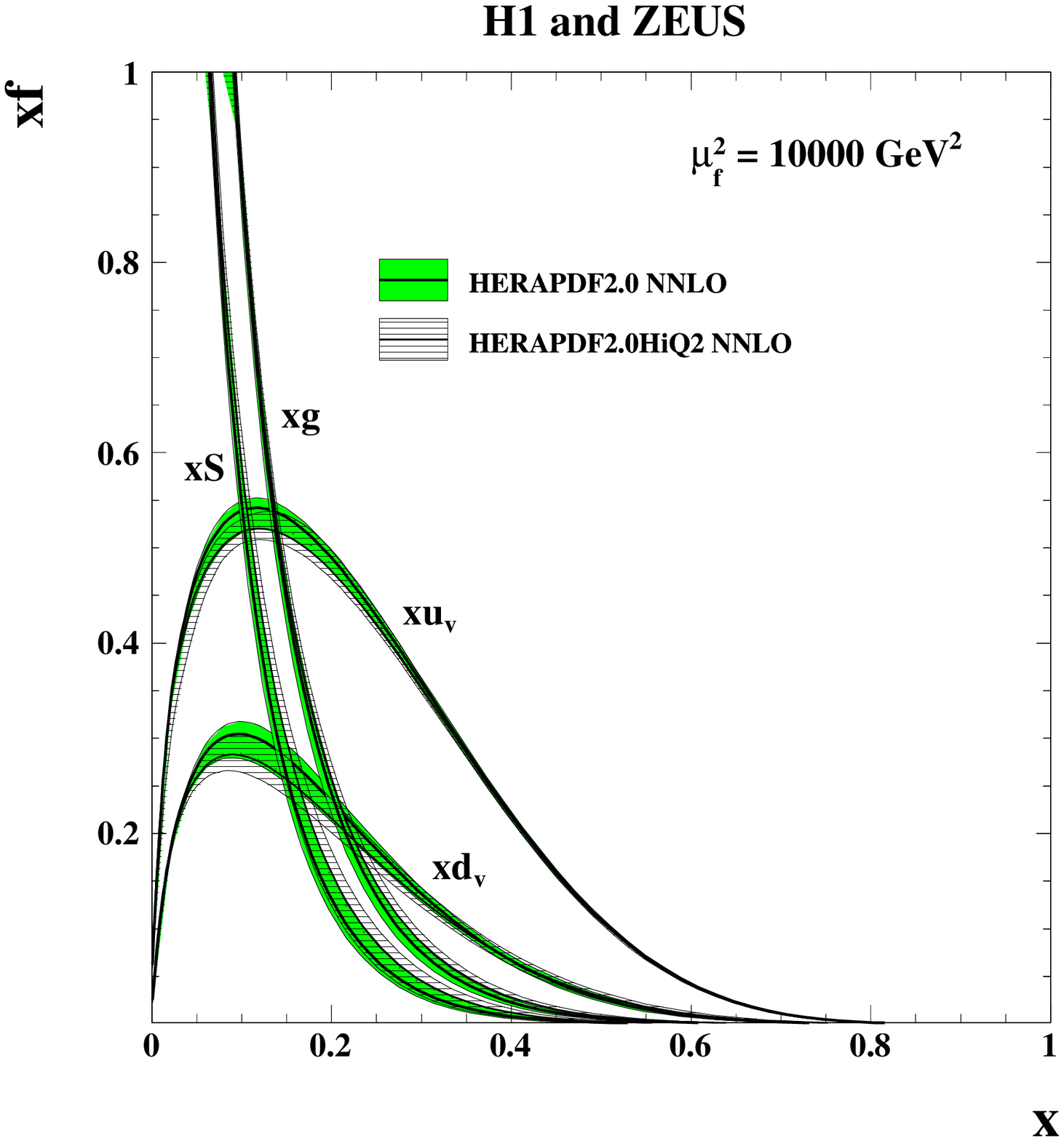,width=0.65\textwidth}}
\caption { 
The parton distribution functions 
$xu_v$, $xd_v$, $xS=2x(\bar{U}+\bar{D})$ and $xg$ of  
HERAPDF2.0 NNLO
at $\mu_{\rm f}^{2} = 10000\,$GeV$^{2}$ 
compared to those of HERAPDF2.0HiQ2 NNLO 
on logarithmic (top)
and linear (bottom) scales. The bands represent the total uncertainties.
}
\label{fig:highscalennlo}
\end{figure}

A comparison of the NNLO PDFs 
of HERAPDF2.0 to those
of
HERAPDF2.0HiQ2 
at the scale $\mu_{\rm f}^2=10\,$GeV$^2$
is shown in Fig.~\ref{fig:nnlo10vs3pt5}.  
The differences in the gluon distributions are pronounced. 
The gluon distribution of HERAPDF2.0HiQ2 NNLO turns over for
$x < 10^{-3}$.
The valence distributions at NNLO also differ between HERAPDF2.0HiQ2 and
HERAPDF2.0, but they are compatible within uncertainties.  
At the high scale of $\mu_{\rm f}^2=10000\,$GeV$^2$, 
the PDFs resulting from the two fits are, as at NLO, very similar, 
see Fig.~\ref{fig:highscalennlo}. 
This demonstrates
that again  no bias is introduced at higher scales 
when low-$x_{\rm B_j}$ and low-$Q^2$ data are included in the fit at NNLO . 

\subsubsection{Comparison of HERAPDF2.0HiQ2 to data}

Figures~\ref{fig:nnloQ210ncepb} and~\ref{fig:nloQ210ncepb} show 
the predictions
of HERAPDF2.0HiQ2 NNLO and NLO 
compared to the data,
which were already presented and compared to HERAPDF2.0 NNLO and NLO in 
Figs.~\ref{fig:nnloQ23pt5ncepb} and~\ref{fig:nloQ23pt5ncepb}.
In the region $3.5 \le Q^2 \le 10$\,GeV$^2$, 
the standard HERAPDF2.0 NLO fit 
compromises between describing the low-$x_{\rm Bj}$ (high-$y$) turnover, 
for which it is too high, 
and the data at slightly higher $x_{\rm Bj}$, for which it is too low. 
In the corresponding HERAPDF2.0HiQ2 
fit, these data are not fitted.  
The resulting fit, when extrapolated 
to the excluded region, is systematically lower 
than the data at lower $x_{\rm Bj}$ and lower $Q^2$, 
but then is significantly
above the data at very low $x_{\rm Bj}$, where the contribution 
from $F_{\rm L}$ becomes
important. 
This implies that the 
pQCD fit evolves more strongly towards lower $x_{\rm Bj}$ 
and $Q^2$ than does the data. 
The situation is not improved at 
NNLO where the fit evolves even more strongly. 
This suggests that the conventional DGLAP resummation may not be fully adequate. 
This observation was also made during 
investigations of the HERA\,I data~\cite{Caola}.


\subsection{HERAPDF2.0FF}

\begin{figure}[tbp]
  \centerline{
  \epsfig{file=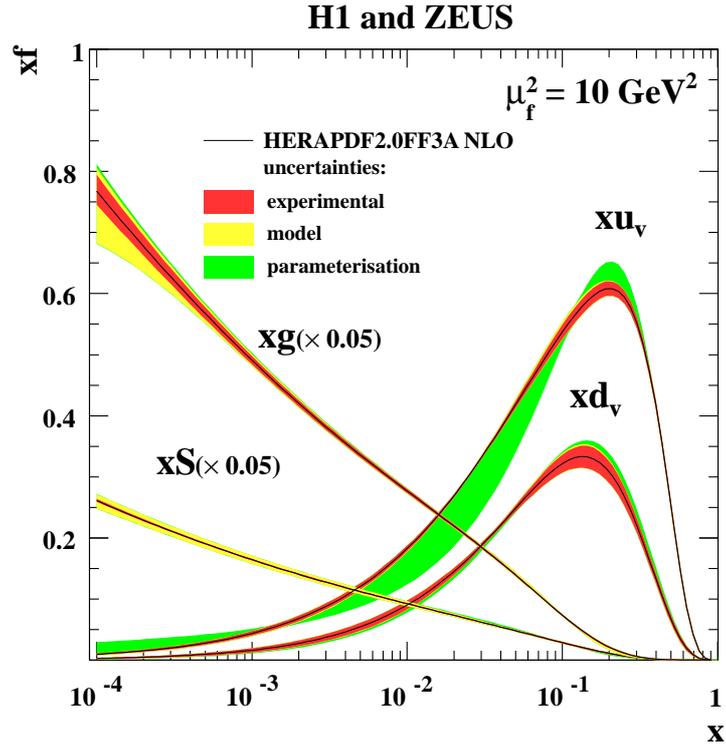,width=0.65\textwidth}}
  \centerline{
  \epsfig{file=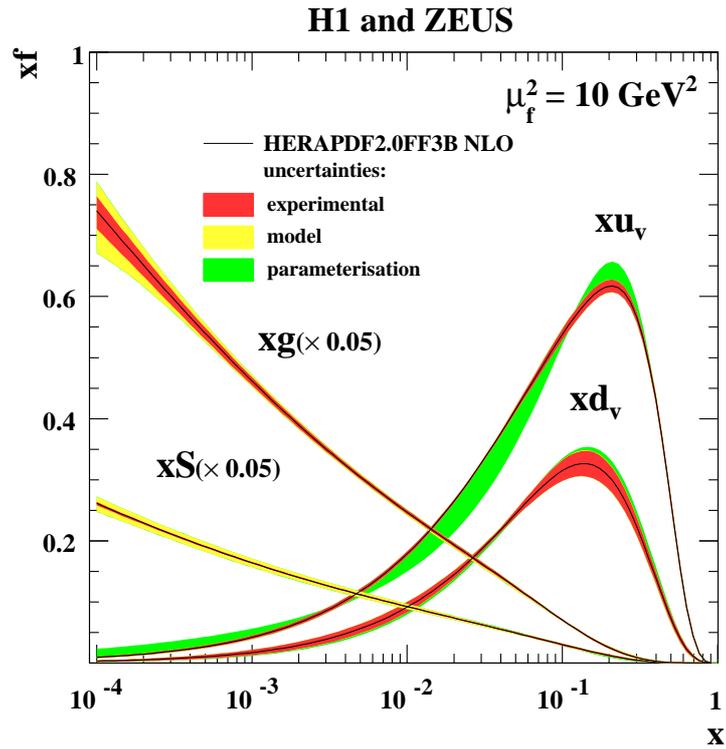,width=0.65\textwidth}}
  \caption{
The parton distribution functions 
$xu_v$, $xd_v$, $xS=2x(\bar{U}+\bar{D})$ and $xg$ of  
of HERAPDF2.0FF3A NLO
and HERAPDF2.0FF3B NLO,
at $\mu_{\rm f}^{2}$ = 10\,GeV$^{2}$.
The experimental, model
and parameterisation uncertainties are shown.}
  \label{fig:PDFFF} 
\end{figure}

Summaries of 
HERAPDF2.0FF3A and HERAPDF2.0FF3B as 
introduced in Section~\ref{sec:althfs} are  shown in
Fig.~\ref{fig:PDFFF}.
The experimental, model  and parameterisation uncertainties
were evaluated as for the standard HERAPDF2.0 NLO, 
see Sections~\ref{sec:exp:unc} and~\ref{sec:assumpt}, 
and are shown separately. 

\begin{figure}[tbp]
\vspace{-0.5cm} 
\centerline{
\epsfig{file=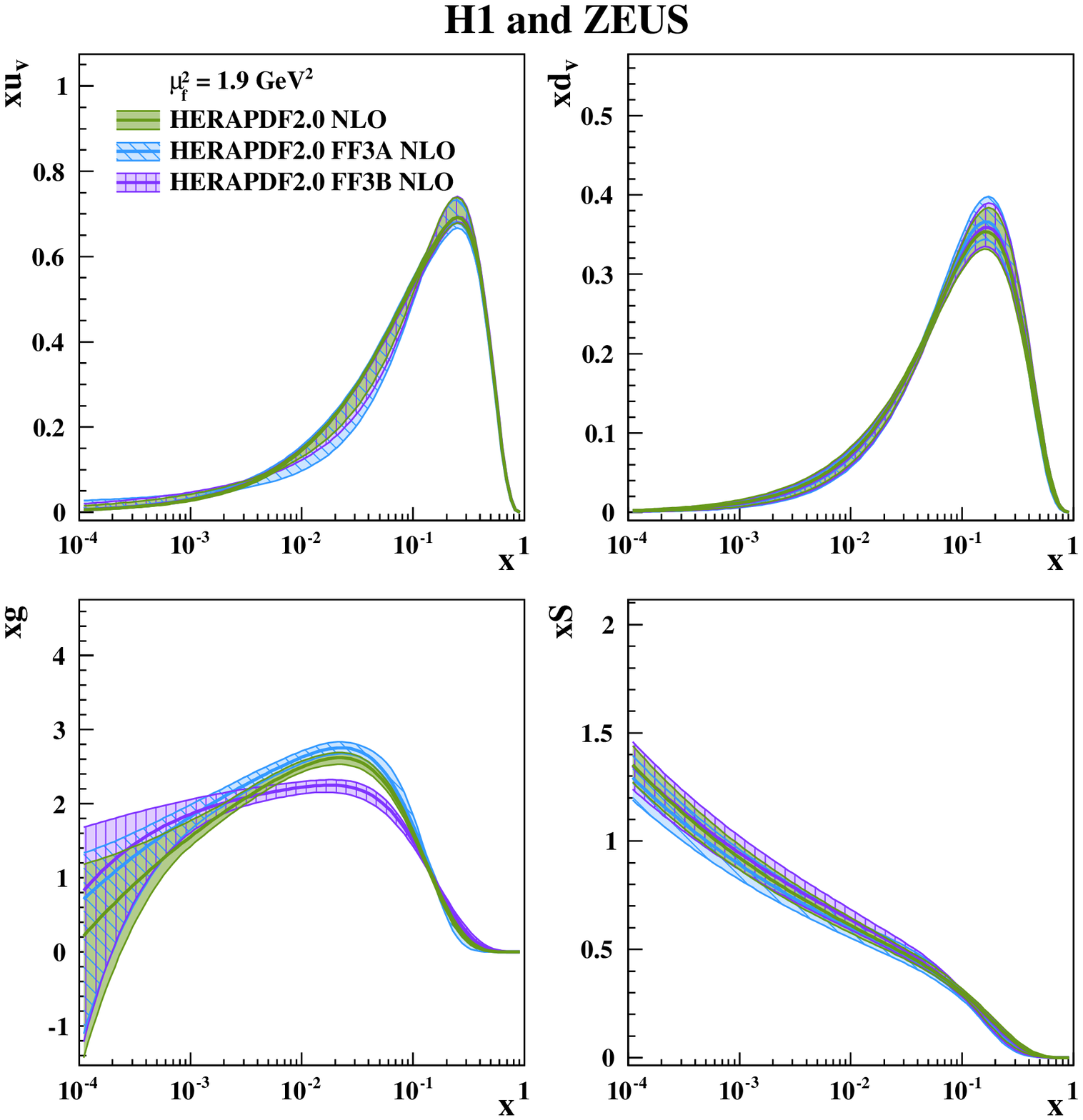,width=0.65\textwidth}}
\vspace*{0.65cm}
\centerline{
\epsfig{file=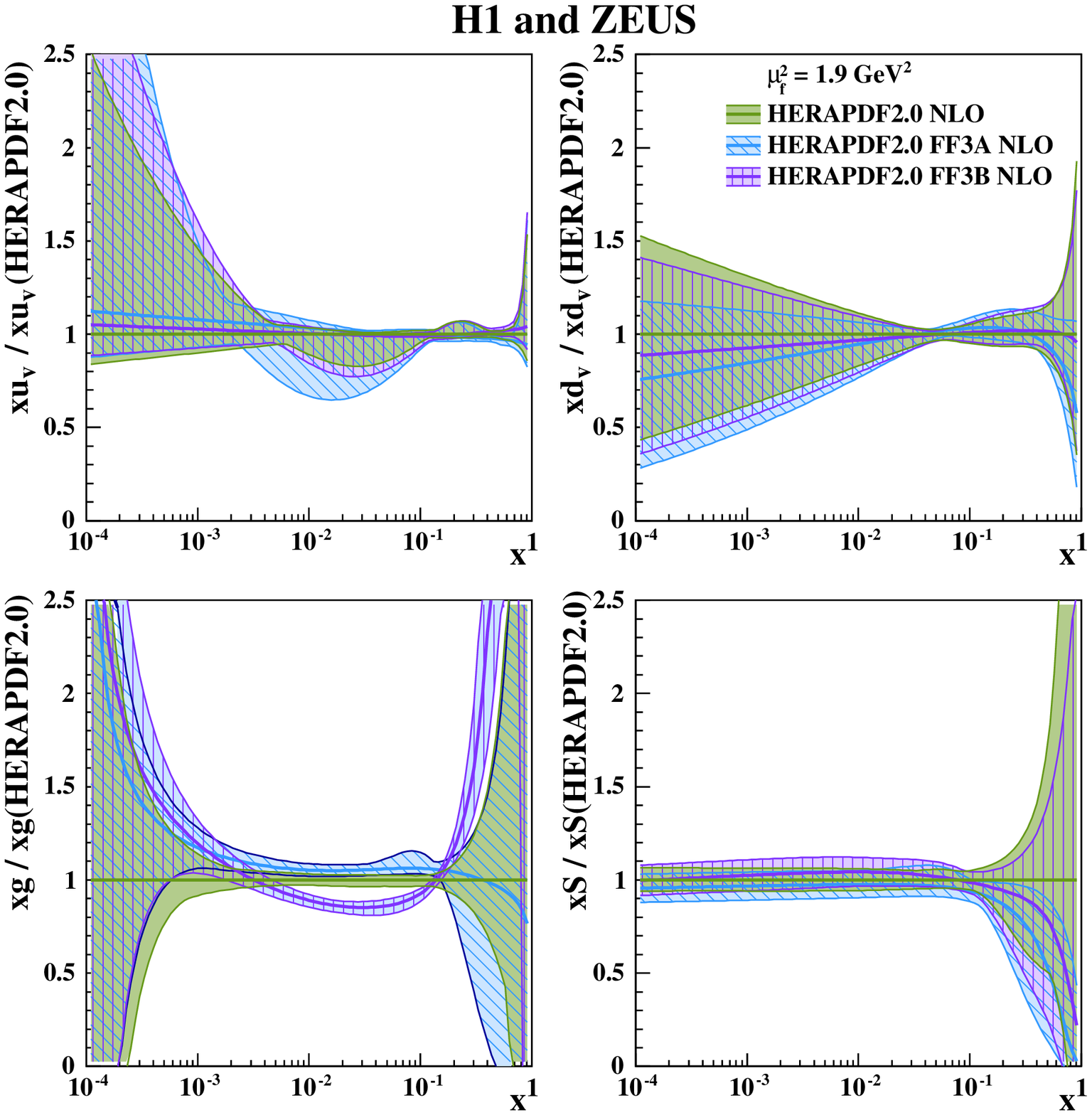,width=0.65\textwidth}}
\caption { 
The parton distribution functions 
$xu_v$, $xd_v$, $xg$ and $xS=2x(\bar{U}+\bar{D})$ of  
HERAPDF2.0FF3A and FF3B
at the starting scale $\mu_{\rm f_{0}}^{2} = 1.9\,$GeV$^{2}$
compared to those of HERAPDF2.0 NLO. 
The top panel
shows the distributions. 
The bottom panel shows the PDFs normalised to HERAPDF2.0 NLO.
The uncertainties are given as differently hatched bands
in both panels.
}
\label{fig:FFNLO-NLO}
\end{figure}

A comparison of the PDFs
of HERAPDF2.0FF3A and HERAPDF2.0FF3B to the standard HERAPDF2.0 NLO
using the RTOPT heavy-flavour scheme is shown
in Fig.~\ref{fig:FFNLO-NLO}.
This comparison is presented at the starting scale $\mu_{\rm f_0}$,
because a meaningful comparison can only be done at scales below the
charm mass. 
There are differences in the valence and in the gluon distributions. 
The latter originate mainly from the different 
${\cal O}(\alpha_s)$ at which the massless contribution 
to $F_{\rm L}$ is calculated and on the $\alpha_s$ evolution scheme.
A comparison of the predictions from HERAPDF2.0FF3B and HERAPDF2.0 NLO
to selected data as already used for Fig.~\ref{fig:5mod}
is shown in Fig.~\ref{fig:FFdata}. 
The predictions are very similar.
However, at low $x_{\rm Bj}$ and low $Q^2$, the $Q^2$ dependence predicted from
HERAPDF2.0FF3B is a bit less steep than the prediction from HERAPDF2.0 NLO.
The predictions of HERAPDF2.0FF3A are also very similar.
The $Q^2$ dependence predicted from HERAPDF2.0FF3A  is however 
slightly steeper than the prediction
from  HERAPDF2.0 NLO at low $x_{\rm Bj}$ and low $Q^2$. 

\begin{figure}[tbp]
\vspace{-0.5cm} 
\centerline{
\epsfig{file=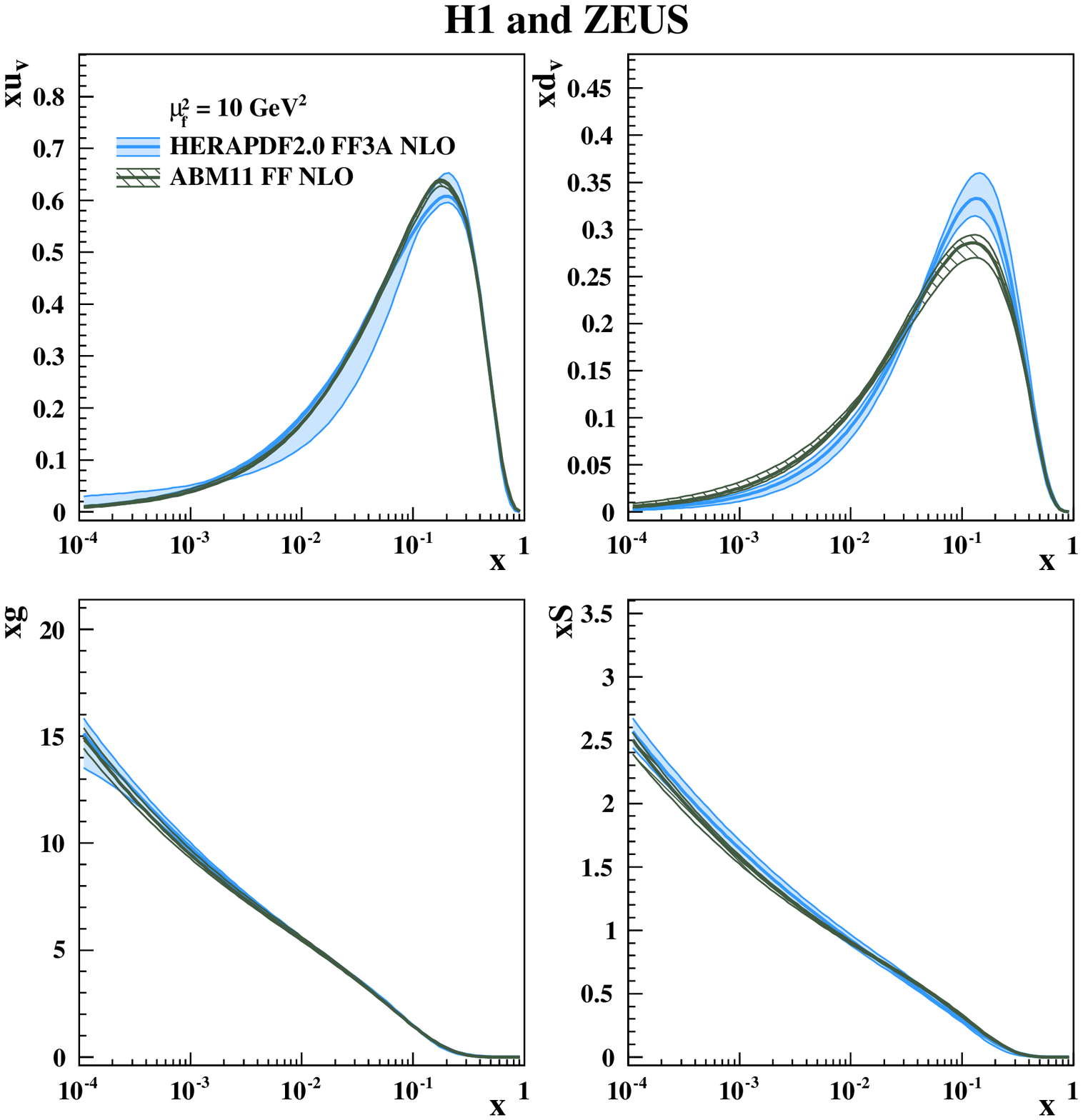,width=0.65\textwidth}}
\vspace*{0.5cm}
\centerline{
\epsfig{file=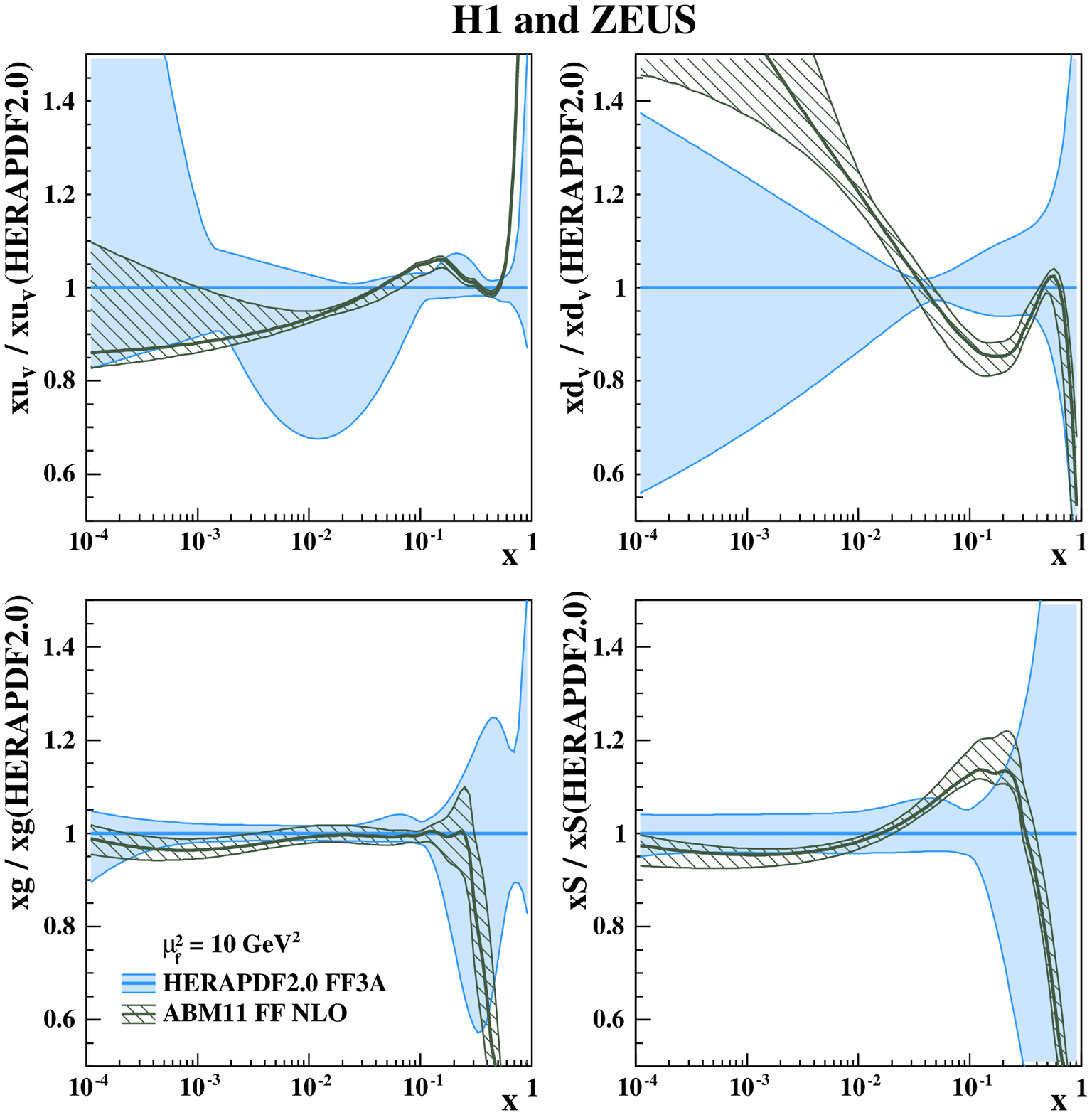,width=0.65\textwidth}}
\caption { 
The parton distribution functions 
$xu_v$, $xd_v$, $xg$ and $xS=2x(\bar{U}+\bar{D})$  of  
HERAPDF2.0FF3A
at $\mu_{\rm f}^{2} = 10\,$GeV$^{2}$
compared to those of ABM11\,FF~\cite{ABM3}. 
The top panel
shows the distributions. 
The bottom panel shows the PDFs normalised to HERAPDF2.0FF3A.
The uncertainties are given as differently hatched bands
in both panels.
}
\label{fig:FFANLO-others}
\end{figure}

A comparison of the PDFs of
HERAPDF2.0FF3A to the PDFs of 
ABM11\,FF~\cite{ABM3} is shown in
Fig.~\ref{fig:FFANLO-others}. 
These two sets of PDFs can be directly compared as they use the same
order for the description of $F_{\rm L}$ and the same $\alpha_s$ evolution. 
The largest difference is observed
for the $xd_v$ distribution.
However, overall the two sets of PDFs are quite similar.

\begin{figure}[tbp]
\vspace{-0.5cm} 
\centerline{
\epsfig{file=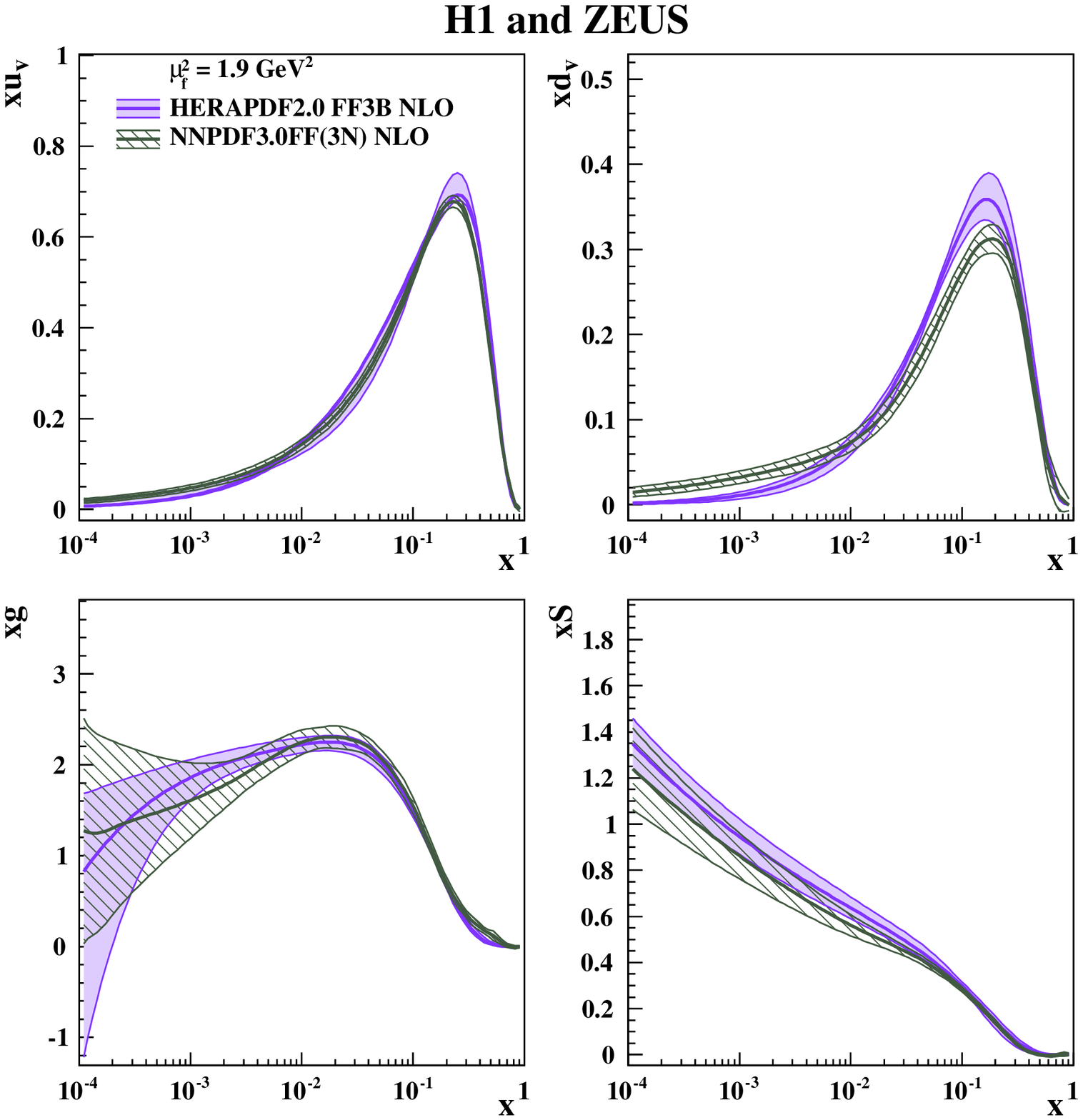,width=0.65\textwidth}}
\vspace*{0.5cm}
\centerline{
\epsfig{file=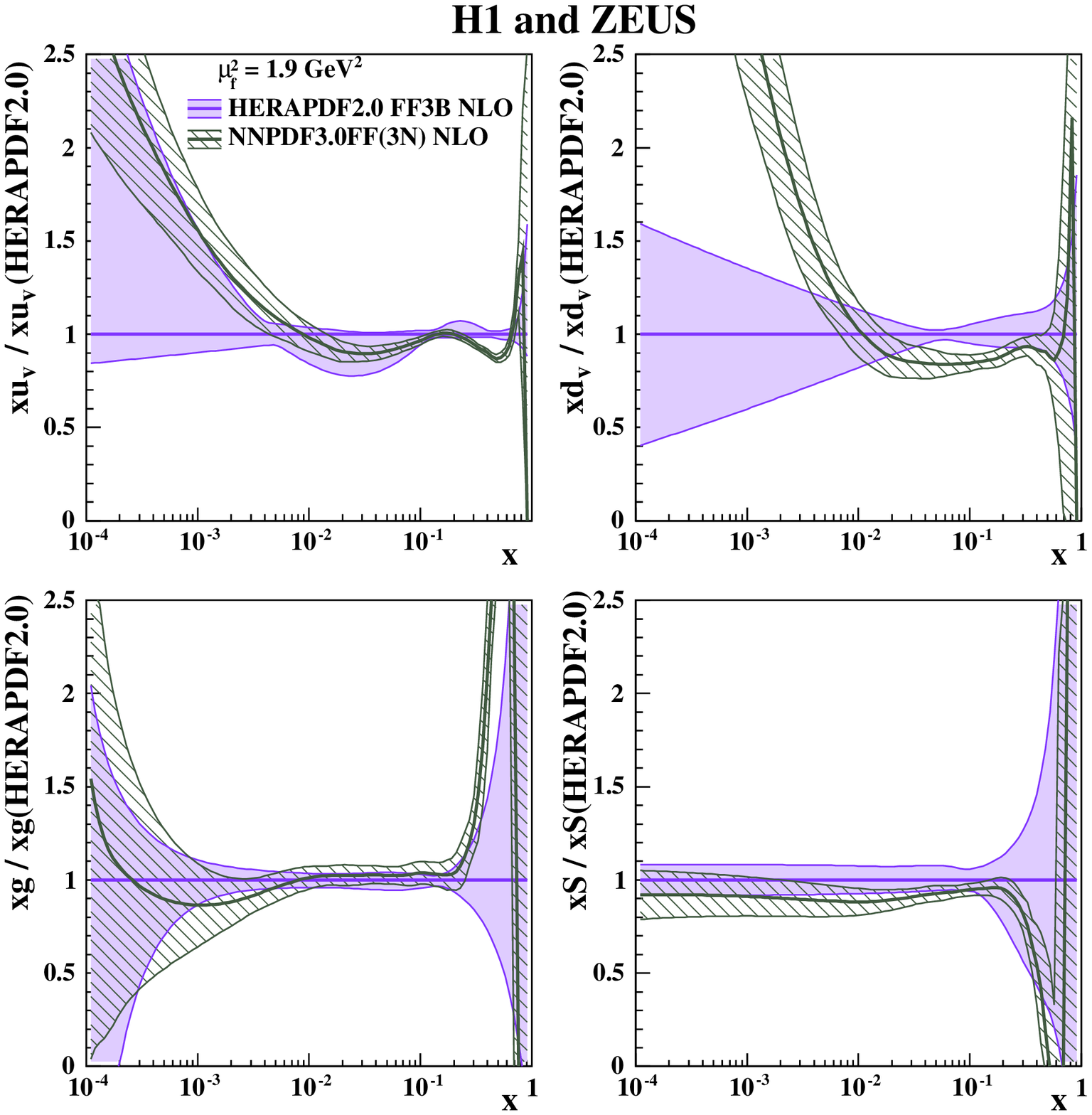,width=0.65\textwidth}}
\caption {  
The parton distribution functions 
$xu_v$, $xd_v$, $xg$ and $xS=2x(\bar{U}+\bar{D})$ of  
HERAPDF2.0FF3B
at the starting scale $\mu_{\rm f_{0}}^{2} = 1.9\,$GeV$^{2}$
compared to those of NNPDF3.0FF\,(3N). 
The top panel
shows the distributions. 
The bottom panel shows the PDFs normalised to HERAPDF2.0FF3B.
The uncertainties are given as differently hatched bands
in both panels.
} 
\label{fig:FFBNLO-others}
\end{figure}

A comparison of the PDFs of
HERAPDF2.0FF3B to the PDFs of
NNPDF3.0\,FF(3N)~\cite{NNPDF3.0} is shown in Fig.~\ref{fig:FFBNLO-others}.
These two sets of PDFs can be directly compared 
at the starting scale
due to their equivalent
treatment of the $F_{\rm L}$ contribution and of the $\alpha_s$ 
evolution.\footnote{
 The NNPDF3.0FF(3N) is based on a fixed number of flavours, NF=3, evolution, 
 but it is calculated from the FONLL-B fit, which is based on a 
 variable NF evolution~\cite{NNPDF3.0,NNPDFadd}.
 Thus, close to the starting scale, the PDFs of NNPDF3.0FF(3N) 
 can be directly 
 compared to the PDFs of HERAPDF2.0FF3B, which are also based on a variable 
 NF evolution.
}
The gluon distributions are quite similar. 
Some differences are observed in the $xu_v$ and $xd_v$ valence distributions.

\subsection{HERAPDF2.0Jets}

\begin{figure}
  \centering
  \setlength{\unitlength}{0.1\textwidth}
  \begin{picture} (9,9)
  \put(0,0){\includegraphics[width=\textwidth]{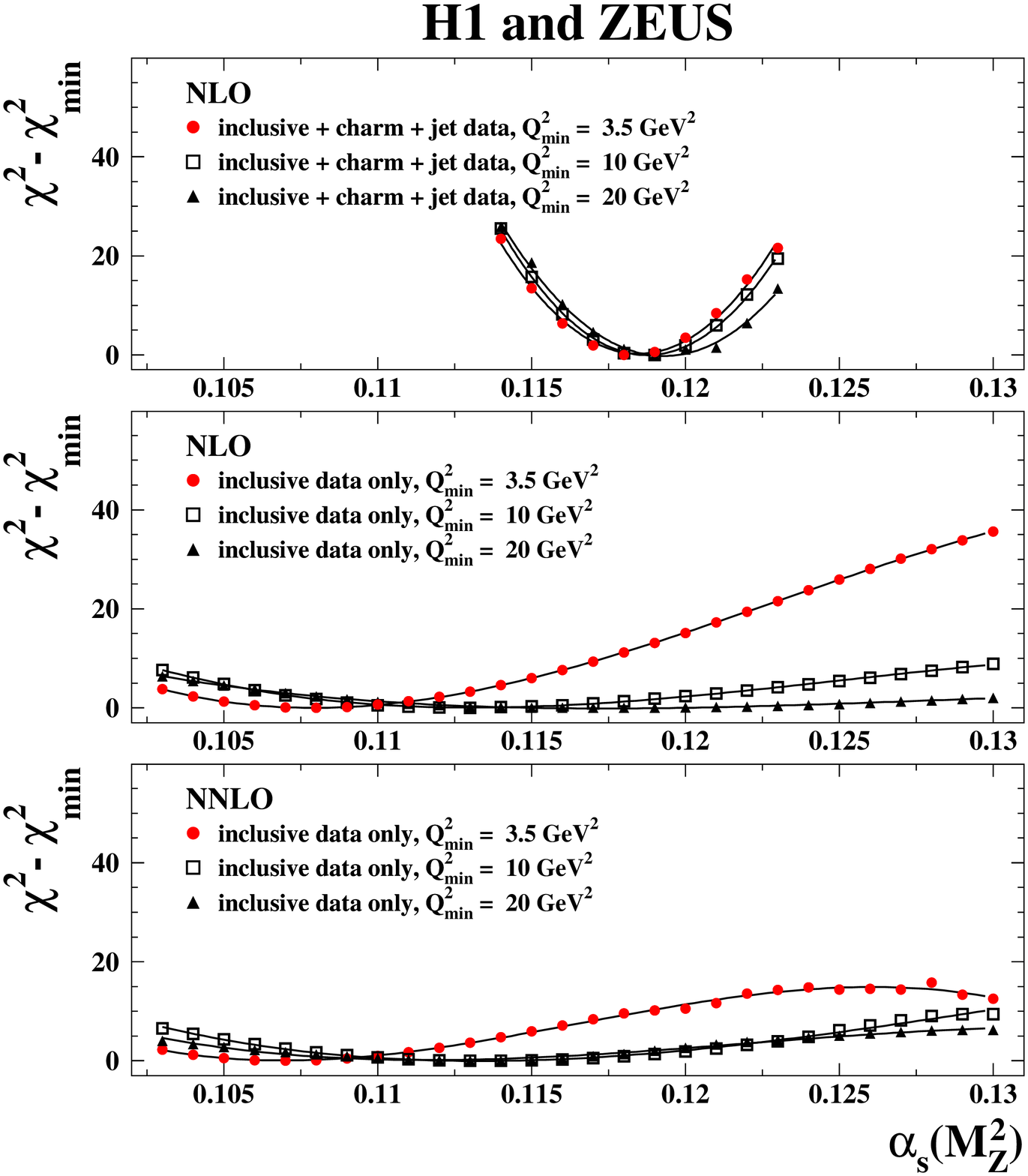}}
  \put (0.3,6.6) {a)}
  \put (0.3,3.8) {b)}
  \put (0.3,0.9) {c)}
  \end{picture}
\caption {$\Delta \chi^2 = \chi^2 - \chi^2_{\rm min}$ vs.\ $\asmz$ for pQCD fits 
with different $Q^2_{\rm min}$ using
data on (a) inclusive, charm and jet production at NLO,
(b) inclusive $ep$ scattering only at NLO and
(c) inclusive $ep$ scattering only at NNLO.
}
\label{fig:alphasscan}
\end{figure}

Data on jet production were included in the analysis as described
in Section~\ref{sec:addjfit}.
This inclusion  was first used to validate the
choice of $\asmz=0.118$ for HERAPDF 
by investigating the dependence of the $\chi^2$s 
of the HERAPDF pQCD fits on $\asmz$.
Three  $\chi^2$ scans vs. the value of $\asmz$ were performed at 
NLO for three values of $Q^2_{\rm min}$. 
The result is depicted in the top panel of Fig.~\ref{fig:alphasscan}.
A distinct minimum at $\asmz \approx 0.118$
is observed, which is basically independent
of $Q^2_{\rm min}$. This validates the choice of  $\asmz = 0.118$
for HERAPDF2.0 NLO.
Scans at NLO and NNLO were also performed for fits to inclusive data only.
The middle and bottom panels of Fig.~\ref{fig:alphasscan} show that
these scans  yielded similar shallow 
$\chi^2$ dependences and the minima
were strongly dependent on the $Q^2_{\rm min}$. 
This demonstrates
that the inclusive data alone cannot constrain $\asmz$
reasonably. 


\subsubsection{PDFs and measurement of $\mathbold{\asmz}$}

\begin{figure}[tbp]
\vspace{-0.5cm} 
\centerline{
\epsfig{file=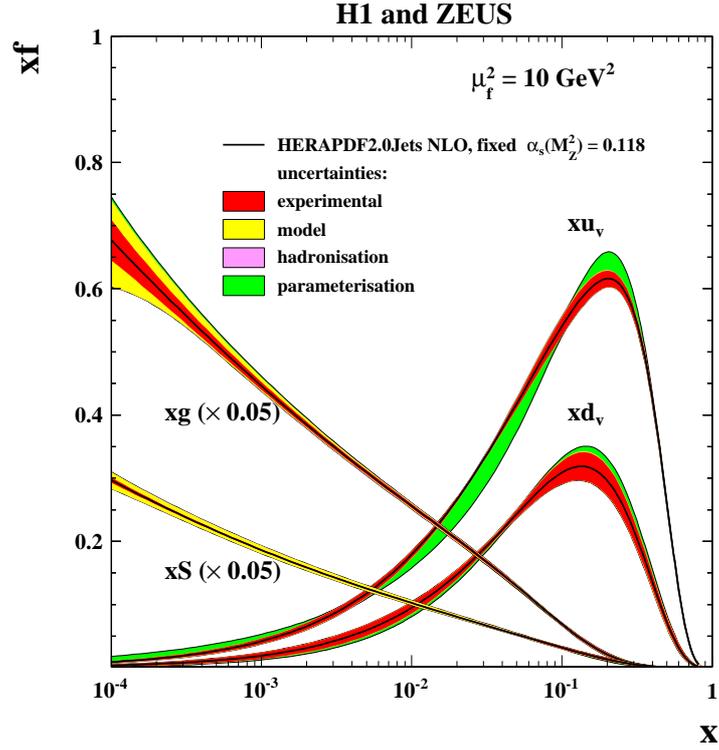,width=0.65\textwidth}}
\centerline{
\epsfig{file=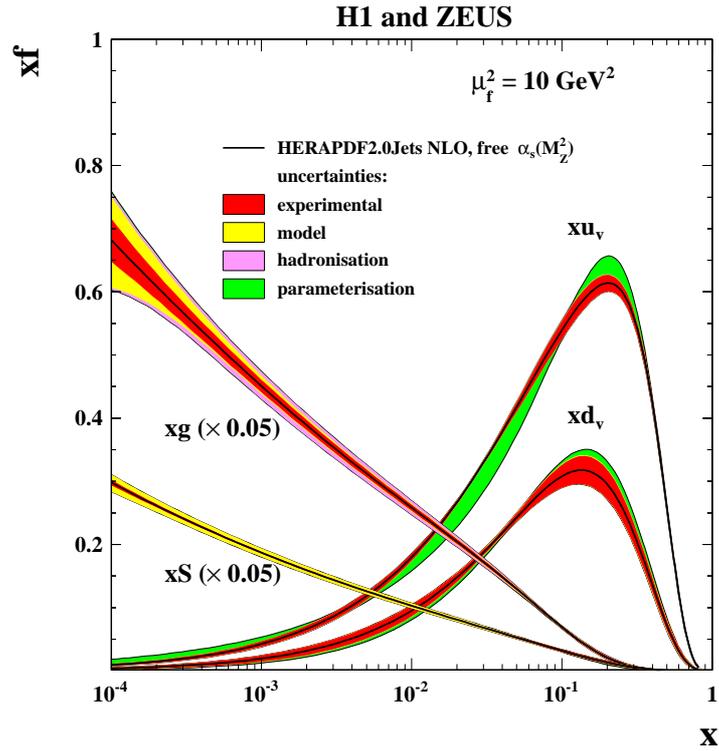,width=0.65\textwidth}}
\caption { 
The parton distribution functions 
$xu_v$, $xd_v$, $xS=2x(\bar{U}+\bar{D})$ and $xg$  of
HERAPDF2.0Jets NLO
at $\mu_{\rm f}^{2} = 10\,$GeV$^{2}$ 
with fixed $\asmz=0.118$ (top) and free $\asmz$ (bottom).
The experimental, model and parameterisation 
uncertainties are shown. 
The hadronisation uncertainty is also included, but it is
only visible for the fit with free $\asmz$.
}
\label{fig:jetsasmzfixfree}
\end{figure}

The PDFs resulting from a fit with
free $\asmz$, HERAPDF2.0Jets, and from a fit with fixed $\asmz=0.118$
are shown in
Fig.~\ref{fig:jetsasmzfixfree}.
A full uncertainty analysis was 
performed in both cases, including model and parameterisation uncertainties as 
well as additional hadronisation uncertainties on the jet data. 
The PDFs are very similar, because the HERAPDF2.0Jets fit with free $\asmz$ yields
a value which is very close to the value used for the fit with fixed $\asmz$.
The jet data determine the value of $\asmz$  very well 
in the HERAPDF2.0Jets fit. 
Thus, the uncertainty on $\asmz$ in this fit does not significantly increase 
the uncertainty on the gluon PDF with respect to the fit with $\asmz$ fixed. 
The difference in the $\asmz$ free fit is mostly due to 
extra uncertainty coming from the hadronisation corrections. 

\begin{figure}[tbp]
\vspace{-0.5cm} 
\centerline{
\epsfig{file=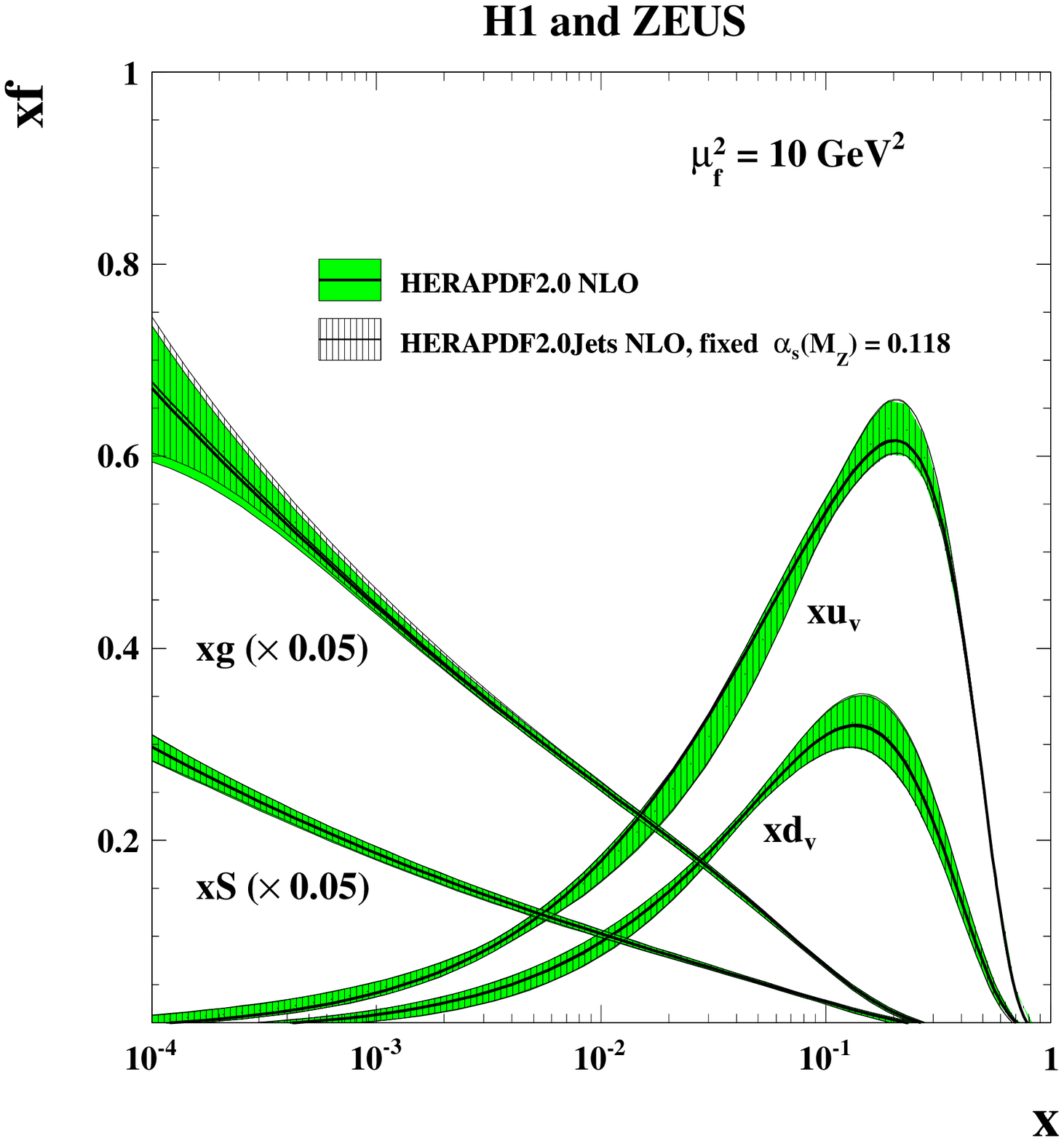,width=0.65\textwidth, height=0.44\textheight}}
\centerline{
\epsfig{file=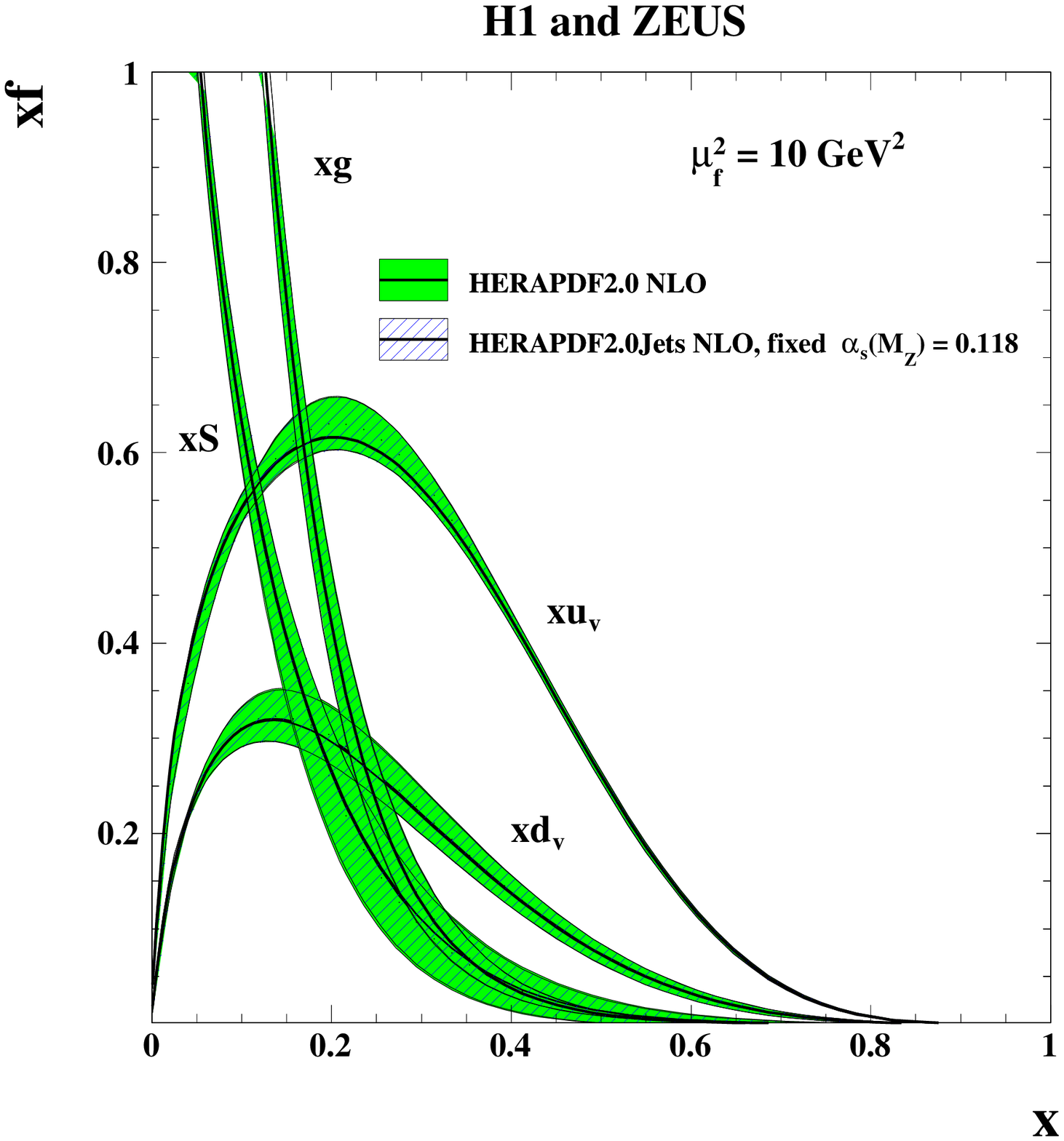,width=0.65\textwidth, height=0.44\textheight}}
\caption { 
The parton distribution functions 
$xu_v$, $xd_v$, $xS=2x(\bar{U}+\bar{D})$ and $xg$  of
HERAPDF2.0Jets NLO 
at $\mu_{\rm f}^{2} = 10\,$GeV$^{2}$ 
compared to those of HERAPDF2.0 NLO
on logarithmic (top) and linear (bottom) scales. 
The fits were done with fixed $\asmz=0.118$.
The bands represent the total uncertainties.
}
\label{fig:alfixjets}
\end{figure}

The PDFs from the HERAPDF2.0Jets fit with  $\asmz=0.118$ fixed 
are also very similar to the standard PDFs from HERAPDF2.0 NLO. 
This is demonstrated in Fig.~\ref{fig:alfixjets}.
This is again the result of the choice of $\asmz=0.118$ for HERAPDF2.0
which is also the preferred value for HERAPDF2.0Jets.
Consequently, there is only a small reduction of the uncertainty 
on the gluon distribution observed for HERAPDF2.0Jets.

The $\chi^2$ of the HERAPDF2.0Jets fit with free $\asmz$ is the same 
as for the fit with fixed $\asmz=0.118$, see Table~\ref{tab:chi2}. 
This is again due the fact that the value of $\asmz$ obtained 
from the fit is very close to
the value previously fixed. 
The 
strong coupling constant obtained is 
\begin{eqnarray}
\nonumber
\asmz =0.1183 \pm 0.0009{\rm (exp)} \pm 0.0005{\rm (model/parameterisation)} \\
\nonumber
 ~~~~~~ \pm 0.0012{ \rm (hadronisation)}~~ ^{+0.0037}_{-0.0030}{\rm (scale)}~~.
\end{eqnarray}
The uncertainty on $\asmz$ due to scale uncertainties 
was evaluated by varying the renormalisation and factorisation 
scales by a factor of two,
both separately and simultaneously, 
and taking the maximal positive and negative deviations.
The uncertainties were assumed to be 
50\,\% correlated and 50\,\% uncorrelated
between bins and data sets.
This resulted in an asymmetric uncertainty of  $+0.0037$ and $-0.0030$.
The result on $\asmz$ is compatible with the world average~\cite{PDG12} 
and 
it is competitive with other 
determinations at NLO.

\subsubsection{Comparison of HERAPDF2.0Jets to data}
\label{sec:comp:fit:jets} 

\begin{figure}[tbp]
\vspace{-0.3cm} 
\centerline{
\epsfig{file=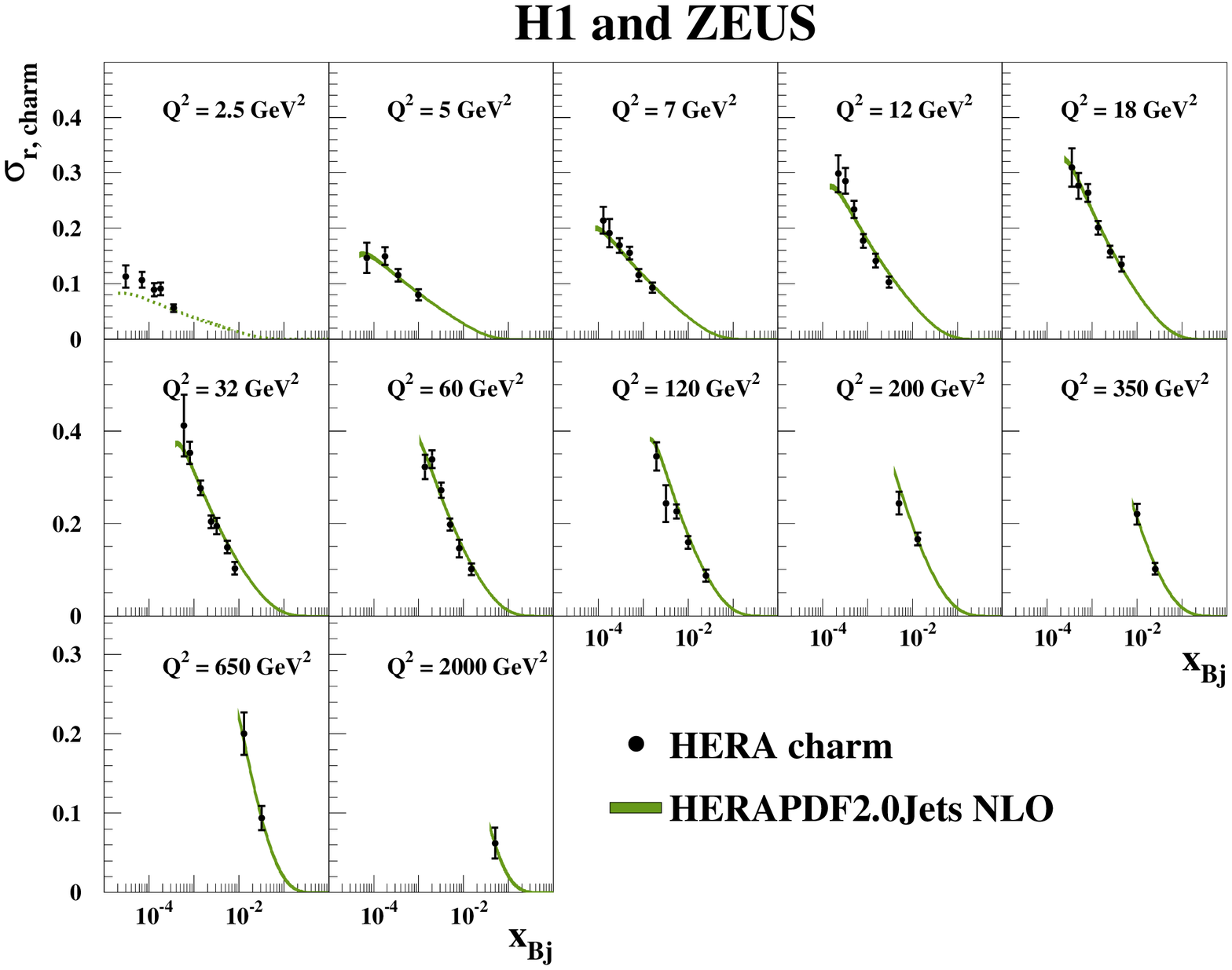 ,width=0.9\textwidth}}
\vspace{0.5cm}
\caption {The HERA reduced cross sections for charm production  
with overlaid predictions of the HERAPDF2.0Jets NLO fit.
The bands represent the total uncertainty on the predictions
excluding scale uncertainties.
Dotted lines indicate extrapolation into 
kinematic regions not included in the
fit.
}
\label{fig:charm-data}
\end{figure}

The predictions of HERAPDF2.0Jets with free $\asmz$ are shown together with the
charm input data~\cite{HERAccombi} in 
Fig.~\ref{fig:charm-data}. The description of the data
is excellent. 

\begin{figure}
  \centering
  \setlength{\unitlength}{0.1\textwidth}
  \begin{picture} (9,12)
  \put(0,0){\includegraphics[width=0.9\textwidth]{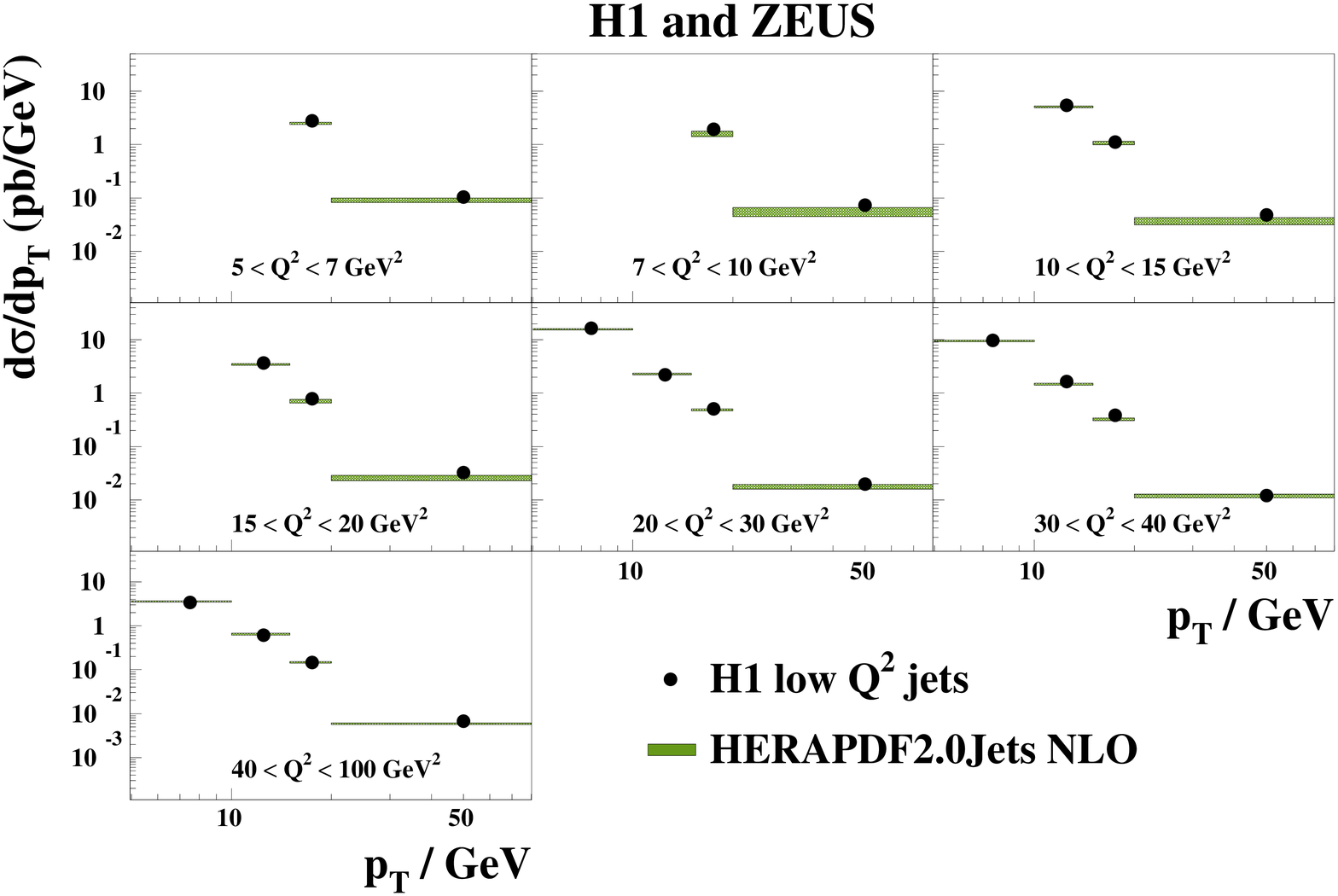}}
  \put(0,7){\includegraphics[width=0.9\textwidth]{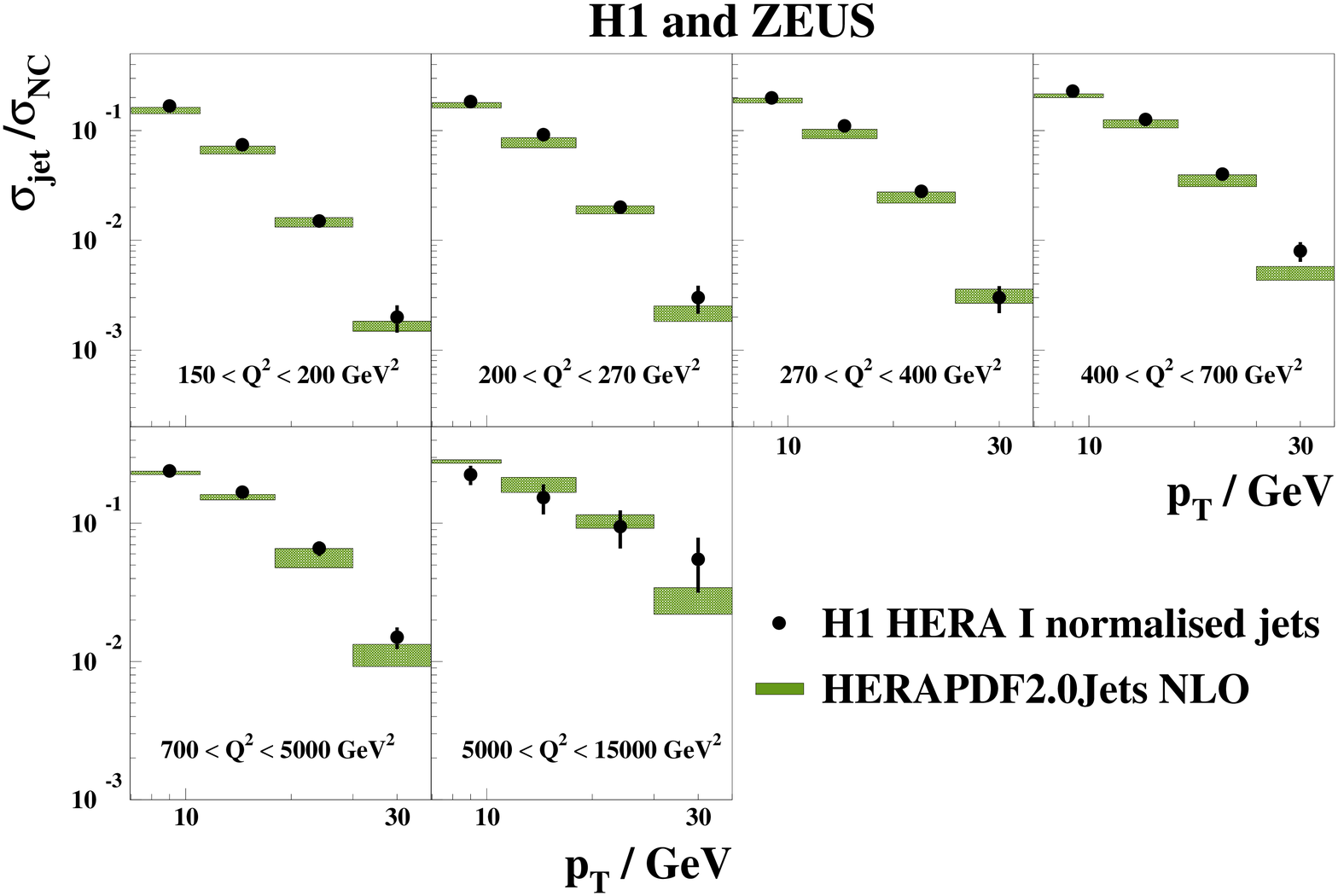}}
  \put (0.1,7.2) {a)}
  \put (0.1,0.2) {b)}
  \end{picture}
\caption{
a) Differential jet cross sections, ${\rm d}\sigma/{\rm d}p_T$, normalised to 
   NC inclusive cross sections, in bins of $Q^2$ 
   between 150 and 15000\,GeV$^2$ as measured by H1.
b) Differential jet cross sections, ${\rm d}\sigma/{\rm d}p_T$, 
   in bins of $Q^2$ between 5 and 100\,GeV$^2$ as measured by H1. 
Also shown are predictions from HERAPDF2.0Jets. 
The bands represent the total uncertainties on the predictions
excluding scale uncertainties.
}
\label{fig:h1old-jet-data}
\end{figure}

\begin{figure}
  \centering
  \setlength{\unitlength}{0.1\textwidth}
  \begin{picture} (9,11)
  \put(0,0){\includegraphics[width=0.9\textwidth]{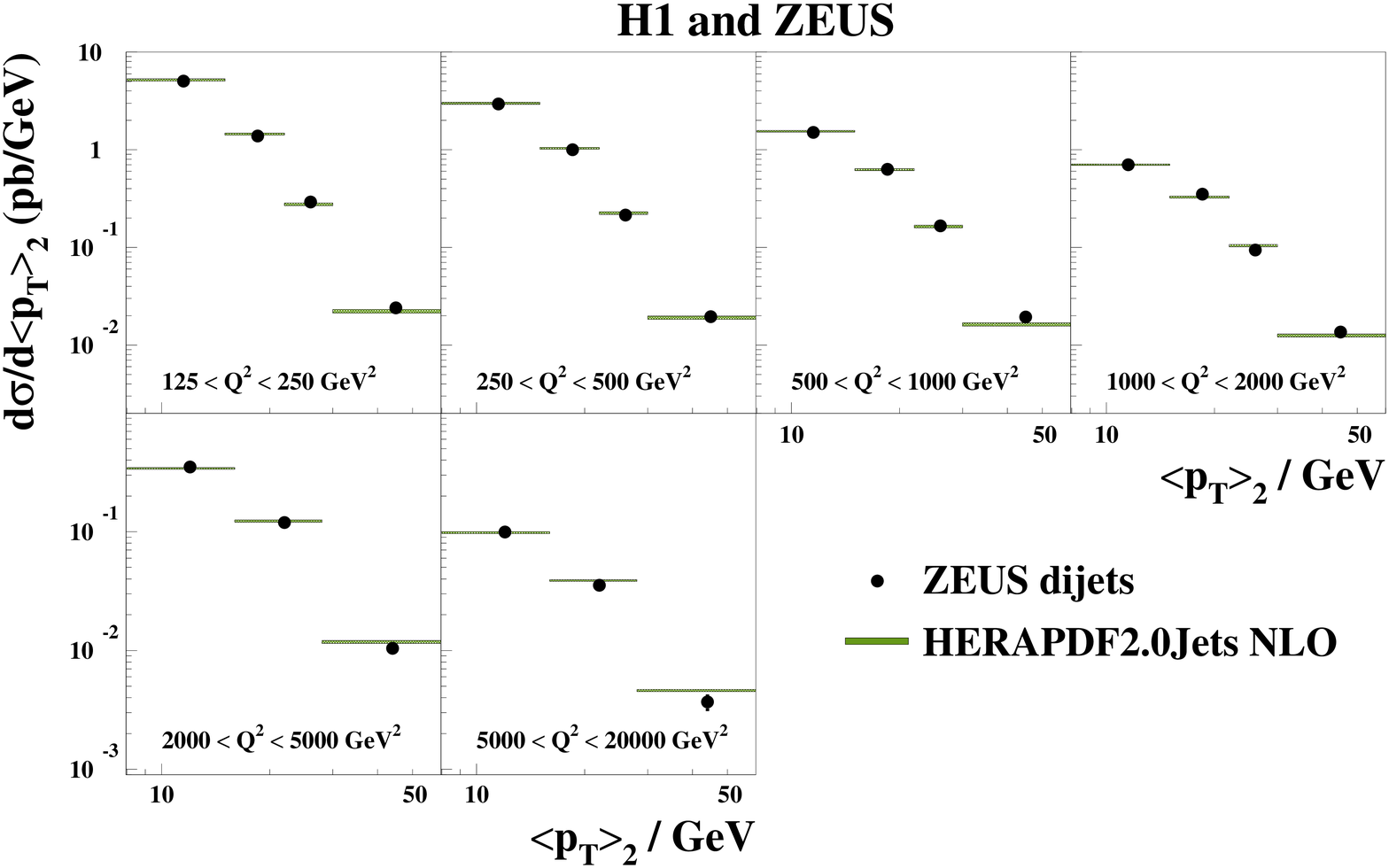}}
  \put(0,6.5){\includegraphics[width=0.9\textwidth]{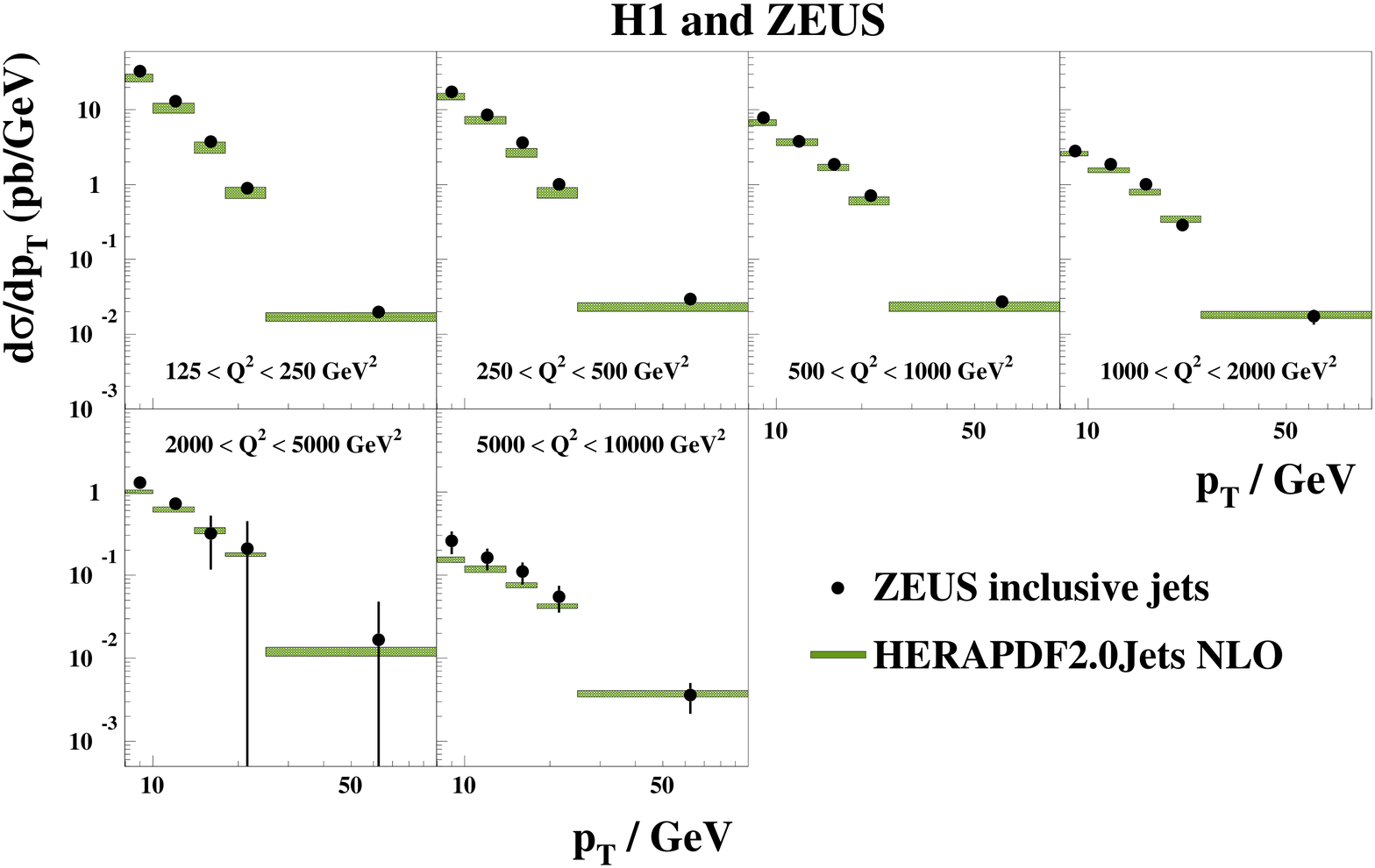}}
  \put (0.1,6.7) {a)}
  \put (0.1,0.2) {b)}
  \end{picture}
\caption {
a) Differential jet cross sections, ${\rm d}\sigma/{\rm d}p_T$, 
   in bins of $Q^2$ between 125 and 10000\,GeV$^2$ 
   as measured by ZEUS.
b) Differential dijet cross sections, $d\sigma/{\rm d} \langle p_T \rangle_2$, 
   in bins of $Q^2$ between 125 and 20000\,GeV$^2$ as measured by ZEUS.
   The variable $\langle p_T \rangle_2$ denotes the average 
   $p_T$ of the two jets. 
Also shown are predictions from HERAPDF2.0Jets. 
The bands represent the total uncertainty on the predictions
excluding scale uncertainties.
}
\label{fig:zeus-jet-data}
\end{figure}

\begin{figure}
\centerline{
\epsfig{file=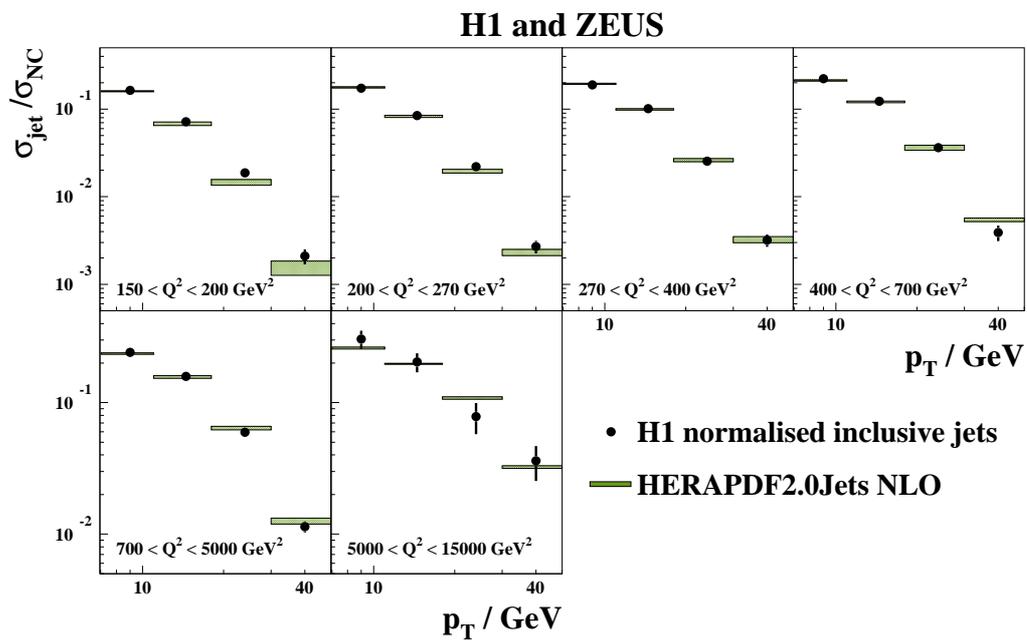 ,width=0.95\textwidth}}
\caption{
Differential jet cross sections, ${\rm d}\sigma/{\rm d}p_T$. 
All cross sections are normalised to NC inclusive cross sections. 
Also shown are predictions from HERAPDF2.0Jets. 
The bands represent the total uncertainties on the predictions
excluding scale uncertainties.
}
\label{fig:jet-data}
\end{figure}

\begin{figure}
\centerline{
\epsfig{file=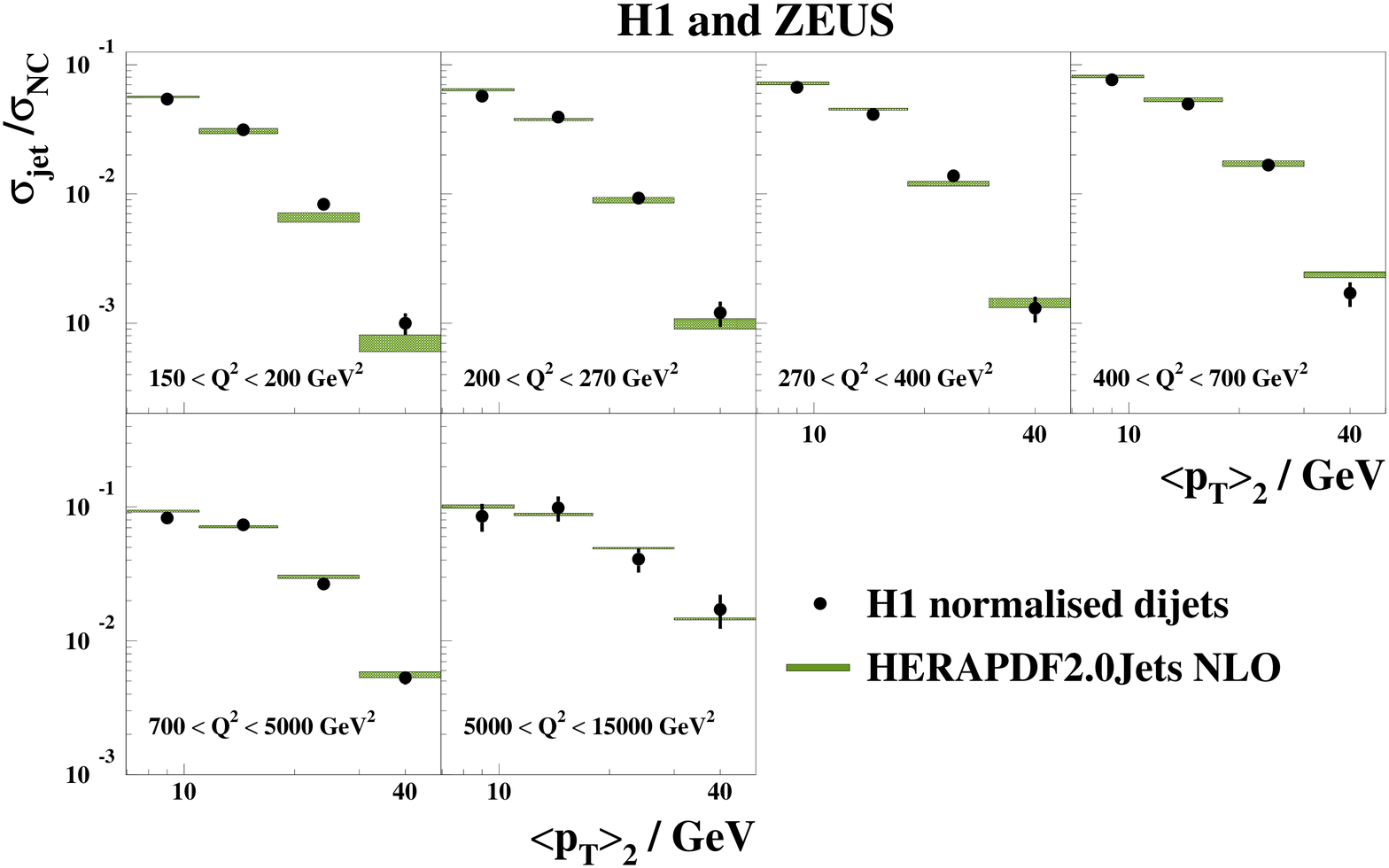 ,width=0.95\textwidth}}
\caption{
Differential dijet cross sections, 
   ${\rm d}\sigma/{\rm d}\langle p_T \rangle_2$, 
   in bins of $Q^2$ between 150 and 15000\,GeV$^2$ as measured by H1.
   The variable $\langle p_T \rangle_2$ 
   denotes the average $p_T$ of the two jets.
All cross sections are normalised to NC inclusive cross sections. 
Also shown are predictions from HERAPDF2.0Jets. 
The bands represent the total uncertainties on the predictions
excluding scale uncertainties.
}
\label{fig:jet-data2}
\end{figure}

\begin{figure}
\centerline{
\epsfig{file=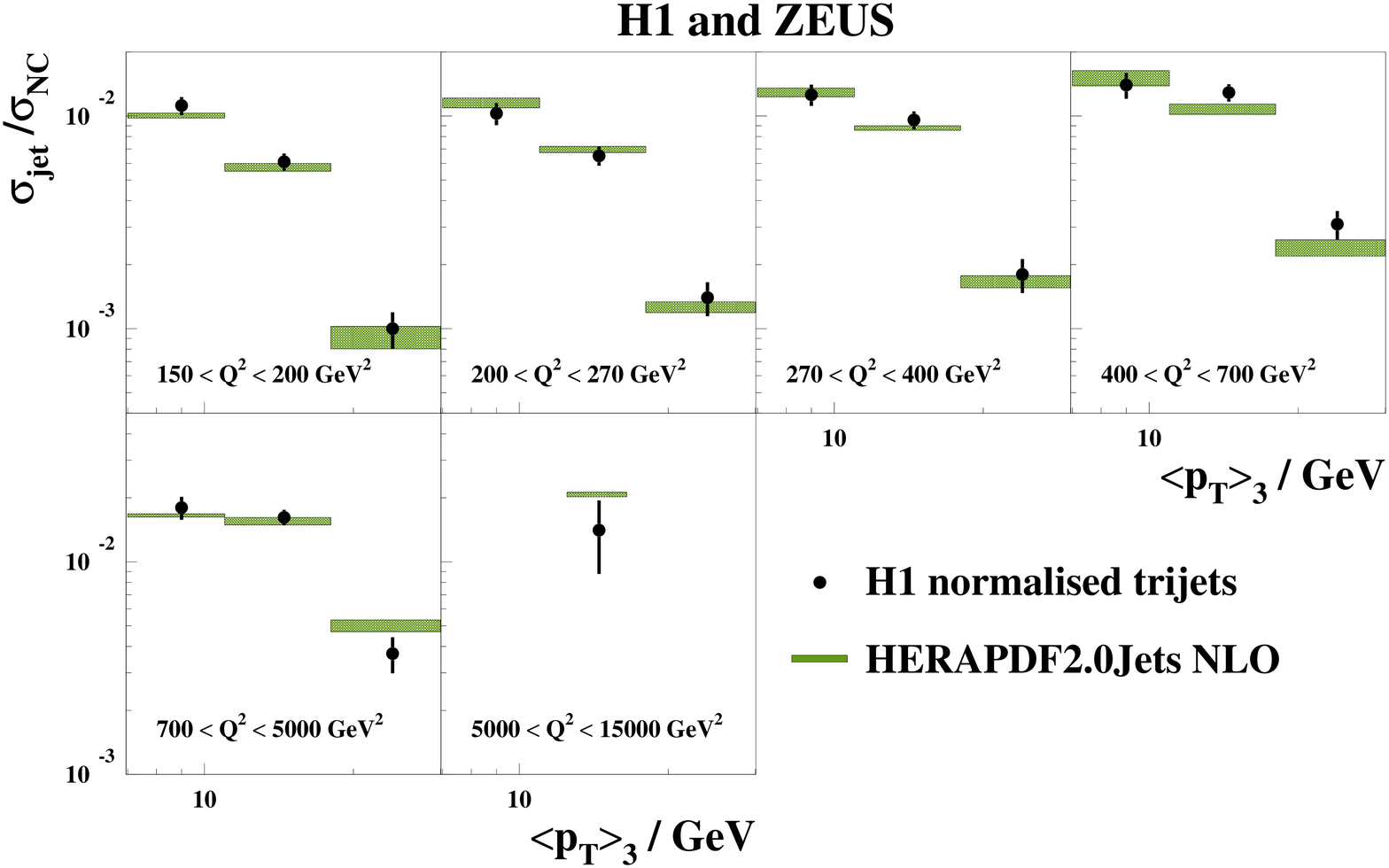 ,width=0.95\textwidth}}
\caption{
Differential trijet cross sections, 
   ${\rm d}\sigma/{\rm d} \langle p_T \rangle_3$, 
   in bins of $Q^2$ between 150 and 15000\,GeV$^2$ as measured by H1.
   The variable  $\langle p_T \rangle_3$
   denotes the average $p_T$ of the three jets.
All cross sections are normalised to NC inclusive cross sections. 
Also shown are predictions from HERAPDF2.0Jets. 
The bands represent the total uncertainties on the predictions
excluding scale uncertainties.
}
\label{fig:jet-data3}
\end{figure}

Comparisons of the predictions of HERAPDF2.0Jets
to the data on jet production used as input 
are shown in
Figs.~\ref{fig:h1old-jet-data}, \ref{fig:zeus-jet-data}
and~\ref{fig:jet-data}~--~\ref{fig:jet-data3}.
All analyses were performed using the assumption of massless 
jets, i.e. the transverse energy, $E_T$, and the transverse momentum
of a jet, $p_T$, are equivalent. 
For inclusive jet analyses, each jet is entered separately with its
$p_T$. For dijet and trijet analyses, the average 
of the transverse momenta is used as $p_T$.
These different definitions of $p_T$ were also used to set the
the renormalisation scale to
$\mu_{\rm r}^2 = (Q^2 + p_{T}^2)/2$ for calculating predictions.
The factorisation scale was chosen as 
$\mu_{\rm f}^2 = Q^2$. 
Scale uncertainties were not considered for
the comparisons to data.

Data from H1 on differential
cross sections, ${\rm d} \sigma / {\rm d} p_T$, 
at low $Q^2$~\cite{h1lowq2jets} and high $Q^2$~\cite{h1highq2oldjets}
are presented in Fig.~\ref{fig:h1old-jet-data} together with the
predictions of HERAPDF2.0Jets.
The high-$Q^2$ data are normalised to the inclusive NC cross sections.
Each event causes as many entries as there are jets.
Data from ZEUS on differential
cross-sections, ${\rm d}\sigma/{\rm d}p_T$, at high $Q^2$
for inclusive~\cite{zeus9697jets} and dijet~\cite{zeusdijets} production
are shown in Fig.~\ref{fig:zeus-jet-data} together with the predictions
of HERAPDF2.0Jets.
Finally, H1 inclusive-jet, dijet and trijet 
cross sections normalised to inclusive NC 
cross sections~\cite{h1highq2newjets} 
are presented in Figs.~\ref{fig:jet-data}~--~\ref{fig:jet-data3}.
The description of all the data on jet production by HERAPDF2.0Jets NLO
is excellent.

\section{Electroweak effects and scaling violations}
\label{sec:legplots}

The precise data and the predictions from HERAPDF2.0
were used to examine both electroweak effects and scaling violations.

\subsection{Electroweak unification}

The combined reduced cross sections were integrated to obtain the
differential cross sections ${\rm d}\sigma/{\rm d}Q^2$.
The integration over $x_{\rm Bj}$ of the double-differential
cross-sections ${\rm d}^2\sigma/{\rm d}Q^2{\rm d}x_{\rm Bj}$ was performed 
in the region $0 < y < 0.9$, using the shapes as predicted
HERAPDF2.0 NLO.
All correlated and uncorrelated uncertainties were taken into account.
The cross-sections ${\rm d}\sigma/{\rm d}Q^2$
are shown in Fig.~\ref{fig:EWuni}  
for NC and CC  $e^-p$ and $e^+p$ scattering
together with
predictions from HERAPDF2.0 NLO. 
Whereas the NC cross sections are three orders of magnitude larger 
at low $Q^2 \approx 100$\,GeV$^2$, where they are dominated 
by virtual photon exchange, 
the NC and CC cross sections become similar in magnitude 
at $Q^2 \approx 10000\,$GeV$^2$, 
i.e.\ at around the mass-scale squared of the electroweak bosons,
demonstrating the success of
electroweak unification in the Standard Model with impressive precision. 
The data also clearly demonstrate that 
the NC $e^-p$ and NC $e^+p$ cross sections 
are the same when photon exchange is dominant but they start 
to differ at  $Q^2 \approx 10000\,$GeV$^2$ when $\gamma$--$Z$ 
interference becomes important.

\subsection{The structure function {$xF_3^{\gamma Z}$}}

\begin{figure}[tbp]
\vspace{-0.3cm} 
\centerline{
\epsfig{file=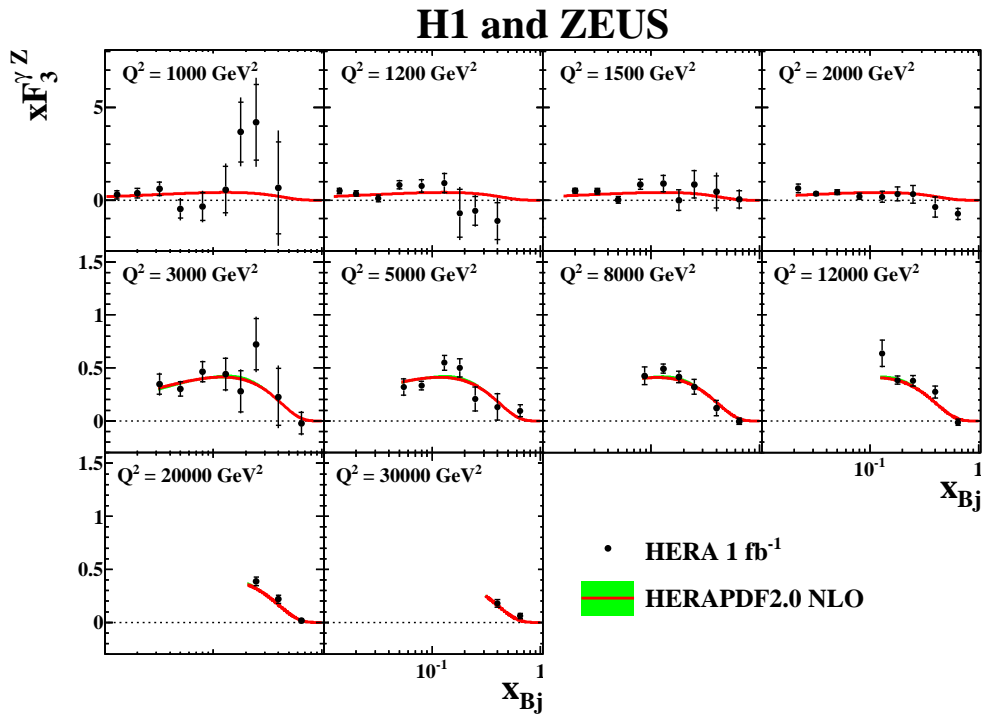 ,width=0.9\textwidth}}
\vspace{0.5cm}
\caption {The structure function $xF_3^{\gamma Z}$ for ten values of $Q^2$
together with predictions from HERAPDF2.0 NLO.
The bands represent the total uncertainties on the predictions.
}
\label{fig:xF3:2d}
\end{figure}

Figures~\ref{fig:nloQ23pt5ncemep} and~\ref{fig:nnloQ23pt5ncemep} 
show the reduced cross sections for both $e^+p$ and $e^-p$
inclusive NC scattering and predictions from HERAPDF2.0 at NLO and NNLO
as a function of $Q^2$ for selected values of $x_{\rm Bj}$.
The differences in the cross sections at high $Q^2$ are clearly visible
and well described by HERAPDF2.0, both at NLO and at NNLO.
The predictions at NNLO have slightly lower uncertainties than at NLO.
As described in Section~\ref{xsecns}, 
the structure function $xF_{3}^{\gamma Z}$ can be
extracted by subtracting the NC $e^+p$ from the NC $e^-p$ cross sections.
This directly probes the valence structure of the proton. 
Equations~\ref{strf} and~\ref{eqn:txf3} were used to obtain $xF_{3}^{\gamma Z}$
for $Q^2 \geq 1000$\,GeV$^2$. The result is shown in
Fig.~\ref{fig:xF3:2d} in bins of $Q^2$ together 
with the predictions of HERAPDF2.0 NLO. 
The values are listed in Table~\ref{tab:xF3inbins:1}.
The subtraction yields precise results above $Q^2$ of 3000\,GeV$^2$.


The valence-quark distributions and hence $xF_{3}^{\gamma Z}$ depend
only minimally on the scale, i.e.\ only small corrections
are needed to translate all values of $xF_{3}^{\gamma Z}$ to a common scale of
$1000\,$GeV$^{2}$. This was done  using HERAPDF2.0 NLO.
The translation factors were close to unity for most points. 
The largest factors
of up to 1.6 
were obtained for points at the highest $Q^2$ and $x_{\rm Bj}$ where
$xF_{3}^{\gamma Z}$ is very small.

\begin{figure}[tbp]
\vspace{-0.3cm} 
\centerline{
\epsfig{file=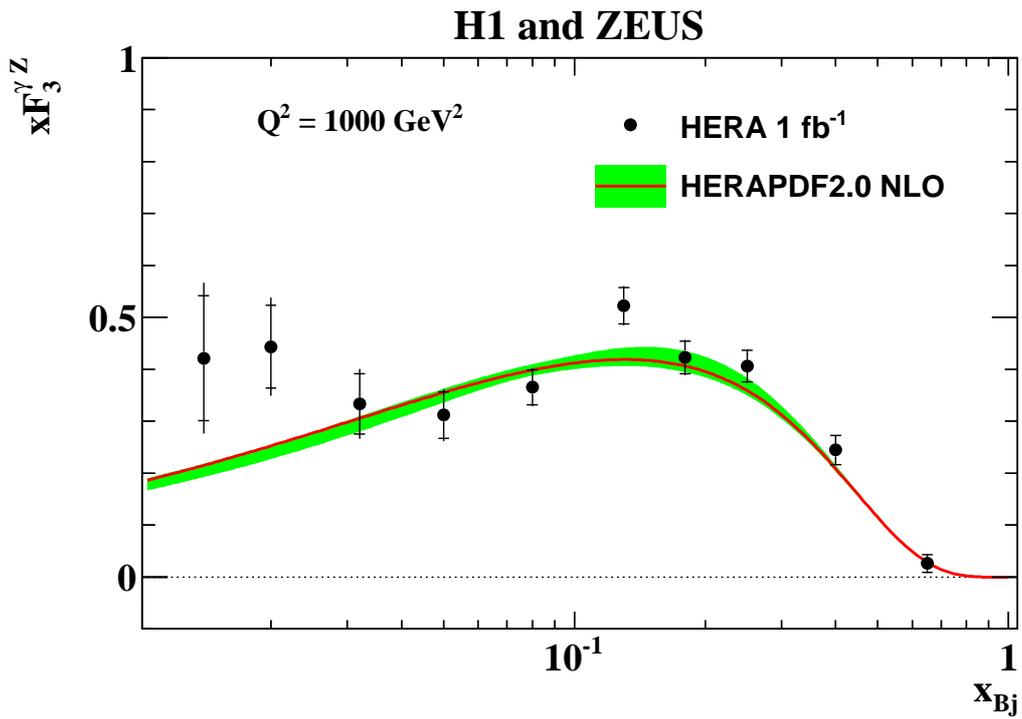 ,width=0.9\textwidth}}
\vspace{0.5cm}
\caption {The structure function $xF_3^{\gamma Z}$ averaged over
$Q^2 \ge 1000\,$GeV$^2$ at the scale $Q^2=1000\,$GeV$^{2}$ 
together with the prediction from HERAPDF2.0 NLO.
The band represents the total uncertainty on the prediction.
}
\label{fig:xF3:1d}
\end{figure}

The  translated $xF_{3}^{\gamma Z}$ values were averaged
using the method described in Section~\ref{sec:comb}.
A full covariance matrix was built using the information on 
the individual sources of uncertainty. 
The averaging of the  $xF_{3}^{\gamma Z}$ values 
has a $\chi^2 /\rm{d.o.f.} = 58.8 / 57$ 
demonstrating the consistency of the data for different
values of $Q^2$.
The result is presented in Fig.~\ref{fig:xF3:1d} together with the
prediction of HERAPDF2.0 NLO. 
The values are listed in Table~\ref{tab:xF3ave}.
The data are well described  by the HERAPDF2.0 NLO prediction.

An integration of  $F_{3}^{\gamma Z}$ was performed
using the averaged cross-section values.
For each bin, the shape prediction of HERAPDF2.0 NLO was used. 
The correlated and uncorrelated uncertainties were taken into account. 
Two intervals,  I1 : $0.016 < x_{\rm Bj} < 0.725$ and 
 I2 : $0 < x_{\rm Bj} < 1$, were considered. 
An integration of the prediction of HERAPDF2.0 NLO was also performed.
The integration was performed in bins
equidistant in log$(x_{\rm Bj})$. The integral boundaries for I1
were derived from the maximum $y$ and kinematic boundaries. 
The results are: 
\begin{eqnarray} \label{eqn:xf3int}     
 \rm{I1:~HERAPDF2.0:} 1.165^{+0.042}_{-0.053}~~~~~\rm{Data:} 1.314 \pm 0.057\rm(stat) 
                                                            \pm 0.057\rm(syst) \\
 \rm{I2:~HERAPDF2.0:} 1.588^{+0.078}_{-0.100}~~~~~\rm{Data:} 1.790 \pm 0.078\rm(stat) 
                                                            \pm 0.078\rm(syst) 
\end{eqnarray}  
The values from HERAPDF2.0 and data agree within uncertainties. 
For I2, they are also close to the QPM prediction of 5/3 from the integration
of Eq.~\ref{eqn:xf3gz_simple}.


\subsection{Helicity effects in CC interactions}

Figures~\ref{fig:scaling-CC-NLO} and~\ref{fig:scaling-CC-NNLO} present
the reduced cross sections for CC inclusive $e^+p$ and $e^-p$ scattering.
The $e^+p$ cross sections are affected strongly by the helicity factor
$(1-y)^2$, see Eq.\ref{ccupdo}. 
Therefore, the contribution of the valence quarks 
is supressed at high $y$ which translates to high $Q^2$ for fixed $x_{\rm Bj}$.
The $e^-p$ cross section is almost unaffected, because the helicity 
factor applies to the
anti-quarks which as part of the sea are already supressed at 
high $x_{\rm Bj}$.

\subsection{Scaling violations}

Scaling violations, i.e.\ the dependence of the structure
functions on $Q^2$ at fixed $x_{\rm Bj}$, are a consequence of the 
strong interactions between the partons in the nucleon.
The larger the kinematic range, the more clearly these violations 
are demonstrated.
They have been used to extract the gluon content of the proton.


Figures~\ref{fig:nloQ23pt5scal} and~\ref{fig:nnloQ23pt5scal} show
the inclusive NC $e^+p$ and $e^-p$ HERA data together 
with fixed-target data~\cite{bcdms,nmc}
and the predictions of HERAPDF2.0 NLO and NNLO, respectively.
The data presented span more than four orders of magnitude, 
both in $Q^2$ and $x_{\rm Bj}$. 
The scaling violations are clearly visible and
are well described by HERAPDF2.0, both at NLO and NNLO.
The scaling violations were also already clearly visible
in Fig.~\ref{fig:5mod}, in which a close-up for a particular
kinematic range was presented.

\begin{figure}[tbp]
\vspace{-0.3cm} 
\centerline{
\epsfig{file=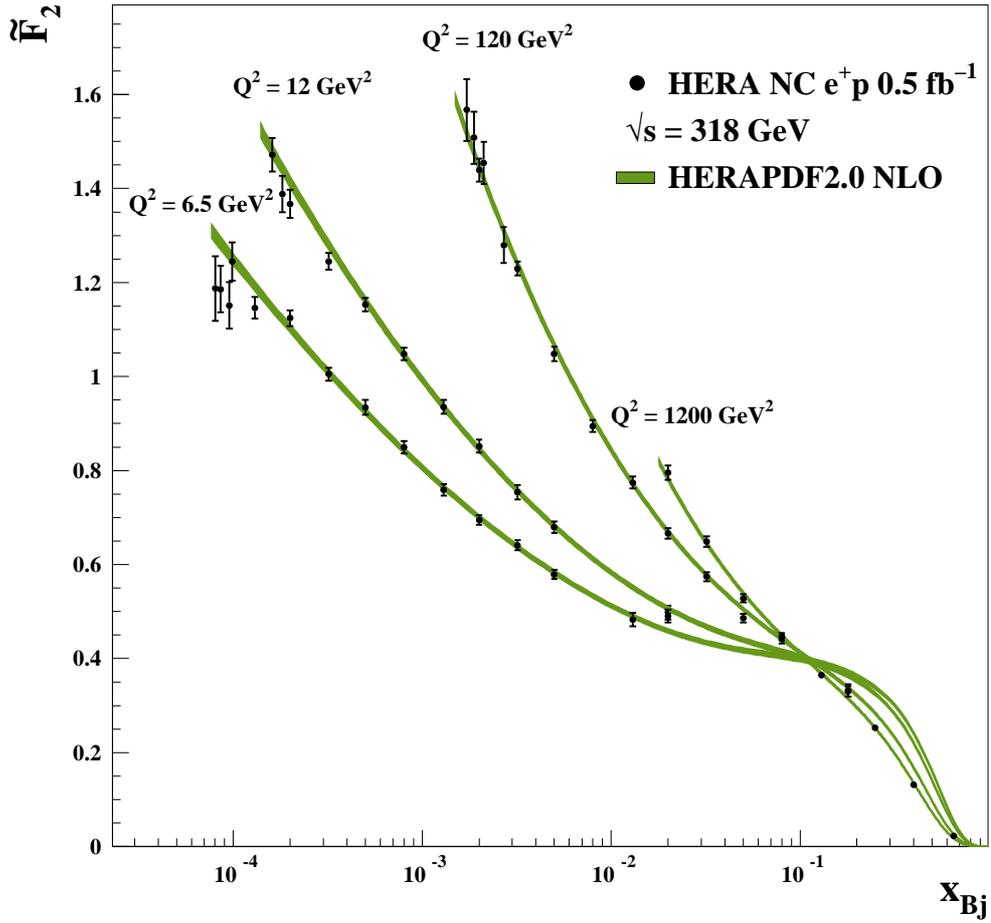 ,width=0.9\textwidth}}
\vspace{0.5cm}
\caption {The structure function $\tilde{F_2}$ 
as extracted from the measured reduced cross sections 
for four values of $Q^2$
together with the predictions of HERAPDF2.0 NLO. 
The bands represent the total uncertainty on the predictions.
}
\label{fig:f2}
\end{figure}

The structure function $\tilde{F_2}$, see Eq.~\ref{ncsi}, can be
displayed  as a function of $x_{\rm Bj}$ at fixed $Q^2$.
For the part of the phase space where both $x\tilde{F_3}$ and
$\tilde{F_{\rm L}}$ are small, the simple expression 
\begin{equation}
  \tilde{F_2} = \ncred \cdot 
                \frac{\tilde{F_2}^{\rm predicted}}{\ncred^{\rm predicted}}
                   =  \ncred \cdot (1+C_F) ~~  
\end{equation}
can be used to extract the values of $\tilde{F_2}$. 
Selected values and HERAPDF2.0 NLO predictions for $\tilde{F_2}$,
for which the correction $|C_F| < 0.1$, 
are shown in Fig.~\ref{fig:f2}.

The function $\tilde{F_2}$ rises toward low $x_{\rm Bj}$ at fixed $Q^2$.
The scaling violations manifest themselves by the rise becoming steeper  
as $Q^2$ increases. 
In the conventional framework of pQCD, this implies an increasing gluon density.
The predictions of HERAPDF2.0 NLO describe the data well. 




\section{Summary and Conclusions}
The H1 and ZEUS collaborations measured inclusive 
$e^{\pm}p$ scattering cross sections at HERA from
1994 to 2007, collecting a total integrated luminosity of about 1\,fb$^{-1}$.
The data were taken in two different beam
configurations, called HERA\,I and HERA\,II, at four different
centre-of-mass energies and with two different detectors changing and
improving over time. All inclusive data were combined to create
one consistent set of NC and CC cross-section measurements 
for unpolarised $e^{\pm}p$ scattering, spanning
six orders of magnitude in both negative four-momentum-transfer
squared, $Q^2$, and Bjorken $x$.
The data from many measurements made independently by the two
collaborations proved to be consistent with a  $\chi^2$ per
degree of freedom being 1.04 for the combination. 
Combined cross sections are provided for values of 
$Q^2$ between $Q^2=0.045$\,GeV$^2$
and $Q^2=50000$\,GeV$^2$ and values of $x_{\rm Bj}$ 
between $x_{\rm Bj}=6\times10^{-7}$ and $x_{\rm Bj}=0.65$.
They are the most precise measurements 
ever published for $ep$ scattering over such a large kinematic range 
and have been used to illustrate scaling violation.
The precision of the data has also been exploited to illustrate
electroweak unification and extract $xF_3^{\gamma Z}$
above $Q^2 = 1000\,$GeV$^2$.

The inclusive cross sections were used as input 
to a QCD analysis within
the DGLAP formalism. In order to constrain the heavy-quark
mass parameters, additional information from data on 
charm and beauty production at HERA was used.
The resulting parton distribution functions are denoted 
HERAPDF2.0 and are available at LO, NLO and NNLO.
They were calculated for a series of fixed values of 
$\asmz$ around the central
value of 0.118.
HERAPDF2.0 has small experimental uncertainties due to the
high precision and
coherence of the input data. 
Parameterisation and model uncertainties have also been estimated.
HERAPDF2.0 makes precise predictions which describe the input data well.

The heavy-flavour scheme used for HERAPDF2.0 is RTOPT, a
variable-flavour number scheme. Two variants HERAPDF2.0\,FF3A and FF3B, 
using  fixed-flavour number schemes, are also available at NLO.

The perturbative QCD fits yielding HERAPDF2.0 are based on data 
with  $Q^2$ above 3.5\,GeV$^2$. Their $\chi^2/{\rm d.o.f.}$ values are 
around 1.2. An extensive investigation included fits 
with different $Q^2_{\rm min}$, below which data were excluded.
For $Q^2_{\rm min} = 10\,$GeV$^2$, a full set of PDFs named
HERAPDF2.0HiQ2 is also released.
These fits have an improved $\chi^2/{\rm d.o.f.}$ of about 1.15.
However, the resulting PDFs  do not describe the data in the
excluded low-$Q^2$ region well.
HERAPDF2.0 shows  tensions between data and fit, independent of the 
heavy-flavour scheme used, 
at low $Q^2$, i.e.\ below $Q^2 = 15\,$GeV$^2$, and at high $Q^2$, i.e.\ above 
$Q^2 = 150\,$GeV$^2$. 
Comparisons between the behaviour
of the fits with different $Q^2_{\rm min}$ values 
indicate that the NLO theory evolves
faster than the data towards lower $Q^2$ and $x$. Fits at NNLO do not 
improve the agreement.
HERAPDF2.0 NNLO and NLO have a similar fit quality.

A measurement of $\asmz$ was made using a perturbative QCD fit for which
the inclusive cross sections were augmented
with selected jet- and charm-production cross sections 
as measured by both the H1 and ZEUS collaborations.
The value obtained is
$\asmz =0.1183 \pm 0.0009 {\rm(exp)} \pm 0.0005{\rm (model/parameterisation)} 
\pm 0.0012{\rm (hadronisation)} ^{+0.0037}_{-0.0030}{\rm (scale)}$. 
This value
is in excellent agreement with the value of the world average  
$\asmz = 0.1185$~\cite{PDG14}. 
The set of PDFs obtained from the analysis
with free $\asmz$ is released as HERAPDF2.0Jets.

The precision data on inclusive $ep$ scattering  presented in this paper
are one of the main legacies of HERA. 


\section{Acknowledgements}

We are grateful to the HERA machine group whose outstanding
efforts have made these experiments possible.
We appreciate the contributions to the construction, 
maintenance and operation of the H1 and ZEUS detectors of 
many people who are not listed as authors.
We thank our funding agencies for financial 
support, the DESY technical staff for continuous assistance and the 
DESY directorate for their support and for the hospitality they 
extended to the non-DESY members of the collaborations. 
We would like to give credit to all partners contributing to the 
EGI computing infrastructure for their support.

\clearpage
\bibliography{desy15-039}

\clearpage
\begin{table}
\begin{center}
\begin{scriptsize}
\begin{tabular}{|lr|ll|rr|c|c|c|c|c|}
\hline
\multicolumn{2}{|c|}{Data Set} &
\multicolumn{2}{|c|}{$x_{\rm Bj}$ Grid} &
\multicolumn{2}{|c|}{$Q^2 [$GeV$^2$] Grid} &
${\cal L}$ & $e^+/e^-$ & $\sqrt{s}$ & $x_{\rm Bj}$,$Q^2$ from & Ref. \\
\multicolumn{2}{|c|}{ } & from & to & from & to &
pb$^{-1}$ &  & GeV &  equations & \\
\hline
\multicolumn{11}{|l|} {HERA I $E_p=820$\,GeV and $E_p=920$\,GeV data sets} \\
\hline
H1~svx-mb\,[2] & $95$-$00$ & $0.000 005$  & $0.02$  & $0.2$ & $12$  & $2.1$ &$e^+p$   & $301$, $319$ & \ref{eq:emeth},\ref{eq:yh},\ref{eq:sigma2} &\cite{Collaboration:2009bp} \\
H1~low~$Q^2$\,[2] & $96$-$00$ & $0.000 2$  & $0.1$   & $12$  & $150$ & $22$ &$e^+p$   & $301$, $319$ & \ref{eq:emeth},\ref{eq:yh},\ref{eq:sigma2} &\cite{Collaboration:2009kv}\\
H1~NC             & $94$-$97$ & $0.0032$  &$0.65$   &$150$  &$30000$   & $35.6$  & $e^+p$ & $301$           &   \ref{eq:esigma}       & \cite{Adloff:1999ah}\\
H1~CC             & $94$-$97$ & $0.013$   &$0.40$   &$300$  &$15000$   & $35.6$  & $e^+p$ & $301$            &   \ref{yjb}     & \cite{Adloff:1999ah}\\
H1~NC             & $98$-$99$ & $0.0032$  &$0.65$   &$150$  &$30000$   & $16.4$  & $e^-p$ & $319$            &    \ref{eq:esigma}     & \cite{Adloff:2000qj}\\
H1~CC             & $98$-$99$ & $0.013$   &$0.40$   &$300$  &$15000$   & $16.4$  & $e^-p$ & $319$             &  \ref{yjb}       & \cite{Adloff:2000qj}\\
H1~NC HY          & $98$-$99$ & $0.0013$  &$0.01$   &$100$  &$800$   & $16.4$  & $e^-p$ & $319$               &  \ref{eq:emeth}     & \cite{Adloff:2003uh}\\
H1~NC             & $99$-$00$ & $0.0013$ &$0.65$   &$100$  &$30000$   & $65.2$  & $e^+p$ & $319$              &  \ref{eq:esigma}  & \cite{Adloff:2003uh}\\
H1~CC             & $99$-$00$ & $0.013$   &$0.40$   &$300$  &$15000$   & $65.2$  & $e^+p$ & $319$              &   \ref{yjb}    & \cite{Adloff:2003uh} \\
\hline
ZEUS~BPC            & $95$ & $0.000 002$    & $0.000 06$ &$0.11$  & $0.65$   & $1.65$  &  $e^+p$  & $300$ & \ref{eq:emeth} &  \cite{Breitweg:1997hz} \\
ZEUS~BPT            & $97$ & $0.000 000 6$    & $0.001$   & $0.045$ & $0.65$  & $3.9$  & $e^+p$   &  $300$  & \ref{eq:emeth}, \ref{eq:esigma} &     \cite{Breitweg:2000yn}\\
ZEUS~SVX            & $95$ & $0.000 012$  & $0.0019$  & $0.6$ & $17$  & $0.2$  & $e^+p$   &  $300$    &  \ref{eq:emeth} & \cite{Breitweg:1998dz}\\
ZEUS~NC\,[2]\,high/low\,$Q^2 $ & $96$-$97$ & $0.000 06$  &$0.65$& $2.7$ & $30000$  & $30.0$  & $e^+p$    & $300$     & \ref{eq:ptmeth}  & \cite{Chekanov:2001qu}\\
ZEUS~CC             & $94$-$97$ & $0.015$  & $0.42$  & $280$  & $17000$   &$47.7$   &  $e^+p$   & $300$         &  \ref{yjb}   & \cite{zeuscc97}\\
ZEUS~NC             & $98$-$99$ & $0.005$  & $0.65$  & $200$  & $30000$  &$15.9$   & $e^-p$    & $318$          &  \ref{qxda} & \cite{Chekanov:2002ej} \\
ZEUS~CC             & $98$-$99$ & $0.015$  & $0.42$  & $280$  & $30000$  &$16.4$   & $e^-p$    & $318$          &  \ref{yjb}   & \cite{Chekanov:2002zs} \\
ZEUS~NC             & $99$-$00$ & $0.005$  & $0.65$  & $200$  & $30000$  &$63.2$   & $e^+p$   & $318$         & \ref{qxda}   & \cite{Chekanov:2003yv} \\
ZEUS~CC             & $99$-$00$ & $0.008$  & $0.42$  & $280$  & $17000$  &$60.9$   &  $e^+p$   &$318$           &  \ref{yjb}  & \cite{Chekanov:2003vw}\\
\hline
\multicolumn{11}{|l|} {HERA II $E_p=920$\,GeV data sets} \\ \hline
H1~NC~~$^{1.5p}$    & $03$-$07$ & $0.0008$  &$0.65$   &$60$   &$30000$   & $182$  & $e^+p$ & $319$           &  \ref{eq:emeth}, \ref{eq:esigma}       & \cite{H1allhQ2}$^1$\\
H1~CC~~$^{1.5p}$             & $03$-$07$ & $0.008$   &$0.40$   &$300$  &$15000$   & $182$  & $e^+p$ & $319$           &    \ref{yjb}     & \cite{H1allhQ2}$^1$\\
H1~NC~~$^{1.5p}$              & $03$-$07$ & $0.0008$  &$0.65$   &$60$   &$50000$   & $151.7$  & $e^-p$ & $319$           &   \ref{eq:emeth}, \ref{eq:esigma}      & \cite{H1allhQ2}$^1$\\
H1~CC~~$^{1.5p}$              & $03$-$07$ & $0.008$   &$0.40$   &$300$  &$30000$   & $151.7$  & $e^-p$ & $319$           &   \ref{yjb}      & \cite{H1allhQ2}$^1$\\
H1~NC med $Q^2$~~$^{*y.5}$  & $03$-$07$ & $0.000 0986$   &$0.005$   &$8.5$   &$90$   & $ 97.6$  & $e^+p$ & $319$   &   \ref{eq:emeth}      & \cite{H1FL2}\\
H1~NC low $Q^2 $~~$^{*y.5}$  & $03$-$07$ & $0.000 029$   &$0.000 32$   &$2.5$   &$12$   & $5.9 $  & $e^+p$ & $319$           &  \ref{eq:emeth}       & \cite{H1FL2}\\

\hline
ZEUS~NC             & $06$-$07$ & $0.005$   &$0.65$   &$200$   &$30000$   & $ 135.5$  & $e^+p$ & $318$           & \ref{eq:emeth},\ref{yjb},\ref{qxda}    & \cite{ZEUS2NCp}\\
ZEUS~CC~~$^{1.5p}$               & $06$-$07$ & $0.0078$   &$0.42$   &$280$   &$30000$   & $ 132$    & $e^+p$ & $318$           &   \ref{yjb}     & \cite{ZEUS2CCp}\\
ZEUS~NC~~$^{1.5}$              & $05$-$06$ & $0.005$  &$0.65$   &$200$   &$30000$   & $ 169.9$  & $e^-p$ & $318$           &  \ref{qxda}      & \cite{ZEUS2NCe}\\
ZEUS~CC~~$^{1.5}$              & $04$-$06$ & $0.015$   &$0.65$   &$280$   &$30000$   & $ 175$    & $e^-p$ & $318$           &   \ref{yjb}     & \cite{ZEUS2CCe}\\
ZEUS~NC nominal~~$^{*y}$     & $06$-$07$ & $0.000092$   &$0.008343$   &$7$   &$110$   & $ 44.5$  & $e^+p$ & $318$           &  \ref{eq:emeth}    & \cite{ZEUSFL}\\
ZEUS~NC satellite~~$^{*y}$    & $06$-$07$ & $0.000071$   &$0.008343$   &$5$   &$110$   & $ 44.5$  & $e^+p$ & $318$           &  \ref{eq:emeth}  & \cite{ZEUSFL} \\\hline
\multicolumn{11}{|l|} {HERA II $E_p=575$\,GeV data sets} \\ \hline
H1~NC high $Q^2 $          & $07$ & $0.00065$   &$0.65$   &$35$   &$800$   & $ 5.4$  & $e^+p$ & $252$           &  \ref{eq:emeth}, \ref{eq:esigma}       & \cite{H1FL1}\\
H1~NC low $Q^2 $           & $07$ & $0.000 0279$   &$0.0148$   &$1.5$   &$90$   & $5.9 $  & $e^+p$ & $252$           &  \ref{eq:emeth}       & \cite{H1FL2}\\
\hline
ZEUS~NC nominal            & $07$ & $0.000147$   &$0.013349$   &$7$   &$110$   & $ 7.1$  & $e^+p$ & $251$           &   \ref{eq:emeth}    & \cite{ZEUSFL}\\
ZEUS~NC satellite          & $07$ & $0.000125$   &$0.013349$   &$5$   &$110$   & $ 7.1$  & $e^+p$ & $251$           &  \ref{eq:emeth}     & \cite{ZEUSFL}\\
\hline
\multicolumn{11}{|l|} {HERA II $E_p=460$\,GeV data sets} \\ \hline
H1~NC high $Q^2 $          & $07$ & $0.00081$   &$0.65$   &$35$   &$800$   & $ 11.8$  & $e^+p$ & $225$           &   \ref{eq:emeth}, \ref{eq:esigma}      & \cite{H1FL1}\\
H1~NC low $Q^2 $           & $07$ & $0.000 0348$   &$0.0148$   &$1.5$   &$90$   & $12.2$  & $e^+p$ & $225$    &    \ref{eq:emeth}     & \cite{H1FL2}\\
\hline
ZEUS~NC nominal            & $07$ & $0.000184$   &$0.016686$   &$7$   &$110$   & $ 13.9$  & $e^+p$ & $225$        &   \ref{eq:emeth}     & \cite{ZEUSFL}\\
ZEUS~NC satellite          & $07$ & $0.000143$   &$0.016686$   &$5$   &$110$   & $ 13.9$  & $e^+p$ & $225$           &  \ref{eq:emeth}  & \cite{ZEUSFL}\\
\hline
\end{tabular}
\end{scriptsize}
\end{center}
\caption{\label{tab:data}The 41 data sets from H1 and ZEUS  
used for the combination.
The marker [2] in the column ``Data Set'' indicates that the
data are treated as two data sets in the analysis.
The markers $^{1.5p}$ and $^{1.5}$ in the column ``Data Set''
indicate that the data were already used for HERAPDF1.5, see Appendix~\ref{appendix:A}.
The $p$ in  $^{1.5p}$ denotes that the cross-sections measurements were
preliminary at that time.
The markers $^{*y.5}$ and $^{*y}$ in the column ``Data Set'' are explained 
in Section~\ref{subsec:extrapol}.
The marker $^1$ for \cite{H1allhQ2} indicates that published cross section
were scaled by a factor of 1.018~\cite{H1lumi2}.
Integrated luminosities are quoted as given by the collaborations.
The equations used for the reconstruction of $x_{\rm Bj}$ and $Q^2$  
are given in Section~\ref{diskine}. 
}
\end{table}

\clearpage
\newcolumntype{e}{D{.}{.}{1}}

\begin{table}[tbp]
\renewcommand*{\arraystretch}{1.2}
\centerline{
\begin{tabular}{|l|c|c|c|}
\hline
\multicolumn{1}{|c|}{Variation} &
\multicolumn{1}{c|}{Standard Value} &
\multicolumn{1}{c|}{Lower Limit} &
\multicolumn{1}{c|}{Upper Limit}  \\
\hline
$Q^2_{\rm min}$ [GeV$^2$] & $~~3.5$  & $2.5$   & $5.0$~~~~   \\
$Q^2_{\rm min}$ [GeV$^2$] HiQ2 & $~10.0$  & $7.5$   & $12.5$~~~~   \\
\hline
$M_c$(NLO) [GeV]     & $1.47$    & $1.41$        & $1.53$~~~~   \\
$M_c$ (NNLO) [GeV]     & $1.43$    & $1.37$      & $1.49$~~~~   \\
$M_b$ [GeV]     & $4.5$~~    & $4.25$ & $4.75$~~~~  \\
\hline
$f_s$           & $0.4$      & $0.3$ & $0.5$~~~~   \\
\hline
$\asmz$         & 0.118 & -- & -- \\
\hline
$\mu_{f_{0}}$ [GeV]            & 1.9      & 1.6 & 2.2~~~~   \\
\hline
\end{tabular}}
\caption{Input parameters for HERAPDF2.0 fits and the variations 
considered to evaluate model and parameterisation ($\mu_{f_{0}}$)
uncertainties.  
}
\label{tab:model}
\end{table}

\begin{table}[htbp]
\renewcommand*{\arraystretch}{1.2}
\begin{center}
\begin{tabular}{|l|l|l|l|l|}
\hline
 scheme & $\asmz$ & $F_{\rm L}$ & $m_c$ [GeV]& $m_b$ [GeV] \\
\hline
FF3A & $\alpha_s^{N_F=3} = 0.106375$ & ${\cal O}(\alpha_s^2)$ &
       $m_c^{\rm pole} = 1.44$ & $m_b^{\rm pole} = 4.5$ \\ 
FF3B & $\alpha_s^{N_F=5} = 0.118$ & ${\cal O}(\alpha_s)$   &
       $m_c(m_c) = 1.26$  & $m_b(m_b) = 4.07$  \\
\hline
\end{tabular}
\end{center}
\caption{Input parameters for HERAPDF2.0FF fits.
All other parameters were set
as for the standard HERAPDF2.0 NLO fit.}
\label{tab:FF}
\end{table}

\begin{table} [h]
\renewcommand*{\arraystretch}{1.2}

\centerline{
\begin{tabular}{|l|r|c|c|c|}
\hline
HERAPDF & $Q^2_{\rm min}$[GeV$^2$] & $\chi^2$ & d.o.f. & $\chi^2$/d.o.f \\
\hline
2.0 NLO  & 3.5~~    &    1357 & 1131 & 1.200\\
2.0HiQ2 NLO &10.0~~    &    1156 & 1002 & 1.154\\
\hline
2.0 NNLO & 3.5~~    &    1363 & 1131 & 1.205\\
2.0HiQ2 NNLO & 10.0~~   &    1146 & 1002 & 1.144\\
\hline
2.0 AG NLO  & 3.5~~    &    1359 & 1132 & 1.201\\
2.0HiQ2 AG NLO         &10.0~~    &    1161 & 1003 & 1.158\\
\hline
2.0 AG NNLO & 3.5~~    &    1385 & 1132 & 1.223\\
2.0HiQ2 AG NNLO        & 10.0~~   &    1175 & 1003 & 1.171\\
\hline
2.0 NLO FF3A & 3.5~~    &    1351 & 1131 & 1.195\\
2.0 NLO FF3B & 3.5~~    &    1315 & 1131 & 1.163\\
\hline
2.0Jets $\asmz$ fixed & 3.5~~    & 1568 & 1340 & 1.170\\
2.0Jets $\asmz$ free  & 3.5~~    & 1568 & 1339 & 1.171\\
\hline
\end{tabular}}
\caption{The values of $\chi^2$ per degree of freedom for HERAPDF2.0 and its variants.}
\label{tab:chi2}
\end{table}
\clearpage

\begin{table}[tbp]
\renewcommand*{\arraystretch}{1.2}
\centerline{
\begin{tabular}{|l|l|r|l|l|l|l|l|}
\hline
\multicolumn{1}{|c|}{ } &
\multicolumn{1}{c|}{~~~$A$~~~} & 
\multicolumn{1}{c|}{~~~$B$~~~} &
\multicolumn{1}{c|}{~~~$C$~~~} &
\multicolumn{1}{c|}{~~~$D$~~~} &
\multicolumn{1}{c|}{~~~$E$~~~} &
\multicolumn{1}{c|}{~~~$A'$~~~}&
\multicolumn{1}{c|}{~~~$B'$~~~}  \\
\hline
 $xg$      & 4.34   & $-$0.015  & 9.11 &  &   & 1.048 & $-$0.167\\
$xu_v$     & 4.07   &  0.714  & 4.84 &  & 13.4 &&\\
$xd_v$     & 3.15   &  0.806  & 4.08 &  &   &&\\
$x\bar{U}$ & 0.105 & $-$0.172 & 8.06 & 11.9 &    &&\\
$x\bar{D}$ & 0.176 & $-$0.172 & 4.88 &   &  &&\\
\hline
\end{tabular}}
\caption{Central values of the HERAPDF2.0 parameters at NLO.}
\label{tab:param}\end{table}

\begin{table}[tbp]
\renewcommand*{\arraystretch}{1.2}
\centerline{
\begin{tabular}{|l|l|r|l|l|l|l|l|}
\hline
\multicolumn{1}{|c|}{ } &
\multicolumn{1}{c|}{~~~$A$~~~} & 
\multicolumn{1}{c|}{~~~$B$~~~} &
\multicolumn{1}{c|}{~~~$C$~~~} &
\multicolumn{1}{c|}{~~~$D$~~~} &
\multicolumn{1}{c|}{~~~$E$~~~} &
\multicolumn{1}{c|}{~~~$A'$~~~}&
\multicolumn{1}{c|}{~~~$B'$~~~}  \\
\hline
 $xg$      & 2.27   & $-$0.062  & 5.56 & &    & 0.167 & $-$0.383\\
$xu_v$     & 5.55   &  0.811  & 4.82 & & 9.92 &&\\
$xd_v$     & 6.29   &  1.03~~  & 4.85 &   &  &&\\
$x\bar{U}$ & 0.161 & $-$0.127 & 7.09 & 1.58&    &&\\
$x\bar{D}$ & 0.269 & $-$0.127 & 9.58 &  &   &&\\
\hline
\end{tabular}}
\caption{Central values of the HERAPDF2.0 parameters at NNLO.}
\label{tab:nnloparam}\end{table}

\clearpage
\clearpage
\begin{table}[tbp]
\renewcommand\arraystretch{1.2}
\centerline{
\begin{tabular}{c c r r r r}
\hline
\hline
$Q^2$ & $x_{\rm Bj}$ & $xF_3^{\gamma Z}$ & $\delta_{\rm{stat}}$ & $\delta_{\rm{syst}}$ &$\delta_{\rm{tot}}$\\
$\rm{GeV}^2$ & & & &  &  \\
\hline
1000  &  0.013  &  0.293  &  0.227  &  0.144  &  0.269 \\
1000  &  0.020  &  0.378  &  0.254  &  0.141  &  0.290 \\
1000  &  0.032  &  0.619  &  0.357  &  0.214  &  0.416 \\
1000  &  0.050  &  $-$0.472  &  $-$0.500  &  $-$0.341  &  $-$0.606 \\
1000  &  0.080  &  $-$0.342  &  $-$0.760  &  $-$0.396  &  $-$0.857 \\
1000  &  0.130  &  0.567  &  1.256  &  0.650  &  1.415 \\
1000  &  0.180  &  3.669  &  1.622  &  0.903  &  1.853 \\
1000  &  0.250  &  4.189  &  2.044  &  1.265  &  2.404 \\
1000  &  0.400  &  0.657  &  2.477  &  1.886  &  3.113 \\
1200  &  0.014  &  0.497  &  0.142  &  0.107  &  0.178 \\
1200  &  0.020  &  0.362  &  0.137  &  0.087  &  0.162 \\
1200  &  0.032  &  0.089  &  0.178  &  0.107  &  0.208 \\
1200  &  0.050  &  0.826  &  0.227  &  0.139  &  0.266 \\
1200  &  0.080  &  0.763  &  0.329  &  0.192  &  0.382 \\
1200  &  0.130  &  0.919  &  0.509  &  0.261  &  0.573 \\
1200  &  0.180  &  $-$0.709  &  $-$1.288  &  $-$0.618  &  $-$1.429 \\
1200  &  0.250  &  $-$0.574  &  $-$0.763  &  $-$0.377  &  $-$0.851 \\
1200  &  0.400  &  $-$1.128  &  $-$0.996  &  $-$0.767  &  $-$1.258 \\
1500  &  0.020  &  0.511  &  0.121  &  0.081  &  0.146 \\
1500  &  0.032  &  0.487  &  0.149  &  0.070  &  0.164 \\
1500  &  0.050  &  0.009  &  0.193  &  0.100  &  0.218 \\
1500  &  0.080  &  0.852  &  0.268  &  0.135  &  0.300 \\
1500  &  0.130  &  0.897  &  0.443  &  0.187  &  0.481 \\
1500  &  0.180  &  $-$0.001  &  $-$1.013  &  $-$0.407  &  $-$1.092 \\
1500  &  0.250  &  0.855  &  0.725  &  0.300  &  0.785 \\
1500  &  0.400  &  0.444  &  0.871  &  0.583  &  1.048 \\
1500  &  0.650  &  0.042  &  0.456  &  0.267  &  0.528 \\
2000  &  0.022  &  0.630  &  0.234  &  0.103  &  0.255 \\
2000  &  0.032  &  0.340  &  0.103  &  0.055  &  0.116 \\
2000  &  0.050  &  0.426  &  0.134  &  0.055  &  0.145 \\
2000  &  0.080  &  0.211  &  0.180  &  0.078  &  0.196 \\
2000  &  0.130  &  0.181  &  0.296  &  0.110  &  0.315 \\
2000  &  0.180  &  0.335  &  0.374  &  0.142  &  0.400 \\
2000  &  0.250  &  0.316  &  0.483  &  0.179  &  0.515 \\
2000  &  0.400  &  $-$0.371  &  $-$0.542  &  $-$0.236  &  $-$0.591 \\
2000  &  0.650  &  $-$0.739  &  $-$0.296  &  $-$0.166  &  $-$0.340 \\
\hline
\hline
\end{tabular}}
\caption{Structure function $xF_3^{\gamma Z}$ for different values of $Q^2$ and $x_{\rm Bj}$;
$\delta_{\rm stat}$, $\delta_{\rm syst}$ and $\delta_{\rm tot}$ represent the statistical, systematic and total uncertainties, respectively.
}
\label{tab:xF3inbins:1}
\end{table}
\clearpage

\begin{table}[tbp]
\renewcommand\arraystretch{1.2}
\centerline{
\begin{tabular}{c c r r r r}
\hline
\hline
$Q^2$ & $x_{\rm Bj}$ & $xF_3^{\gamma Z}$ & $\delta_{\rm{stat}}$ & $\delta_{\rm{syst}}$ &$\delta_{\rm{tot}}$\\
$\rm{GeV}^2$ & & & &  &  \\
\hline
3000  &  0.032  &  0.347  &  0.096  &  0.049  &  0.108 \\
3000  &  0.050  &  0.303  &  0.068  &  0.033  &  0.075 \\
3000  &  0.080  &  0.463  &  0.095  &  0.041  &  0.104 \\
3000  &  0.130  &  0.440  &  0.150  &  0.059  &  0.161 \\
3000  &  0.180  &  0.279  &  0.194  &  0.073  &  0.208 \\
3000  &  0.250  &  0.723  &  0.241  &  0.102  &  0.262 \\
3000  &  0.400  &  0.227  &  0.268  &  0.128  &  0.297 \\
3000  &  0.650  &  $-$0.022  &  $-$0.106  &  $-$0.053  &  $-$0.118 \\
5000  &  0.055  &  0.320  &  0.078  &  0.033  &  0.084 \\
5000  &  0.080  &  0.333  &  0.041  &  0.019  &  0.045 \\
5000  &  0.130  &  0.548  &  0.072  &  0.027  &  0.077 \\
5000  &  0.180  &  0.500  &  0.087  &  0.030  &  0.092 \\
5000  &  0.250  &  0.207  &  0.115  &  0.035  &  0.120 \\
5000  &  0.400  &  0.132  &  0.124  &  0.046  &  0.132 \\
5000  &  0.650  &  0.096  &  0.055  &  0.027  &  0.062 \\
8000  &  0.087  &  0.425  &  0.084  &  0.029  &  0.089 \\
8000  &  0.130  &  0.493  &  0.043  &  0.016  &  0.046 \\
8000  &  0.180  &  0.415  &  0.056  &  0.018  &  0.059 \\
8000  &  0.250  &  0.321  &  0.070  &  0.022  &  0.074 \\
8000  &  0.400  &  0.120  &  0.072  &  0.025  &  0.077 \\
8000  &  0.650  &  $-$0.004  &  $-$0.031  &  $-$0.013  &  $-$0.034 \\
12000  &  0.130  &  0.637  &  0.125  &  0.038  &  0.131 \\
12000  &  0.180  &  0.385  &  0.040  &  0.013  &  0.042 \\
12000  &  0.250  &  0.379  &  0.049  &  0.013  &  0.050 \\
12000  &  0.400  &  0.272  &  0.056  &  0.019  &  0.059 \\
12000  &  0.650  &  $-$0.012  &  $-$0.027  &  $-$0.009  &  $-$0.028 \\
20000  &  0.250  &  0.388  &  0.040  &  0.013  &  0.042 \\
20000  &  0.400  &  0.218  &  0.040  &  0.012  &  0.041 \\
20000  &  0.650  &  0.016  &  0.019  &  0.009  &  0.021 \\
30000  &  0.400  &  0.178  &  0.036  &  0.008  &  0.037 \\
30000  &  0.650  &  0.060  &  0.025  &  0.009  &  0.026 \\
\hline
\hline
\end{tabular}}
\captcont{Continued.}
\end{table}

\clearpage
\begin{table}[tbp]
\begin{center}

\renewcommand\arraystretch{1.2}
\centerline{
\begin{tabular}{c c c c c c}
\hline
\hline
$Q^2$ & $x_{\rm Bj}$ & $xF_3^{\gamma Z}$ & $\delta_{\rm{stat}}$ & $\delta_{\rm{syst}}$ &$\delta_{\rm{tot}}$\\
$\rm{GeV}^2$ & & & & & \\
\hline
1000  &  0.014  &  0.422  &  0.120  &  0.082  &  0.146 \\
1000  &  0.020  &  0.443  &  0.080  &  0.051  &  0.094 \\
1000  &  0.032  &  0.334  &  0.058  &  0.034  &  0.067 \\
1000  &  0.050  &  0.312  &  0.045  &  0.023  &  0.050 \\
1000  &  0.080  &  0.365  &  0.033  &  0.016  &  0.037 \\
1000  &  0.130  &  0.523  &  0.035  &  0.014  &  0.038 \\
1000  &  0.180  &  0.423  &  0.032  &  0.011  &  0.034 \\
1000  &  0.250  &  0.407  &  0.031  &  0.011  &  0.032 \\
1000  &  0.400  &  0.245  &  0.028  &  0.009  &  0.029 \\
1000  &  0.650  &  0.026  &  0.017  &  0.007  &  0.018 \\
\hline
\hline
\end{tabular}}
\caption{Structure function $xF_3^{\gamma Z}$ averaged over $Q^2 \ge 1000$\,GeV$^2$ at the scale
         1000\,GeV$^2$;
$\delta_{\rm stat}$, $\delta_{\rm syst}$ and $\delta_{\rm tot}$ represent the statistical, systematic and total uncertainties, respectively.
}
\label{tab:xF3ave}

\end{center}
\end{table}

\clearpage

\begin{figure}
  \centering
  \setlength{\unitlength}{0.1\textwidth}
  \begin{picture} (10,10)
  \put(0,0){\includegraphics[width=\textwidth]{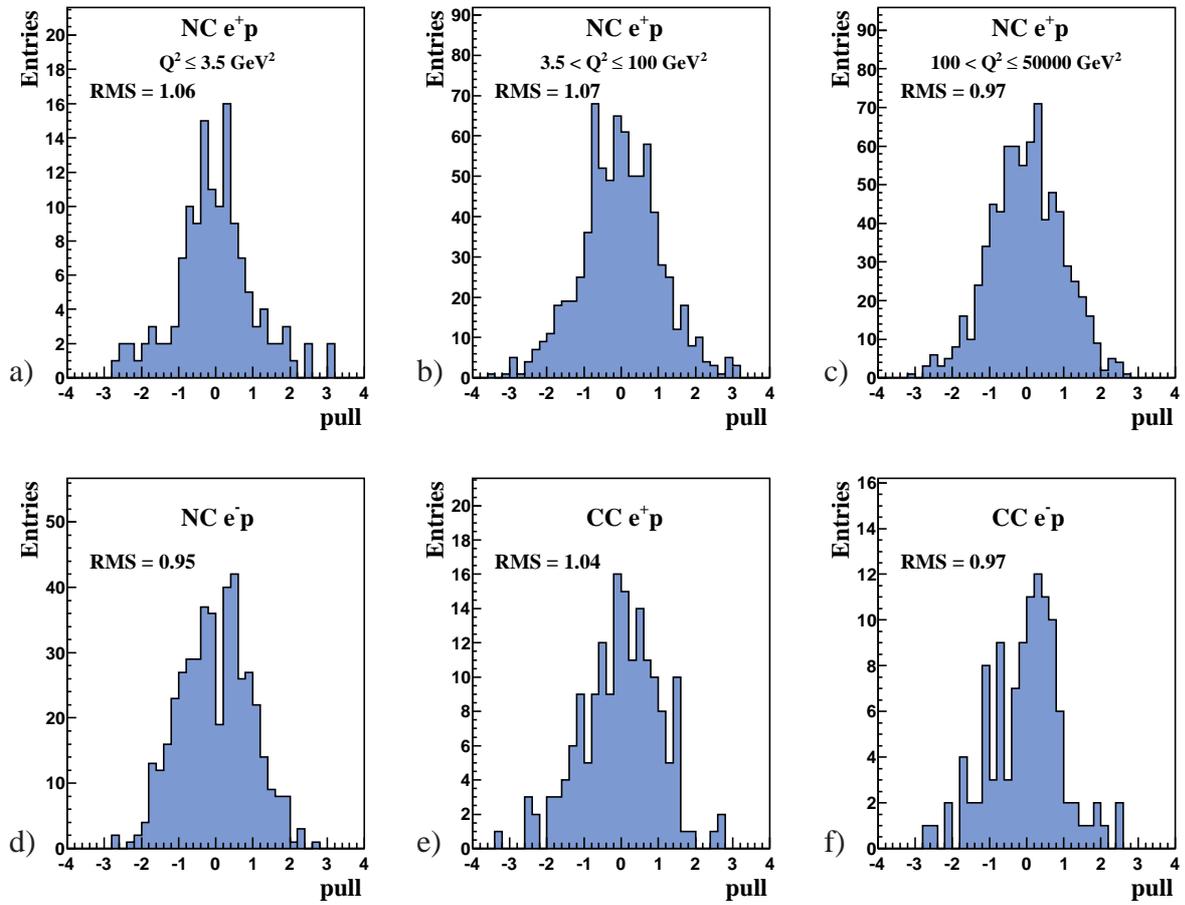}}
  \put (0.0,4.4) {a)}
  \put (3.35,4.4) {b)}
  \put (6.7,4.4) {c)}
  \put (0.0,0.5) {d)}
  \put (3.35,0.5) {e)}
  \put (6.7,0.5) {f)}
  \end{picture}
\caption{Distributions of pulls $\rm p$ for:
a) NC $e^+p$ for $Q^2 \le 3.5$\,GeV$^2$;
b) NC $e^+p$ for $3.5 < Q^2 \le 100$\,GeV$^2$;
c) NC $e^+p$ for $100 < Q^2 \le 50000$\,GeV$^2$;
d) NC $e^-p$ for $60 \le Q^2 \le 50000$\,GeV$^2$;
e) CC $e^+p$ for $300 \le Q^2 \le 30000$\,GeV$^2$;  and 
f) CC $e^-p$ for $300 \le Q^2 \le 30000$\,GeV$^2$.  
There are no entries outside the histogram ranges. 
The root mean square, RMS, of each distribution is given.
}
\label{fig:pulls}
\end{figure}

\clearpage

\clearpage

\begin{figure}[tbp]
\vspace{-0.5cm} 
\centerline{
\epsfig{file=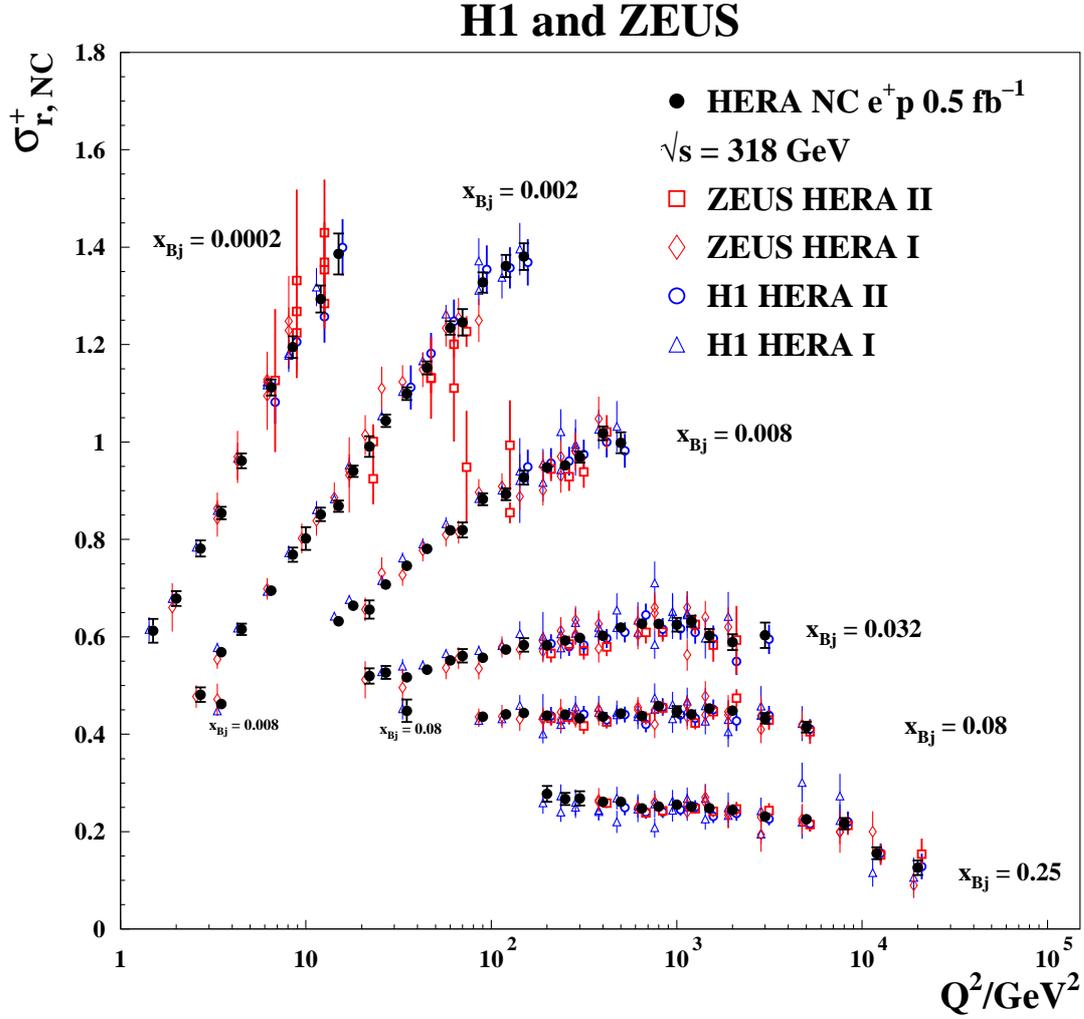 ,width=\linewidth}}
\caption {The combined HERA data for the inclusive 
NC $e^+p$ reduced 
cross sections as a function of 
$Q^2$ for six selected values of $x_{\rm Bj}$ compared to the individual 
H1 and ZEUS data.
The individual measurements are displaced horizontally 
for better visibility.
Error bars represent the total uncertainties.
The two labelled entries at $x_{\rm Bj}=0.008$ and $0.08$
come from data which were 
taken at $\sqrt{s}=300$\,GeV and $y<0.35$ and were translated 
to $\sqrt{s}=318$\,GeV, see Section~\ref{subsec:extrapol}.
}
\label{fig:quality:NCepp}

\end{figure}
\clearpage
\begin{figure}[tbp]
\vspace{-0.5cm} 
\centerline{
\epsfig{file=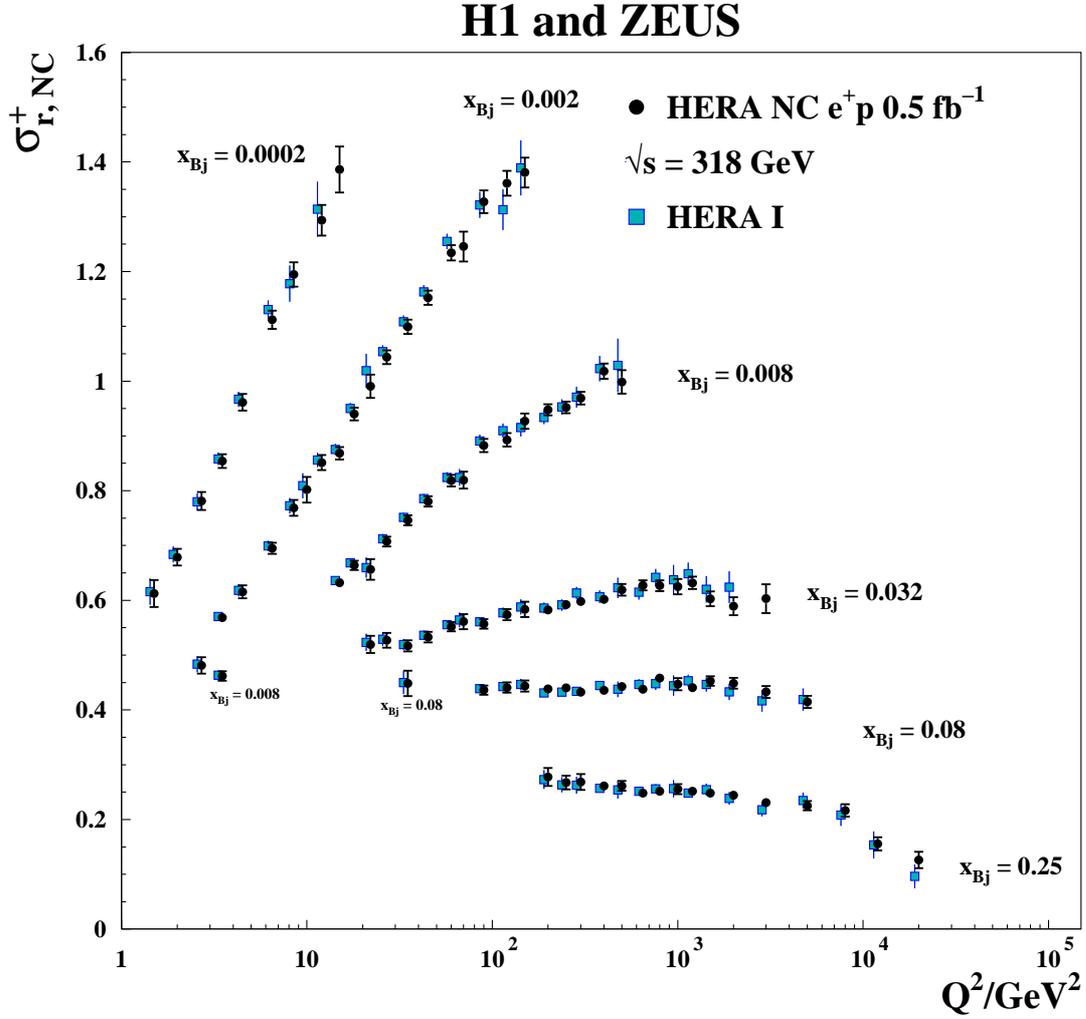  ,width=\linewidth}}
\caption {The combined HERA data for the inclusive NC $e^+p$ reduced 
cross sections as a function of 
$Q^2$ for six selected values of $x_{\rm Bj}$ compared to the results from
HERA\,I alone~\cite{HERAIcombi}. 
The two measurements are displaced horizontally for better visibility.
Error bars represent the total uncertainties.
The two labelled entries at $x_{\rm Bj}=0.008$ and $0.08$
come from data which were 
taken at $\sqrt{s}=300$\,GeV and $y<0.35$ and were translated 
to $\sqrt{s}=318$\,GeV, see Section~\ref{subsec:extrapol}.
}
\label{fig:Hera1:NCepp}

\end{figure}
\clearpage

\begin{figure}[tbp]
\vspace{-0.5cm} 
\centerline{
\epsfig{file=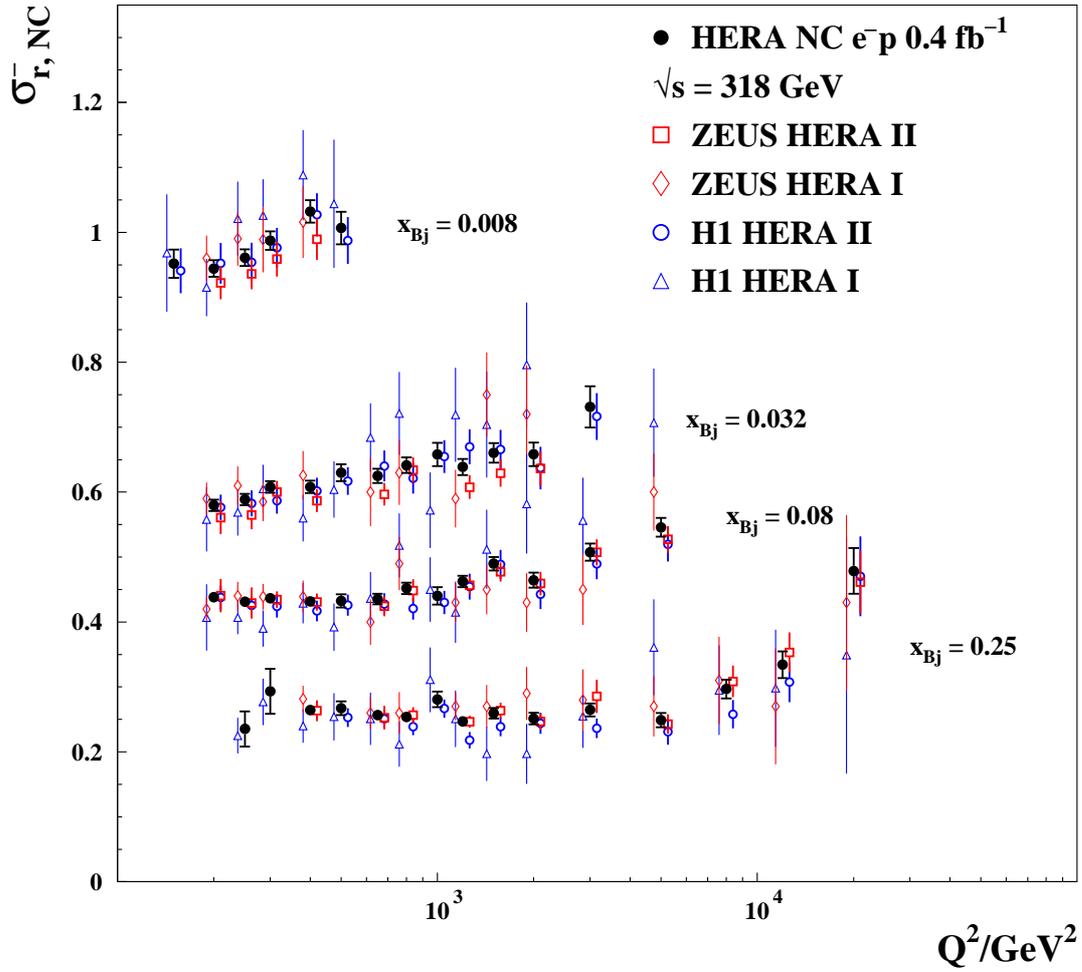  ,width=\linewidth}}
\caption {The combined HERA data for the inclusive 
NC $e^-p$ reduced 
cross sections as a function of 
$Q^2$ for four selected values of $x_{\rm Bj}$ compared to the individual 
H1 and ZEUS data.
The individual measurements are displaced 
horizontally for better visibility.
Error bars represent the total uncertainties.
}
\label{fig:quality:NCemp}

\end{figure}
\clearpage

\begin{figure}[tbp]
\vspace{-0.5cm} 
\centerline{
\epsfig{file=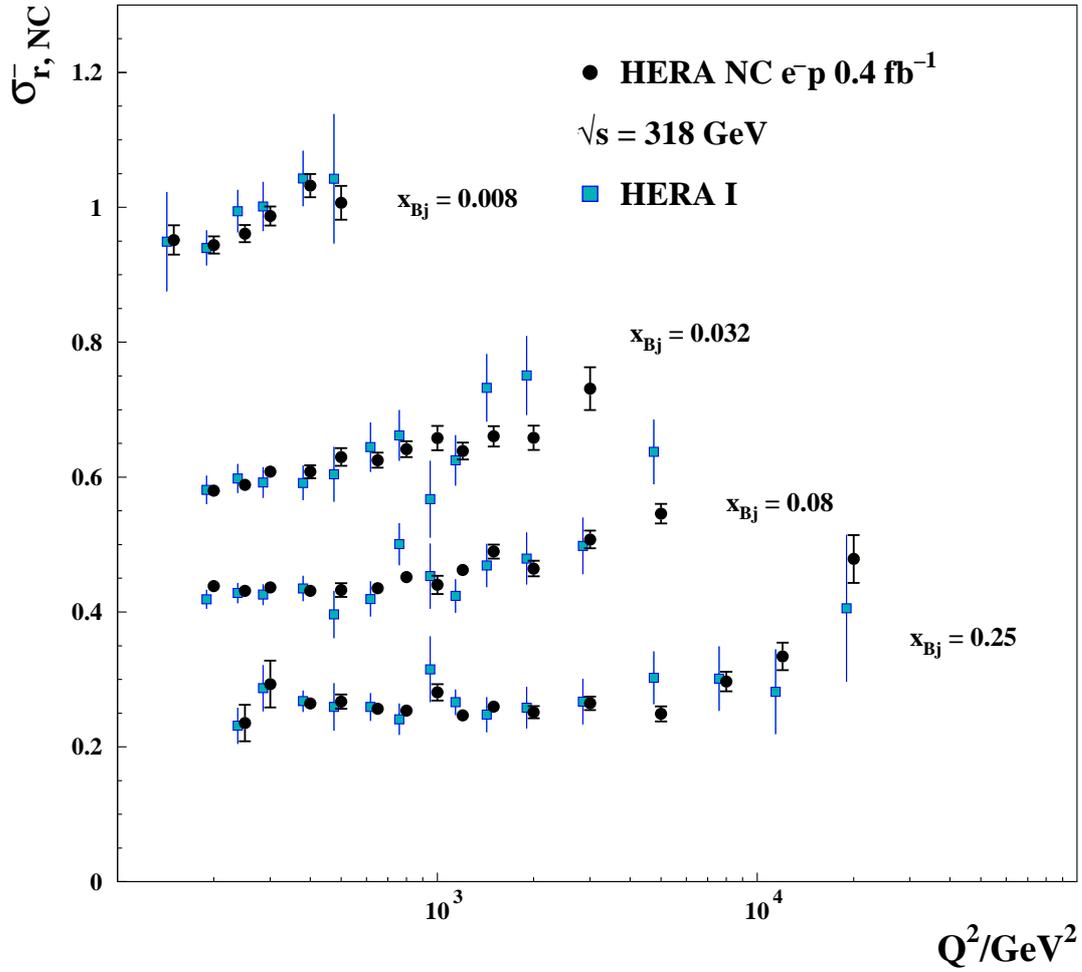  ,width=\linewidth}}
\caption {The combined HERA data for the inclusive 
NC $e^-p$ reduced 
cross section as a function of 
$Q^2$ for four selected values of $x_{\rm Bj}$ compared 
to the results from
HERA\,I alone~\cite{HERAIcombi}. 
The two measurements are displaced horizontally for better visibility.
Error bars represent the total uncertainties.
}
\label{fig:Hera1:NCemp}
\end{figure}

\clearpage

\begin{figure}[tbp]
\vspace{-0.5cm} 
\centerline{
\epsfig{file=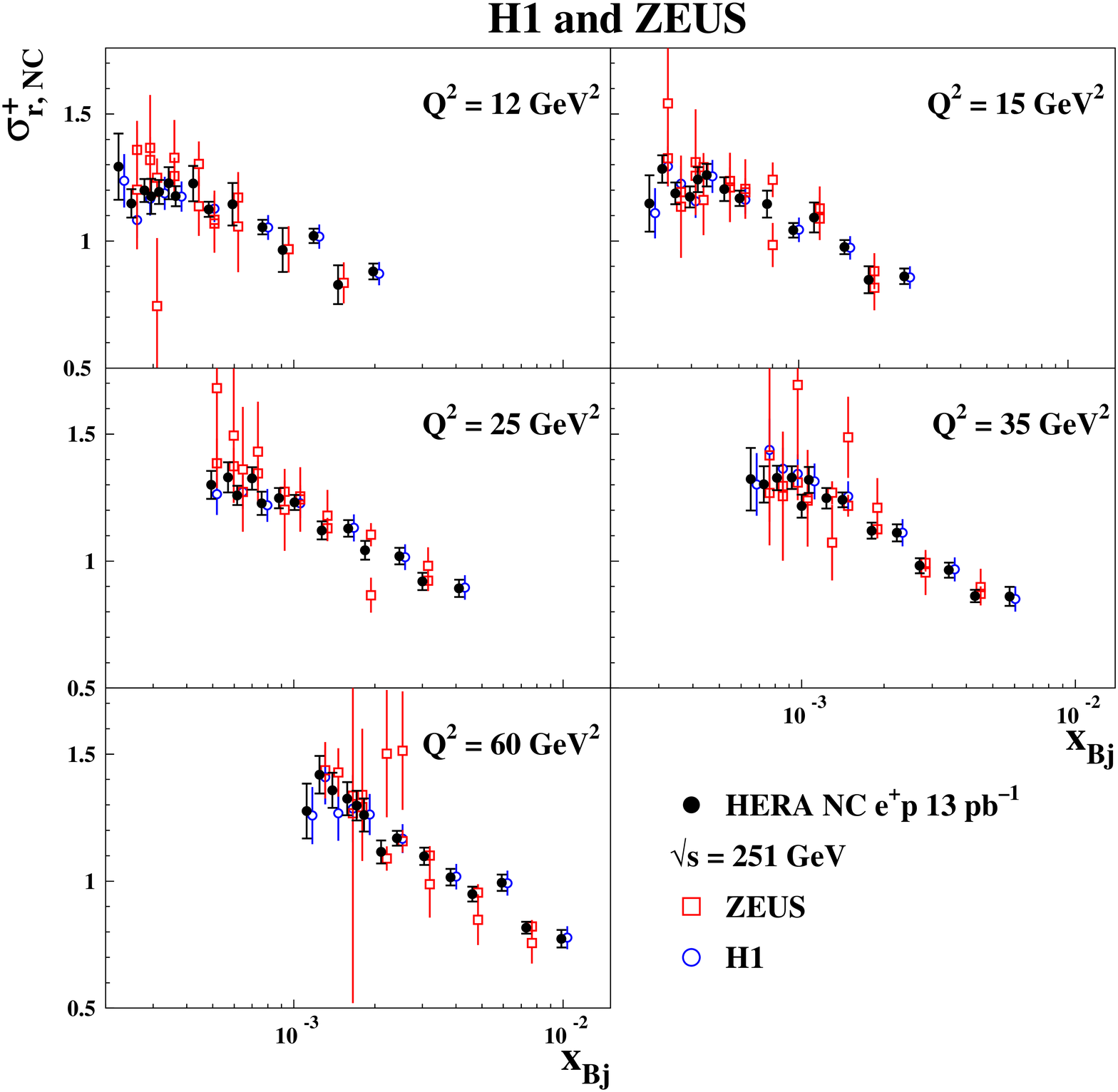  ,width=\linewidth}}
\caption {The combined HERA data for the inclusive 
NC $e^+p$ reduced 
cross sections  at $\sqrt{s} = 251$\,GeV  as a function 
of $x_{\rm Bj}$ for five selected
values of $Q^2$ compared to the individual 
H1 and ZEUS data.
The individual measurements are displaced horizontally 
for better visibility.
The ZEUS points at the same $x_{\rm Bj}$ and $Q^2$ values 
are from two different data sets. 
Error bars represent the total uncertainties.
}
\label{fig:quality:575}
\end{figure}
\clearpage

\begin{figure}[tbp]
\vspace{-0.5cm} 
\centerline{
\epsfig{file=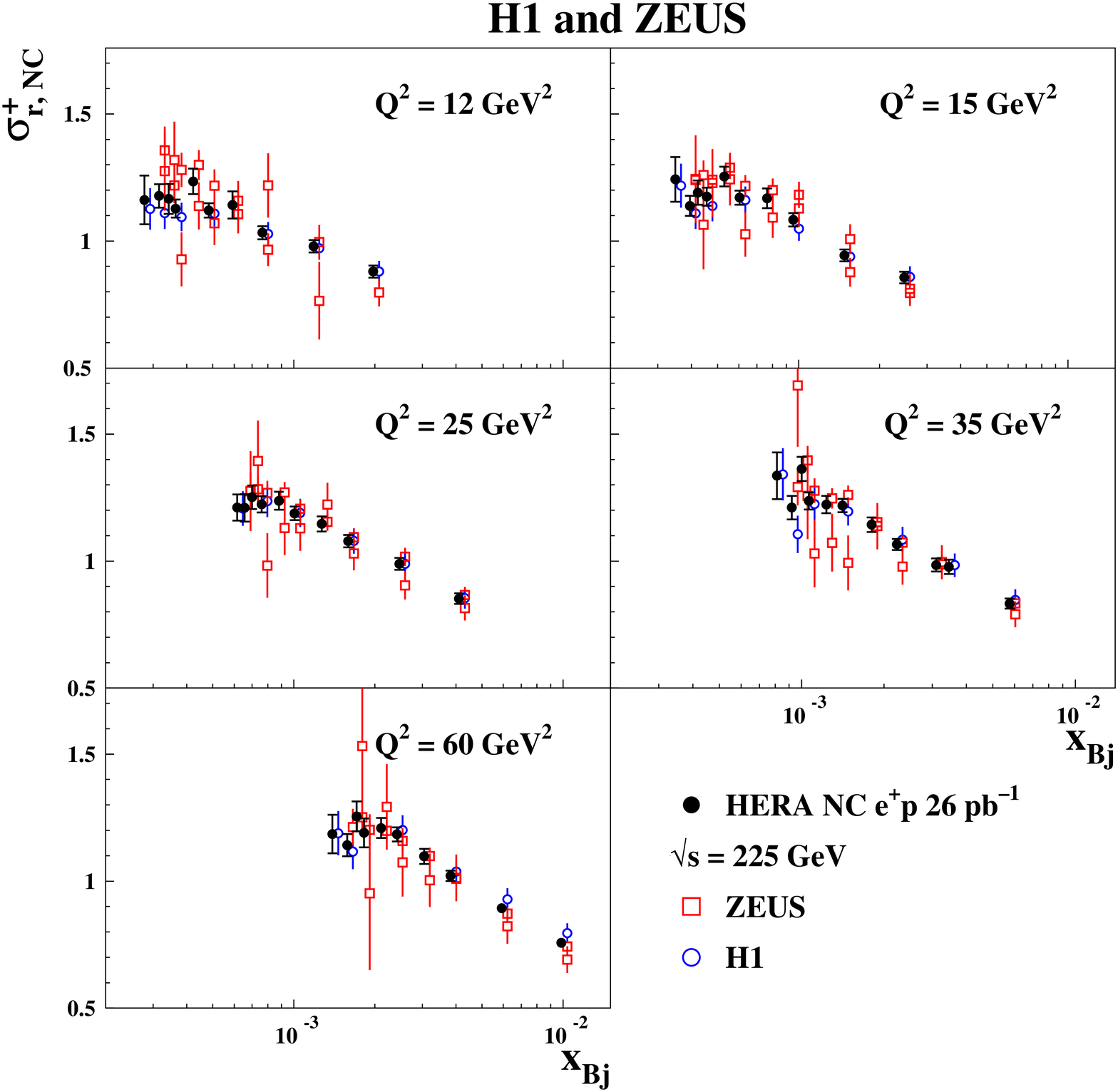  ,width=\linewidth}}
\caption {The combined HERA data for the inclusive 
NC $e^+p$ reduced 
cross sections  at $\sqrt{s} = 225$\,GeV  as a function 
of $x_{\rm Bj}$ for five selected
values of $Q^2$ compared to the individual
H1 and ZEUS data.
The individual measurements are displaced horizontally for 
better visibility.
The ZEUS points at the same $x_{\rm Bj}$ and $Q^2$ values 
are from two different data sets. 
Error bars represent the total uncertainties.
}
\label{fig:quality:460}
\end{figure}
\clearpage


\begin{figure}[tbp]
\vspace{-0.3cm} 
\centerline{
\epsfig{file=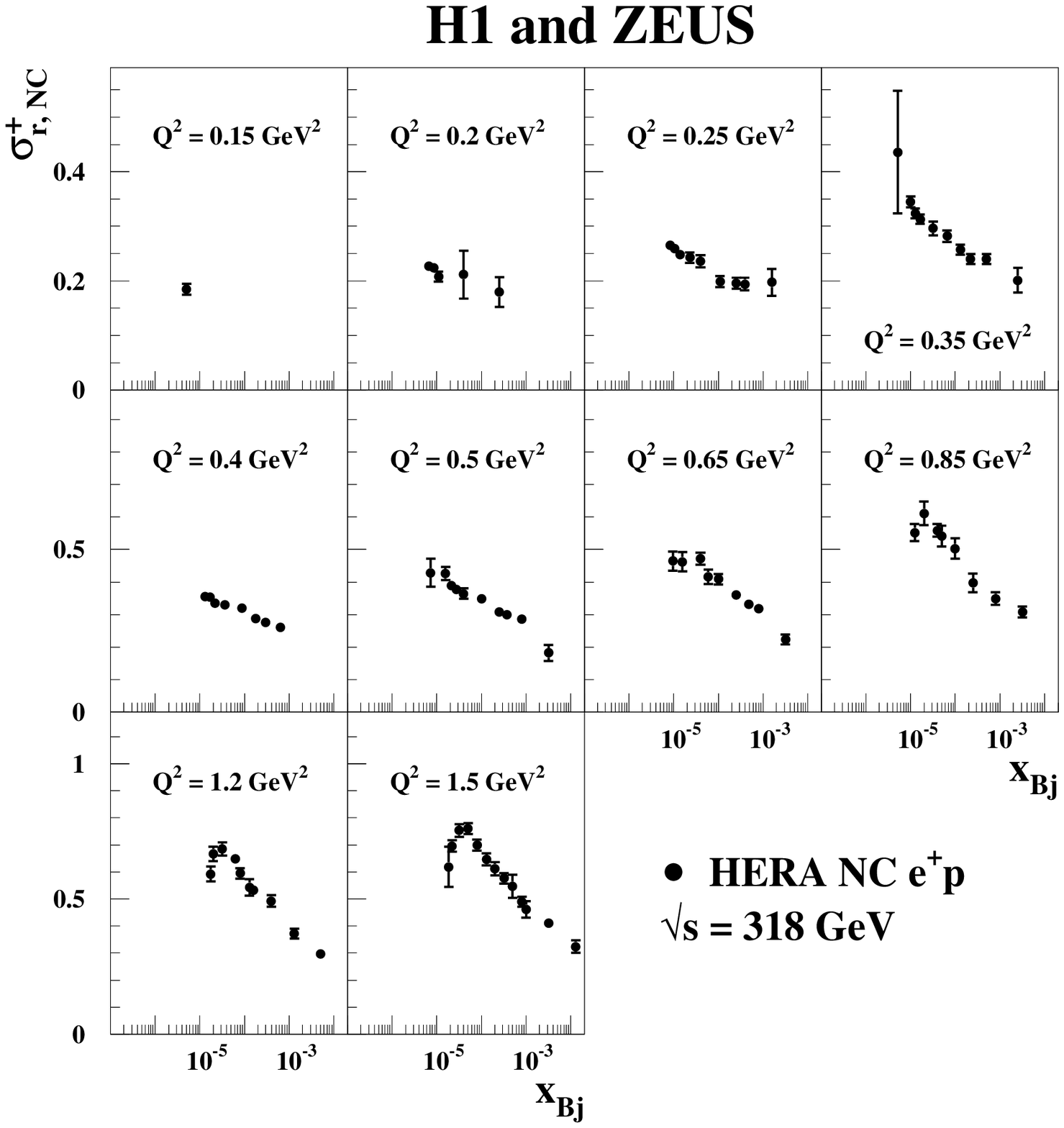   ,width=0.9\textwidth}}
\vspace{0.5cm}
\caption {The combined HERA data for the inclusive NC $e^+p$ 
reduced cross sections at $\sqrt{s} = 318$\,GeV at very low $Q^2$.
Error bars represent the total uncertainties.
}
\label{fig:NCvlQ2-920}
\end{figure}
\clearpage

\begin{figure}[tbp]
\vspace{-0.3cm} 
\centerline{
\epsfig{file=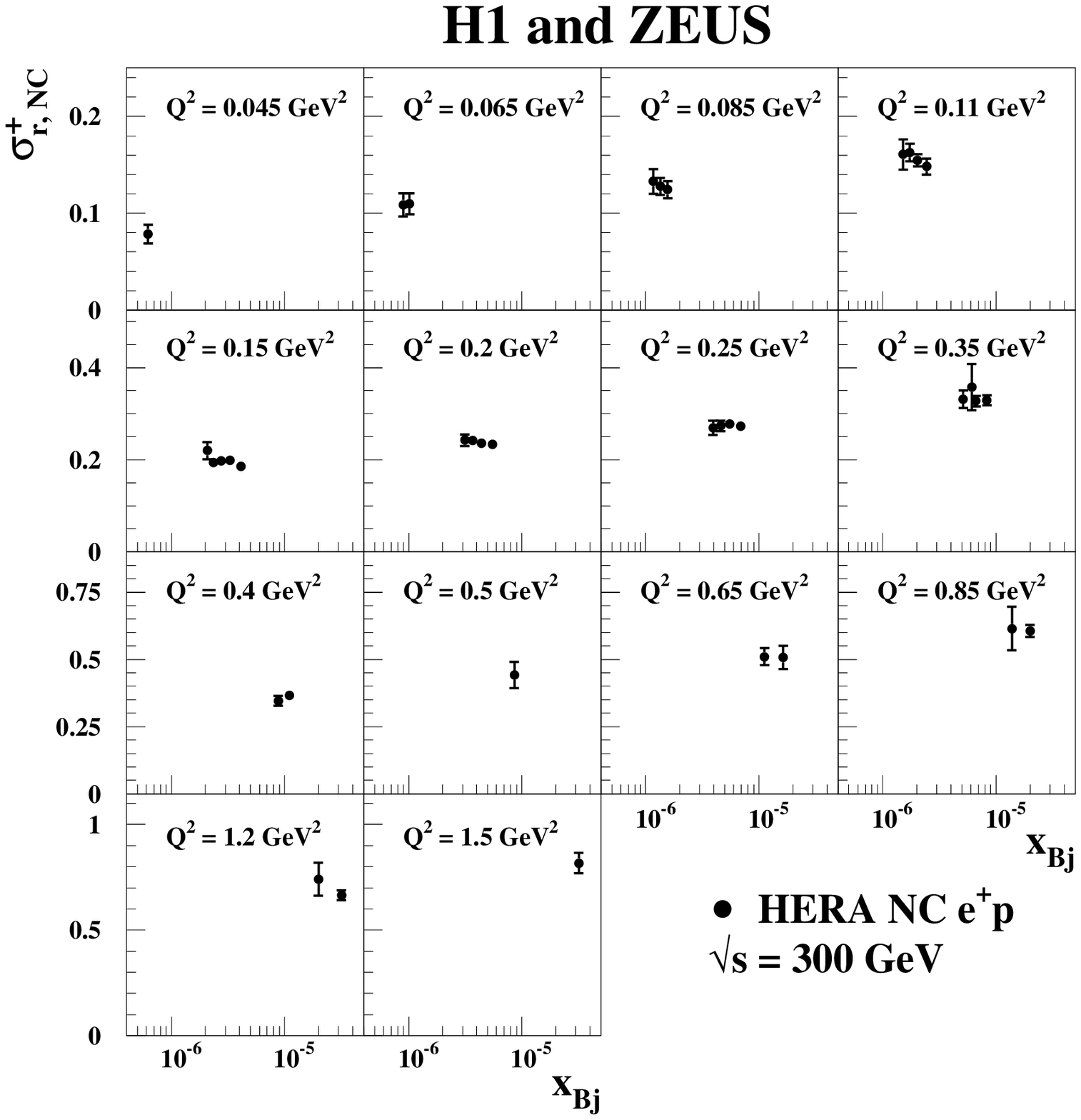   ,width=0.9\textwidth}}
\vspace{0.5cm}
\caption {The combined HERA data for the inclusive 
NC $e^+p$ reduced cross sections at $\sqrt{s} = 300$~GeV at very low $Q^2$.
Error bars represent the total uncertainties.
}
\label{fig:NCvlQ2-820}
\end{figure}
\clearpage

\begin{figure}[tbp]
\vspace{-0.5cm} 
\centerline{
\epsfig{file=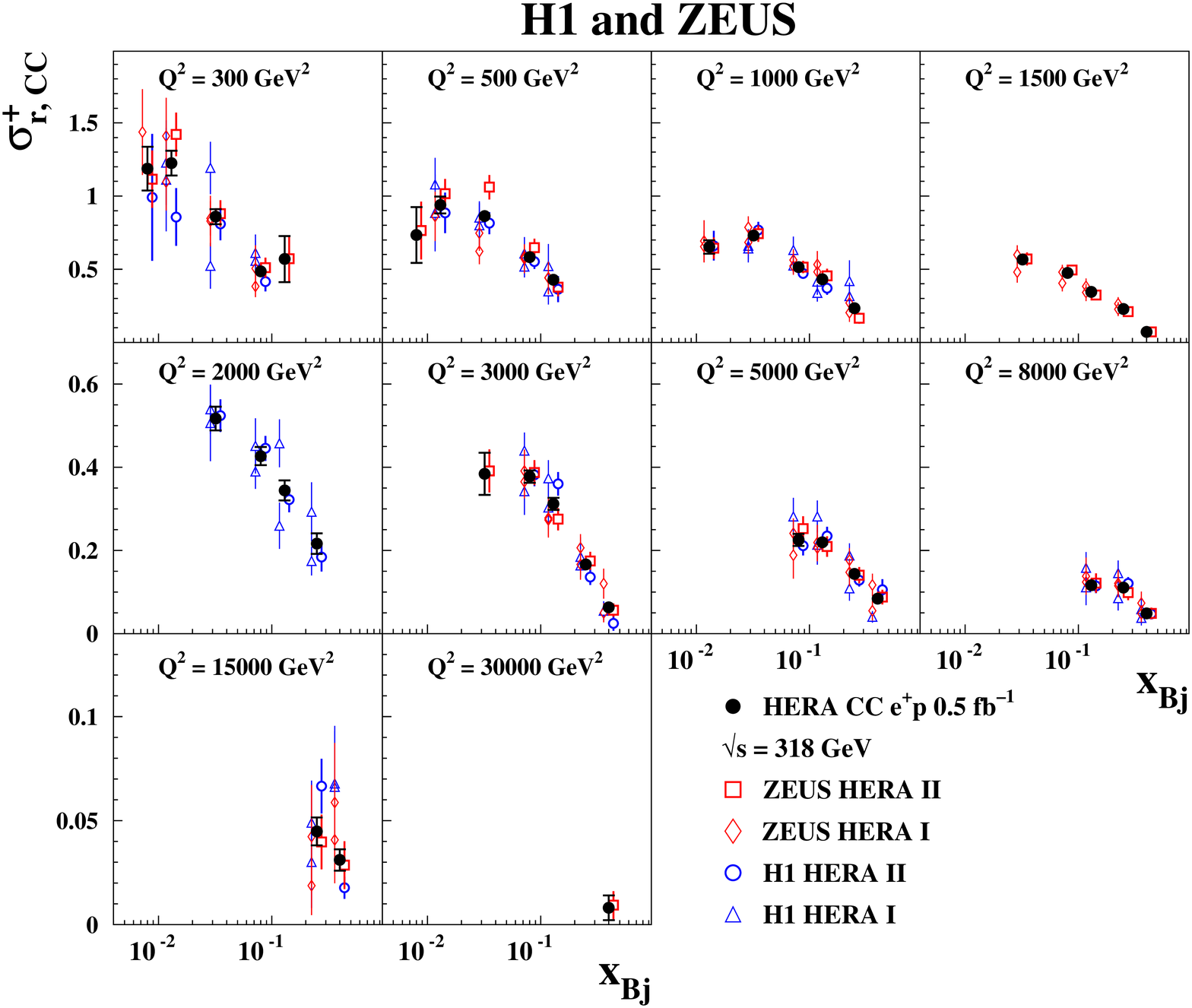  ,width=\linewidth}}
\caption {The combined HERA data for the inclusive 
CC $e^+p$ reduced 
cross sections as a function of $x_{\rm Bj}$ for the 10 different 
values of $Q^2$ compared to the individual
H1 and ZEUS data.
The individual measurements are displaced horizontally for better visibility.
Error bars represent the total uncertainties.
}
\label{fig:quality:CCepp}
\end{figure}
\clearpage

\begin{figure}[tbp]
\vspace{-0.5cm} 
\centerline{
\epsfig{file=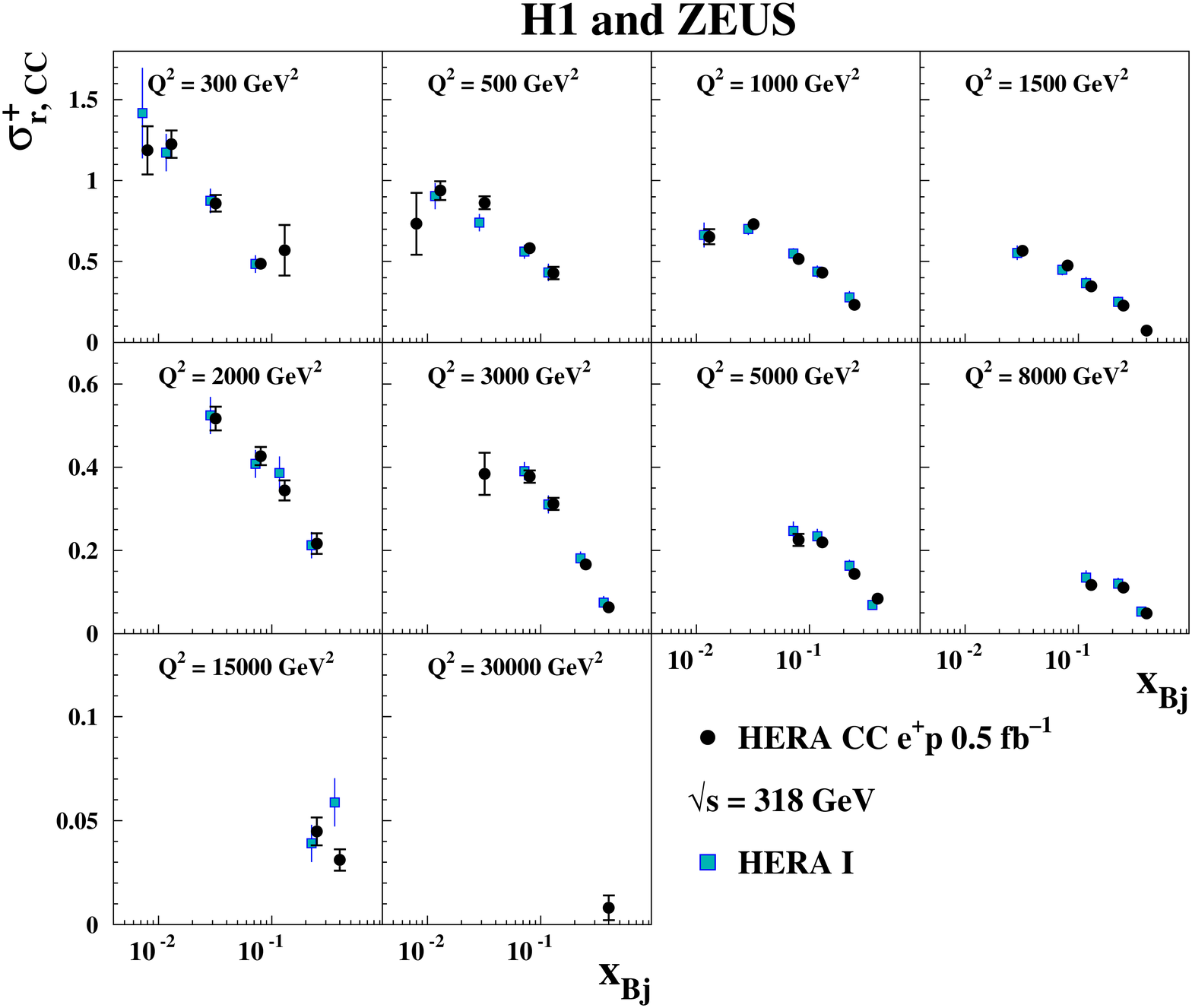  ,width=\linewidth}}
\caption {The combined HERA data for the inclusive CC 
$e^+p$ reduced 
cross sections as a function of $x_{\rm Bj}$ for the 10 different 
values of $Q^2$ compared
to the results from
HERA\,I alone~\cite{HERAIcombi}. 
The individual measurements are displaced horizontally for better visibility.
Error bars represent the total uncertainties.
}
\label{fig:Hera1:CCepp}
\end{figure}
\clearpage

\begin{figure}[tbp]
\vspace{-0.5cm} 
\centerline{
\epsfig{file=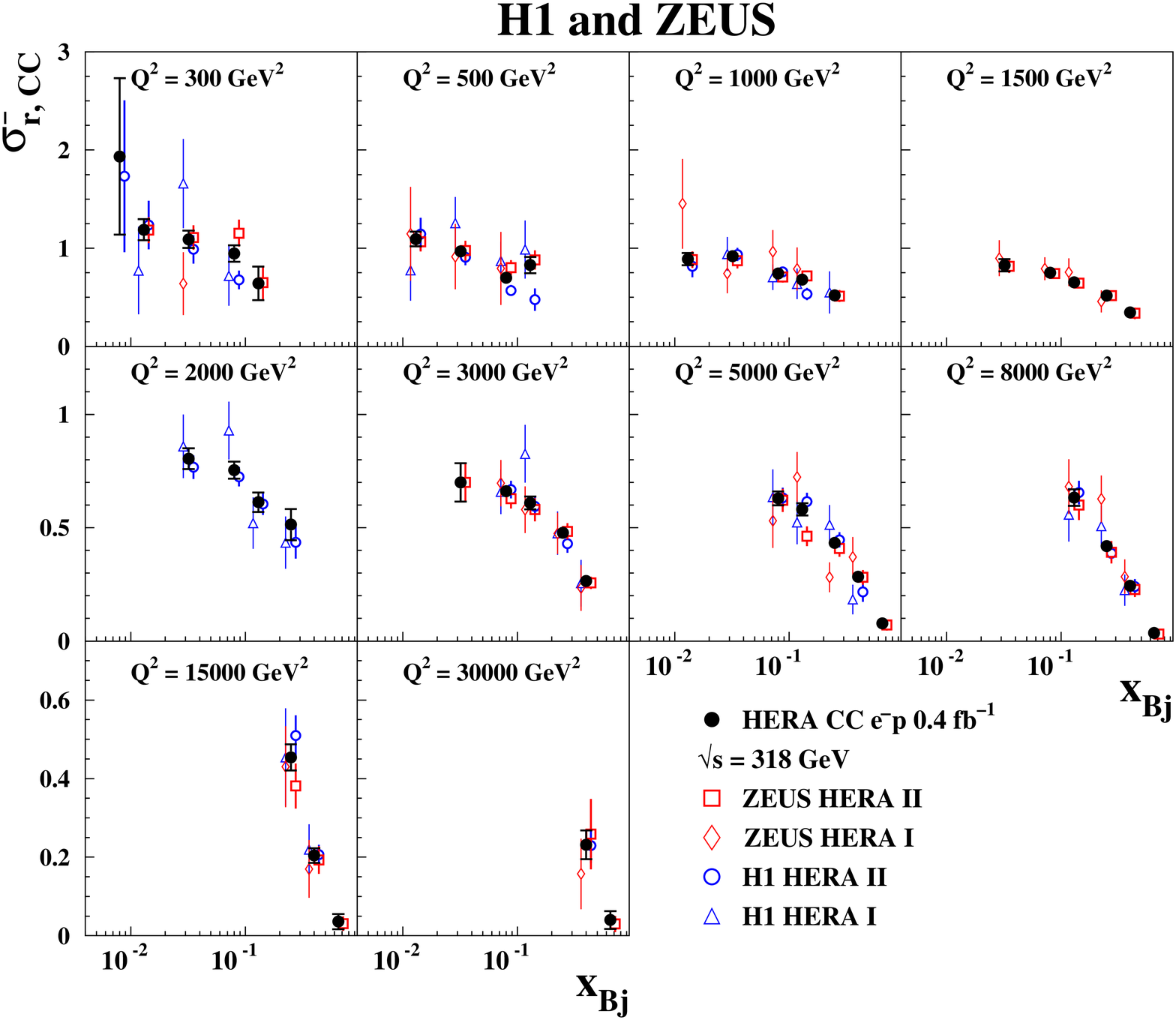  ,width=\linewidth}}
\caption {The combined HERA data for the inclusive 
CC $e^-p$ reduced 
cross sections as a function of $x_{\rm Bj}$ for the 10 different
values of $Q^2$ compared to the individual 
H1 and ZEUS data.
The individual measurements are displaced horizontally 
for better visibility.
Error bars represent the total uncertainties.
}
\label{fig:quality:CCemp}
\end{figure}
\clearpage

\begin{figure}[tbp]
\vspace{-0.5cm} 
\centerline{
\epsfig{file=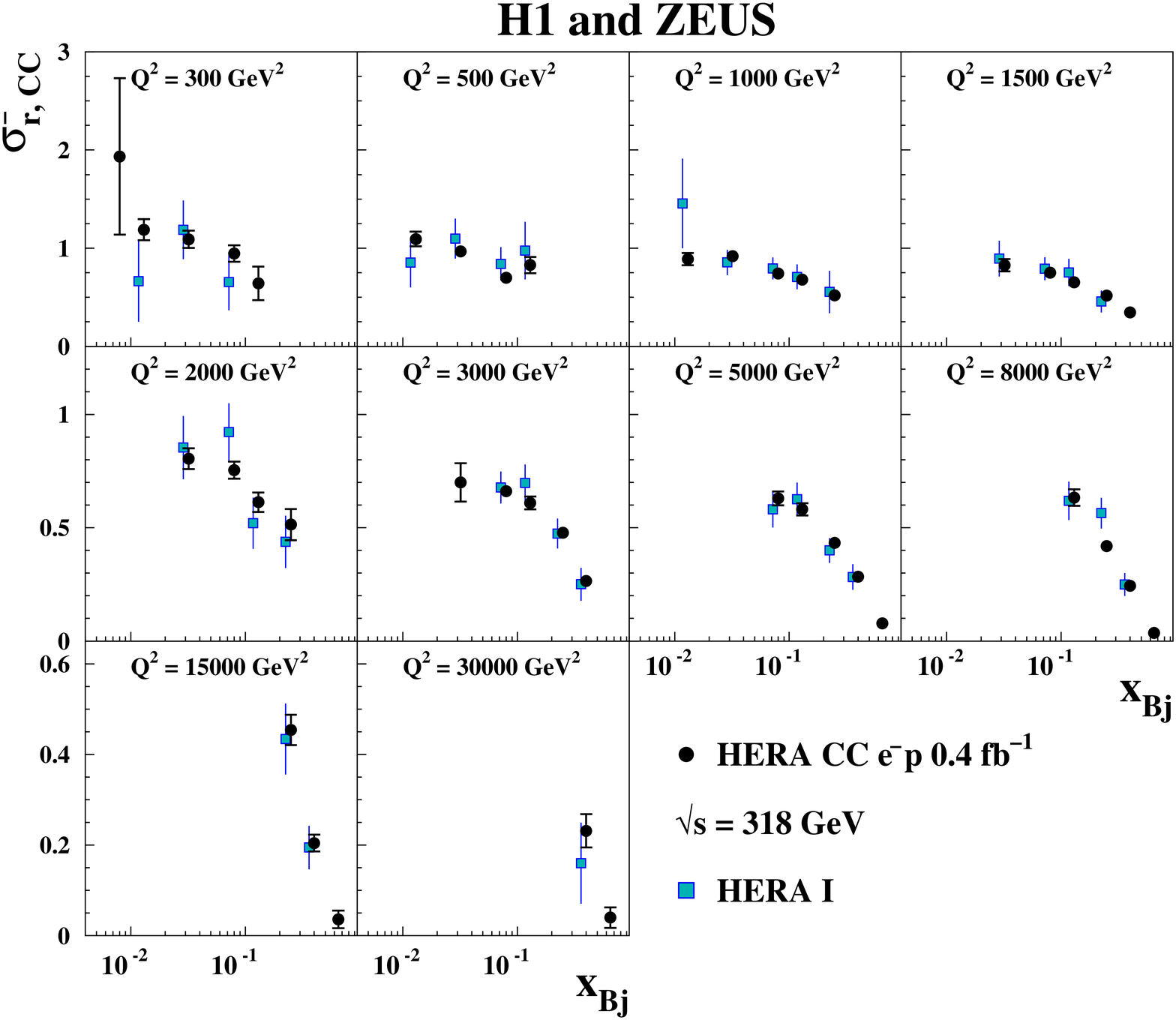  ,width=\linewidth}}
\caption {The combined HERA data for the inclusive 
CC $e^-p$ reduced 
cross sections as a function of $x_{\rm Bj}$ for the 10 different
values of $Q^2$ compared 
to the results from
HERA\,I alone~\cite{HERAIcombi}. 
The individual measurements are displaced horizontally for better visibility.
Error bars represent the total uncertainties.
}
\label{fig:Hera1:CCemp}
\end{figure}
\clearpage


\clearpage

\clearpage
\clearpage

\clearpage

\clearpage

\clearpage

\clearpage

\clearpage

\begin{figure}[tbp]
\vspace{-0.3cm} 
\centerline{
\epsfig{file=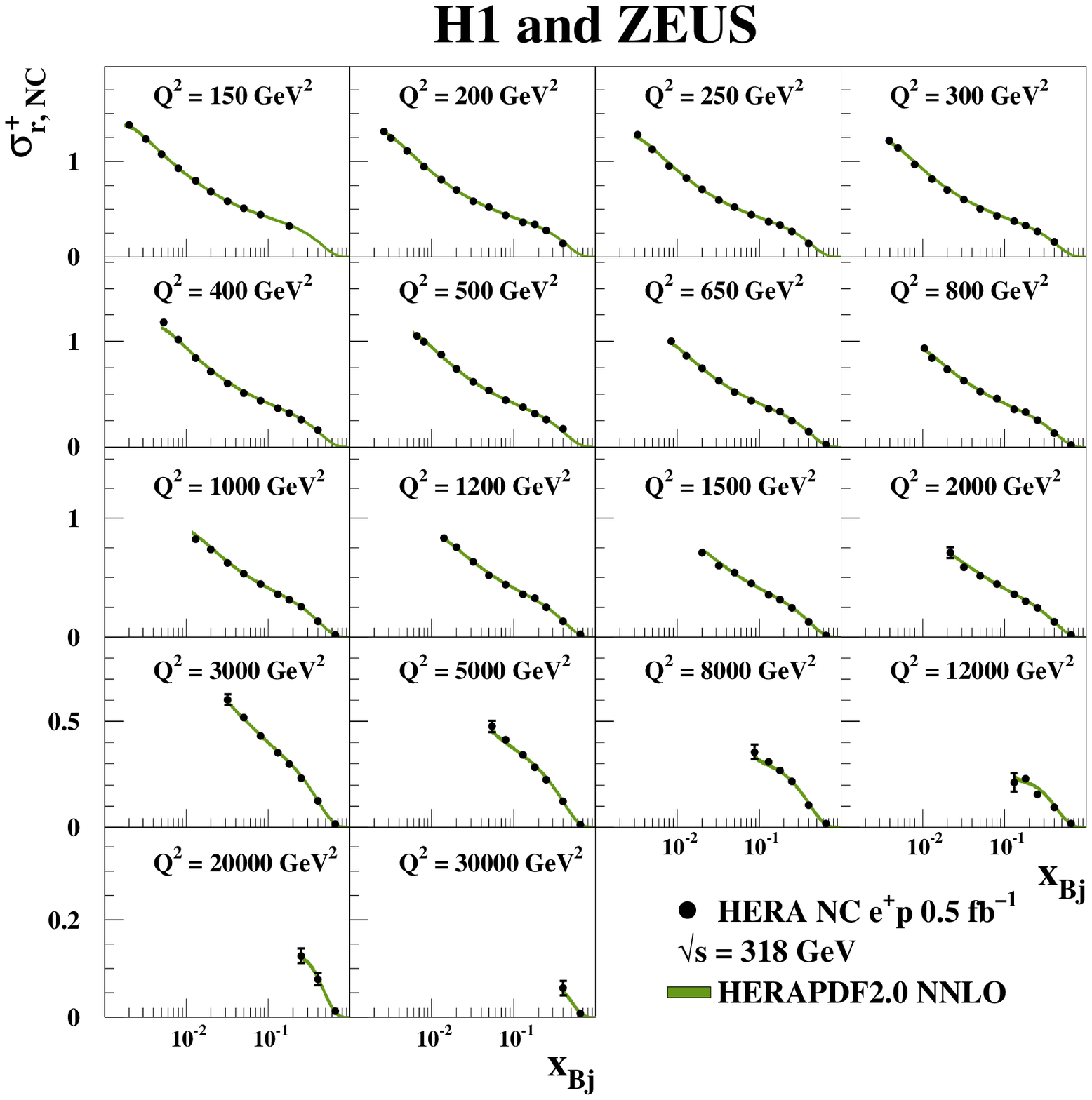   ,width=0.9\textwidth}}
\vspace{0.5cm}
\caption {The combined high-$Q^2$ HERA inclusive NC $e^+p$ 
          reduced cross sections  
          at $\sqrt{s} = 318$\,GeV with overlaid predictions from 
          HERAPDF2.0 NNLO.
          The bands represent the total uncertainties on the predictions.
}
\label{fig:nnloQ23pt5ncepc}
\end{figure}
\clearpage

\begin{figure}[tbp]
\vspace{-0.3cm} 
\centerline{
\epsfig{file=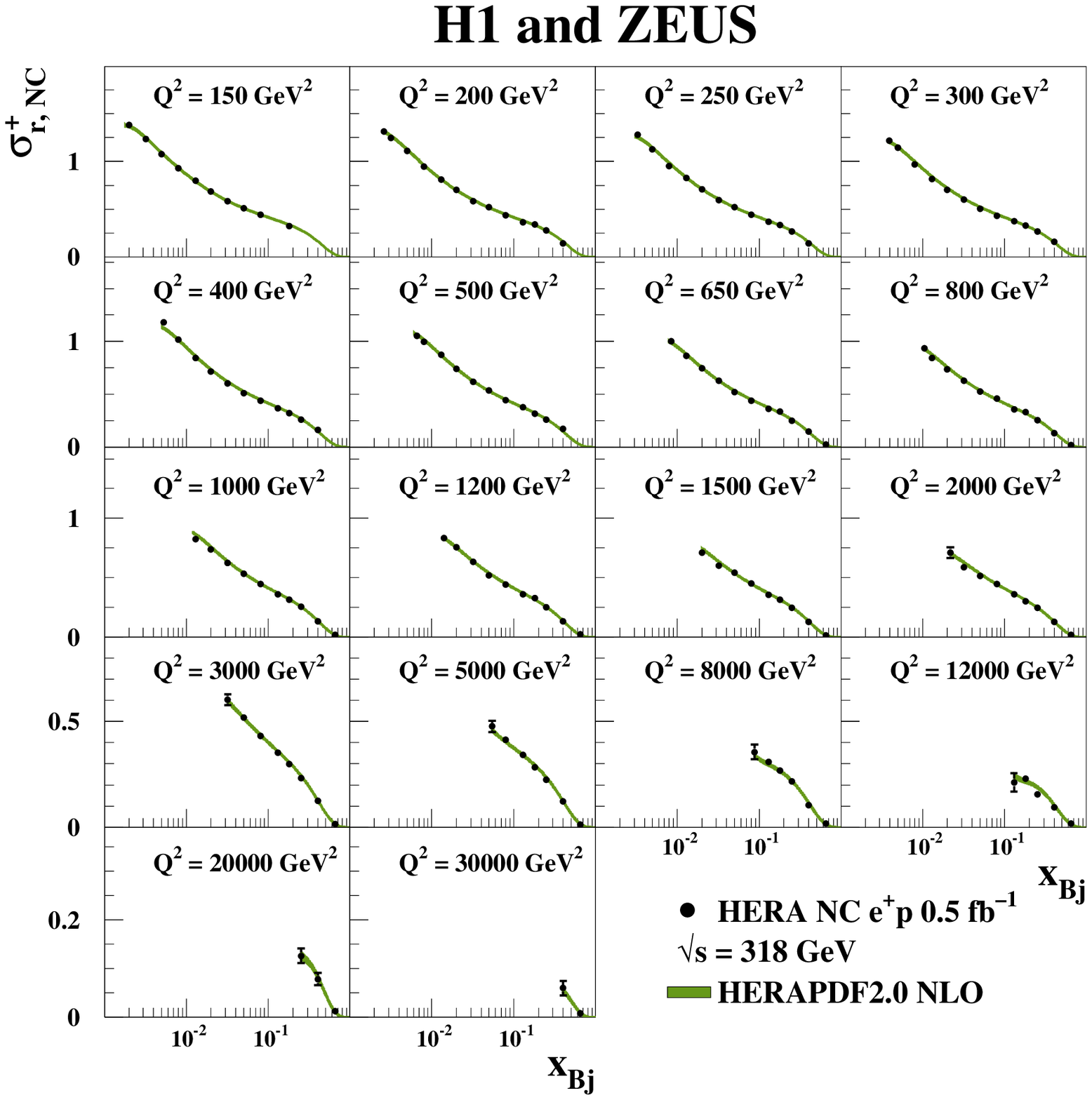   ,width=0.9\textwidth}}
\vspace{0.5cm}
\caption {The combined high-$Q^2$ HERA  
inclusive NC $e^+p$ reduced cross sections  
at $\sqrt{s} = 318$\,GeV with overlaid predictions of the HERAPDF2.0 NLO.
The   bands represent the total uncertainties on the predictions.
}
\label{fig:nloQ23pt5ncepc}
\end{figure}
\clearpage

\begin{figure}[tbp]
\vspace{-0.3cm} 
\centerline{
\epsfig{file=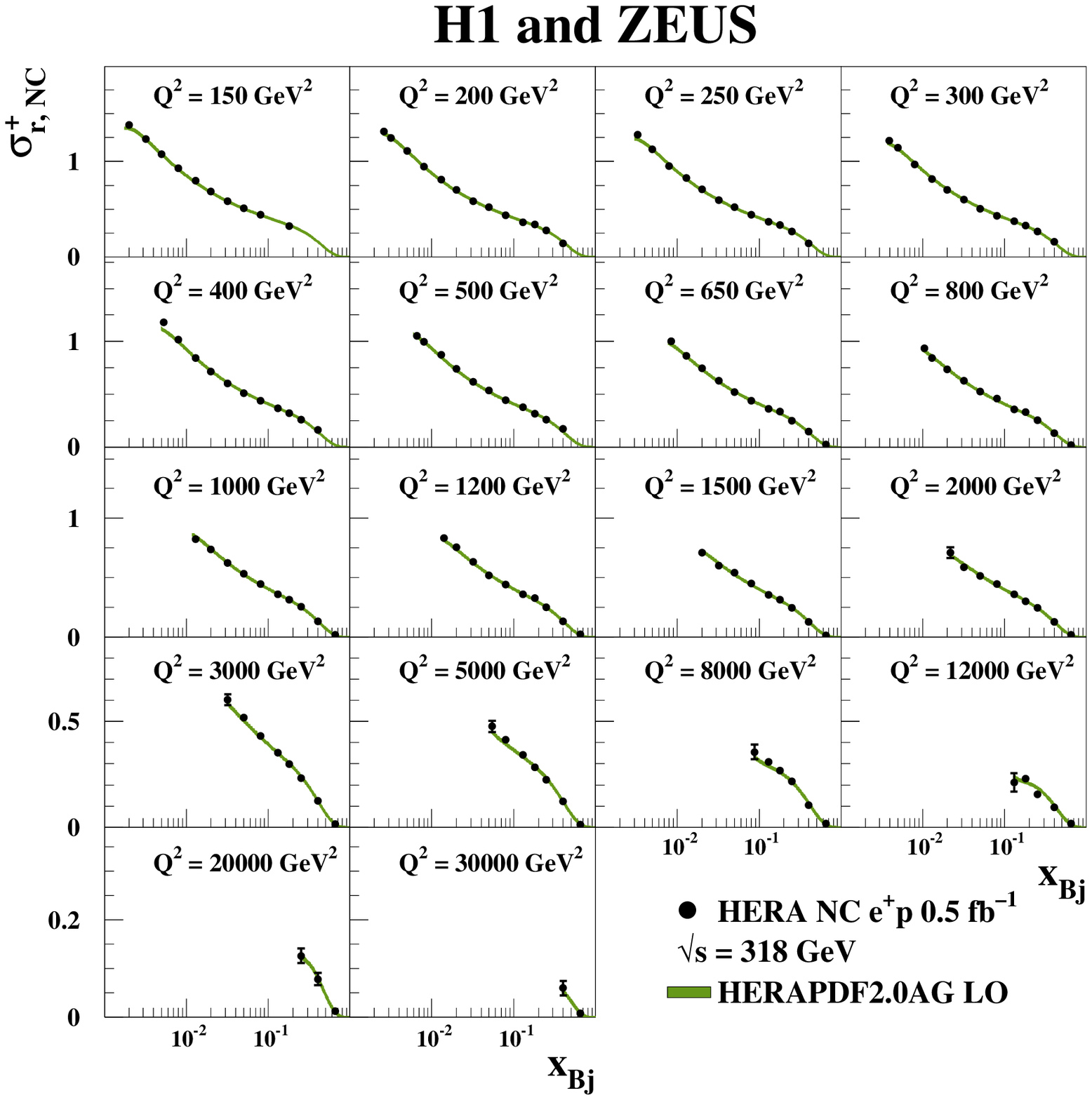   ,width=0.9\textwidth}}
\vspace{0.5cm}
\caption {The combined high-$Q^2$ HERA 
inclusive NC $e^+p$ reduced cross sections  
at $\sqrt{s} = 318$\,GeV with overlaid predictions of the HERAPDF2.0AG LO.
The bands represent the experimental uncertainties on the predictions.
}
\label{fig:loQ23pt5ncepc}
\end{figure}
\clearpage


\begin{figure}[tbp]
\vspace{-0.3cm} 
\centerline{
\epsfig{file=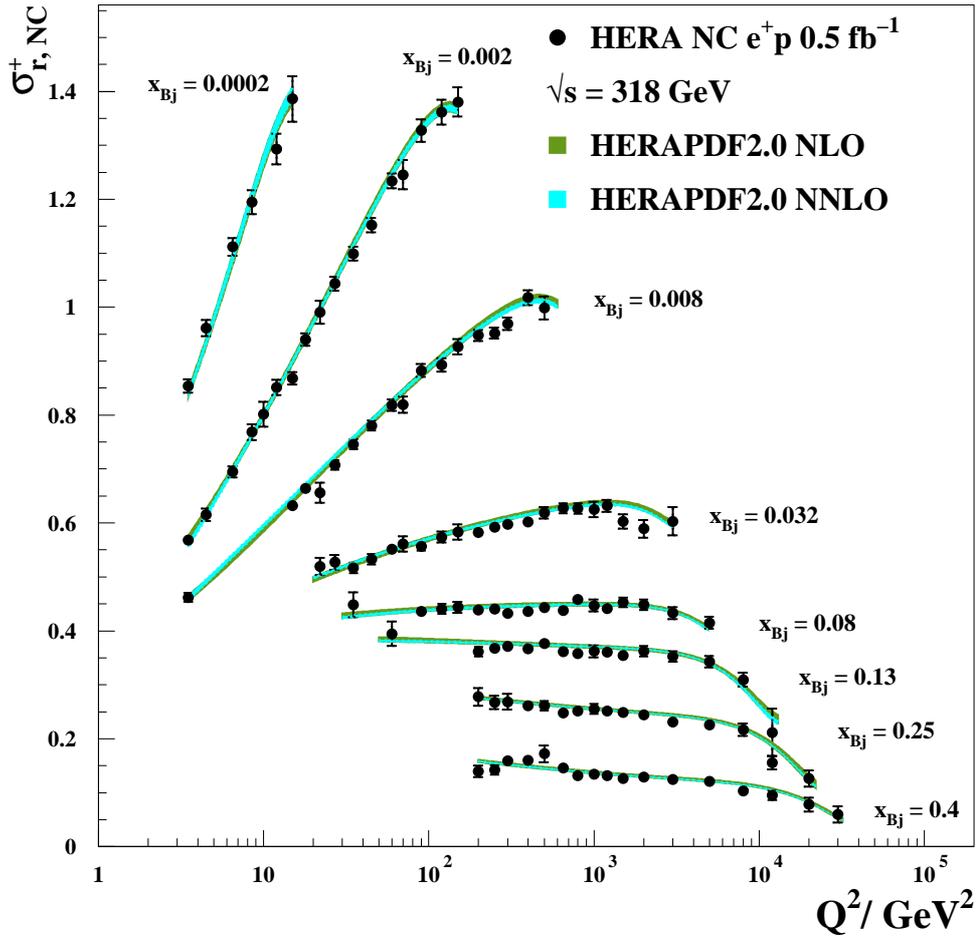 ,width=0.9\textwidth}}
\vspace{0.5cm}
\caption {The combined high-$Q^2$ HERA 
     inclusive NC $e^+p$ reduced cross sections  as 
     partially shown 
     already in Fig.~\ref{fig:Hera1:NCepp}
     with overlaid predictions of HERAPDF2.0 NLO and NNLO.
     The two differently shaded bands represent the total 
     uncertainties on the two predictions.   
}
\label{fig:5mod}
\end{figure}


\begin{figure}[tbp]
\vspace{-0.3cm} 
\centerline{
\epsfig{file=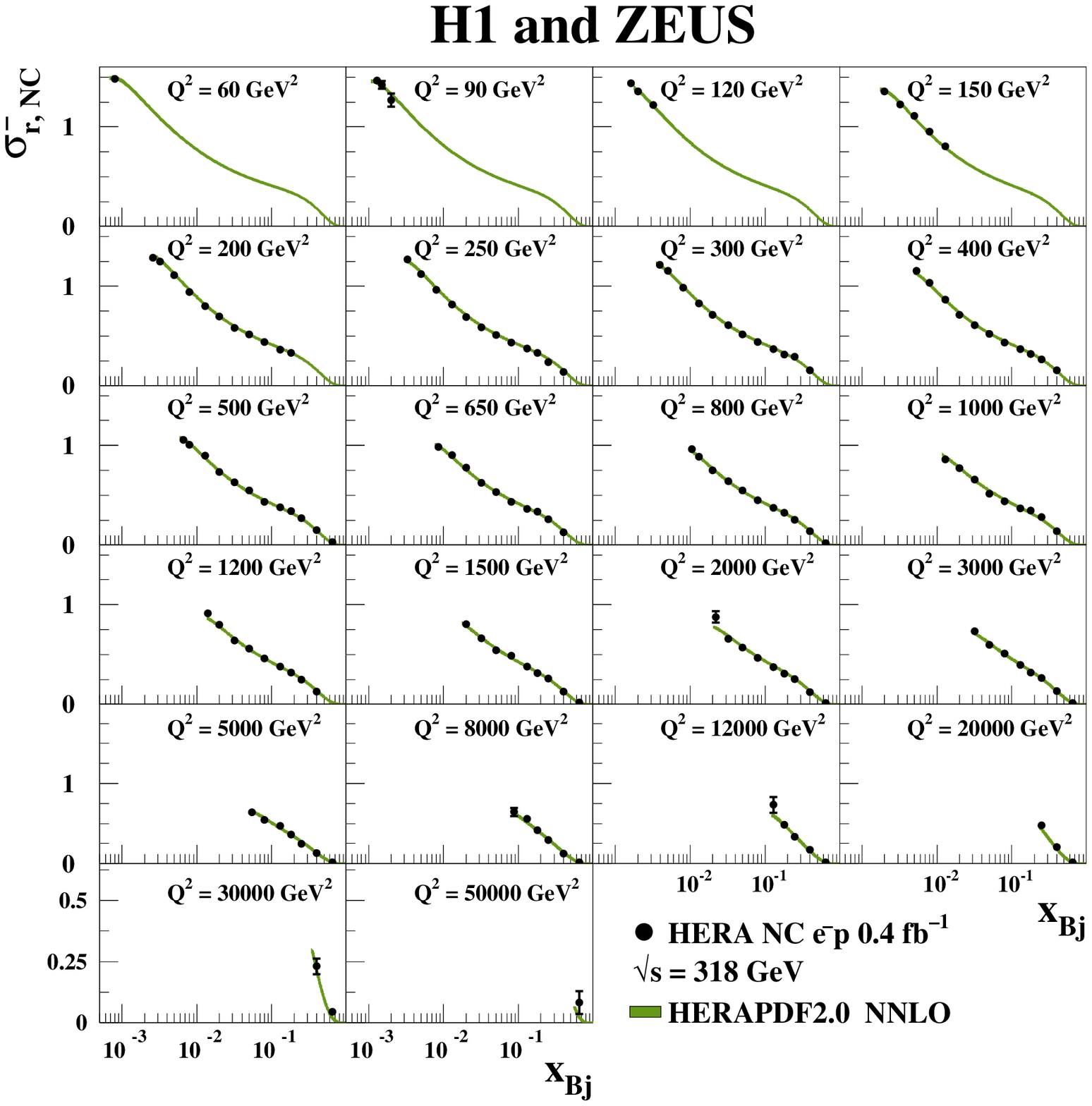   ,width=0.9\textwidth}}
\vspace{0.5cm}
\caption {The combined HERA inclusive NC $e^-p$ reduced cross sections  
at $\sqrt{s} = 318$\,GeV with overlaid predictions from HERAPDF2.0 NNLO.
     The bands represent the total uncertainties on the predictions.   
}
\label{fig:nnloQ23pt5ncem}
\end{figure}
\clearpage

\begin{figure}[tbp]
\vspace{-0.3cm} 
\centerline{
\epsfig{file=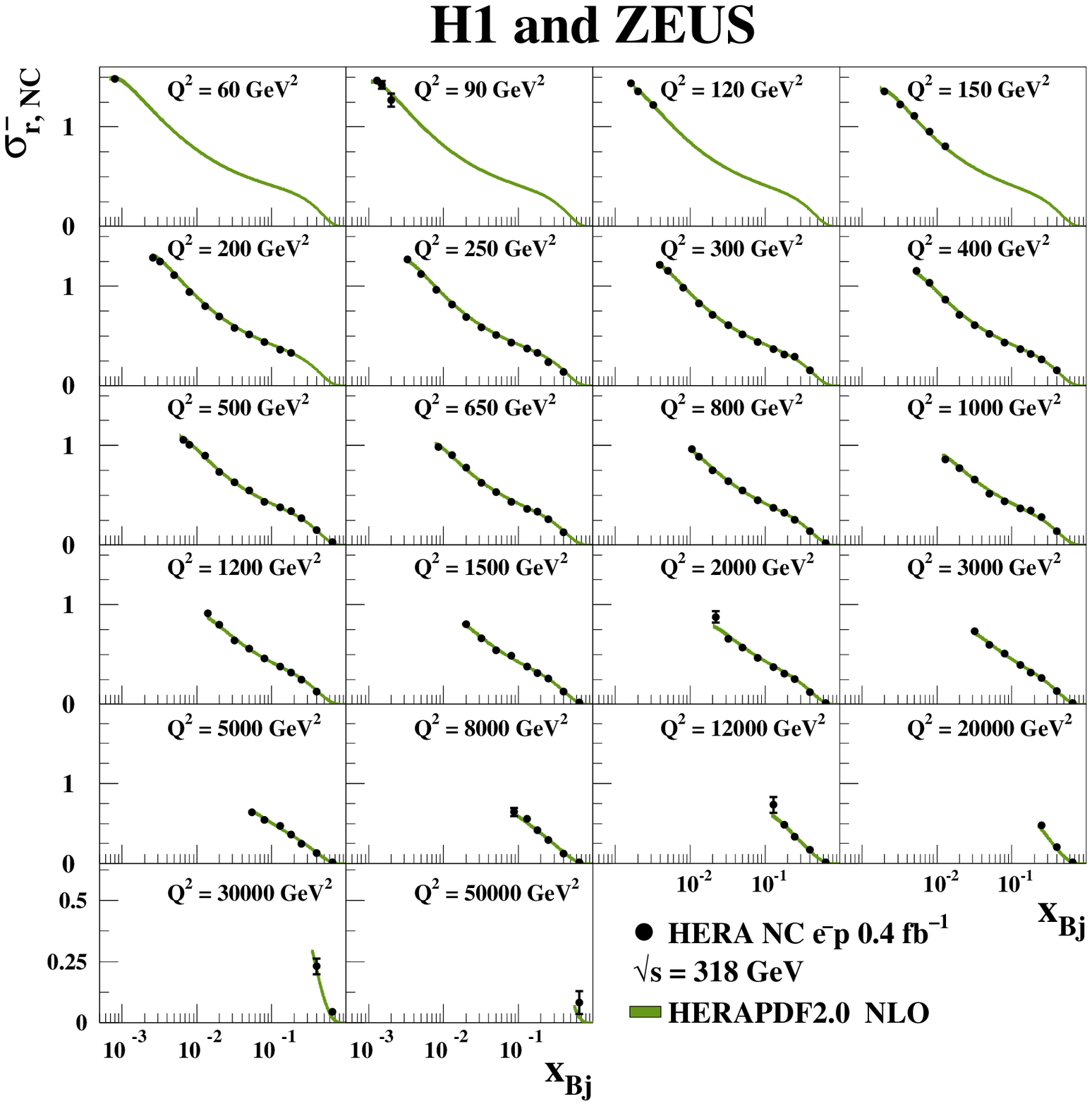   ,width=0.9\textwidth}}
\vspace{0.5cm}
\caption {The combined HERA inclusive NC $e^-p$ reduced cross sections  
at $\sqrt{s} = 318$\,GeV with overlaid predictions from HERAPDF2.0 NLO.
     The bands represent the total uncertainties on the predictions.   
}
\label{fig:nloQ23pt5ncem}
\end{figure}
\clearpage

\begin{figure}[tbp]
\vspace{-0.3cm} 
\centerline{
\epsfig{file=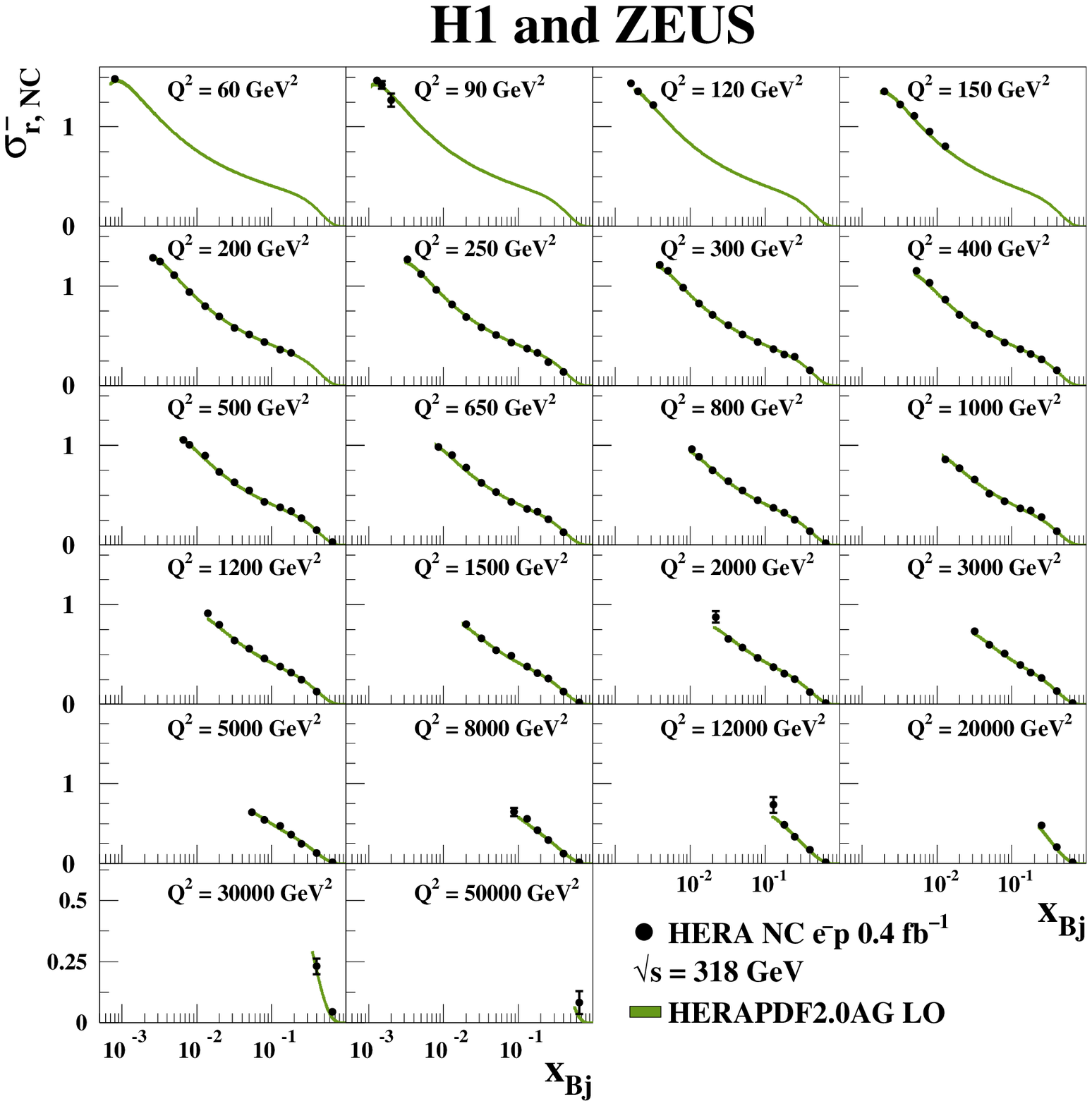   ,width=0.9\textwidth}}
\vspace{0.5cm}
\caption {The combined HERA inclusive NC $e^-p$ reduced cross sections  
at $\sqrt{s} = 318$\,GeV with overlaid predictions from HERAPDF2.0AG LO.
     The bands represent the experimental uncertainties 
     on the predictions.   
}
\label{fig:loQ23pt5ncem}
\end{figure}
\clearpage



\begin{figure}[tbp]
\vspace{-0.3cm} 
\centerline{
\epsfig{file=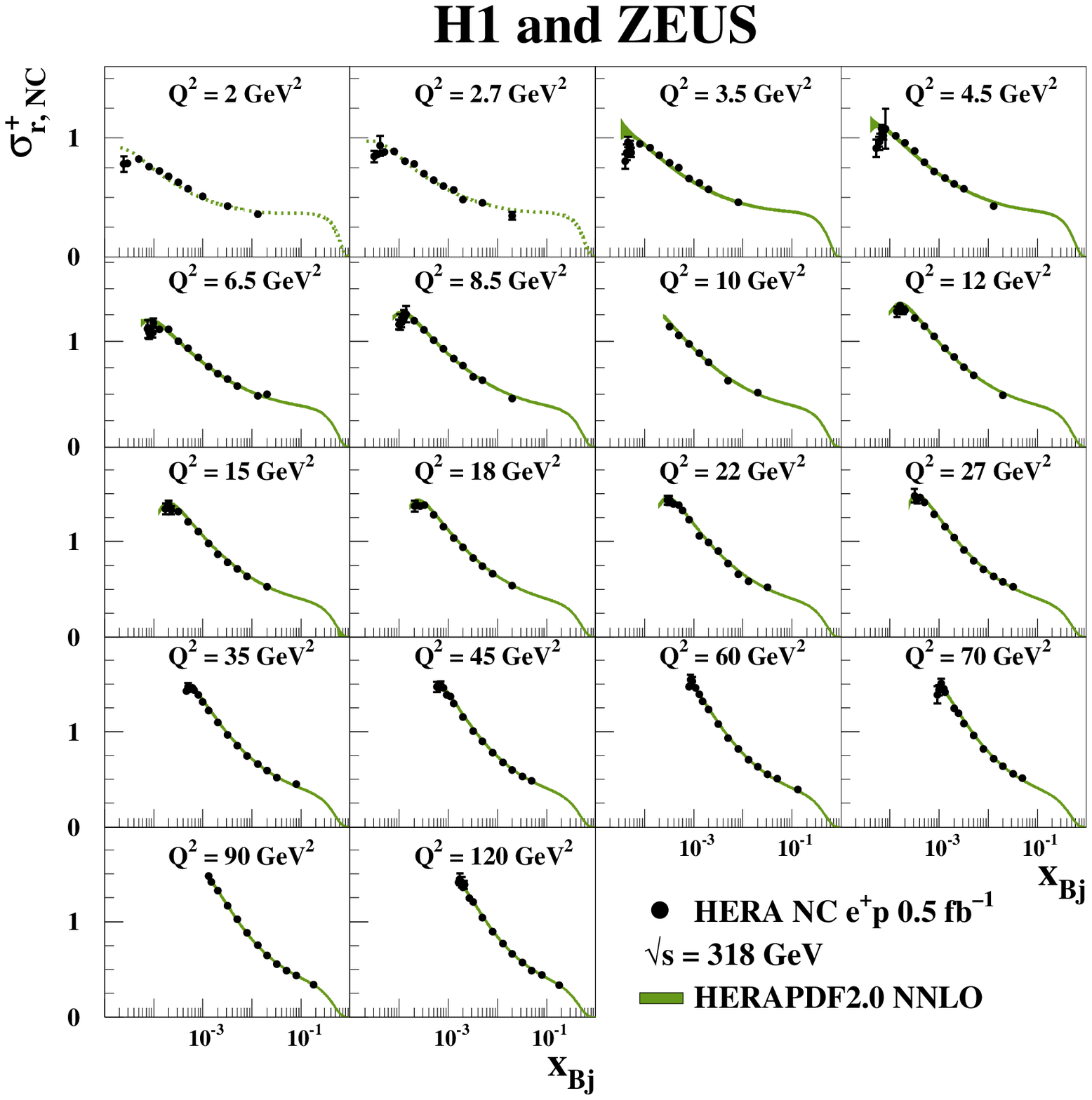   ,width=0.9\textwidth}}
\vspace{0.5cm}
\caption {The combined low-$Q^2$ HERA inclusive NC $e^+p$ 
reduced cross sections  
at $\sqrt{s} = 318$\,GeV with overlaid predictions from HERAPDF2.0 NNLO. 
     The bands represent the total uncertainties on the predictions.   
Dotted lines indicate extrapolation into 
kinematic regions not included in the
fit.
}
\label{fig:nnloQ23pt5ncepb}
\end{figure}
\clearpage

\begin{figure}[tbp]
\vspace{-0.3cm} 
\centerline{
\epsfig{file=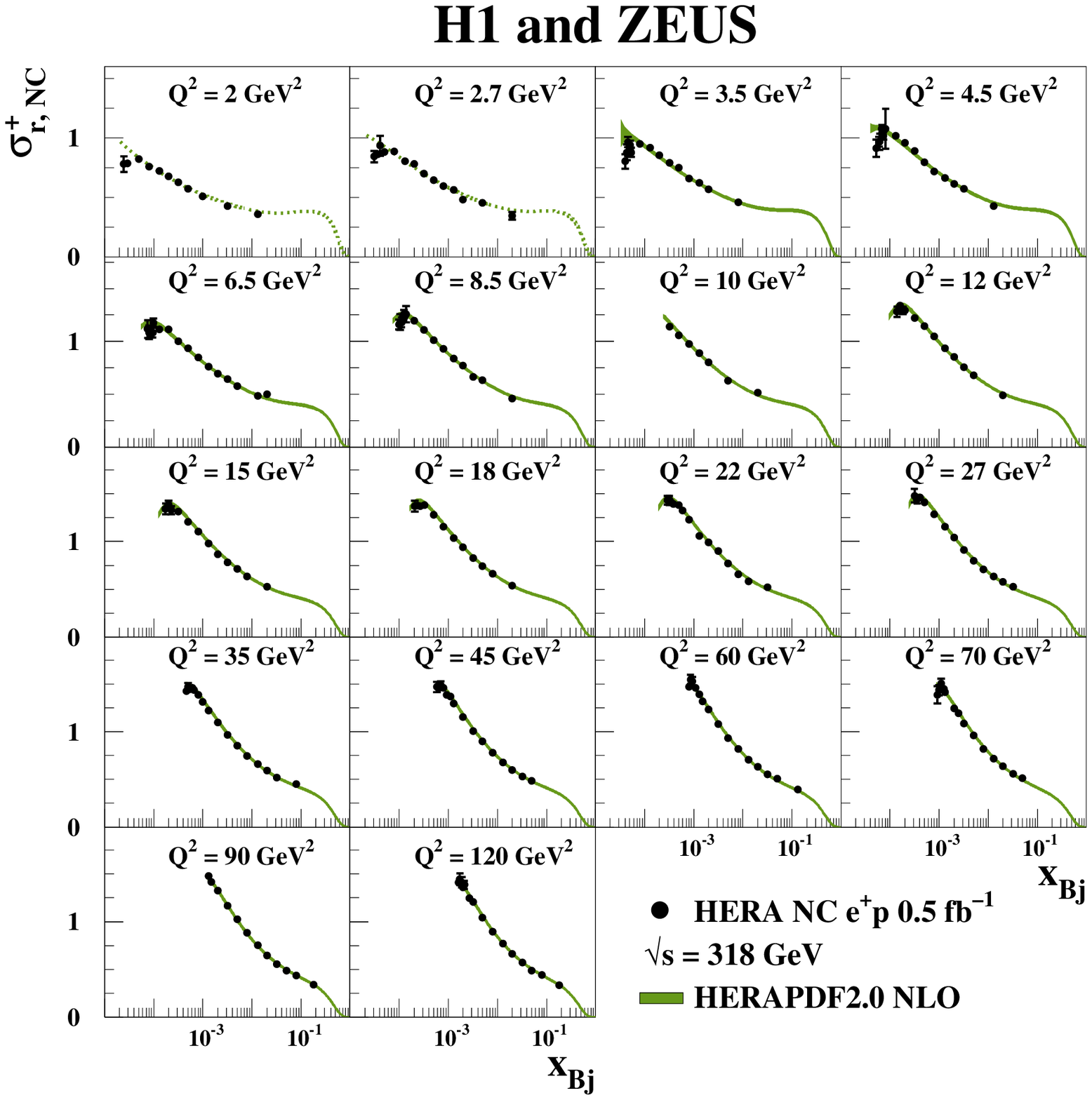   ,width=0.9\textwidth}}
\vspace{0.5cm}
\caption {The combined low-$Q^2$ HERA inclusive NC $e^+p$ reduced cross sections  
at $\sqrt{s} =318$\,GeV with overlaid predictions from HERAPDF2.0 NLO.
     The bands represent the total uncertainties on the predictions.   
Dotted lines indicate extrapolation into kinematic regions not included in the
fit.
}
\label{fig:nloQ23pt5ncepb}
\end{figure}
\clearpage

\begin{figure}[tbp]
\vspace{-0.3cm} 
\centerline{
\epsfig{file=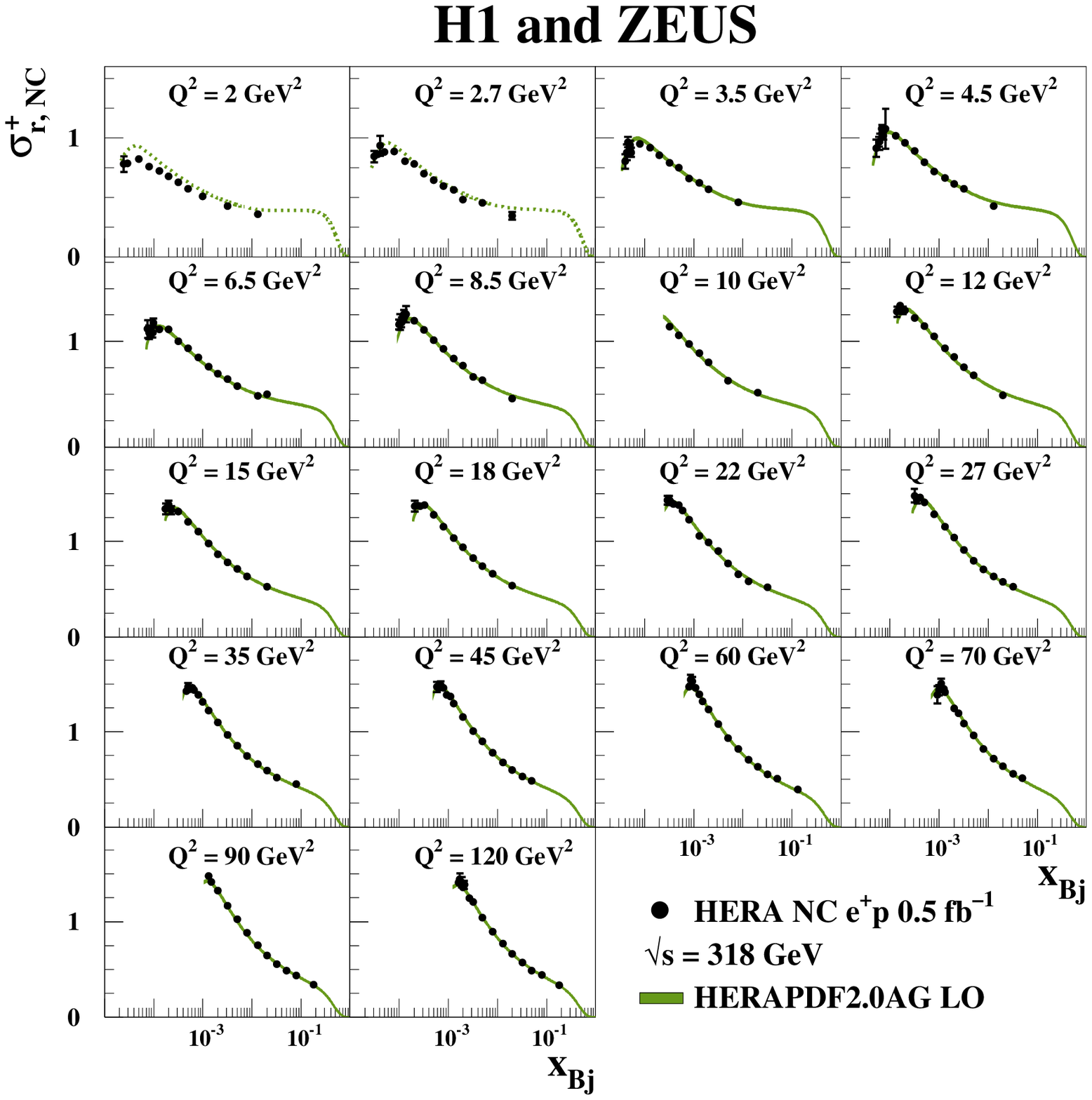   ,width=0.9\textwidth}}
\vspace{0.5cm}
\caption {The combined low-$Q^2$ HERA inclusive NC $e^+p$ reduced cross sections  
at $\sqrt{s} = 318$\,GeV with overlaid predictions from HERAPDF2.0AG LO.
     The bands represent the experimental uncertainties on the predictions.   
Dotted lines indicate extrapolation 
into kinematic regions not included in the
fit.
}
\label{fig:loQ23pt5ncepb}
\end{figure}

\clearpage


\begin{figure}[tbp]
\vspace{-0.3cm} 
\centerline{
\epsfig{file=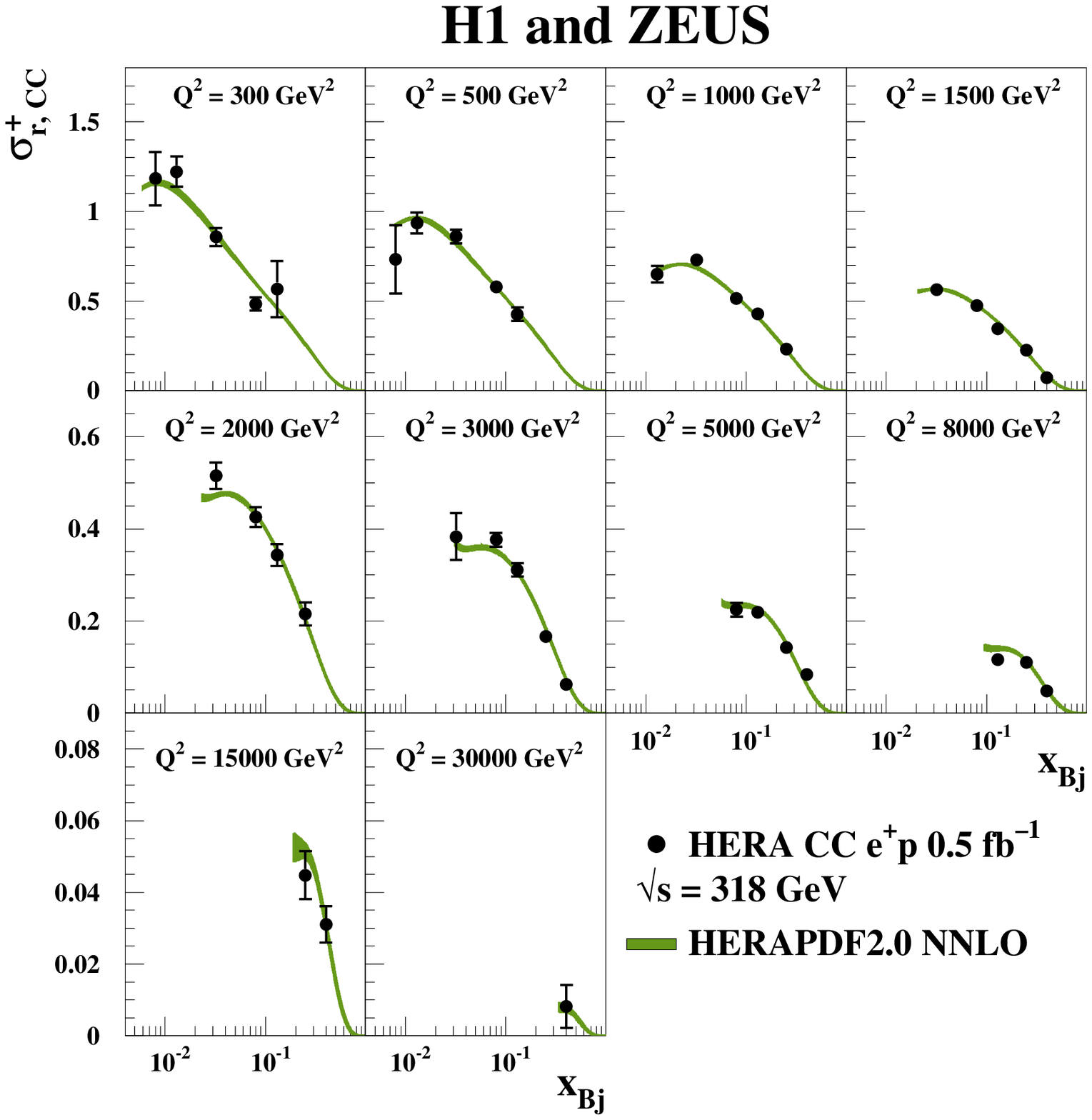 ,width=0.9\textwidth}}
\vspace{0.5cm}
\caption{The combined  HERA inclusive CC $e^+p$ reduced cross sections  
at $\sqrt{s} = 318$\,GeV with overlaid predictions from HERAPDF2.0 NNLO.
     The bands represent the total uncertainties on the predictions.   
}
\label{fig:nnloQ23pt5ccep}
\end{figure}

\clearpage

\begin{figure}[tbp]
\vspace{-0.3cm} 
\centerline{
\epsfig{file=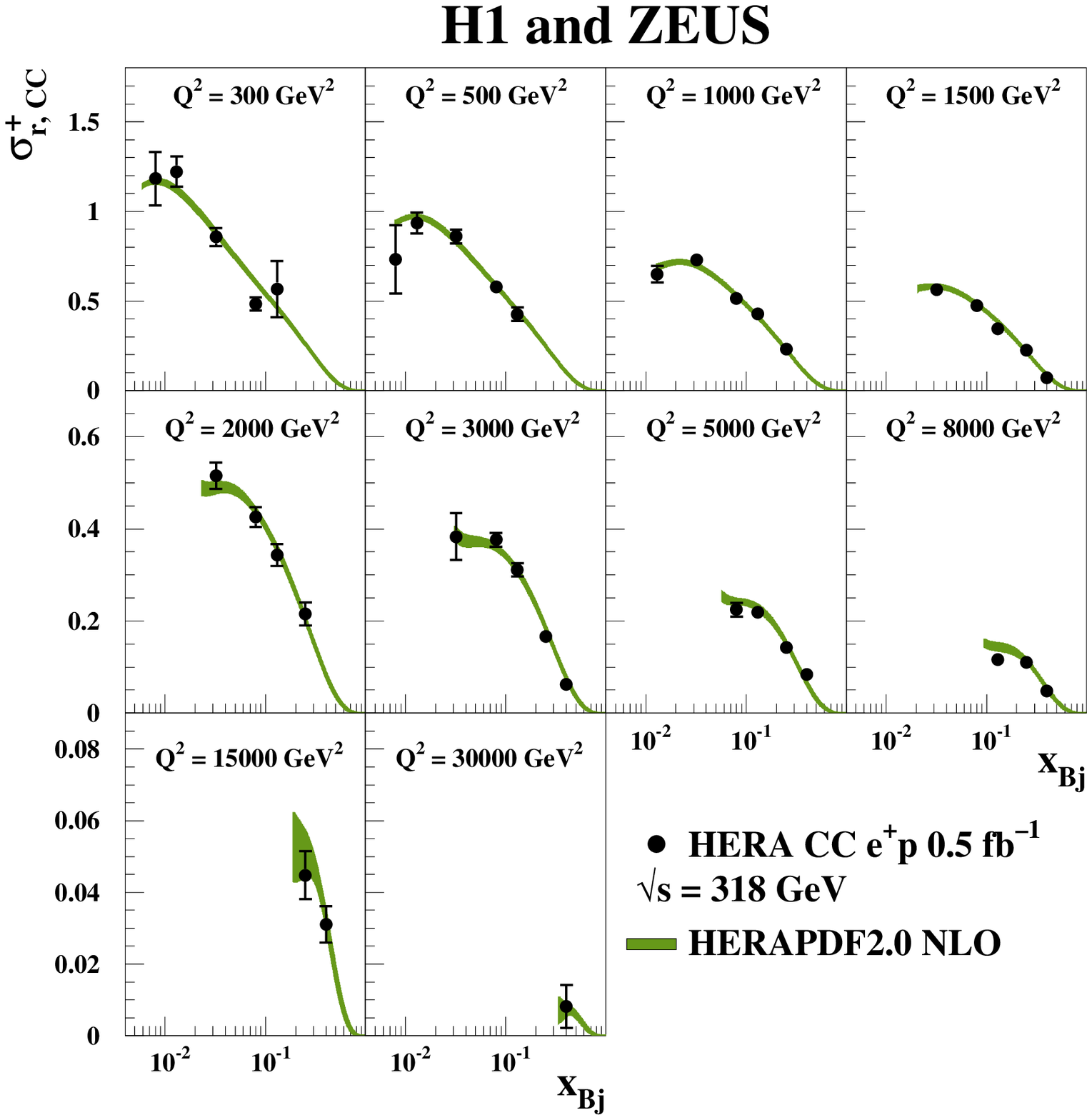   ,width=0.9\textwidth}}
\vspace{0.5cm}
\caption {The combined  HERA  inclusive CC $e^+p$ reduced cross sections  
at $\sqrt{s} = 318$ GeV with overlaid predictions from HERAPDF2.0 NLO.
     The bands represent the total uncertainties on the predictions.   
}
\label{fig:nloQ23pt5ccep}
\end{figure}

\clearpage

\begin{figure}[tbp]
\vspace{-0.3cm} 
\centerline{
\epsfig{file=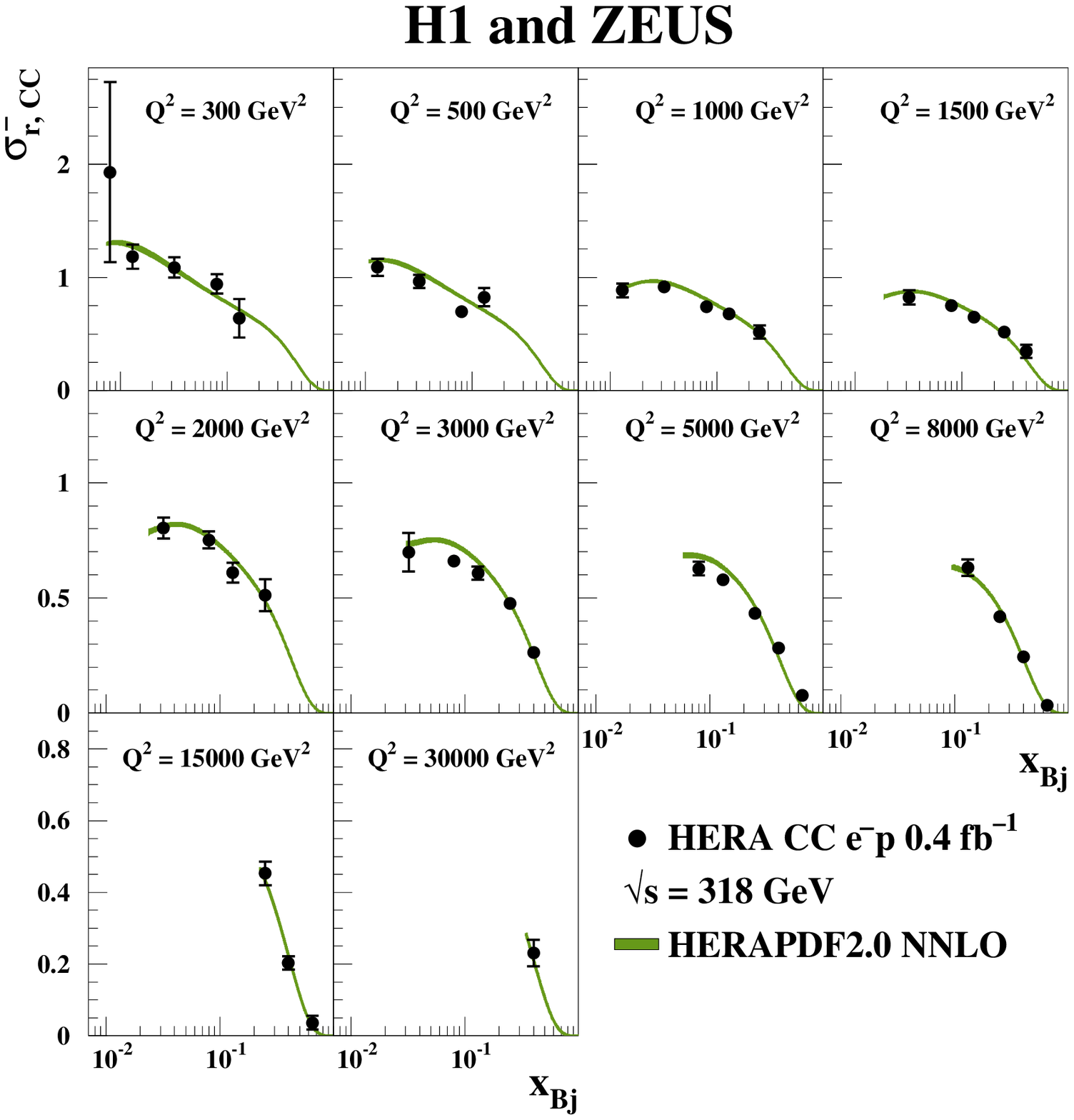   ,width=0.9\textwidth}}
\vspace{0.5cm}
\caption {The combined  HERA inclusive CC $e^-p$ reduced cross sections  
at $\sqrt{s} = 318$ GeV with overlaid predictions from HERAPDF2.0 NNLO.
     The bands represent the total uncertainties on the predictions.   
}
\label{fig:nnloQ23pt5ccem}
\end{figure}

\clearpage

\begin{figure}[tbp]
\vspace{-0.3cm} 
\centerline{
\epsfig{file=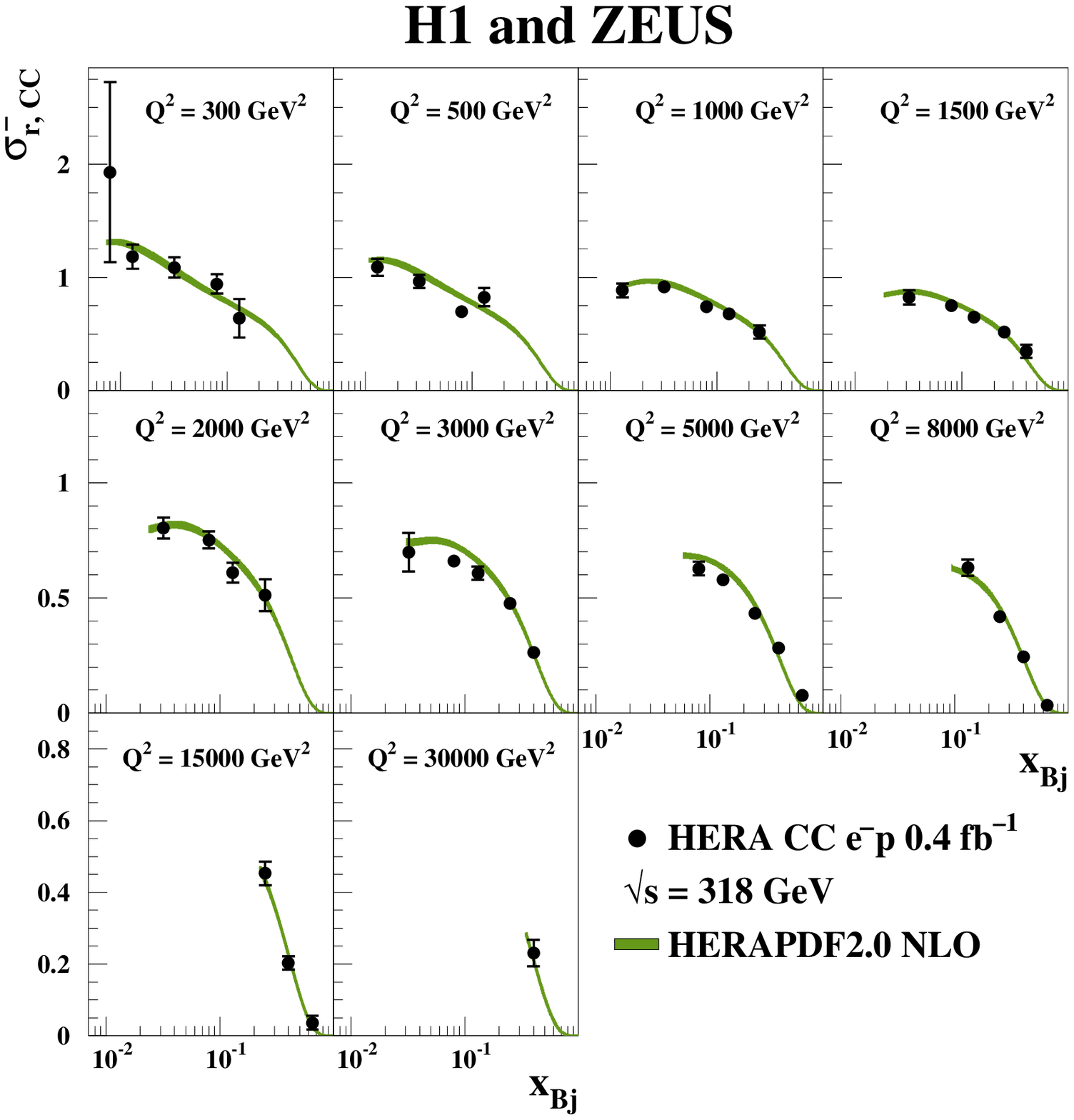   ,width=0.9\textwidth}}
\vspace{0.5cm}
\caption {The combined  HERA inclusive CC $e^-p$ 
reduced cross sections  
at $\sqrt{s} = 318$\,GeV 
with overlaid predictions of the HERAPDF2.0 NLO.
     The bands represent the total uncertainties on the predictions.   
}
\label{fig:nloQ23pt5ccem}
\end{figure}


%
\begin{figure}[tbp]
\vspace{-0.3cm} 
\centerline{
\epsfig{file=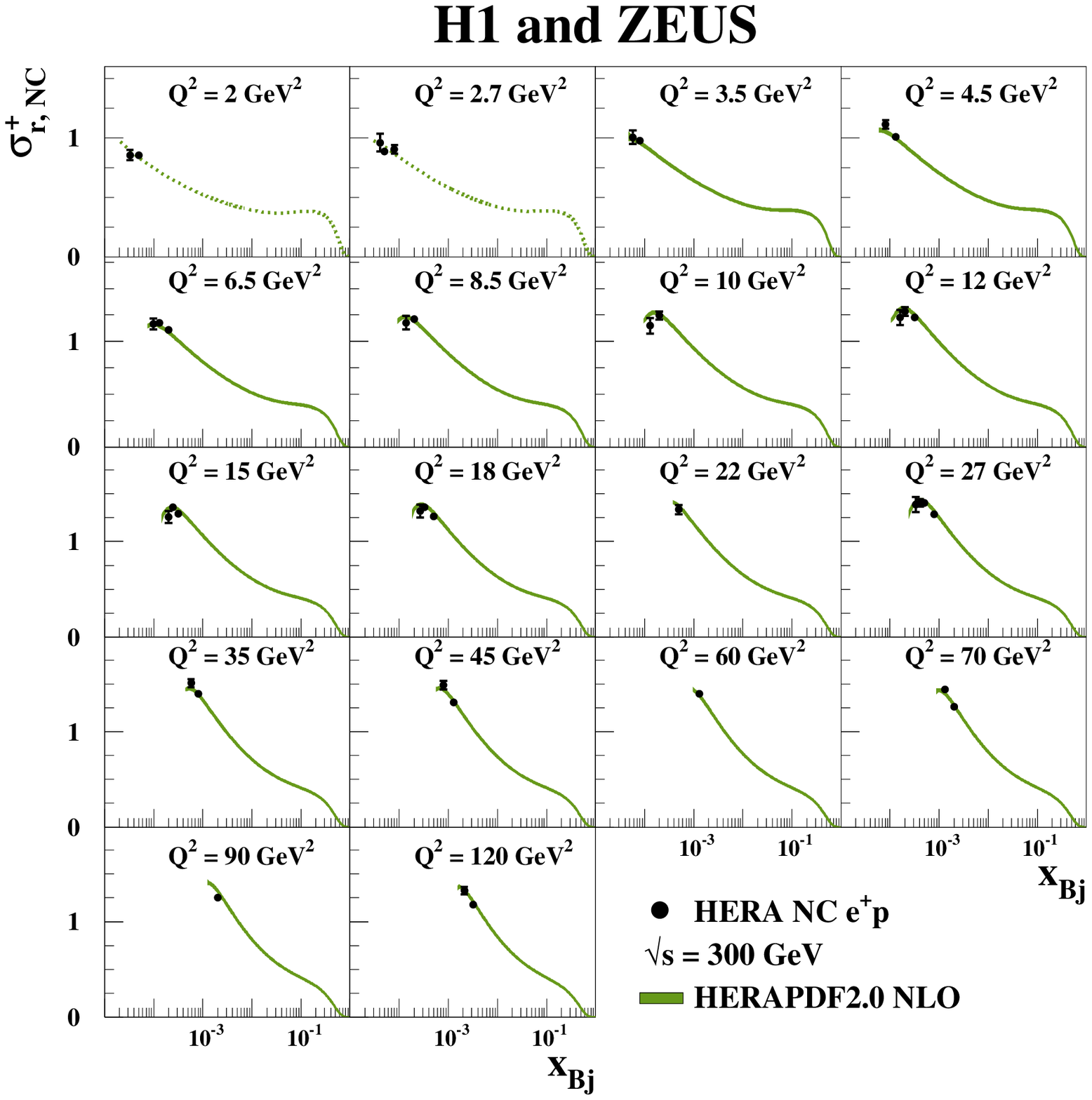   ,width=0.9\textwidth}}
\vspace{0.5cm}
\caption {The combined low-$Q^2$ HERA 
inclusive NC $e^+p$ reduced cross sections  
at $\sqrt{s} = 300$\,GeV with overlaid predictions 
of HERAPDF2.0 NLO. 
     The bands represent the total uncertainties on the predictions.   
Dotted lines indicate extrapolation into kinematic regions 
not included in the fit.
}
\label{fig:nloQ23pt5ncepb820}
\end{figure}

\clearpage

\begin{figure}[tbp]
\vspace{-0.3cm} 
\centerline{
\epsfig{file=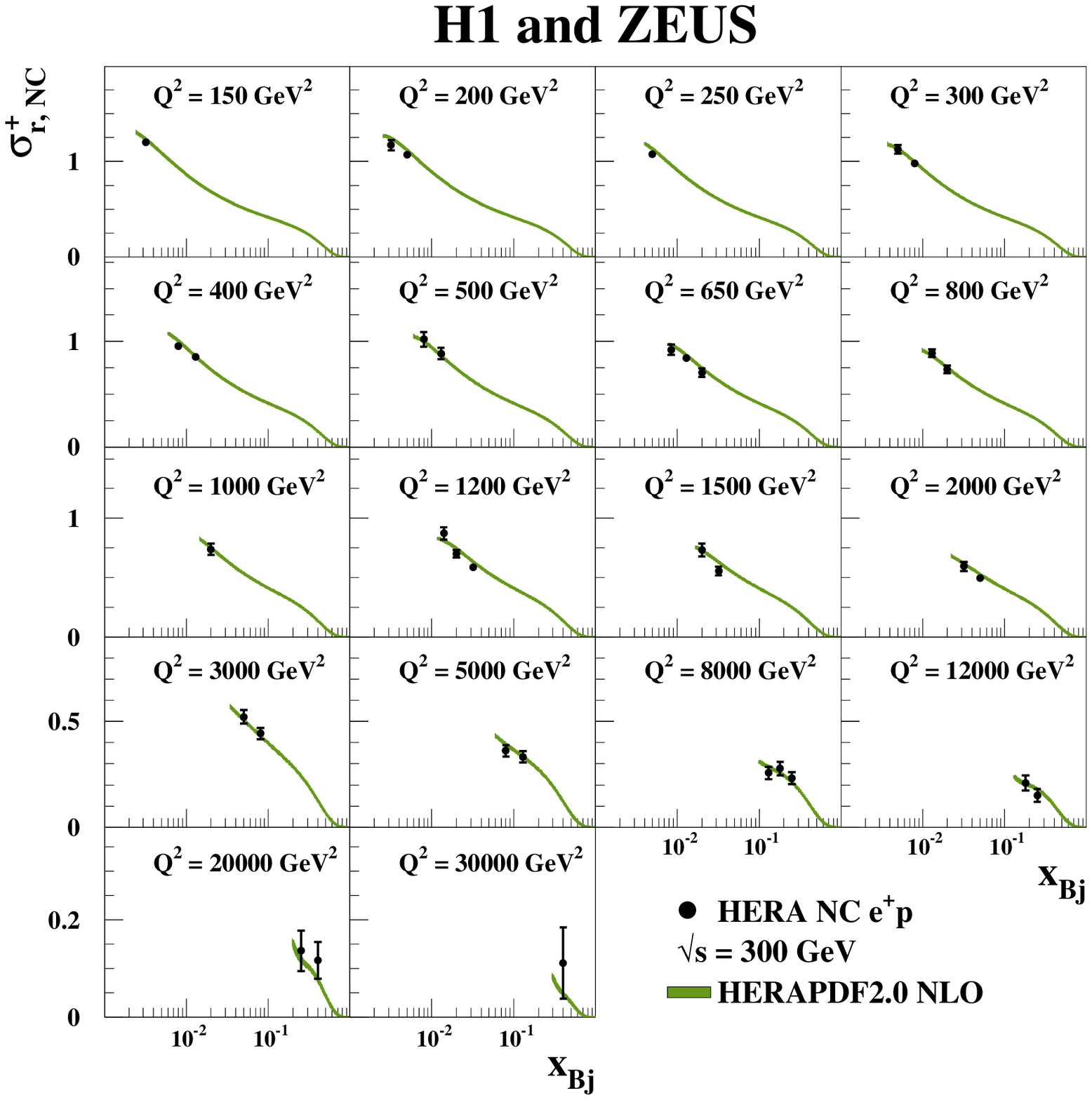   ,width=0.9\textwidth}}
\vspace{0.5cm}
\caption {The combined high-$Q^2$ HERA inclusive NC $e^+p$ 
reduced cross sections  at $\sqrt{s} = 300$\,GeV with overlaid 
predictions of HERAPDF2.0 NLO.
The bands represent the total uncertainties on the predictions.   
}
\label{fig:nloQ23pt5ncepc820}
\end{figure}

\clearpage

%
\begin{figure}[tbp]
\vspace{-0.3cm} 
\centerline{
\epsfig{file=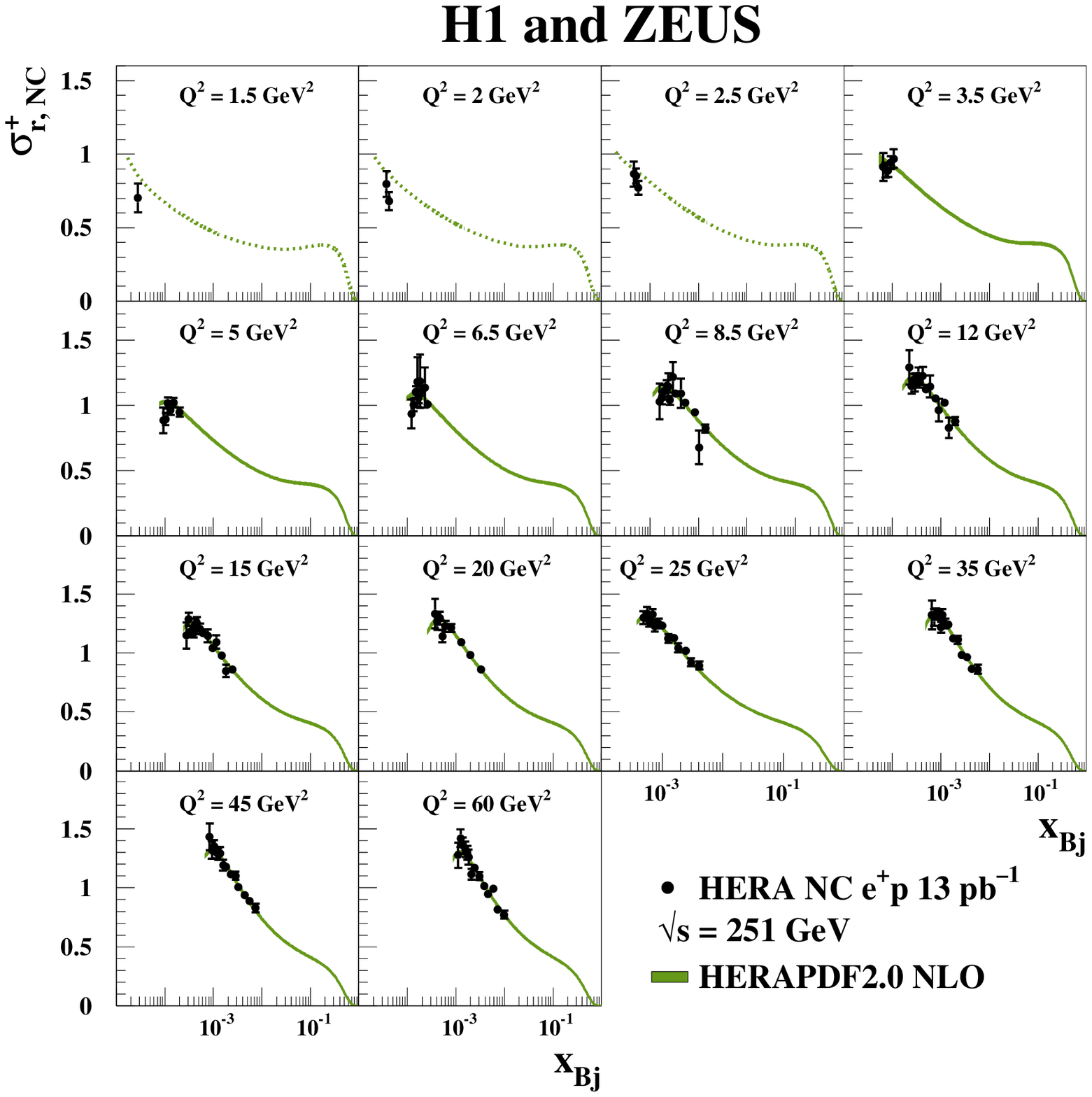   ,width=0.9\textwidth}}
\vspace{0.5cm}
\caption {The combined low-$Q^2$ HERA inclusive NC $e^+p$ 
reduced cross sections  
at $\sqrt{s} = 251$\,GeV with overlaid predictions from HERAPDF2.0 NLO.
The bands represent the total uncertainties on the predictions.   
Dotted lines indicate extrapolation into kinematic regions 
not included in the
fit.
}
\label{fig:nloQ23pt5ncepb575}
\end{figure}

\clearpage

\begin{figure}[tbp]
\vspace{-0.3cm} 
\centerline{
\epsfig{file=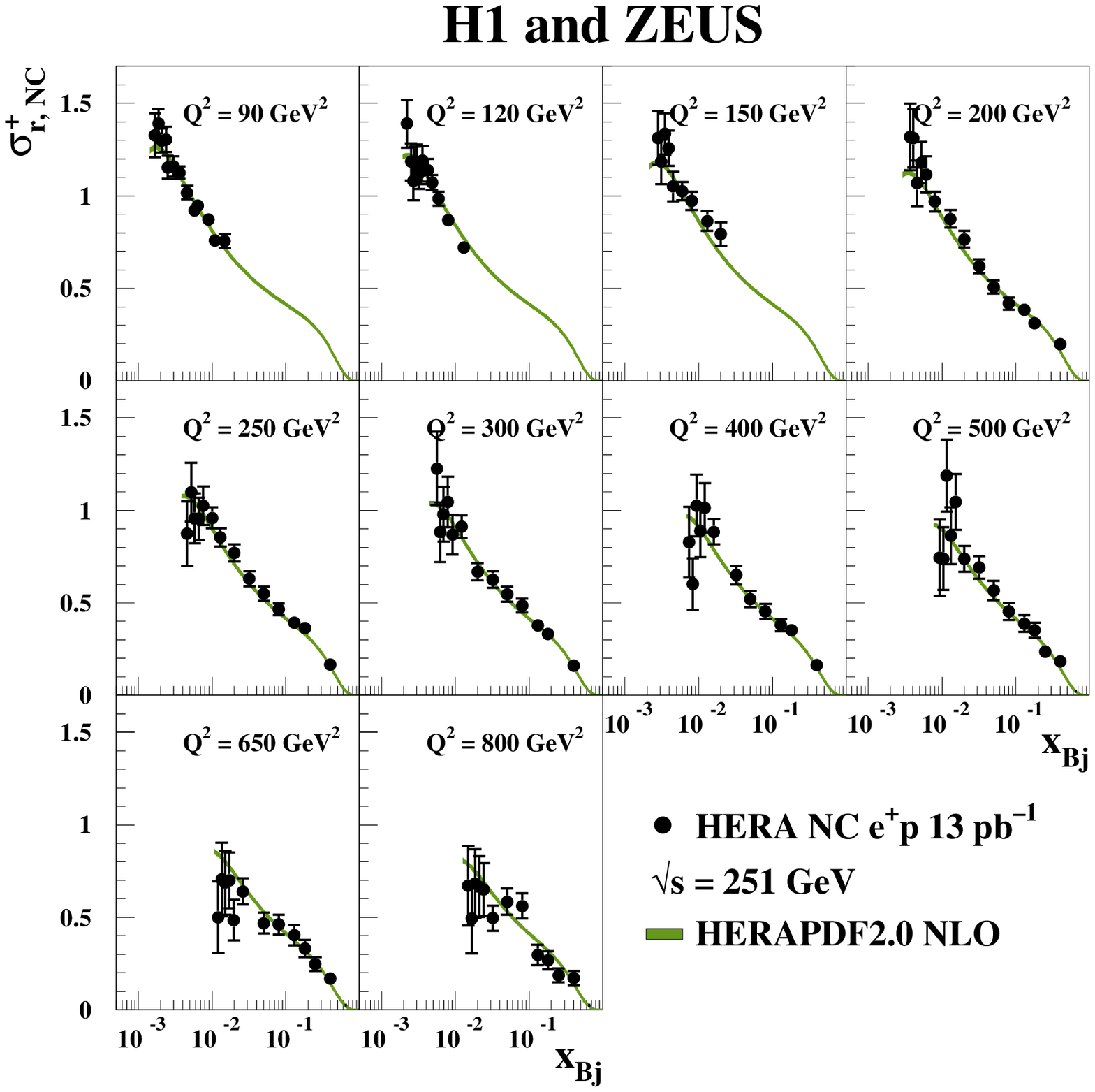   ,width=0.9\textwidth}}
\vspace{0.5cm}
\caption {The combined high-$Q^2$ HERA inclusive NC $e^+p$ 
reduced cross sections  
at $\sqrt{s} = 251$\,GeV with overlaid predictions 
from HERAPDF2.0 NLO.
The bands represent the total uncertainties on the predictions.   
}
\label{fig:nloQ23pt5ncepc575}
\end{figure}

\clearpage

\begin{figure}[tbp]
\vspace{-0.3cm} 
\centerline{
\epsfig{file=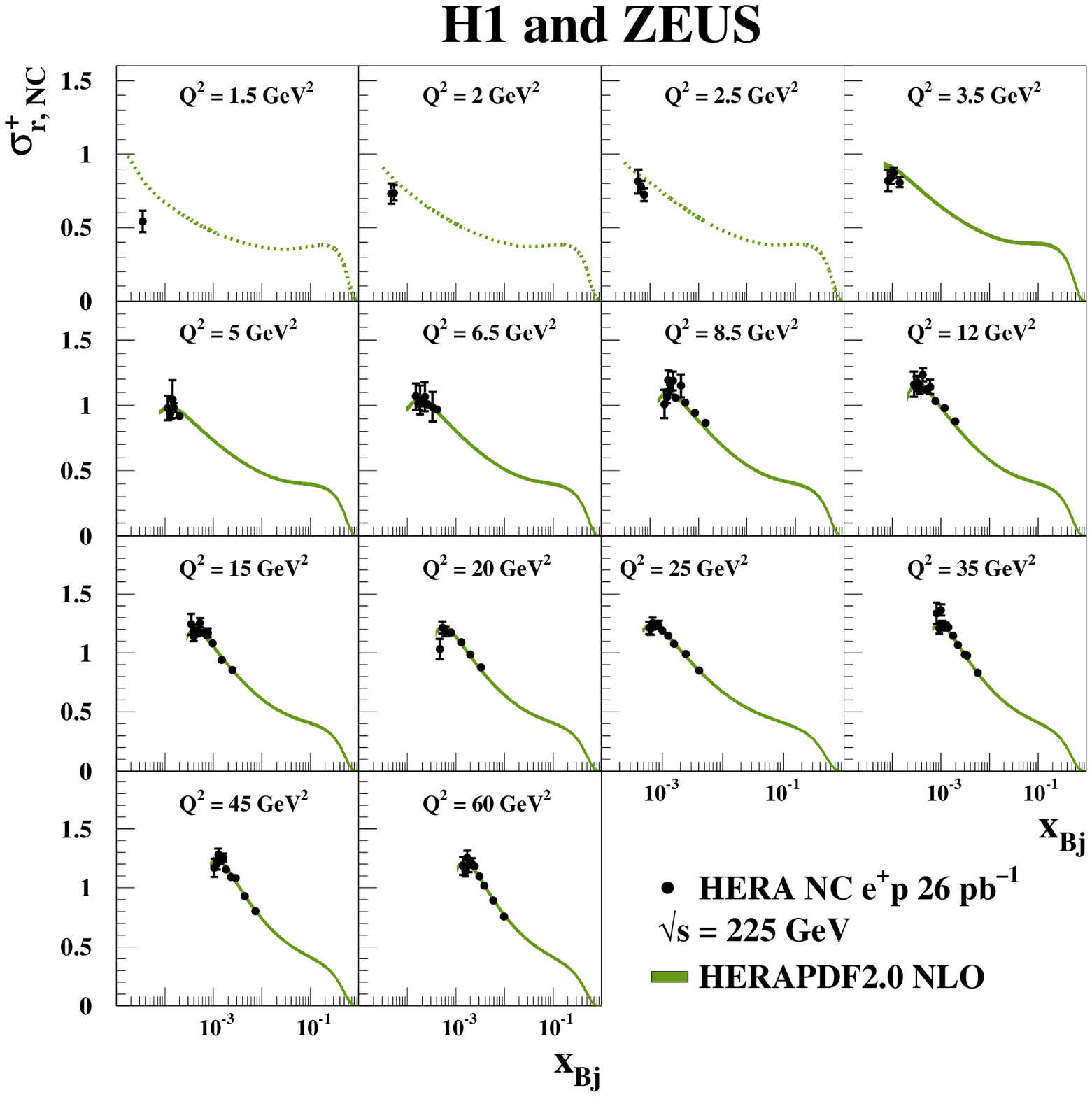   ,width=0.9\textwidth}}
\vspace{0.5cm}
\caption {The combined low-$Q^2$ HERA 
inclusive NC $e^+p$ reduced cross sections  
at $\sqrt{s} = 225$\,GeV with overlaid predictions from HERAPDF2.0 NLO.
     The bands represent the total uncertainties on the predictions.   
Dotted lines indicate extrapolation into kinematic regions not included in the
fit.
}
\label{fig:nloQ23pt5ncepb460}
\end{figure}

\clearpage

\begin{figure}[tbp]
\vspace{-0.3cm} 
\centerline{
\epsfig{file=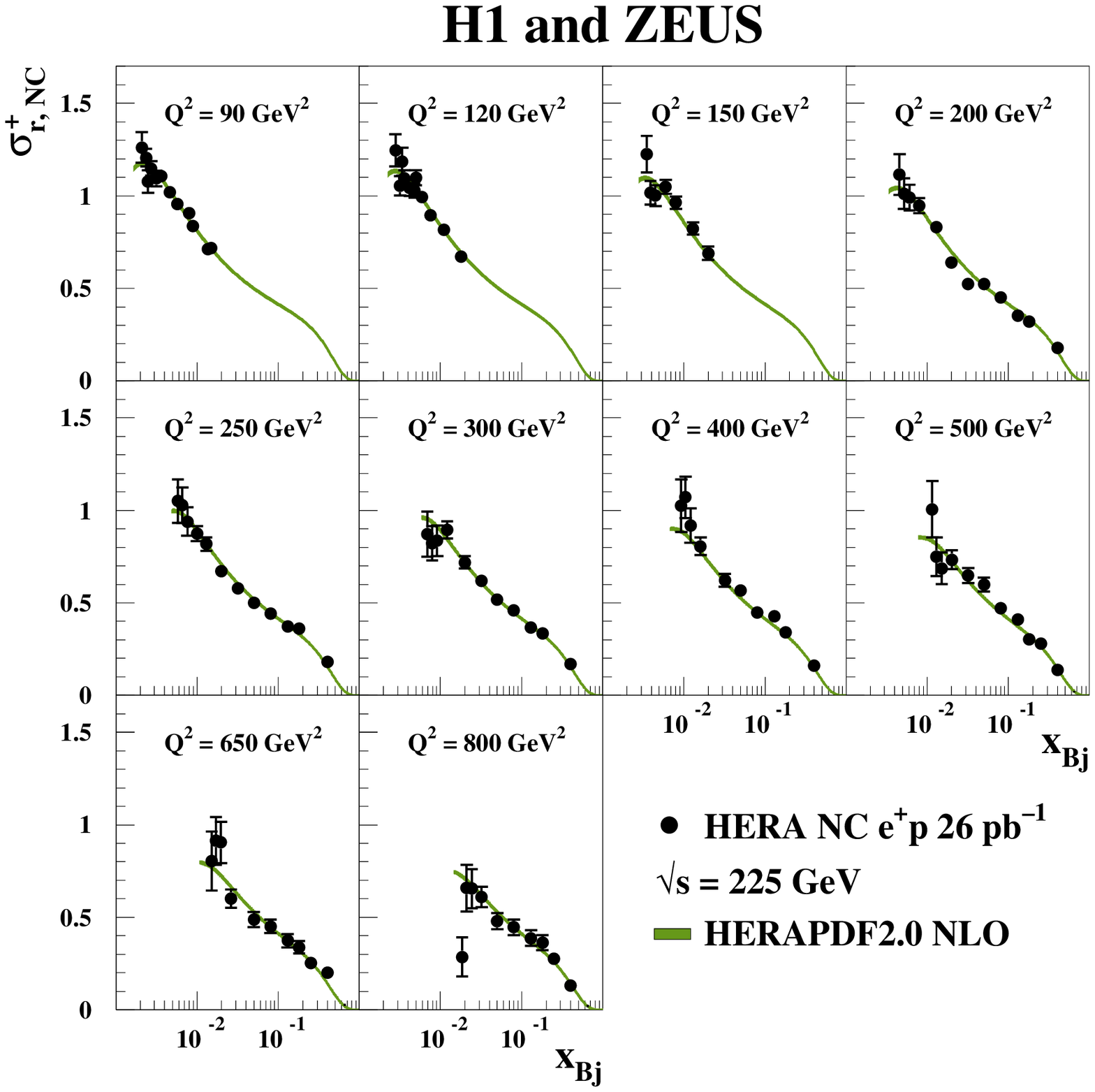   ,width=0.9\textwidth}}
\vspace{0.5cm}
\caption {The combined high-$Q^2$ HERA inclusive NC $e^+p$ reduced cross sections  
at $\sqrt{s} = 225$ GeV with overlaid predictions from HERAPDF2.0 NLO.
     The bands represent the total uncertainties on the predictions.   
}
\label{fig:nloQ23pt5ncepc460}
\end{figure}


\clearpage

\clearpage

\clearpage

\clearpage

\clearpage

\clearpage



\clearpage

\clearpage 

\clearpage

\clearpage

\begin{figure}[tbp]
\vspace{-0.3cm} 
\centerline{
\epsfig{file=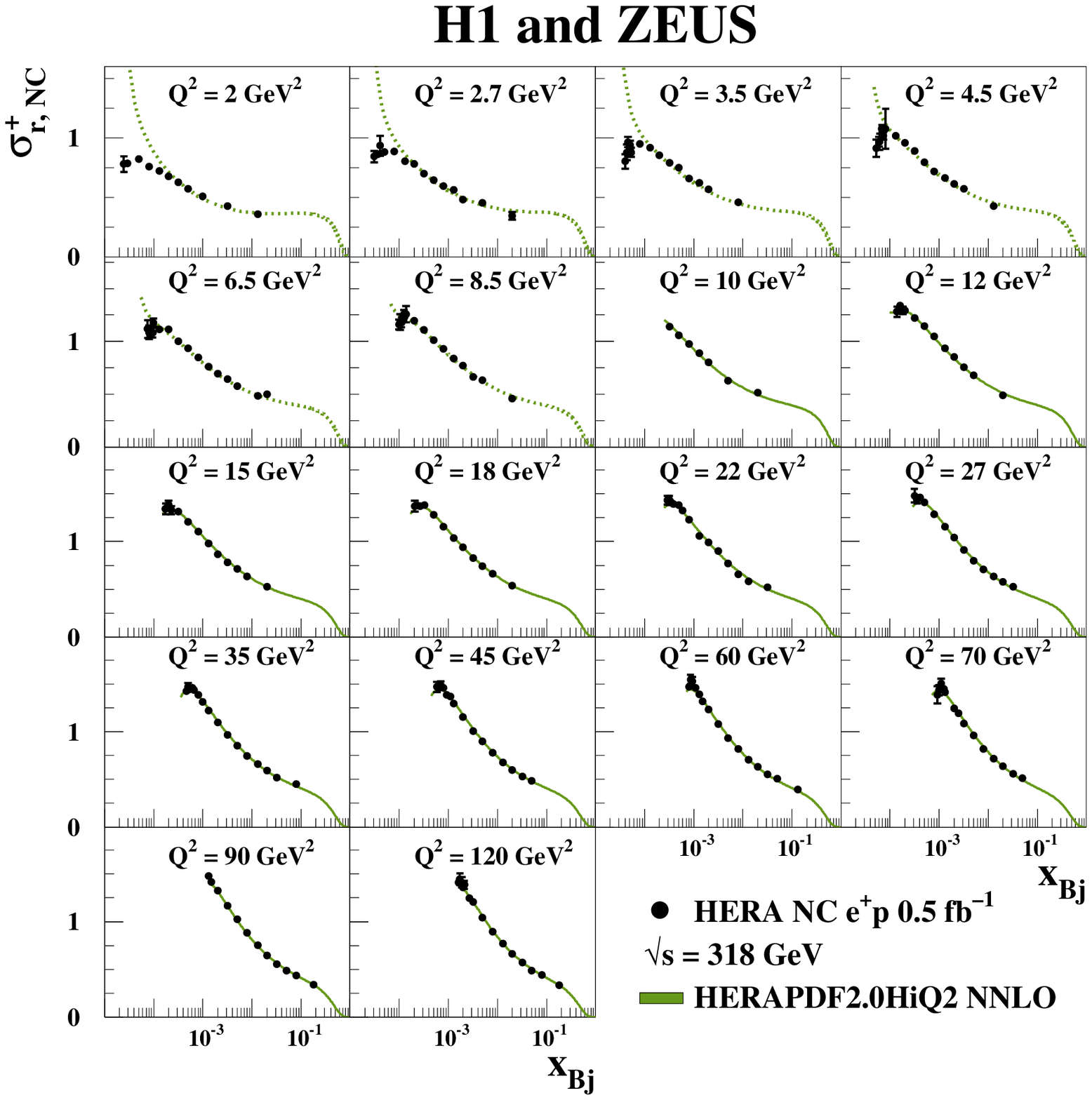   ,width=0.9\textwidth}}
\vspace{0.5cm}
\caption {The combined low-$Q^2$ HERA data on inclusive NC $e^+p$ reduced cross sections  
at $\sqrt{s} = 318$ GeV with overlaid predictions from HERAPDF2.0HiQ2 NNLO 
The   bands represent the total uncertainty on the predictions.
Dotted lines indicate extrapolation 
into kinematic regions not included in the
fit.
}
\label{fig:nnloQ210ncepb}
\end{figure}
\clearpage

\begin{figure}[tbp]
\vspace{-0.3cm} 
\centerline{
\epsfig{file=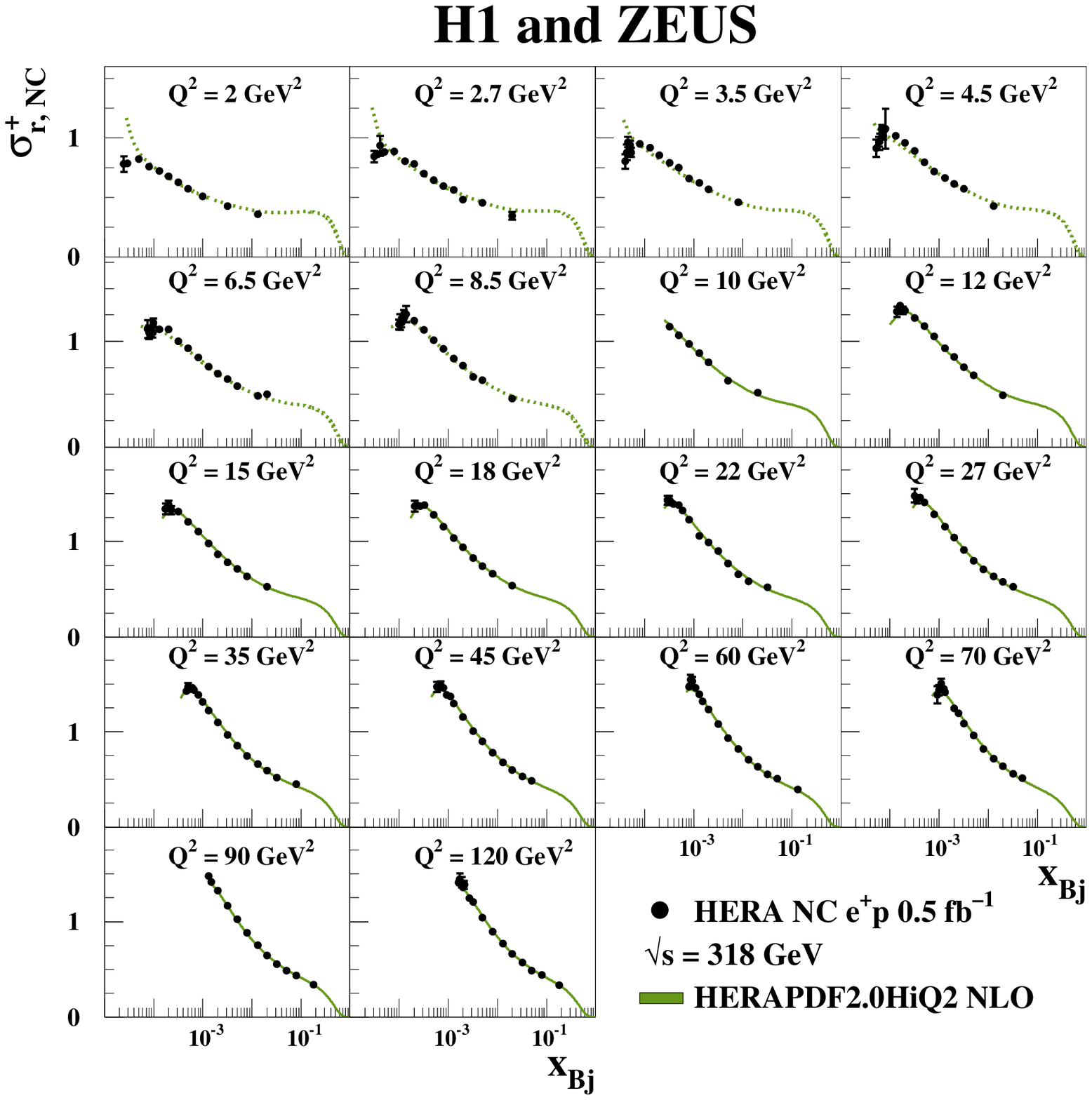   ,width=0.9\textwidth}}
\vspace{0.5cm}
\caption {The combined low-$Q^2$ HERA data on 
          inclusive NC $e^+p$ reduced cross sections  
          at $\sqrt{s} = 318$ GeV with 
          overlaid predictions from HERAPDF2.0HiQ2 NLO. 
The  bands represent the total uncertainty on the predictions.
Dotted lines indicate extrapolation into 
kinematic regions not included in the
fit.
}
\label{fig:nloQ210ncepb}
\end{figure}
\clearpage

\clearpage

\clearpage

\clearpage

\begin{figure}[tbp]
\vspace{-0.3cm} 
\centerline{
\epsfig{file=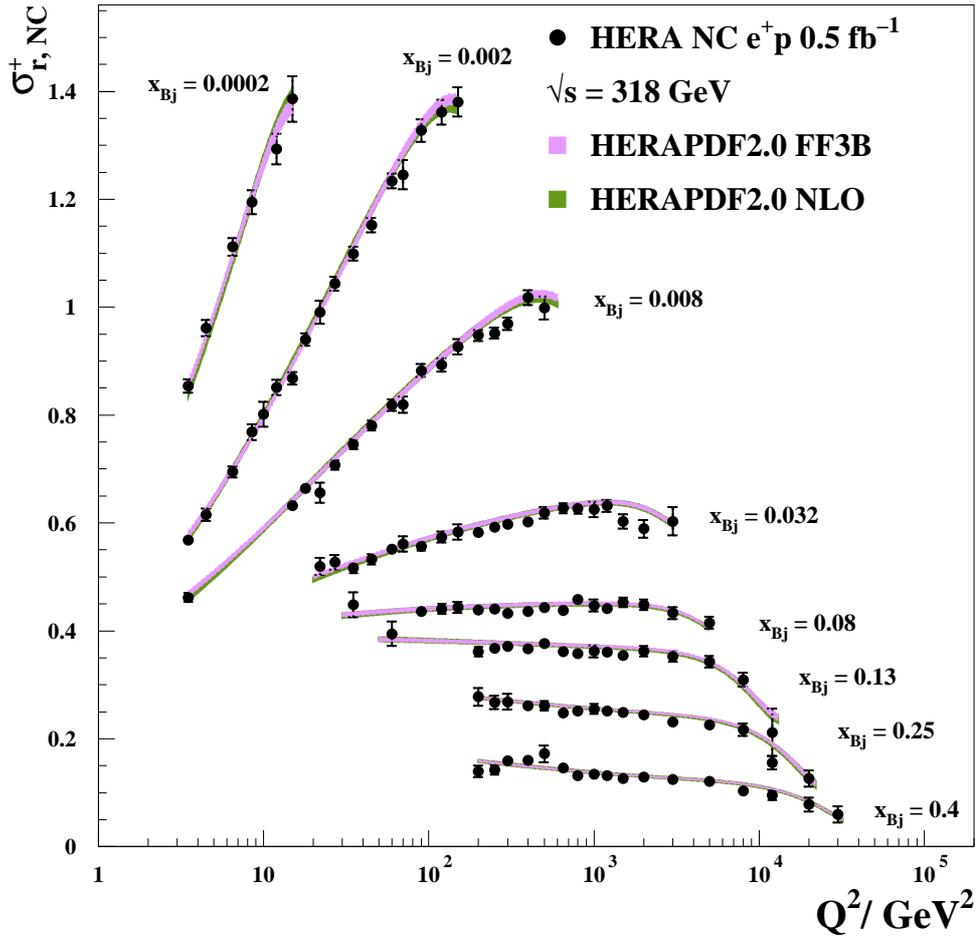 ,width=0.9\textwidth}}
\vspace{0.5cm}
\caption { 
     Selected combined HERA 
     inclusive NC $e^+p$ reduced cross sections compared to 
     predictions of HERAPDF2.0 NLO and HERAPDF2.0FF3B.
     The two differently shaded bands represent the total 
     uncertainties on the two predictions.   
} 
\label{fig:FFdata}
\end{figure}

\clearpage

\clearpage

\clearpage



\clearpage

\clearpage

\clearpage


\clearpage

\clearpage


\clearpage

\clearpage

\clearpage

\clearpage
\begin{figure}[tbp]
\vspace{-0.3cm} 
\centerline{
\epsfig{file=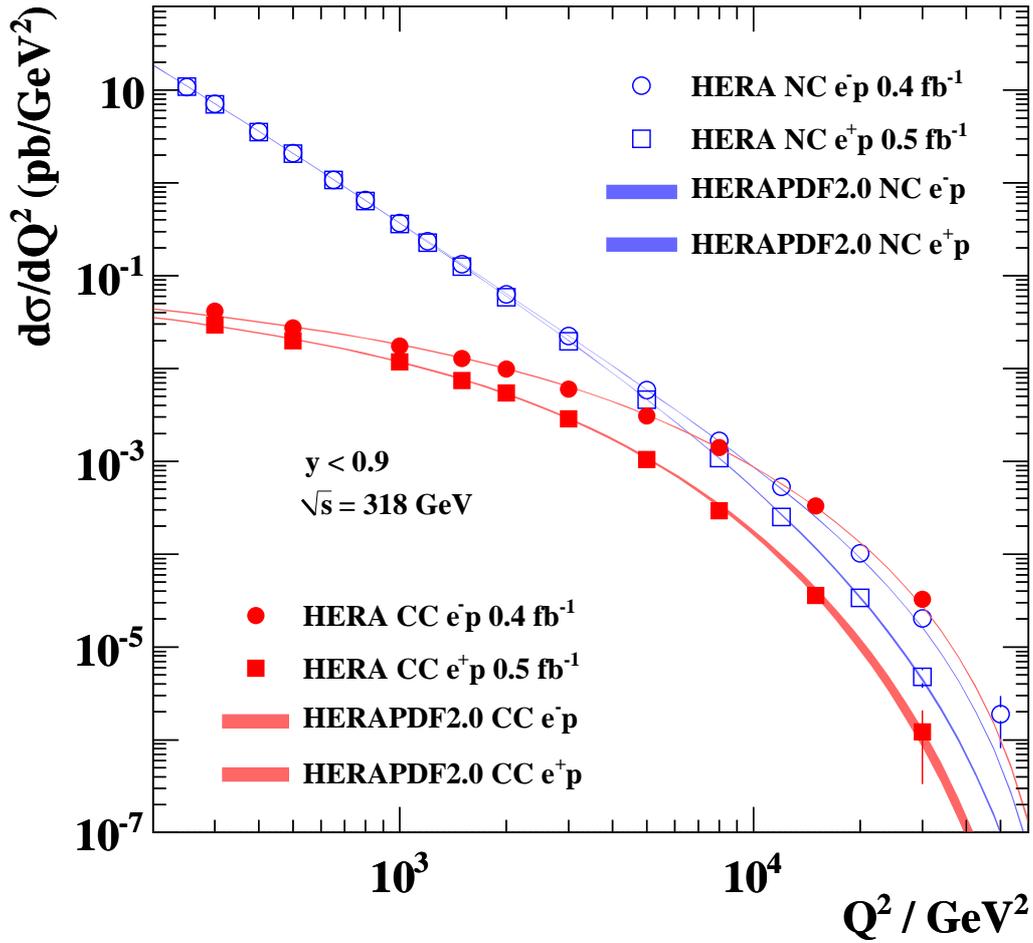 ,width=0.9\textwidth}}
\vspace{0.5cm}
\caption {The combined HERA NC and CC  $e^-p$ and $e^+p$ cross sections,
${\rm d}\sigma/{\rm d}Q^2$, 
together with predictions from HERAPDF2.0 NLO.
The bands represent the total uncertainty on the predictions.
}
\label{fig:EWuni}
\end{figure}

\clearpage


\begin{figure}[tbp]
\vspace{-0.3cm} 
\centerline{
\epsfig{file=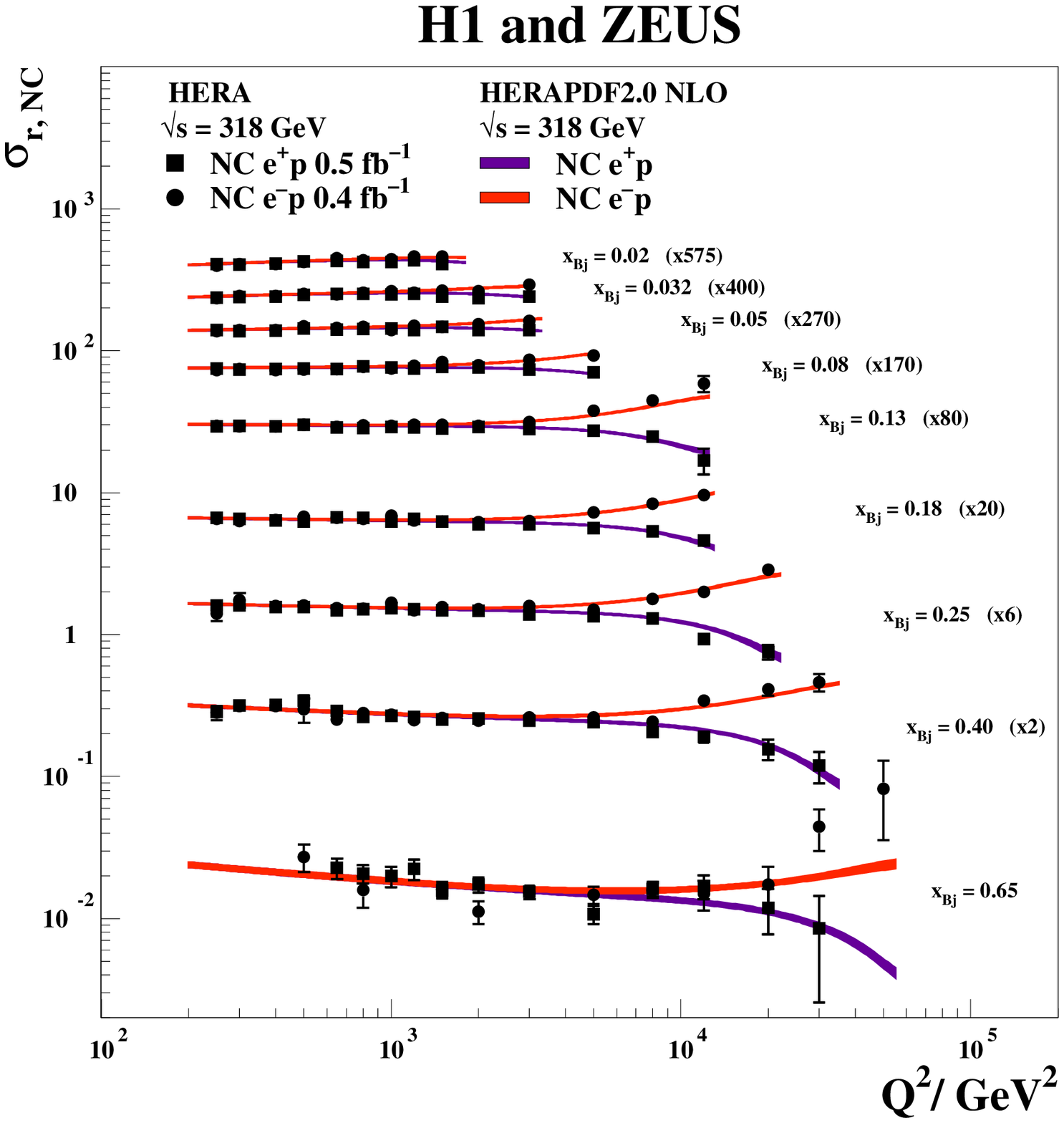   ,width=0.9\textwidth}}
\vspace{0.5cm}
\caption {The combined HERA data for the inclusive NC $e^+p$  
and $e^-p$ reduced cross sections 
as a function
of $Q^2$ for selected values of $x_{\rm Bj}$
at $\sqrt{s} = 318$\,GeV 
with overlaid predictions of HERAPDF2.0 NLO. 
The bands represent the total uncertainties of the predictions.
}
\label{fig:nloQ23pt5ncemep}
\end{figure}
\clearpage

\begin{figure}[tbp]
\vspace{-0.3cm} 
\centerline{
\epsfig{file=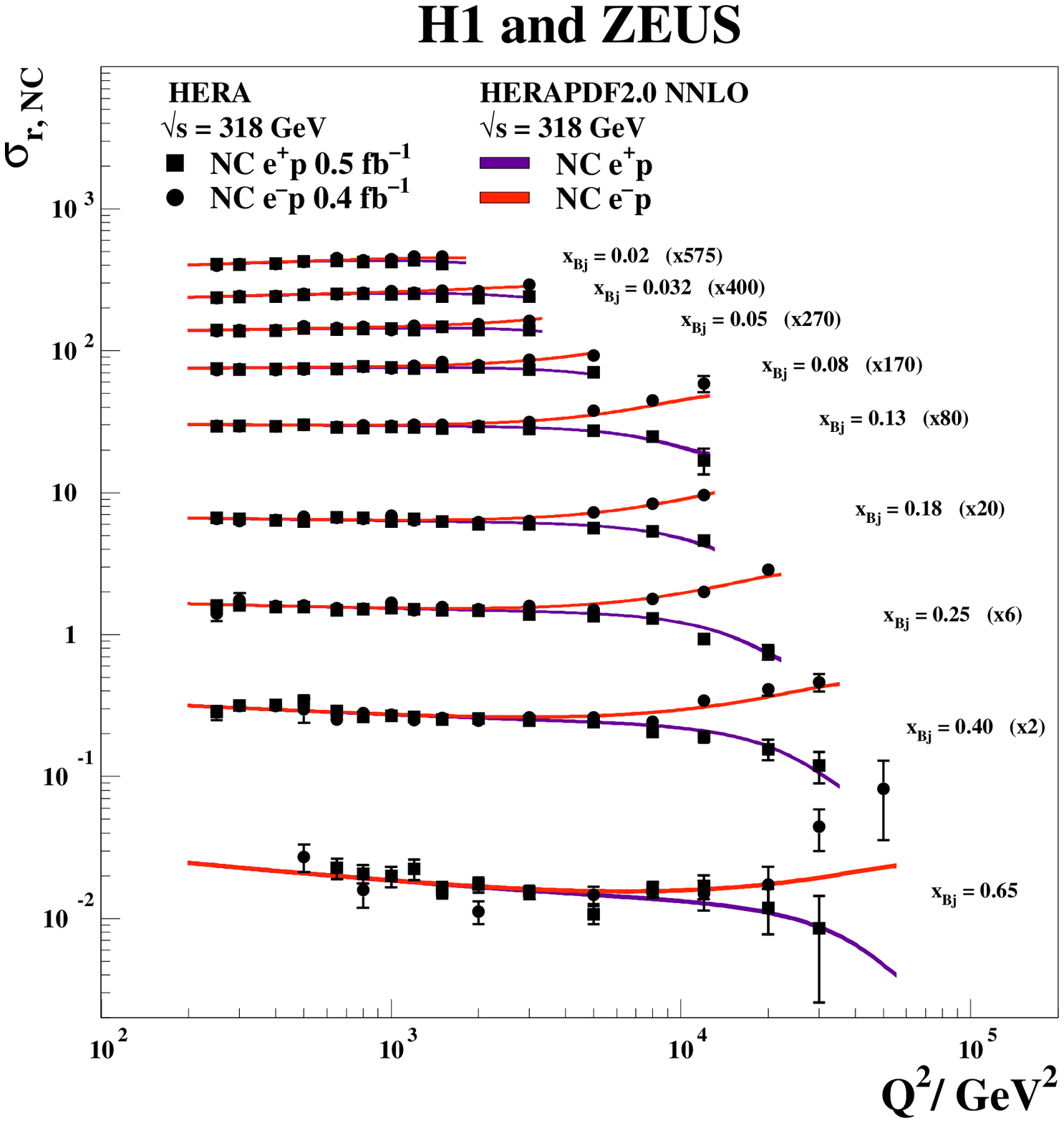   ,width=0.9\textwidth}}
\vspace{0.5cm}
\caption {The combined HERA data for the inclusive NC $e^+p$  
and $e^-p$ reduced cross sections 
as a function
of $Q^2$ for selected values of $x_{\rm Bj}$
at $\sqrt{s} = 318$\,GeV 
with overlaid predictions of HERAPDF2.0 NNLO. 
The bands represent the total uncertainties of the predictions.
}
\label{fig:nnloQ23pt5ncemep}
\end{figure}
\clearpage


\clearpage


\clearpage
\begin{figure}[tbp]
\vspace{-0.3cm} 
\centerline{
\epsfig{file=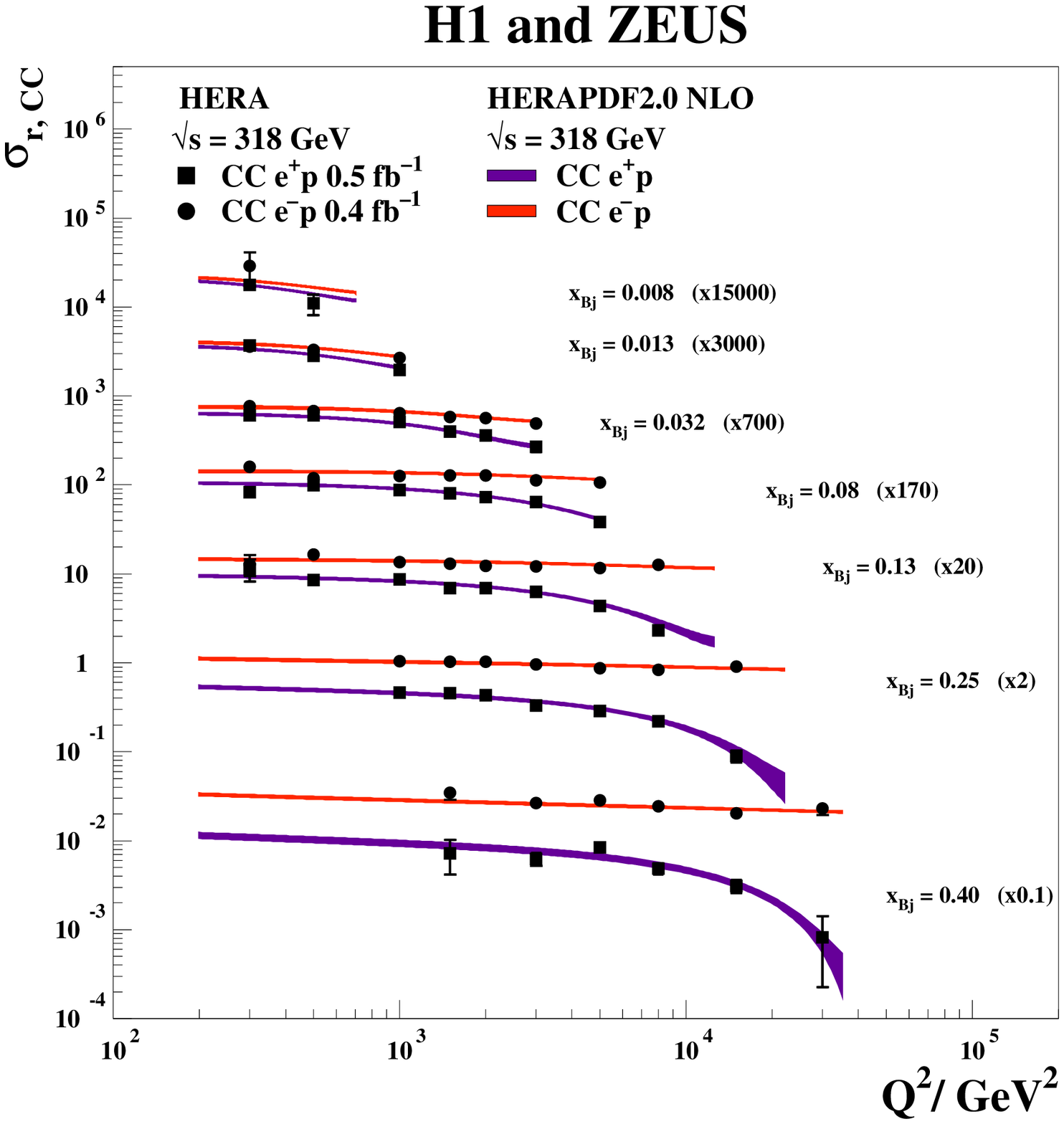   ,width=0.9\textwidth}}
\vspace{0.5cm}
\caption {The combined HERA data for  inclusive CC $e^+p$  
and $e^-p$ reduced cross sections at $\sqrt{s} = 318$\,GeV 
with overlaid predictions of HERAPDF2.0 NLO.
The bands represent the total uncertainties on the predictions.
}
\label{fig:scaling-CC-NLO}
\end{figure}

\clearpage
\begin{figure}[tbp]
\vspace{-0.3cm} 
\centerline{
\epsfig{file=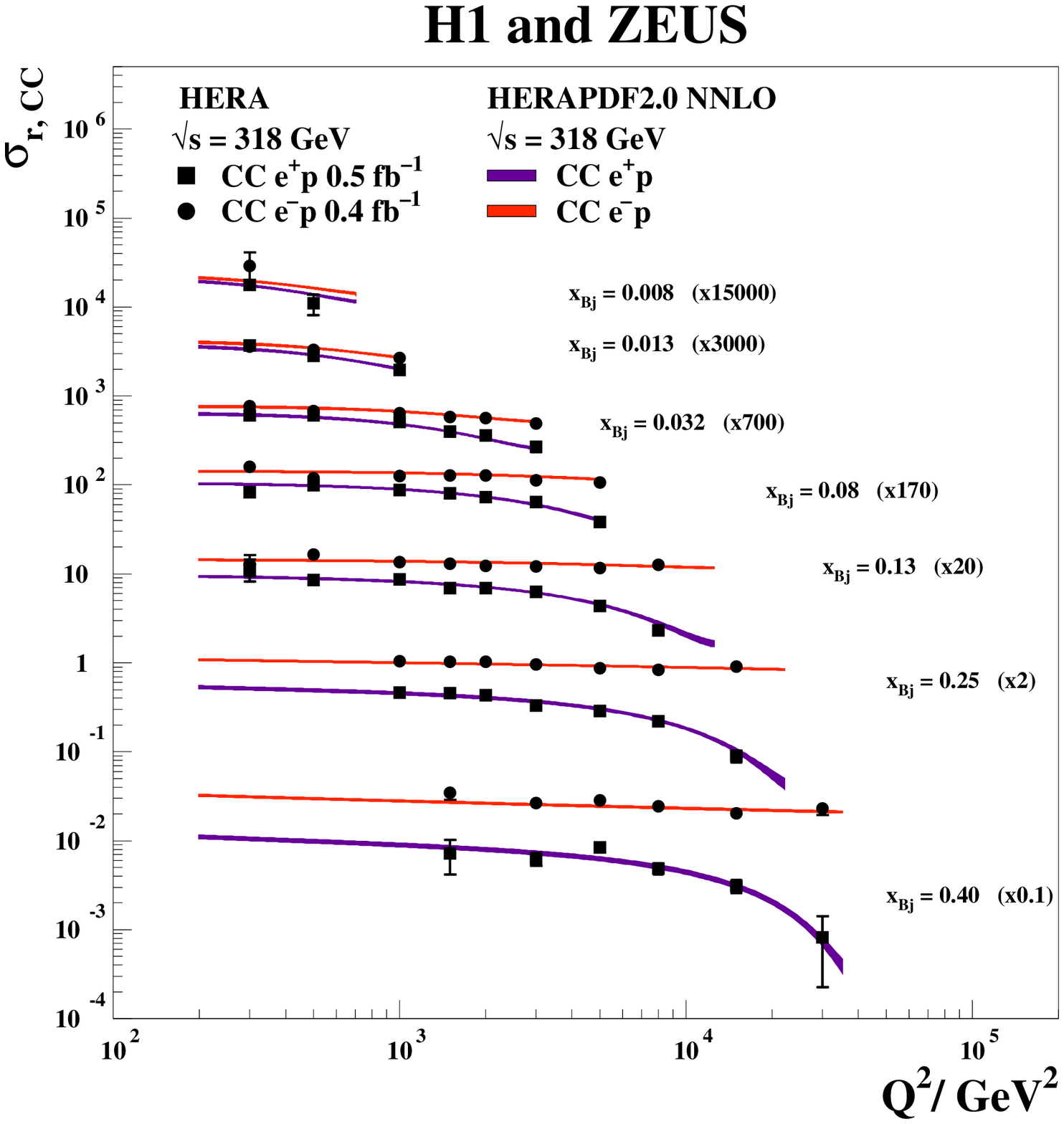   ,width=0.9\textwidth}}
\vspace{0.5cm}
\caption {The combined HERA data for inclusive CC $e^+p$  
and $e^-p$ reduced cross sections at $\sqrt{s} = 318$\,GeV 
with overlaid predictions of  
HERAPDF2.0 NNLO.
The bands represent the total uncertainty on the predictions.
}
\label{fig:scaling-CC-NNLO}
\end{figure}
\clearpage


\begin{figure}[tbp]
\vspace{-0.3cm} 
\centerline{
\epsfig{file=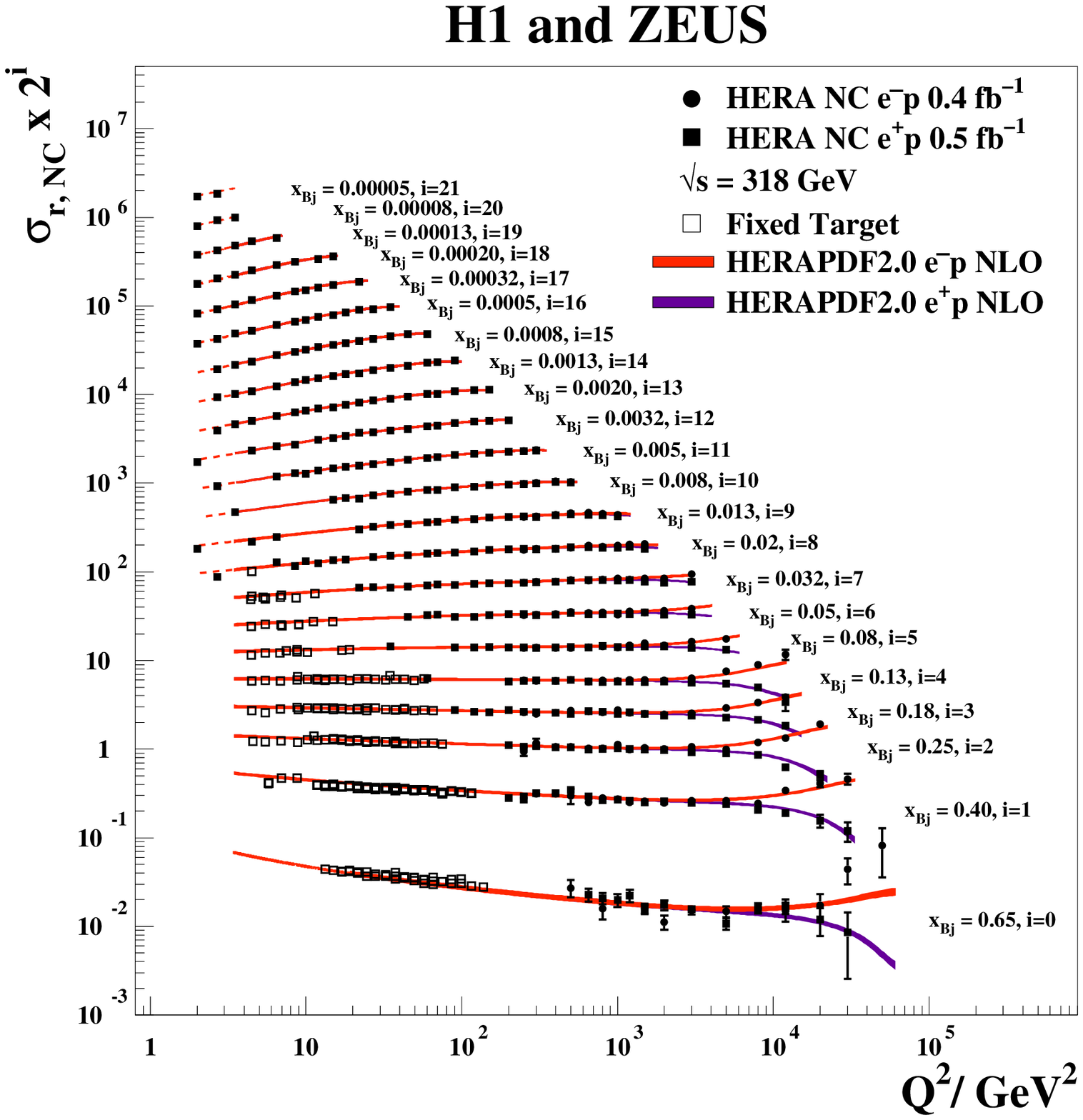 ,width=0.9\textwidth}}
\vspace{0.5cm}
\caption {The combined HERA data for the inclusive NC $e^+p$ 
          and $e^-p$ reduced cross sections 
          together with fixed-target data~\cite{bcdms,nmc} 
          and the predictions of HERAPDF2.0 NLO. 
          The bands represent the total uncertainties on 
          the predictions.
Dashed lines indicate extrapolation into 
kinematic regions not included in the
fit.
}
\label{fig:nloQ23pt5scal}
\end{figure}

\clearpage

\begin{figure}[tbp]
\vspace{-0.3cm} 
\centerline{
\epsfig{file=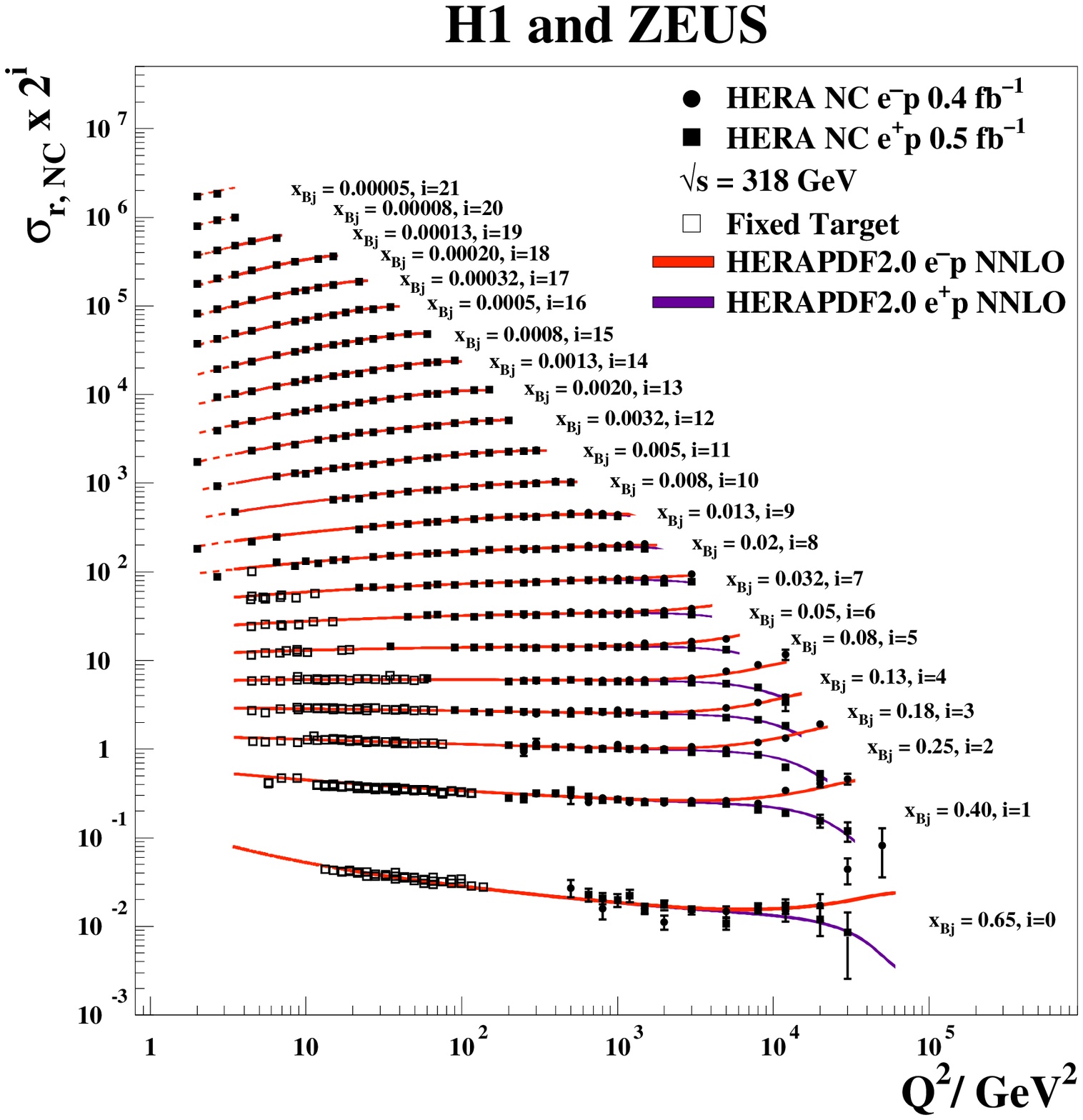   ,width=0.9\textwidth}}
\vspace{0.5cm}
\caption {The combined HERA data for the inclusive NC $e^+p$ 
          and $e^-p$ reduced cross sections 
          together with fixed-target data~\cite{bcdms,nmc}  
          and the predictions of HERAPDF2.0 NNLO. 
          The bands represent the total uncertainties on 
          the predictions.
Dashed lines indicate extrapolation into 
kinematic regions not included in the
fit.
}
\label{fig:nnloQ23pt5scal}
\end{figure}
\clearpage


\clearpage

\appendix
\refstepcounter{section}
\section*{Appendix \Alph{section} -- HERAPDF1.5}
\label{appendix:A}

HERAPDF1.5 NLO and NNLO were released in 2010~\cite{HERAPDF15}. 
They were obtained from all HERA\,I data sets and 
the selected HERA\,II data sets marked in Table~\ref{tab:data}.
Some cross-section measurements were preliminary at the time; this is
also marked in Table~\ref{tab:data}.
All these data sets were combined as described in Section~\ref{sec:comb}.
However, only the three procedural uncertainties described in
Sections~\ref{subsec:procerr1} and~\ref{subsec:procerr2} were considered.
Table \ref{tab:16} provides a comparison between the main settings for
HERAPDF2.0 and 1.5.
\begin{table} [h]
\renewcommand*{\arraystretch}{1.2}
\centerline{
\begin{tabular}{|l|c|c|c|c|}
\hline
                       &\multicolumn{2}{c|}{HERAPDF2.0} 
                       & \multicolumn{2}{c|}{HERAPDF1.5}\\
\hline
                       &   NNLO  &  NLO   &  NNLO &  NLO   \\
\hline
                        &  \multicolumn{2}{c|}{}       
                        &  \multicolumn{2}{c|}{}       \\
Data as in Table~\ref{tab:data}               
                       & \multicolumn{2}{c|}{combination} 
                       & \multicolumn{2}{c|}{preliminary combination}\\
Uncertainties:         & \multicolumn{2}{c|}{}
                       & \multicolumn{2}{c|}{}\\ 
Experimental           & \multicolumn{2}{c|} {Hessian}
                       & \multicolumn{2}{c|} {Hessian}\\
Procedural 
                       &  \multicolumn{2}{c|}{7} 
                       &  \multicolumn{2}{c|}{3}\\
\hline
Parameterisation  
 & \multicolumn{2}{c|} {as in Equations~\ref{eq:xgpar} to~\ref{eq:xdbarpar}} 
 & \multicolumn{2}{c|} {as in Equations~\ref{eq:xgpar} to~\ref{eq:xdbarpar}} \\
Number of Parameters   &  14  & 14 
                       &  14$^{**}$  & 10 $^*$       \\
~~~ -- Variations      &  15 [$D_{u_{v}}$] &  15 [$D_{u_{v}}$]       
                       &  none & 11 [$D_{u_{v}}$], 12  [$D_{\bar{U}}$]      \\
\hline
$\mu^2_{\rm f_{0}}$ [GeV$^2$] 
                       &  1.9 & 1.9        
                       &  1.9 & 1.9\\
~~~ -- Variations    
                       &  1.6, 2.2$^{a}$ & 1.6, 2.2$^{b}$          
                       &  1.5, 2.5$^{c}$ & 1.5$^{d}$, 2.5$^{c}$ \\
\hline
$M_c$ [GeV] 
                       &  1.43 & 1.47 
                       &  1.4  & 1.4 \\       
~~~ -- Variations    
                       &   1.37$^{e}$, 1.49  & 1.41, 1.53
                       &   1.35$^{f}$, 1.65  & 1.35$^{f}$, 1.65 \\
\hline
$M_b$ [GeV] 
                       &  4.5 & 4.5           
                       &  4.75 & 4.75   \\
~~~ -- Variations    
                       &  4.25, 4.75   &  4.25, 4.75          
                       &  4.30, 5.00   &  4.30, 5.00 \\
\hline
$f_s$ [GeV] 
                       &  0.40 & 0.40          
                       &  0.31 & 0.31    \\
~~~ -- Variations    
                       &  0.30, 0.50  &  0.30, 0.50           
                       &  0.23, 0.38  &  0.23, 0.38    \\
\hline
$Q^2_{\rm min}$ [GeV$^2$] of Data 
                       &  3.5 & 3.5 & 3.5 & 3.5 \\        
~~~ -- Variations    
                       &  2.5, 5.0 &  2.5, 5.0 &  2.5, 5.0 &  2.5, 5.0 \\
\hline
Fixed $\alpha_s$  
                       &  0.118 & 0.118 & 0.1176 & 0.1176 \\        
\hline
\end{tabular}}
\caption{\label{tab:16}Settings for HERAPDF2.0 and HERAPDF1.5. \newline
        $^{*}$:  Setting was chosen exactly as for
                    HERAPDF1.0. \newline
        $^{**}$: Parameter number 14 was $D_{u_{v}}$ 
                                 and not $D_{\bar U}$. \newline
        $^{a}$: $M_c=1.49$\,GeV to assure
                 $\mu^2_{\rm f_{0}}<M_c^2$ \newline
        $^{b}$: $M_c=1.53$\,GeV to assure
                 $\mu^2_{\rm f_{0}}<M_c^2$ \newline
        $^{c}$: $M_c=1.6$\,GeV to assure
                 $\mu^2_{\rm f_{0}}<M_c^2$ \newline
        $^{d}$: For $\mu^2_{\rm f_{0}}=1.5$\,GeV$^2$, also $A'_{g}$ and $B_g'$
                 were introduced (as for HERAPDF1.0 NLO). \newline
        $^{e}$: $\mu^2_{\rm f_{0}}=1.6$\,GeV$^2$ to assure 
                 $\mu^2_{\rm f_{0}} < M_c^2$ \newline
        $^{f}$: $\mu^2_{\rm f_{0}}=1.8$\,GeV$^2$ to assure 
                 $\mu^2_{\rm f_{0}} < M_c^2$ \newline
}
\label{tab:HERAPD15}
\end{table}

%


\clearpage

\refstepcounter{section}
\section*{Appendix  \Alph{section} -- PDFs released} 
\label{appendix:B}

The following sets of PDFs are released~\cite{fullcorr} and available on LHAPDF:

(https://lhapdf.hepforge.org/pdfsets.html).


\begin{itemize}
\item HERAPDF2.0
  \begin{itemize}
  \item based on the combination of all inclusive data
        from the H1 and ZEUS collaborations;
  \item with $Q^2_{\rm min}=3.5\,$GeV$^2$;
  \item at NLO and NNLO;
  \item using the RTOPT variable-flavour-number scheme;
  \item with $\asmz=0.118$;
  \item 14 eigenvector pairs give the Hessian experimental uncertainties;  
  \item for the NLO and NNLO fits, grids of 13 variations
        are released to describe the model and parameterisation uncertainties; 
  \item grids with 
        alternative values of $\asmz$ are released
        for $\asmz=0.110$ to $\asmz=0.130$ in steps of 0.001;
  \end{itemize}

\item HERAPDF2.0HiQ2
  \begin{itemize}
  \item as HERAPDF2.0, but
           with $Q^2_{\rm min}=10.0\,$GeV$^2$;
  \item only at $\asmz=0.118$;
  \end{itemize}

\item HERAPDF2.0AG
  \begin{itemize}
  \item based on the combination of all inclusive data
        from the H1 and ZEUS collaborations;
  \item with $Q^2_{\rm min}=3.5\,$GeV$^2$;
  \item at LO, NLO and NNLO;
  \item using the alternative gluon parameterisation 
        as defined in Section~\ref{sec:altparam};
  \item with $\asmz=0.130$ for LO and $\asmz=0.118$ for NLO and NNLO;
  \item experimental uncertainties provided at LO;
  \item no uncertainties provided at NLO and NNLO;
  \end{itemize}

\item HERAPDF2.0FF3A and FF3B
  \begin{itemize}
  \item based on the combination of all inclusive data
        from the H1 and ZEUS collaborations;
  \item with $Q^2_{\rm min}=3.5\,$GeV$^2$;
  \item at NLO;
  \item using the fixed-flavour-number schemes as decribed in 
             Table~\ref{tab:FF};
  \item with $\asmz^{N_F=3} = 0.106375$,
             equivalent to $\asmz^{N_F=5} = 0.118$ for FF3A,
             and with $\asmz=0.118$ for FF3B;
  \item 14 eigenvector pairs give the Hessian experimental uncertainties;
  \item grids of 13 variations are released to describe the model 
        and parameterisation uncertainties; 
  \end{itemize}

\item HERAPDF2.0Jets
  \begin{itemize}
  \item based on the combination of all inclusive data
        from the H1 and ZEUS collaborations
        and selected data on charm and jet production;
  \item with $Q^2_{\rm min}=3.5\,$GeV$^2$;
  \item at NLO;
  \item with free $\asmz$;
  \item 15 eigenvector pairs give Hessian experimental 
        uncertainties including the uncertainty on $\asmz$;
  \item grids of 15 variations are released
        to describe the model, parameterisation 
        and hadronisation uncertainties.  
  \end{itemize}
\end{itemize}

\clearpage

\refstepcounter{section}
\section*{Appendix  \Alph{section} -- Data tables}
\label{appendix:C}

Tables \ref{tab615-318a1}--\ref{tab3515a1} summarise the combined cross
section measurements and uncertainties. The full information about
correlations between cross-section measurements is available elsewhere 
\cite{fullcorr}. The new values supersede those published
previously~\cite{HERAIcombi}.

%
%

%

 
\clearpage
\begin{table}
\begin{center}
\begin{scriptsize}\renewcommand\arraystretch{1.1}

\begin{tabular}[H]{ c l c r r r r r r r r r r r }
\hline \hline
$Q^2$ &  $x_{\rm Bj}$ & $\sigma_{r, \rm NC}^{+ }$ & $\delta_{\rm stat}$ & $\delta_{\rm uncor}$ & $\delta_{\rm cor}$ & $\delta_{\rm rel}$ & $\delta_{\gamma p}$ & $\delta_{\rm had}$ &  $\delta_{1}$ & $\delta_{2}$ & $\delta_{3}$ & $\delta_{4}$ &$\delta_{\rm tot}$ \\
${\rm GeV^2}$ & & & \% & \% & \% & \% & \% & \% & \% & \% & \% & \% & \%     \\
\hline
0.15 & $0.502 \times 10^{-5}$ & 0.185 & 3.79 & 1.50 & 3.62 & 1.39 & 0.35 & $-$0.21 & $-$0.17 & $-$0.01 & 0.00 & 0.01 & 5.65 \\
0.2 & $0.669 \times 10^{-5}$ & 0.227 & 1.65 & 0.78 & 1.70 & 0.86 & 0.57 & 0.00 & 0.00 & 0.00 & 0.00 & 0.01 & 2.70 \\
0.2 & $0.849 \times 10^{-5}$ & 0.223 & 1.61 & 0.61 & 2.19 & 1.06 & 0.55 & $-$0.26 & $-$0.06 & 0.00 & 0.00 & 0.00 & 3.04 \\
0.2 & $0.110 \times 10^{-4}$ & 0.208 & 2.79 & 1.50 & 2.83 & 1.01 & 0.34 & $-$0.08 & $-$0.18 & 0.00 & 0.00 & 0.01 & 4.38 \\
0.2 & $0.398 \times 10^{-4}$ & 0.211 & 14.93 & 11.96 & 5.18 & 0.33 & 4.70 & 2.93 & 1.57 & $-$0.03 & $-$0.02 & 0.15 & 20.64 \\
0.2 & $0.251 \times 10^{-3}$ & 0.180 & 13.49 & 6.17 & 3.00 & 0.32 & 1.39 & $-$1.67 & 1.19 & 0.01 & 0.02 & 0.02 & 15.34 \\
0.25 & $0.836 \times 10^{-5}$ & 0.265 & 1.46 & 0.73 & 1.92 & 1.17 & 0.63 & $-$0.23 & 0.45 & 0.00 & 0.00 & 0.01 & 2.89 \\
0.25 & $0.106 \times 10^{-4}$ & 0.260 & 1.29 & 0.66 & 1.84 & 1.11 & 0.63 & $-$0.10 & 0.32 & 0.00 & 0.00 & 0.01 & 2.69 \\
0.25 & $0.138 \times 10^{-4}$ & 0.249 & 1.27 & 0.72 & 1.85 & 1.24 & 0.61 & $-$0.22 & 0.08 & 0.00 & 0.00 & 0.00 & 2.74 \\
0.25 & $0.230 \times 10^{-4}$ & 0.243 & 1.41 & 1.50 & 2.37 & 2.23 & 0.38 & $-$0.60 & 0.43 & 0.00 & $-$0.02 & 0.01 & 3.94 \\
0.25 & $0.398 \times 10^{-4}$ & 0.236 & 3.32 & 1.54 & 2.79 & 0.50 & 1.03 & 0.29 & 0.21 & 0.00 & 0.01 & 0.02 & 4.76 \\
0.25 & $0.110 \times 10^{-3}$ & 0.199 & 3.96 & 1.50 & 2.50 & 0.77 & 0.32 & 0.06 & $-$0.58 & 0.00 & 0.00 & 0.01 & 5.02 \\
0.25 & $0.251 \times 10^{-3}$ & 0.196 & 3.75 & 1.44 & 3.26 & $-$0.23 & 0.35 & 0.51 & $-$0.21 & 0.01 & 0.02 & 0.02 & 5.22 \\
0.25 & $0.394 \times 10^{-3}$ & 0.194 & 4.16 & 1.50 & 3.61 & 1.65 & 0.46 & $-$0.29 & $-$0.20 & 0.00 & 0.00 & 0.02 & 5.97 \\
0.25 & $0.158 \times 10^{-2}$ & 0.198 & 11.00 & 5.29 & 2.41 & 0.29 & 0.01 & $-$1.76 & 1.08 & 0.01 & 0.03 & 0.02 & 12.61 \\
0.35 & $0.512 \times 10^{-5}$ & 0.436 & 22.08 & 12.79 & 1.83 & 0.22 & 1.82 & $-$0.44 & 1.33 & $-$0.03 & 0.00 & 0.03 & 25.68 \\
0.35 & $0.100 \times 10^{-4}$ & 0.345 & 1.63 & 0.80 & 1.80 & 1.11 & 0.60 & $-$0.38 & 0.12 & 0.00 & 0.00 & 0.01 & 2.87 \\
0.35 & $0.127 \times 10^{-4}$ & 0.324 & 1.44 & 0.79 & 1.86 & 1.16 & 0.58 & $-$0.16 & $-$0.03 & 0.00 & 0.00 & 0.01 & 2.81 \\
0.35 & $0.165 \times 10^{-4}$ & 0.313 & 1.21 & 0.63 & 1.88 & 1.19 & 0.63 & 0.06 & 0.52 & 0.00 & 0.00 & 0.01 & 2.74 \\
0.35 & $0.320 \times 10^{-4}$ & 0.296 & 1.14 & 0.72 & 2.54 & 2.55 & 0.72 & $-$0.77 & 1.05 & $-$0.01 & $-$0.02 & 0.01 & 4.12 \\
0.35 & $0.662 \times 10^{-4}$ & 0.282 & 2.62 & 1.50 & 1.72 & 0.42 & 0.42 & $-$0.06 & 1.19 & 0.00 & 0.01 & 0.01 & 3.72 \\
0.35 & $0.130 \times 10^{-3}$ & 0.257 & 2.53 & 1.43 & 1.54 & 0.81 & 0.59 & $-$0.11 & 0.70 & 0.00 & 0.00 & 0.02 & 3.50 \\
0.35 & $0.220 \times 10^{-3}$ & 0.240 & 2.67 & 1.50 & 2.01 & 1.13 & 0.53 & $-$0.31 & 0.14 & 0.00 & 0.00 & 0.01 & 3.88 \\
0.35 & $0.500 \times 10^{-3}$ & 0.240 & 2.52 & 1.42 & 1.81 & 1.49 & 0.44 & $-$0.46 & 0.28 & 0.00 & 0.00 & 0.01 & 3.79 \\
0.35 & $0.251 \times 10^{-2}$ & 0.201 & 10.00 & 4.55 & 1.54 & 0.55 & $-$0.32 & $-$0.44 & 0.34 & 0.00 & 0.02 & 0.04 & 11.12 \\
0.4 & $0.133 \times 10^{-4}$ & 0.355 & 1.97 & 0.88 & 2.04 & 1.36 & 0.67 & $-$0.31 & 0.64 & 0.00 & 0.00 & 0.01 & 3.42 \\
0.4 & $0.170 \times 10^{-4}$ & 0.354 & 1.66 & 0.83 & 1.87 & 1.15 & 0.61 & $-$0.05 & 0.29 & 0.00 & 0.01 & 0.01 & 2.96 \\
0.4 & $0.220 \times 10^{-4}$ & 0.334 & 1.45 & 0.78 & 1.86 & 1.28 & 0.64 & $-$0.34 & 0.38 & 0.00 & 0.00 & 0.01 & 2.91 \\
0.4 & $0.368 \times 10^{-4}$ & 0.330 & 1.26 & 0.77 & 2.52 & 2.58 & 0.65 & $-$0.84 & 1.12 & $-$0.01 & $-$0.02 & 0.01 & 4.19 \\
0.4 & $0.883 \times 10^{-4}$ & 0.320 & 2.72 & 1.50 & 1.57 & 1.03 & 0.48 & $-$0.23 & 0.41 & 0.00 & 0.00 & 0.01 & 3.69 \\
0.4 & $0.176 \times 10^{-3}$ & 0.287 & 2.79 & 1.50 & 1.76 & 0.68 & 0.49 & $-$0.16 & 0.22 & 0.00 & 0.01 & 0.01 & 3.73 \\
0.4 & $0.294 \times 10^{-3}$ & 0.277 & 2.75 & 1.50 & 1.66 & 1.03 & 0.47 & $-$0.30 & 0.74 & 0.00 & 0.00 & 0.01 & 3.81 \\
0.4 & $0.631 \times 10^{-3}$ & 0.260 & 2.74 & 1.50 & 2.06 & 1.33 & 0.43 & $-$0.32 & 0.54 & 0.00 & 0.00 & 0.01 & 4.04 \\
0.5 & $0.732 \times 10^{-5}$ & 0.428 & 5.55 & 5.74 & 4.16 & $-$0.02 & 4.04 & 0.31 & 1.90 & $-$0.05 & $-$0.01 & 0.05 & 10.05 \\
0.5 & $0.158 \times 10^{-4}$ & 0.426 & 3.53 & 1.48 & 2.38 & 1.22 & 0.46 & $-$0.20 & 0.70 & 0.00 & 0.00 & 0.00 & 4.75 \\
0.5 & $0.212 \times 10^{-4}$ & 0.390 & 2.06 & 0.80 & 2.01 & 1.07 & 0.58 & 0.21 & 0.33 & 0.00 & 0.00 & 0.01 & 3.25 \\
0.5 & $0.276 \times 10^{-4}$ & 0.377 & 1.71 & 0.76 & 2.00 & 1.38 & 0.64 & $-$0.18 & 0.31 & 0.00 & 0.00 & 0.01 & 3.15 \\
0.5 & $0.398 \times 10^{-4}$ & 0.364 & 1.48 & 0.81 & 2.60 & 2.64 & 0.85 & $-$0.97 & 1.03 & $-$0.01 & $-$0.01 & 0.01 & 4.40 \\
0.5 & $0.100 \times 10^{-3}$ & 0.348 & 1.74 & 1.45 & 1.57 & 0.65 & 0.54 & $-$0.07 & 0.73 & 0.00 & 0.01 & 0.01 & 2.98 \\
0.5 & $0.251 \times 10^{-3}$ & 0.308 & 1.87 & 1.43 & 1.62 & 0.36 & 0.38 & 0.01 & 0.02 & 0.00 & 0.01 & 0.01 & 2.91 \\
0.5 & $0.368 \times 10^{-3}$ & 0.300 & 2.03 & 1.50 & 1.68 & 0.83 & 0.43 & $-$0.12 & 0.35 & 0.00 & 0.00 & 0.01 & 3.19 \\
0.5 & $0.800 \times 10^{-3}$ & 0.287 & 2.05 & 1.38 & 1.54 & 0.98 & 0.37 & $-$0.27 & 0.09 & 0.00 & 0.00 & 0.01 & 3.11 \\
0.5 & $0.320 \times 10^{-2}$ & 0.182 & 11.38 & 6.39 & 1.30 & 0.38 & $-$0.40 & $-$0.77 & $-$0.48 & $-$0.01 & 0.01 & 0.08 & 13.16 \\
0.65 & $0.952 \times 10^{-5}$ & 0.464 & 4.02 & 2.90 & 2.57 & 0.22 & 2.36 & $-$0.10 & 1.52 & $-$0.04 & 0.00 & 0.04 & 6.25 \\
0.65 & $0.158 \times 10^{-4}$ & 0.462 & 3.10 & 5.44 & 1.56 & 0.40 & 0.34 & $-$0.41 & 0.15 & $-$0.02 & 0.01 & 0.05 & 6.48 \\
0.65 & $0.398 \times 10^{-4}$ & 0.472 & 2.73 & 0.68 & 2.37 & 1.26 & 0.55 & 0.15 & 0.46 & 0.00 & 0.00 & 0.00 & 3.96 \\
0.65 & $0.598 \times 10^{-4}$ & 0.416 & 1.99 & 0.74 & 3.16 & 2.93 & 0.60 & $-$0.67 & 1.56 & 0.00 & $-$0.01 & 0.01 & 5.13 \\
0.65 & $0.100 \times 10^{-3}$ & 0.409 & 2.09 & 1.47 & 2.04 & 0.93 & 0.66 & $-$0.19 & 1.59 & 0.00 & 0.00 & 0.01 & 3.81 \\
0.65 & $0.251 \times 10^{-3}$ & 0.361 & 2.14 & 1.39 & 1.69 & 0.88 & 0.32 & $-$0.32 & 1.13 & 0.00 & 0.00 & 0.01 & 3.41 \\
0.65 & $0.478 \times 10^{-3}$ & 0.332 & 2.29 & 1.50 & 1.93 & 0.88 & 0.43 & $-$0.36 & 0.94 & 0.00 & 0.00 & 0.01 & 3.63 \\
0.65 & $0.800 \times 10^{-3}$ & 0.318 & 2.14 & 1.35 & 1.64 & 0.71 & 0.23 & $-$0.39 & 0.56 & 0.00 & 0.01 & 0.03 & 3.18 \\
0.65 & $0.320 \times 10^{-2}$ & 0.224 & 5.82 & 3.18 & 1.27 & 0.39 & $-$0.10 & $-$0.56 & 0.54 & 0.00 & 0.02 & 0.06 & 6.81 \\

\hline \hline

\end{tabular}
\caption{\label{tab615-318a1}
HERA combined reduced  cross sections $\sigma^{+ }_{r,{\rm NC}}$ for NC $e^{+}p$ scattering at $\sqrt{s} = 318 $~GeV; 
$\delta_{\rm stat}$, $\delta_{\rm uncor}$ and $\delta_{\rm cor}$  represent the statistical, uncorrelated
systematic and correlated systematic uncertainties, respectively; 
$\delta_{\rm rel}$, $\delta_{\gamma p}$, $\delta_{\rm had}$ and $\delta_{1}$ -- $\delta_{4}$
are the correlated sources of uncertainties arising from the combination procedure.
The total uncertainty $\delta_{\rm tot}$ is 
calculated by adding $\delta_{\rm stat}$, $\delta_{\rm uncor}$, $\delta_{\rm cor}$ and the procedural uncertainties in quadrature.
The uncertainties are quoted in percent relative to $\sigma^{+ }_{r,{\rm NC}}$.
}

\end{scriptsize}
\end{center}
\end{table}

\clearpage
\begin{table}
\begin{center}
\begin{scriptsize}\renewcommand\arraystretch{1.1}

\begin{tabular}[H]{ c l c r r r r r r r r r r r }
\hline \hline
$Q^2$ &  $x_{\rm Bj}$ & $\sigma_{r, \rm NC}^{+ }$ & $\delta_{\rm stat}$ & $\delta_{\rm uncor}$ & $\delta_{\rm cor}$ & $\delta_{\rm rel}$ & $\delta_{\gamma p}$ & $\delta_{\rm had}$ &  $\delta_{1}$ & $\delta_{2}$ & $\delta_{3}$ & $\delta_{4}$ &$\delta_{\rm tot}$ \\
${\rm GeV^2}$ & & & \% & \% & \% & \% & \% & \% & \% & \% & \% & \% & \%     \\
\hline

0.85 & $0.124 \times 10^{-4}$ & 0.552 & 2.57 & 2.52 & 2.21 & 0.27 & 1.92 & 0.02 & 1.03 & $-$0.04 & 0.00 & 0.04 & 4.76 \\
0.85 & $0.200 \times 10^{-4}$ & 0.611 & 1.96 & 5.36 & 1.63 & 0.49 & $-$0.13 & $-$0.40 & $-$0.04 & $-$0.02 & 0.01 & 0.05 & 5.97 \\
0.85 & $0.398 \times 10^{-4}$ & 0.558 & 2.03 & 1.86 & 1.77 & 0.44 & $-$0.45 & $-$0.39 & $-$1.05 & $-$0.01 & 0.00 & 0.03 & 3.51 \\
0.85 & $0.500 \times 10^{-4}$ & 0.541 & 2.93 & 4.52 & 2.36 & 0.62 & $-$0.28 & $-$0.09 & $-$0.25 & 0.00 & 0.01 & 0.06 & 5.93 \\
0.85 & $0.100 \times 10^{-3}$ & 0.503 & 2.65 & 3.78 & 3.73 & 0.43 & $-$1.31 & 0.49 & $-$0.55 & $-$0.05 & 0.00 & 0.03 & 6.14 \\
0.85 & $0.251 \times 10^{-3}$ & 0.397 & 6.01 & 2.98 & 2.33 & 0.38 & $-$0.76 & $-$0.54 & 0.96 & $-$0.02 & 0.02 & 0.04 & 7.23 \\
\hline \hline

\end{tabular}\captcont{Continued.}


\end{scriptsize}
\end{center}
\end{table}

\clearpage
\begin{table}
\begin{center}
\begin{scriptsize}\renewcommand\arraystretch{1.1}

\begin{tabular}[H]{ c l c r r r r r r r r r r r }
\hline \hline
$Q^2$ &  $x_{\rm Bj}$ & $\sigma_{r, \rm NC}^{+ }$ & $\delta_{\rm stat}$ & $\delta_{\rm uncor}$ & $\delta_{\rm cor}$ & $\delta_{\rm rel}$ & $\delta_{\gamma p}$ & $\delta_{\rm had}$ &  $\delta_{1}$ & $\delta_{2}$ & $\delta_{3}$ & $\delta_{4}$ &$\delta_{\rm tot}$ \\
${\rm GeV^2}$ & & & \% & \% & \% & \% & \% & \% & \% & \% & \% & \% & \%     \\
\hline

3.5 & $0.406 \times 10^{-4}$ & 0.806 & 6.14 & 4.17 & 1.18 & 1.09 & $-$0.25 & $-$0.46 & $-$0.04 & $-$0.75 & $-$0.01 & $-$0.15 & 7.65 \\
3.5 & $0.432 \times 10^{-4}$ & 0.881 & 3.08 & 2.83 & 3.31 & 0.70 & $-$4.07 & 0.56 & $-$2.62 & $-$0.18 & $-$0.01 & $-$0.05 & 7.26 \\
3.5 & $0.460 \times 10^{-4}$ & 0.965 & 3.05 & 2.99 & 1.10 & 0.35 & $-$0.21 & $-$0.41 & $-$0.05 & $-$0.15 & $-$0.01 & $-$0.22 & 4.45 \\
3.5 & $0.512 \times 10^{-4}$ & 0.940 & 2.16 & 2.25 & 1.53 & 0.52 & $-$1.61 & 0.05 & $-$1.16 & $-$0.07 & 0.01 & 0.01 & 4.04 \\
3.5 & $0.531 \times 10^{-4}$ & 0.880 & 3.10 & 2.64 & 0.91 & 0.48 & $-$0.20 & $-$0.30 & $-$0.03 & $-$0.01 & $-$0.01 & $-$0.21 & 4.22 \\
3.5 & $0.800 \times 10^{-4}$ & 0.952 & 1.25 & 1.55 & 0.88 & 0.43 & $-$0.26 & $-$0.09 & $-$0.08 & $-$0.09 & 0.01 & 0.03 & 2.24 \\
\hline \hline

\end{tabular}\captcont{Continued.}


\end{scriptsize}
\end{center}
\end{table}

\clearpage
\begin{table}
\begin{center}
\begin{scriptsize}\renewcommand\arraystretch{1.1}

\begin{tabular}[H]{ c l c r r r r r r r r r r r }
\hline \hline
$Q^2$ &  $x_{\rm Bj}$ & $\sigma_{r, \rm NC}^{+ }$ & $\delta_{\rm stat}$ & $\delta_{\rm uncor}$ & $\delta_{\rm cor}$ & $\delta_{\rm rel}$ & $\delta_{\gamma p}$ & $\delta_{\rm had}$ &  $\delta_{1}$ & $\delta_{2}$ & $\delta_{3}$ & $\delta_{4}$ &$\delta_{\rm tot}$ \\
${\rm GeV^2}$ & & & \% & \% & \% & \% & \% & \% & \% & \% & \% & \% & \%     \\
\hline

10 & $0.320 \times 10^{-3}$ & 1.138 & 1.92 & 1.12 & 0.92 & 0.66 & $-$0.06 & $-$0.04 & $-$0.02 & $-$0.01 & 0.01 & $-$0.07 & 2.50 \\
10 & $0.500 \times 10^{-3}$ & 1.058 & 1.78 & 1.19 & 0.86 & 0.34 & 0.09 & 0.16 & 0.03 & $-$0.01 & 0.06 & 0.05 & 2.34 \\
10 & $0.800 \times 10^{-3}$ & 0.977 & 1.84 & 1.31 & 0.84 & 0.30 & 0.07 & 0.05 & 0.02 & $-$0.02 & 0.09 & 0.15 & 2.44 \\
10 & $0.130 \times 10^{-2}$ & 0.888 & 2.13 & 1.21 & 1.01 & 0.11 & $-$0.16 & 0.01 & 0.02 & $-$0.06 & 0.22 & $-$0.02 & 2.66 \\
10 & $0.200 \times 10^{-2}$ & 0.802 & 2.04 & 1.77 & 0.89 & 0.42 & $-$0.14 & $-$0.42 & 0.05 & $-$0.04 & 0.13 & 0.02 & 2.91 \\
10 & $0.500 \times 10^{-2}$ & 0.628 & 1.12 & 0.77 & 0.91 & 0.41 & $-$0.11 & $-$0.48 & $-$0.02 & $-$0.03 & 0.10 & 0.10 & 1.76 \\
\hline \hline

\end{tabular}\captcont{Continued.}


\end{scriptsize}
\end{center}
\end{table}

\clearpage
\begin{table}
\begin{center}
\begin{scriptsize}\renewcommand\arraystretch{1.1}

\begin{tabular}[H]{ c l c r r r r r r r r r r r }
\hline \hline
$Q^2$ &  $x_{\rm Bj}$ & $\sigma_{r, \rm NC}^{+ }$ & $\delta_{\rm stat}$ & $\delta_{\rm uncor}$ & $\delta_{\rm cor}$ & $\delta_{\rm rel}$ & $\delta_{\gamma p}$ & $\delta_{\rm had}$ &  $\delta_{1}$ & $\delta_{2}$ & $\delta_{3}$ & $\delta_{4}$ &$\delta_{\rm tot}$ \\
${\rm GeV^2}$ & & & \% & \% & \% & \% & \% & \% & \% & \% & \% & \% & \%     \\
\hline
27 & $0.314 \times 10^{-3}$ & 1.480 & 2.47 & 3.86 & 0.91 & 0.58 & $-$0.22 & $-$0.47 & $-$0.04 & $-$0.23 & $-$0.01 & $-$0.20 & 4.74 \\
27 & $0.355 \times 10^{-3}$ & 1.441 & 0.88 & 2.64 & 0.91 & 0.54 & $-$0.19 & $-$0.35 & $-$0.03 & 0.07 & $-$0.01 & $-$0.24 & 3.01 \\
27 & $0.410 \times 10^{-3}$ & 1.462 & 0.73 & 2.16 & 0.90 & 0.66 & $-$0.20 & $-$0.34 & $-$0.02 & $-$0.07 & $-$0.01 & $-$0.21 & 2.58 \\
27 & $0.500 \times 10^{-3}$ & 1.411 & 0.70 & 0.74 & 0.87 & 0.64 & $-$0.43 & $-$0.23 & $-$0.02 & 0.04 & $-$0.07 & $-$0.16 & 1.57 \\
27 & $0.800 \times 10^{-3}$ & 1.289 & 0.44 & 0.67 & 0.74 & 0.49 & $-$0.12 & $-$0.16 & $-$0.01 & 0.02 & $-$0.02 & $-$0.13 & 1.22 \\
27 & $0.130 \times 10^{-2}$ & 1.154 & 0.47 & 0.65 & 0.76 & 0.48 & $-$0.07 & $-$0.12 & 0.00 & $-$0.01 & $-$0.03 & $-$0.13 & 1.22 \\
\hline \hline

\end{tabular}\captcont{Continued.}


\end{scriptsize}
\end{center}
\end{table}

\clearpage
\begin{table}
\begin{center}
\begin{scriptsize}\renewcommand\arraystretch{1.1}

\begin{tabular}[H]{ c l c r r r r r r r r r r r }
\hline \hline
$Q^2$ &  $x_{\rm Bj}$ & $\sigma_{r, \rm NC}^{+ }$ & $\delta_{\rm stat}$ & $\delta_{\rm uncor}$ & $\delta_{\rm cor}$ & $\delta_{\rm rel}$ & $\delta_{\gamma p}$ & $\delta_{\rm had}$ &  $\delta_{1}$ & $\delta_{2}$ & $\delta_{3}$ & $\delta_{4}$ &$\delta_{\rm tot}$ \\
${\rm GeV^2}$ & & & \% & \% & \% & \% & \% & \% & \% & \% & \% & \% & \%     \\
\hline
70 & $0.922 \times 10^{-3}$ & 1.385 & 2.57 & 5.67 & 1.10 & 0.12 & $-$0.16 & $-$0.29 & $-$0.04 & $-$0.29 & 0.01 & $-$0.80 & 6.39 \\
70 & $0.100 \times 10^{-2}$ & 1.434 & 2.18 & 2.55 & 0.86 & $-$0.08 & $-$0.11 & $-$0.20 & $-$0.02 & 0.08 & 0.01 & $-$0.93 & 3.60 \\
70 & $0.110 \times 10^{-2}$ & 1.506 & 1.88 & 2.10 & 0.81 & $-$0.08 & $-$0.11 & $-$0.17 & $-$0.01 & 0.48 & 0.01 & $-$0.97 & 3.14 \\
70 & $0.124 \times 10^{-2}$ & 1.445 & 1.48 & 1.80 & 0.80 & $-$0.15 & $-$0.10 & $-$0.16 & $-$0.01 & 0.24 & 0.01 & $-$0.95 & 2.66 \\
70 & $0.130 \times 10^{-2}$ & 1.414 & 1.32 & 1.32 & 0.82 & $-$0.16 & $-$0.08 & $-$0.12 & 0.00 & 0.45 & 0.00 & $-$1.09 & 2.37 \\
70 & $0.200 \times 10^{-2}$ & 1.246 & 1.15 & 1.36 & 0.78 & $-$0.03 & $-$0.09 & $-$0.12 & 0.00 & 0.22 & 0.01 & $-$0.96 & 2.19 \\
\hline \hline

\end{tabular}\captcont{Continued.}


\end{scriptsize}
\end{center}
\end{table}

\clearpage
\begin{table}
\begin{center}
\begin{scriptsize}\renewcommand\arraystretch{1.1}

\begin{tabular}[H]{ c l c r r r r r r r r r r r }
\hline \hline
$Q^2$ &  $x_{\rm Bj}$ & $\sigma_{r, \rm NC}^{+ }$ & $\delta_{\rm stat}$ & $\delta_{\rm uncor}$ & $\delta_{\rm cor}$ & $\delta_{\rm rel}$ & $\delta_{\gamma p}$ & $\delta_{\rm had}$ &  $\delta_{1}$ & $\delta_{2}$ & $\delta_{3}$ & $\delta_{4}$ &$\delta_{\rm tot}$ \\
${\rm GeV^2}$ & & & \% & \% & \% & \% & \% & \% & \% & \% & \% & \% & \%     \\
\hline
200 & $0.260 \times 10^{-2}$ & 1.312 & 1.09 & 1.56 & 0.84 & 0.08 & $-$0.03 & $-$0.14 & 0.00 & $-$0.05 & 0.11 & 0.24 & 2.10 \\
200 & $0.320 \times 10^{-2}$ & 1.248 & 0.92 & 0.72 & 0.92 & 0.52 & $-$0.57 & $-$0.16 & 0.01 & $-$0.04 & 0.09 & 0.19 & 1.69 \\
200 & $0.500 \times 10^{-2}$ & 1.110 & 0.54 & 0.55 & 0.79 & 0.17 & $-$0.11 & $-$0.08 & 0.00 & $-$0.05 & 0.15 & 0.26 & 1.17 \\
200 & $0.800 \times 10^{-2}$ & 0.948 & 0.50 & 0.42 & 0.76 & 0.11 & $-$0.07 & $-$0.09 & 0.01 & $-$0.05 & 0.11 & 0.31 & 1.07 \\
200 & $0.130 \times 10^{-1}$ & 0.809 & 0.49 & 0.35 & 0.75 & 0.05 & $-$0.06 & $-$0.17 & 0.00 & $-$0.06 & 0.06 & 0.34 & 1.04 \\
200 & $0.200 \times 10^{-1}$ & 0.701 & 0.59 & 0.30 & 0.76 & $-$0.03 & $-$0.04 & $-$0.20 & 0.00 & $-$0.06 & 0.01 & 0.36 & 1.09 \\
\hline \hline

\end{tabular}\captcont{Continued.}


\end{scriptsize}
\end{center}
\end{table}

\clearpage
\begin{table}
\begin{center}
\begin{scriptsize}\renewcommand\arraystretch{1.1}

\begin{tabular}[H]{ c l c r r r r r r r r r r r }
\hline \hline
$Q^2$ &  $x_{\rm Bj}$ & $\sigma_{r, \rm NC}^{+ }$ & $\delta_{\rm stat}$ & $\delta_{\rm uncor}$ & $\delta_{\rm cor}$ & $\delta_{\rm rel}$ & $\delta_{\gamma p}$ & $\delta_{\rm had}$ &  $\delta_{1}$ & $\delta_{2}$ & $\delta_{3}$ & $\delta_{4}$ &$\delta_{\rm tot}$ \\
${\rm GeV^2}$ & & & \% & \% & \% & \% & \% & \% & \% & \% & \% & \% & \%     \\
\hline

650 & $0.850 \times 10^{-2}$ & 1.000 & 1.91 & 1.67 & 0.78 & $-$0.04 & $-$0.01 & $-$0.08 & 0.00 & $-$0.05 & 0.16 & 0.25 & 2.68 \\
650 & $0.130 \times 10^{-1}$ & 0.862 & 1.02 & 0.63 & 0.79 & 0.15 & $-$0.12 & $-$0.09 & 0.00 & $-$0.05 & 0.10 & 0.27 & 1.48 \\
650 & $0.200 \times 10^{-1}$ & 0.746 & 1.05 & 0.55 & 0.75 & 0.09 & $-$0.07 & $-$0.07 & 0.01 & $-$0.05 & 0.08 & 0.30 & 1.45 \\
650 & $0.320 \times 10^{-1}$ & 0.627 & 1.09 & 0.56 & 0.76 & 0.15 & $-$0.06 & $-$0.11 & 0.00 & $-$0.05 & 0.05 & 0.26 & 1.48 \\
650 & $0.500 \times 10^{-1}$ & 0.520 & 1.12 & 0.49 & 0.76 & 0.13 & $-$0.04 & $-$0.21 & 0.00 & $-$0.06 & 0.02 & 0.33 & 1.50 \\
650 & $0.800 \times 10^{-1}$ & 0.438 & 1.18 & 0.48 & 0.76 & 0.10 & $-$0.07 & $-$0.17 & 0.00 & $-$0.06 & $-$0.02 & 0.32 & 1.53 \\
\hline \hline

\end{tabular}\captcont{Continued.}


\end{scriptsize}
\end{center}
\end{table}

\clearpage
\begin{table}
\begin{center}
\begin{scriptsize}\renewcommand\arraystretch{1.1}

\begin{tabular}[H]{ c l c r r r r r r r r r r r }
\hline \hline
$Q^2$ &  $x_{\rm Bj}$ & $\sigma_{r, \rm NC}^{+ }$ & $\delta_{\rm stat}$ & $\delta_{\rm uncor}$ & $\delta_{\rm cor}$ & $\delta_{\rm rel}$ & $\delta_{\gamma p}$ & $\delta_{\rm had}$ &  $\delta_{1}$ & $\delta_{2}$ & $\delta_{3}$ & $\delta_{4}$ &$\delta_{\rm tot}$ \\
${\rm GeV^2}$ & & & \% & \% & \% & \% & \% & \% & \% & \% & \% & \% & \%     \\
\hline
2000 & $0.219 \times 10^{-1}$ & 0.710 & 5.58 & 2.40 & 1.43 & 0.96 & $-$1.31 & $-$0.22 & 0.00 & $-$0.04 & 0.03 & 0.15 & 6.45 \\
2000 & $0.320 \times 10^{-1}$ & 0.589 & 2.30 & 1.02 & 1.09 & 0.41 & $-$0.15 & $-$0.04 & 0.00 & $-$0.04 & 0.07 & 0.08 & 2.78 \\
2000 & $0.500 \times 10^{-1}$ & 0.516 & 2.17 & 0.65 & 0.79 & 0.11 & $-$0.05 & $-$0.08 & 0.00 & $-$0.05 & $-$0.04 & 0.29 & 2.42 \\
2000 & $0.800 \times 10^{-1}$ & 0.449 & 1.94 & 0.54 & 0.75 & 0.01 & $-$0.04 & $-$0.15 & 0.00 & $-$0.05 & 0.05 & 0.34 & 2.18 \\
2000 & 0.130 & 0.362 & 2.36 & 0.66 & 0.76 & 0.17 & $-$0.06 & $-$0.15 & 0.01 & $-$0.06 & $-$0.07 & 0.30 & 2.60 \\
2000 & 0.180 & 0.300 & 2.50 & 0.64 & 0.77 & 0.07 & $-$0.06 & $-$0.12 & 0.01 & $-$0.05 & 0.05 & 0.35 & 2.72 \\
\hline \hline

\end{tabular}\captcont{Continued.}


\end{scriptsize}
\end{center}
\end{table}

\clearpage
\begin{table}
\begin{center}
\begin{scriptsize}\renewcommand\arraystretch{1.1}

\begin{tabular}[H]{ c l c r r r r r r r r r r r }
\hline \hline
$Q^2$ &  $x_{\rm Bj}$ & $\sigma_{r, \rm NC}^{+ }$ & $\delta_{\rm stat}$ & $\delta_{\rm uncor}$ & $\delta_{\rm cor}$ & $\delta_{\rm rel}$ & $\delta_{\gamma p}$ & $\delta_{\rm had}$ &  $\delta_{1}$ & $\delta_{2}$ & $\delta_{3}$ & $\delta_{4}$ &$\delta_{\rm tot}$ \\
${\rm GeV^2}$ & & & \% & \% & \% & \% & \% & \% & \% & \% & \% & \% & \%     \\
\hline
0.045 & $0.621 \times 10^{-6}$ & 0.078 & 5.06 & 1.50 & 7.30 & 3.16 & 1.25 & $-$0.75 & $-$7.46 & 0.00 & $-$0.02 & 0.02 & 12.20 \\
0.065 & $0.897 \times 10^{-6}$ & 0.109 & 2.99 & 1.50 & 6.28 & 3.03 & 1.42 & $-$0.90 & $-$7.40 & 0.00 & $-$0.02 & 0.02 & 10.83 \\
0.065 & $0.102 \times 10^{-5}$ & 0.110 & 4.70 & 1.50 & 5.99 & 3.78 & 1.54 & $-$0.79 & $-$4.23 & $-$0.01 & $-$0.02 & 0.01 & 9.77 \\
0.085 & $0.117 \times 10^{-5}$ & 0.133 & 2.96 & 1.50 & 5.55 & 2.86 & 1.49 & $-$0.83 & $-$6.30 & 0.00 & $-$0.02 & 0.01 & 9.62 \\
0.085 & $0.134 \times 10^{-5}$ & 0.128 & 2.28 & 1.50 & 4.03 & 2.02 & 1.20 & $-$0.53 & $-$4.09 & 0.00 & $-$0.02 & 0.01 & 6.79 \\
0.085 & $0.156 \times 10^{-5}$ & 0.125 & 4.79 & 1.50 & 4.55 & 1.08 & 1.14 & 0.07 & $-$1.47 & 0.00 & 0.00 & 0.02 & 7.11 \\
\hline \hline

\end{tabular}

\caption{\label{tab615-300a1}
HERA combined reduced  cross sections $\sigma^{+ }_{r,{\rm NC}}$ for NC $e^{+}p$ scattering at $\sqrt{s} = 300 $~GeV.
 The uncertainties are quoted in percent relative to $\sigma^{+ }_{r,{\rm NC}}$.
Other details as for Table~\ref{tab615-318a1}.}

\end{scriptsize}
\end{center}
\end{table}

\clearpage
\begin{table}
\begin{center}
\begin{scriptsize}\renewcommand\arraystretch{1.1}

\begin{tabular}[H]{ c l c r r r r r r r r r r r }
\hline \hline
$Q^2$ &  $x_{\rm Bj}$ & $\sigma_{r, \rm NC}^{+ }$ & $\delta_{\rm stat}$ & $\delta_{\rm uncor}$ & $\delta_{\rm cor}$ & $\delta_{\rm rel}$ & $\delta_{\gamma p}$ & $\delta_{\rm had}$ &  $\delta_{1}$ & $\delta_{2}$ & $\delta_{3}$ & $\delta_{4}$ &$\delta_{\rm tot}$ \\
${\rm GeV^2}$ & & & \% & \% & \% & \% & \% & \% & \% & \% & \% & \% & \%     \\
\hline

15 & $0.200 \times 10^{-3}$ & 1.258 & 3.21 & 3.61 & 1.00 & 0.54 & $-$0.11 & $-$0.14 & $-$0.01 & $-$0.01 & $-$0.48 & $-$0.15 & 4.99 \\
15 & $0.246 \times 10^{-3}$ & 1.360 & 1.17 & 1.52 & 1.43 & 0.76 & $-$1.11 & $-$0.46 & $-$0.13 & 0.01 & $-$0.39 & $-$0.10 & 2.82 \\
15 & $0.320 \times 10^{-3}$ & 1.290 & 0.94 & 1.36 & 1.00 & 0.68 & $-$0.63 & $-$0.22 & $-$0.05 & 0.00 & $-$0.19 & $-$0.10 & 2.17 \\
18 & $0.268 \times 10^{-3}$ & 1.318 & 3.26 & 3.67 & 0.99 & 0.54 & $-$0.11 & $-$0.17 & $-$0.01 & $-$0.01 & $-$0.46 & $-$0.15 & 5.06 \\
18 & $0.328 \times 10^{-3}$ & 1.362 & 1.25 & 1.67 & 1.09 & 0.62 & $-$0.67 & $-$0.47 & 0.06 & 0.00 & $-$0.18 & $-$0.08 & 2.58 \\
18 & $0.500 \times 10^{-3}$ & 1.266 & 0.93 & 1.01 & 0.95 & 0.71 & $-$0.44 & $-$0.29 & $-$0.08 & 0.00 & $-$0.14 & $-$0.07 & 1.89 \\
22 & $0.500 \times 10^{-3}$ & 1.335 & 2.57 & 1.22 & 1.18 & 1.36 & $-$0.85 & $-$0.43 & $-$0.08 & 0.02 & $-$0.07 & 0.04 & 3.50 \\
27 & $0.335 \times 10^{-3}$ & 1.389 & 4.11 & 3.92 & 0.98 & 0.54 & $-$0.12 & $-$0.15 & $-$0.01 & $-$0.01 & $-$0.45 & $-$0.14 & 5.81 \\
27 & $0.410 \times 10^{-3}$ & 1.405 & 1.16 & 2.10 & 1.15 & 0.63 & $-$0.70 & $-$0.40 & $-$0.04 & $-$0.01 & $-$0.08 & $-$0.11 & 2.85 \\
27 & $0.500 \times 10^{-3}$ & 1.407 & 1.15 & 1.46 & 0.92 & 0.77 & $-$0.59 & $-$0.31 & $-$0.02 & 0.00 & $-$0.21 & $-$0.07 & 2.32 \\
27 & $0.800 \times 10^{-3}$ & 1.286 & 2.46 & 0.83 & 0.87 & 0.77 & $-$0.48 & $-$0.28 & 0.02 & $-$0.02 & 0.02 & 0.04 & 2.89 \\
35 & $0.574 \times 10^{-3}$ & 1.507 & 1.36 & 2.01 & 0.99 & 0.59 & $-$0.47 & $-$0.35 & $-$0.03 & $-$0.03 & $-$0.19 & $-$0.14 & 2.76 \\
35 & $0.800 \times 10^{-3}$ & 1.399 & 0.95 & 0.81 & 0.91 & 0.95 & $-$0.60 & $-$0.25 & $-$0.01 & $-$0.01 & $-$0.15 & $-$0.17 & 1.94 \\
45 & $0.800 \times 10^{-3}$ & 1.486 & 1.81 & 1.45 & 1.44 & 0.80 & $-$1.20 & $-$0.53 & $-$0.06 & $-$0.07 & 0.08 & $-$0.09 & 3.14 \\
45 & $0.130 \times 10^{-2}$ & 1.306 & 0.95 & 0.62 & 0.86 & 0.73 & $-$0.03 & 0.05 & 0.00 & $-$0.02 & $-$0.07 & $-$0.21 & 1.61 \\
60 & $0.130 \times 10^{-2}$ & 1.395 & 1.31 & 0.55 & 0.91 & 1.01 & $-$0.42 & $-$0.23 & 0.00 & $-$0.04 & $-$0.03 & $-$0.27 & 2.04 \\
70 & $0.130 \times 10^{-2}$ & 1.443 & 1.77 & 0.60 & 1.03 & 1.09 & $-$0.44 & $-$0.41 & $-$0.03 & $-$0.06 & 0.01 & $-$0.12 & 2.48 \\
70 & $0.200 \times 10^{-2}$ & 1.262 & 1.72 & 0.33 & 0.90 & 0.53 & $-$0.22 & 0.06 & 0.02 & $-$0.05 & $-$0.04 & $-$0.33 & 2.08 \\
90 & $0.200 \times 10^{-2}$ & 1.252 & 1.98 & 0.35 & 0.98 & 1.10 & $-$0.32 & $-$0.02 & 0.02 & $-$0.03 & $-$0.05 & $-$0.24 & 2.53 \\
120 & $0.212 \times 10^{-2}$ & 1.323 & 2.22 & 0.77 & 1.22 & 1.26 & $-$0.38 & $-$0.33 & $-$0.02 & $-$0.05 & 0.04 & 0.20 & 2.99 \\
120 & $0.320 \times 10^{-2}$ & 1.178 & 1.93 & 0.73 & 0.88 & 0.91 & $-$0.14 & 0.06 & 0.02 & $-$0.03 & $-$0.01 & $-$0.15 & 2.43 \\
150 & $0.320 \times 10^{-2}$ & 1.203 & 1.82 & 0.89 & 1.04 & 0.88 & $-$0.37 & $-$0.24 & $-$0.02 & $-$0.06 & 0.05 & 0.14 & 2.49 \\
200 & $0.320 \times 10^{-2}$ & 1.170 & 3.80 & 2.00 & 1.32 & 1.49 & $-$0.51 & $-$0.25 & 0.00 & $-$0.04 & 0.03 & 0.22 & 4.77 \\
200 & $0.500 \times 10^{-2}$ & 1.072 & 1.89 & 0.88 & 0.92 & 0.77 & $-$0.20 & 0.13 & 0.02 & $-$0.03 & 0.00 & 0.07 & 2.42 \\
250 & $0.500 \times 10^{-2}$ & 1.078 & 2.30 & 1.62 & 1.02 & 1.08 & $-$0.34 & $-$0.04 & 0.02 & $-$0.02 & $-$0.01 & 0.11 & 3.19 \\
300 & $0.500 \times 10^{-2}$ & 1.126 & 2.94 & 2.65 & 0.98 & 0.50 & $-$0.54 & $-$0.29 & 0.03 & $-$0.04 & $-$0.04 & 0.00 & 4.15 \\
300 & $0.800 \times 10^{-2}$ & 0.977 & 2.29 & 1.12 & 0.95 & 0.66 & $-$0.27 & 0.23 & 0.01 & $-$0.05 & 0.07 & 0.05 & 2.83 \\
400 & $0.800 \times 10^{-2}$ & 0.956 & 2.63 & 1.78 & 0.86 & 0.58 & $-$0.32 & 0.08 & 0.02 & $-$0.04 & 0.03 & $-$0.07 & 3.36 \\
400 & $0.130 \times 10^{-1}$ & 0.855 & 2.96 & 0.57 & 0.83 & 0.53 & $-$0.15 & $-$0.02 & 0.02 & $-$0.04 & 0.02 & 0.02 & 3.17 \\
500 & $0.800 \times 10^{-2}$ & 1.021 & 4.21 & 5.00 & 1.06 & 0.13 & $-$0.56 & $-$0.35 & 0.01 & $-$0.03 & $-$0.12 & 0.14 & 6.66 \\
500 & $0.130 \times 10^{-1}$ & 0.885 & 3.34 & 5.00 & 0.99 & 0.35 & $-$0.25 & 0.15 & 0.00 & $-$0.04 & 0.16 & 0.21 & 6.12 \\
650 & $0.850 \times 10^{-2}$ & 0.921 & 4.63 & 1.82 & 1.35 & 1.40 & $-$0.44 & $-$0.33 & $-$0.03 & $-$0.06 & 0.09 & 0.21 & 5.38 \\
650 & $0.130 \times 10^{-1}$ & 0.843 & 2.65 & 1.13 & 0.84 & 0.35 & $-$0.20 & 0.23 & 0.01 & $-$0.04 & 0.06 & 0.03 & 3.04 \\
650 & $0.200 \times 10^{-1}$ & 0.704 & 4.14 & 3.70 & 0.97 & 0.35 & $-$0.19 & 0.02 & 0.00 & $-$0.05 & 0.08 & 0.23 & 5.66 \\
800 & $0.130 \times 10^{-1}$ & 0.890 & 3.45 & 1.49 & 0.92 & 0.82 & $-$0.26 & $-$0.01 & 0.01 & $-$0.04 & 0.00 & 0.10 & 3.96 \\
800 & $0.200 \times 10^{-1}$ & 0.736 & 4.12 & 2.31 & 0.89 & 0.30 & $-$0.19 & 0.26 & 0.02 & $-$0.05 & 0.16 & 0.06 & 4.83 \\
1000 & $0.200 \times 10^{-1}$ & 0.740 & 5.45 & 3.70 & 1.03 & 0.25 & $-$0.38 & 0.13 & 0.00 & $-$0.04 & 0.16 & 0.15 & 6.69 \\
1200 & $0.140 \times 10^{-1}$ & 0.872 & 5.51 & 1.05 & 1.31 & 1.43 & $-$0.44 & $-$0.03 & 0.00 & $-$0.04 & 0.06 & 0.14 & 5.95 \\
1200 & $0.200 \times 10^{-1}$ & 0.703 & 4.19 & 1.13 & 0.81 & 0.42 & $-$0.33 & 0.20 & 0.04 & $-$0.03 & 0.04 & $-$0.01 & 4.45 \\
1200 & $0.320 \times 10^{-1}$ & 0.589 & 3.96 & 1.51 & 0.79 & $-$0.08 & $-$0.15 & $-$0.03 & 0.03 & $-$0.05 & 0.10 & 0.03 & 4.32 \\
1500 & $0.200 \times 10^{-1}$ & 0.734 & 5.99 & 3.87 & 0.93 & $-$0.02 & $-$0.38 & $-$0.34 & 0.04 & $-$0.04 & $-$0.19 & $-$0.05 & 7.21 \\
1500 & $0.320 \times 10^{-1}$ & 0.556 & 6.45 & 2.29 & 0.85 & 0.22 & $-$0.26 & 0.27 & 0.03 & $-$0.04 & 0.18 & 0.06 & 6.92 \\
2000 & $0.320 \times 10^{-1}$ & 0.596 & 6.11 & 2.45 & 0.83 & 0.16 & $-$0.23 & 0.16 & $-$0.01 & $-$0.04 & 0.05 & 0.08 & 6.64 \\
2000 & $0.500 \times 10^{-1}$ & 0.496 & 5.68 & 1.70 & 0.80 & 0.26 & $-$0.13 & 0.10 & 0.00 & $-$0.05 & 0.17 & 0.13 & 6.00 \\
3000 & $0.500 \times 10^{-1}$ & 0.522 & 5.63 & 2.20 & 0.85 & 0.18 & $-$0.29 & 0.29 & 0.03 & $-$0.04 & 0.14 & 0.02 & 6.12 \\
3000 & $0.800 \times 10^{-1}$ & 0.444 & 5.78 & 1.98 & 0.80 & $-$0.02 & $-$0.07 & 0.01 & 0.02 & $-$0.05 & 0.17 & 0.12 & 6.17 \\
5000 & $0.800 \times 10^{-1}$ & 0.362 & 6.86 & 2.22 & 0.85 & 0.38 & $-$0.33 & 0.33 & 0.00 & $-$0.03 & 0.14 & 0.06 & 7.29 \\
5000 & 0.130 & 0.332 & 7.82 & 2.14 & 0.92 & 0.29 & $-$0.10 & $-$0.33 & 0.01 & $-$0.06 & 0.08 & 0.10 & 8.18 \\
8000 & 0.130 & 0.257 & 10.93 & 3.07 & 0.79 & 0.30 & $-$0.26 & 0.05 & 0.01 & $-$0.04 & 0.13 & 0.12 & 11.39 \\
8000 & 0.180 & 0.278 & 10.93 & 3.44 & 0.94 & 0.59 & $-$0.15 & $-$0.28 & $-$0.04 & $-$0.04 & 0.11 & 0.19 & 11.52 \\
8000 & 0.250 & 0.232 & 11.77 & 3.15 & 1.30 & 0.29 & $-$0.20 & 0.03 & 0.11 & $-$0.08 & 0.13 & 0.21 & 12.26 \\
12000 & 0.180 & 0.209 & 17.34 & 2.43 & 0.92 & 0.85 & $-$0.32 & $-$0.05 & 0.00 & $-$0.04 & $-$0.01 & 0.14 & 17.56 \\
12000 & 0.250 & 0.150 & 20.21 & 3.46 & 0.81 & 0.24 & $-$0.17 & 0.04 & $-$0.01 & $-$0.04 & 0.17 & 0.11 & 20.52 \\
20000 & 0.250 & 0.136 & 30.61 & 3.58 & 0.99 & 0.74 & $-$0.40 & $-$0.34 & 0.01 & $-$0.04 & $-$0.15 & 0.17 & 30.85 \\
20000 & 0.400 & 0.116 & 31.61 & 7.66 & 1.23 & 1.07 & $-$0.32 & 0.70 & 0.02 & $-$0.03 & 0.42 & 0.19 & 32.58 \\
30000 & 0.400 & 0.111 & 65.50 & 5.27 & 1.39 & $-$0.41 & $-$0.52 & $-$0.07 & $-$0.01 & $-$0.03 & $-$0.08 & 0.27 & 65.73 \\

\hline \hline

\end{tabular}\captcont{Continued.}


\end{scriptsize}
\end{center}
\end{table}

\clearpage
\begin{table}
\begin{center}
\begin{scriptsize}\renewcommand\arraystretch{1.1}

\begin{tabular}[H]{ c l c r r r r r r r r r r r }
\hline \hline
$Q^2$ &  $x_{\rm Bj}$ & $\sigma_{r, \rm NC}^{+ }$ & $\delta_{\rm stat}$ & $\delta_{\rm uncor}$ & $\delta_{\rm cor}$ & $\delta_{\rm rel}$ & $\delta_{\gamma p}$ & $\delta_{\rm had}$ &  $\delta_{1}$ & $\delta_{2}$ & $\delta_{3}$ & $\delta_{4}$ &$\delta_{\rm tot}$ \\
${\rm GeV^2}$ & & & \% & \% & \% & \% & \% & \% & \% & \% & \% & \% & \%     \\
\hline
1.5 & $0.279 \times 10^{-4}$ & 0.702 & 9.08 & 4.93 & 3.52 & 2.22 & $-$0.37 & $-$0.41 & $-$0.12 & $-$8.43 & 0.08 & 0.80 & 14.00 \\
2 & $0.372 \times 10^{-4}$ & 0.796 & 6.22 & 4.33 & 2.91 & 1.91 & $-$0.31 & $-$0.76 & $-$0.10 & $-$7.05 & 0.07 & 0.66 & 10.97 \\
2 & $0.415 \times 10^{-4}$ & 0.680 & 7.33 & 3.97 & 2.54 & 0.57 & $-$0.19 & $-$0.34 & $-$0.08 & $-$2.55 & 0.08 & 0.49 & 9.12 \\
2.5 & $0.465 \times 10^{-4}$ & 0.865 & 5.31 & 4.12 & 2.76 & 2.11 & $-$0.31 & $-$0.39 & $-$0.08 & $-$6.40 & 0.07 & 0.70 & 9.94 \\
2.5 & $0.519 \times 10^{-4}$ & 0.852 & 3.81 & 3.10 & 1.59 & 1.44 & $-$0.23 & $-$0.50 & $-$0.05 & $-$2.39 & 0.06 & 0.50 & 5.92 \\
2.5 & $0.580 \times 10^{-4}$ & 0.771 & 4.97 & 3.14 & 1.33 & 0.80 & $-$0.13 & $-$0.18 & $-$0.02 & $-$0.48 & 0.06 & 0.40 & 6.12 \\
3.5 & $0.651 \times 10^{-4}$ & 0.913 & 5.43 & 4.04 & 2.82 & 2.13 & $-$0.34 & $-$0.65 & $-$0.10 & $-$6.78 & 0.07 & 0.71 & 10.26 \\
3.5 & $0.727 \times 10^{-4}$ & 0.894 & 3.47 & 2.92 & 1.72 & 1.46 & $-$0.25 & $-$0.37 & $-$0.06 & $-$2.53 & 0.07 & 0.57 & 5.70 \\
3.5 & $0.812 \times 10^{-4}$ & 0.889 & 3.11 & 2.49 & 1.87 & 1.15 & $-$0.24 & $-$0.34 & $-$0.07 & $-$1.72 & 0.07 & 0.56 & 4.91 \\
3.5 & $0.921 \times 10^{-4}$ & 0.944 & 3.35 & 2.66 & 1.48 & 0.58 & $-$0.14 & $-$0.42 & $-$0.04 & $-$0.18 & 0.07 & 0.37 & 4.61 \\
3.5 & $0.106 \times 10^{-3}$ & 0.970 & 5.35 & 3.27 & 2.09 & 0.56 & $-$0.20 & $-$0.01 & $-$0.06 & $-$0.14 & 0.07 & 0.50 & 6.66 \\
5 & $0.931 \times 10^{-4}$ & 0.887 & 6.33 & 3.99 & 3.01 & 1.99 & $-$0.29 & $-$0.59 & $-$0.08 & $-$7.35 & 0.07 & 0.65 & 11.14 \\
5 & $0.104 \times 10^{-3}$ & 0.896 & 3.49 & 2.84 & 1.51 & 1.10 & $-$0.19 & $-$0.42 & $-$0.05 & $-$1.93 & 0.06 & 0.46 & 5.28 \\
5 & $0.116 \times 10^{-3}$ & 1.019 & 2.64 & 2.37 & 1.43 & 1.06 & $-$0.20 & $-$0.22 & $-$0.04 & $-$1.01 & 0.07 & 0.48 & 4.14 \\
5 & $0.132 \times 10^{-3}$ & 0.965 & 2.54 & 2.33 & 1.44 & 1.12 & $-$0.20 & $-$0.23 & $-$0.05 & $-$0.77 & 0.07 & 0.49 & 4.02 \\
5 & $0.152 \times 10^{-3}$ & 1.020 & 2.42 & 2.04 & 1.45 & 0.79 & $-$0.18 & $-$0.21 & $-$0.04 & $-$0.19 & 0.07 & 0.44 & 3.61 \\
5 & $0.201 \times 10^{-3}$ & 0.949 & 2.57 & 2.06 & 1.59 & 0.61 & $-$0.16 & $-$0.20 & $-$0.05 & $-$0.10 & 0.07 & 0.42 & 3.74 \\
6.5 & $0.121 \times 10^{-3}$ & 0.937 & 7.04 & 4.03 & 3.42 & 1.81 & $-$0.21 & $-$0.42 & $-$0.05 & $-$7.69 & 0.06 & 0.56 & 11.85 \\
6.5 & $0.135 \times 10^{-3}$ & 1.005 & 3.22 & 2.79 & 1.46 & 1.57 & $-$0.22 & $-$0.37 & $-$0.05 & $-$1.53 & 0.06 & 0.52 & 5.06 \\
6.5 & $0.151 \times 10^{-3}$ & 1.105 & 2.44 & 2.28 & 1.37 & 1.38 & $-$0.20 & $-$0.39 & $-$0.05 & $-$0.95 & 0.05 & 0.48 & 4.03 \\
6.5 & $0.163 \times 10^{-3}$ & 1.180 & 10.62 & 12.01 & 1.81 & 0.95 & $-$0.15 & $-$0.14 & $-$0.02 & 0.71 & 0.07 & 0.50 & 16.19 \\
6.5 & $0.171 \times 10^{-3}$ & 1.057 & 2.29 & 2.32 & 1.45 & 0.73 & $-$0.15 & $-$0.24 & $-$0.04 & $-$0.29 & 0.07 & 0.43 & 3.69 \\
6.5 & $0.183 \times 10^{-3}$ & 1.184 & 9.84 & 13.94 & 1.98 & 1.28 & $-$0.18 & $-$0.14 & $-$0.02 & 0.81 & 0.05 & 0.30 & 17.25 \\
6.5 & $0.197 \times 10^{-3}$ & 1.112 & 2.04 & 1.96 & 1.33 & 0.89 & $-$0.16 & $-$0.23 & $-$0.03 & $-$0.12 & 0.06 & 0.43 & 3.30 \\
6.5 & $0.228 \times 10^{-3}$ & 1.138 & 9.66 & 9.40 & 1.79 & 0.76 & $-$0.12 & $-$0.11 & $-$0.01 & $-$0.83 & 0.07 & 0.45 & 13.65 \\
6.5 & $0.262 \times 10^{-3}$ & 1.010 & 1.37 & 1.81 & 1.38 & 0.83 & $-$0.16 & $-$0.09 & $-$0.03 & $-$0.05 & 0.07 & 0.45 & 2.83 \\
8.5 & $0.158 \times 10^{-3}$ & 1.031 & 6.90 & 3.99 & 3.87 & 2.57 & $-$0.32 & $-$0.34 & $-$0.09 & $-$9.71 & 0.06 & 0.78 & 13.42 \\
8.5 & $0.177 \times 10^{-3}$ & 1.058 & 3.18 & 2.81 & 1.46 & 1.25 & $-$0.17 & $-$0.39 & $-$0.03 & $-$0.77 & 0.06 & 0.48 & 4.77 \\
8.5 & $0.197 \times 10^{-3}$ & 1.112 & 2.46 & 2.24 & 1.36 & 1.37 & $-$0.21 & $-$0.32 & $-$0.04 & $-$0.76 & 0.06 & 0.56 & 3.98 \\
8.5 & $0.224 \times 10^{-3}$ & 1.154 & 2.11 & 2.05 & 1.26 & 1.26 & $-$0.16 & $-$0.22 & $-$0.03 & 0.00 & 0.05 & 0.46 & 3.48 \\
8.5 & $0.240 \times 10^{-3}$ & 1.146 & 5.87 & 6.26 & 1.44 & 1.31 & $-$0.17 & $-$0.13 & $-$0.03 & 0.43 & 0.06 & 0.63 & 8.84 \\
8.5 & $0.258 \times 10^{-3}$ & 1.041 & 2.02 & 1.93 & 1.53 & 1.07 & $-$0.19 & $-$0.19 & $-$0.03 & $-$0.30 & 0.06 & 0.47 & 3.42 \\
8.5 & $0.299 \times 10^{-3}$ & 1.217 & 6.86 & 6.25 & 1.68 & 1.13 & $-$0.15 & $-$0.10 & $-$0.02 & 0.90 & 0.05 & 0.39 & 9.55 \\
8.5 & $0.342 \times 10^{-3}$ & 1.091 & 1.19 & 1.76 & 1.31 & 0.79 & $-$0.15 & $-$0.08 & $-$0.03 & 0.15 & 0.06 & 0.41 & 2.66 \\
8.5 & $0.433 \times 10^{-3}$ & 1.092 & 6.83 & 7.45 & 1.78 & 0.79 & $-$0.12 & $-$0.08 & $-$0.01 & $-$0.14 & 0.06 & 0.38 & 10.31 \\
8.5 & $0.541 \times 10^{-3}$ & 1.022 & 1.30 & 1.80 & 1.27 & 0.80 & $-$0.12 & 0.05 & $-$0.01 & 0.37 & 0.07 & 0.39 & 2.74 \\
8.5 & $0.838 \times 10^{-3}$ & 0.946 & 1.47 & 1.85 & 1.42 & 0.73 & $-$0.14 & 0.06 & $-$0.03 & 0.18 & 0.07 & 0.44 & 2.89 \\
8.5 & $0.103 \times 10^{-2}$ & 0.678 & 12.90 & 13.65 & 1.77 & 0.48 & $-$0.10 & $-$0.10 & $-$0.01 & $-$0.47 & 0.07 & 0.32 & 18.88 \\
8.5 & $0.140 \times 10^{-2}$ & 0.823 & 2.57 & 2.07 & 1.41 & 0.85 & $-$0.16 & 0.07 & $-$0.03 & 0.06 & 0.07 & 0.46 & 3.72 \\
12 & $0.223 \times 10^{-3}$ & 1.293 & 5.39 & 4.00 & 2.80 & 2.10 & $-$0.30 & $-$0.61 & $-$0.09 & $-$6.55 & 0.07 & 0.69 & 10.06 \\
12 & $0.249 \times 10^{-3}$ & 1.148 & 2.95 & 2.72 & 1.44 & 1.80 & $-$0.22 & $-$0.31 & $-$0.05 & $-$1.38 & 0.06 & 0.62 & 4.89 \\
12 & $0.278 \times 10^{-3}$ & 1.199 & 2.35 & 2.17 & 1.31 & 1.35 & $-$0.17 & $-$0.30 & $-$0.03 & $-$0.26 & 0.05 & 0.49 & 3.77 \\
12 & $0.295 \times 10^{-3}$ & 1.176 & 3.77 & 3.74 & 1.43 & 1.78 & $-$0.23 & $-$0.20 & $-$0.04 & 0.60 & 0.06 & 0.75 & 5.87 \\
12 & $0.316 \times 10^{-3}$ & 1.193 & 2.54 & 2.37 & 1.37 & 1.15 & $-$0.16 & $-$0.27 & $-$0.02 & $-$0.41 & 0.06 & 0.42 & 3.97 \\
12 & $0.343 \times 10^{-3}$ & 1.228 & 3.43 & 3.12 & 1.36 & 1.44 & $-$0.20 & $-$0.14 & $-$0.02 & 0.78 & 0.05 & 0.68 & 5.15 \\
12 & $0.364 \times 10^{-3}$ & 1.176 & 2.12 & 1.96 & 1.35 & 0.90 & $-$0.14 & $-$0.26 & $-$0.02 & 0.10 & 0.06 & 0.39 & 3.35 \\
12 & $0.423 \times 10^{-3}$ & 1.226 & 4.09 & 3.32 & 1.37 & 1.39 & $-$0.17 & $-$0.10 & $-$0.02 & 0.52 & 0.06 & 0.62 & 5.68 \\
12 & $0.483 \times 10^{-3}$ & 1.124 & 1.18 & 1.71 & 1.26 & 0.92 & $-$0.14 & $-$0.04 & $-$0.02 & 0.12 & 0.06 & 0.42 & 2.64 \\
12 & $0.592 \times 10^{-3}$ & 1.145 & 5.23 & 4.72 & 1.64 & 0.89 & $-$0.13 & $-$0.08 & $-$0.01 & 0.33 & 0.06 & 0.39 & 7.31 \\
12 & $0.764 \times 10^{-3}$ & 1.054 & 1.23 & 1.82 & 1.28 & 0.80 & $-$0.12 & 0.06 & $-$0.02 & 0.34 & 0.06 & 0.40 & 2.72 \\
12 & $0.910 \times 10^{-3}$ & 0.965 & 5.51 & 6.78 & 1.80 & 0.86 & $-$0.13 & $-$0.08 & 0.00 & 0.46 & 0.06 & 0.33 & 8.98 \\
12 & $0.118 \times 10^{-2}$ & 1.020 & 1.29 & 1.83 & 1.30 & 0.85 & $-$0.13 & 0.06 & $-$0.01 & 0.25 & 0.07 & 0.42 & 2.77 \\
12 & $0.146 \times 10^{-2}$ & 0.828 & 5.42 & 7.17 & 1.83 & 0.70 & $-$0.12 & $-$0.11 & $-$0.01 & $-$1.04 & 0.07 & 0.53 & 9.27 \\
12 & $0.197 \times 10^{-2}$ & 0.880 & 2.24 & 2.00 & 1.49 & 1.04 & $-$0.18 & 0.07 & $-$0.04 & $-$0.22 & 0.07 & 0.50 & 3.56 \\

\hline \hline

\end{tabular}

\caption{\label{tab615-251a1}
HERA combined reduced  cross sections $\sigma^{+ }_{r,{\rm NC}}$ for NC $e^{+}p$ scattering at $\sqrt{s} = 251 $~GeV.
The uncertainties are quoted in percent relative to $\sigma^{+ }_{r,{\rm NC}}$.
Other details as for Table~\ref{tab615-318a1}. }

\end{scriptsize}
\end{center}
\end{table}

\clearpage
\begin{table}
\begin{center}
\begin{scriptsize}\renewcommand\arraystretch{1.1}

\begin{tabular}[H]{ c l c r r r r r r r r r r r }
\hline \hline
$Q^2$ &  $x_{\rm Bj}$ & $\sigma_{r, \rm NC}^{+ }$ & $\delta_{\rm stat}$ & $\delta_{\rm uncor}$ & $\delta_{\rm cor}$ & $\delta_{\rm rel}$ & $\delta_{\gamma p}$ & $\delta_{\rm had}$ &  $\delta_{1}$ & $\delta_{2}$ & $\delta_{3}$ & $\delta_{4}$ &$\delta_{\rm tot}$ \\
${\rm GeV^2}$ & & & \% & \% & \% & \% & \% & \% & \% & \% & \% & \% & \%     \\
\hline

15 & $0.279 \times 10^{-3}$ & 1.147 & 6.20 & 4.04 & 2.44 & 1.78 & $-$0.24 & $-$0.46 & $-$0.06 & $-$5.39 & 0.07 & 0.60 & 9.67 \\
15 & $0.312 \times 10^{-3}$ & 1.284 & 2.62 & 2.56 & 1.37 & 1.08 & $-$0.19 & $-$0.41 & $-$0.05 & $-$0.89 & 0.07 & 0.55 & 4.21 \\
15 & $0.348 \times 10^{-3}$ & 1.187 & 2.40 & 2.11 & 1.25 & 0.86 & $-$0.16 & $-$0.22 & $-$0.03 & $-$0.06 & 0.06 & 0.46 & 3.58 \\
15 & $0.394 \times 10^{-3}$ & 1.174 & 2.19 & 2.01 & 1.27 & 1.24 & $-$0.17 & $-$0.29 & $-$0.03 & $-$0.11 & 0.06 & 0.53 & 3.52 \\
15 & $0.422 \times 10^{-3}$ & 1.242 & 2.61 & 2.13 & 1.32 & 1.33 & $-$0.18 & $-$0.14 & $-$0.02 & 0.64 & 0.06 & 0.59 & 3.96 \\
15 & $0.455 \times 10^{-3}$ & 1.259 & 2.33 & 2.01 & 1.32 & 0.98 & $-$0.15 & $-$0.24 & $-$0.02 & $-$0.06 & 0.06 & 0.41 & 3.52 \\
15 & $0.529 \times 10^{-3}$ & 1.204 & 2.72 & 1.94 & 1.32 & 1.25 & $-$0.17 & $-$0.12 & $-$0.02 & 0.26 & 0.06 & 0.66 & 3.87 \\
15 & $0.604 \times 10^{-3}$ & 1.168 & 1.22 & 1.46 & 1.26 & 1.13 & $-$0.16 & $-$0.07 & $-$0.02 & 0.07 & 0.05 & 0.48 & 2.60 \\
15 & $0.763 \times 10^{-3}$ & 1.145 & 3.30 & 2.60 & 1.36 & 1.27 & $-$0.15 & $-$0.08 & $-$0.02 & 0.35 & 0.05 & 0.61 & 4.65 \\
15 & $0.955 \times 10^{-3}$ & 1.042 & 1.28 & 1.82 & 1.28 & 0.73 & $-$0.12 & 0.05 & $-$0.01 & 0.52 & 0.06 & 0.37 & 2.75 \\
15 & $0.114 \times 10^{-2}$ & 1.093 & 3.78 & 3.31 & 1.44 & 1.05 & $-$0.15 & $-$0.09 & $-$0.02 & 0.12 & 0.06 & 0.59 & 5.37 \\
15 & $0.148 \times 10^{-2}$ & 0.976 & 1.32 & 1.83 & 1.36 & 1.07 & $-$0.14 & 0.07 & $-$0.01 & 0.07 & 0.06 & 0.44 & 2.89 \\
15 & $0.182 \times 10^{-2}$ & 0.847 & 4.43 & 3.95 & 1.54 & 0.87 & $-$0.14 & $-$0.10 & $-$0.01 & $-$0.64 & 0.07 & 0.62 & 6.26 \\
15 & $0.247 \times 10^{-2}$ & 0.861 & 2.16 & 1.98 & 1.48 & 1.25 & $-$0.17 & 0.08 & $-$0.02 & $-$0.15 & 0.06 & 0.48 & 3.56 \\
20 & $0.372 \times 10^{-3}$ & 1.332 & 6.47 & 4.25 & 2.21 & 1.82 & $-$0.29 & $-$0.38 & $-$0.08 & $-$4.26 & 0.08 & 0.66 & 9.33 \\
20 & $0.415 \times 10^{-3}$ & 1.261 & 3.63 & 2.93 & 1.39 & 1.06 & $-$0.18 & $-$0.37 & $-$0.04 & $-$1.21 & 0.06 & 0.45 & 5.16 \\
20 & $0.464 \times 10^{-3}$ & 1.294 & 3.04 & 2.46 & 1.31 & 0.79 & $-$0.12 & $-$0.24 & $-$0.02 & $-$0.09 & 0.06 & 0.36 & 4.22 \\
20 & $0.526 \times 10^{-3}$ & 1.142 & 3.16 & 2.46 & 1.33 & 1.09 & $-$0.17 & $-$0.35 & $-$0.03 & $-$0.60 & 0.06 & 0.42 & 4.44 \\
20 & $0.607 \times 10^{-3}$ & 1.226 & 2.57 & 2.06 & 1.53 & 1.17 & $-$0.16 & $-$0.25 & $-$0.01 & $-$0.07 & 0.05 & 0.41 & 3.85 \\
20 & $0.805 \times 10^{-3}$ & 1.211 & 1.41 & 1.84 & 1.31 & 0.91 & $-$0.12 & $-$0.10 & $-$0.01 & 0.25 & 0.06 & 0.38 & 2.85 \\
20 & $0.127 \times 10^{-2}$ & 1.091 & 1.39 & 1.84 & 1.33 & 0.93 & $-$0.11 & 0.05 & 0.00 & 0.38 & 0.06 & 0.39 & 2.87 \\
20 & $0.197 \times 10^{-2}$ & 0.982 & 1.44 & 1.85 & 1.35 & 1.04 & $-$0.14 & 0.06 & $-$0.01 & 0.15 & 0.06 & 0.43 & 2.94 \\
20 & $0.329 \times 10^{-2}$ & 0.858 & 2.38 & 2.03 & 1.40 & 1.09 & $-$0.16 & 0.07 & $-$0.02 & $-$0.10 & 0.06 & 0.47 & 3.64 \\
25 & $0.493 \times 10^{-3}$ & 1.300 & 2.79 & 2.55 & 1.30 & 1.03 & $-$0.17 & $-$0.36 & $-$0.04 & $-$0.66 & 0.06 & 0.40 & 4.22 \\
25 & $0.570 \times 10^{-3}$ & 1.330 & 3.15 & 2.69 & 1.38 & 0.74 & $-$0.15 & $-$0.21 & $-$0.04 & $-$0.25 & 0.07 & 0.48 & 4.48 \\
25 & $0.616 \times 10^{-3}$ & 1.259 & 1.77 & 1.83 & 1.21 & 0.83 & $-$0.15 & $-$0.29 & $-$0.03 & $-$0.07 & 0.06 & 0.42 & 2.98 \\
25 & $0.700 \times 10^{-3}$ & 1.325 & 2.27 & 1.88 & 1.25 & 0.90 & $-$0.14 & $-$0.12 & $-$0.02 & 0.18 & 0.07 & 0.45 & 3.37 \\
25 & $0.759 \times 10^{-3}$ & 1.227 & 2.60 & 2.06 & 1.34 & 0.99 & $-$0.17 & $-$0.24 & $-$0.02 & $-$0.16 & 0.07 & 0.45 & 3.76 \\
25 & $0.880 \times 10^{-3}$ & 1.248 & 2.16 & 1.62 & 1.28 & 1.03 & $-$0.14 & $-$0.10 & $-$0.02 & 0.28 & 0.06 & 0.48 & 3.21 \\
25 & $0.101 \times 10^{-2}$ & 1.231 & 1.25 & 1.23 & 1.21 & 0.99 & $-$0.14 & $-$0.07 & $-$0.01 & 0.43 & 0.06 & 0.45 & 2.44 \\
25 & $0.127 \times 10^{-2}$ & 1.120 & 2.00 & 1.70 & 1.32 & 1.09 & $-$0.14 & $-$0.08 & $-$0.01 & $-$0.03 & 0.06 & 0.56 & 3.18 \\
25 & $0.159 \times 10^{-2}$ & 1.128 & 1.55 & 1.87 & 1.28 & 0.75 & $-$0.11 & 0.05 & $-$0.01 & 0.50 & 0.05 & 0.37 & 2.92 \\
25 & $0.184 \times 10^{-2}$ & 1.042 & 2.05 & 2.20 & 1.33 & 1.17 & $-$0.14 & $-$0.08 & $-$0.01 & 0.04 & 0.06 & 0.58 & 3.54 \\
25 & $0.247 \times 10^{-2}$ & 1.020 & 1.58 & 1.88 & 1.48 & 1.25 & $-$0.16 & 0.08 & $-$0.01 & $-$0.10 & 0.06 & 0.46 & 3.16 \\
25 & $0.300 \times 10^{-2}$ & 0.920 & 2.16 & 2.25 & 1.36 & 1.07 & $-$0.15 & $-$0.09 & $-$0.01 & $-$0.62 & 0.06 & 0.70 & 3.69 \\
25 & $0.411 \times 10^{-2}$ & 0.892 & 2.64 & 2.11 & 1.38 & 0.91 & $-$0.10 & 0.05 & 0.00 & 0.49 & 0.06 & 0.36 & 3.81 \\
35 & $0.651 \times 10^{-3}$ & 1.322 & 7.00 & 3.79 & 3.80 & 0.01 & $-$0.05 & $-$0.14 & $-$0.02 & $-$2.86 & 0.18 & $-$0.11 & 9.28 \\
35 & $0.727 \times 10^{-3}$ & 1.302 & 3.89 & 3.42 & 1.64 & $-$0.10 & $-$0.09 & $-$0.39 & $-$0.04 & $-$0.25 & 0.07 & 0.35 & 5.46 \\
35 & $0.812 \times 10^{-3}$ & 1.327 & 2.57 & 2.07 & 1.21 & 0.77 & $-$0.16 & $-$0.29 & $-$0.03 & $-$0.29 & 0.07 & 0.31 & 3.64 \\
35 & $0.921 \times 10^{-3}$ & 1.329 & 2.29 & 1.99 & 1.19 & 0.72 & $-$0.14 & $-$0.27 & $-$0.02 & 0.25 & 0.06 & 0.29 & 3.38 \\
35 & $0.100 \times 10^{-2}$ & 1.217 & 2.64 & 2.20 & 1.23 & 0.60 & $-$0.12 & $-$0.14 & $-$0.02 & $-$0.12 & 0.07 & 0.35 & 3.72 \\
35 & $0.106 \times 10^{-2}$ & 1.319 & 2.72 & 2.11 & 1.32 & 1.05 & $-$0.16 & $-$0.18 & $-$0.02 & $-$0.14 & 0.06 & 0.42 & 3.87 \\
35 & $0.123 \times 10^{-2}$ & 1.247 & 2.33 & 1.70 & 1.24 & 0.70 & $-$0.12 & $-$0.10 & $-$0.01 & 0.40 & 0.06 & 0.31 & 3.26 \\
35 & $0.141 \times 10^{-2}$ & 1.241 & 1.25 & 1.30 & 1.18 & 0.80 & $-$0.13 & $-$0.08 & $-$0.02 & 0.19 & 0.06 & 0.39 & 2.34 \\
35 & $0.180 \times 10^{-2}$ & 1.120 & 1.83 & 1.34 & 1.27 & 0.88 & $-$0.13 & $-$0.08 & 0.00 & 0.13 & 0.06 & 0.45 & 2.79 \\
35 & $0.223 \times 10^{-2}$ & 1.111 & 1.68 & 1.89 & 1.29 & 0.88 & $-$0.12 & 0.05 & $-$0.01 & 0.33 & 0.05 & 0.40 & 3.02 \\
35 & $0.270 \times 10^{-2}$ & 0.982 & 1.71 & 1.80 & 1.29 & 0.93 & $-$0.13 & $-$0.07 & $-$0.01 & $-$0.32 & 0.06 & 0.50 & 3.01 \\
35 & $0.345 \times 10^{-2}$ & 0.964 & 1.83 & 1.91 & 1.30 & 0.80 & $-$0.10 & 0.05 & 0.00 & 0.53 & 0.06 & 0.36 & 3.13 \\
35 & $0.430 \times 10^{-2}$ & 0.862 & 1.64 & 1.39 & 1.32 & 0.96 & $-$0.13 & $-$0.08 & $-$0.01 & $-$0.67 & 0.06 & 0.62 & 2.85 \\
35 & $0.575 \times 10^{-2}$ & 0.861 & 2.99 & 2.20 & 1.65 & 1.39 & $-$0.20 & 0.09 & $-$0.03 & $-$0.54 & 0.06 & 0.54 & 4.37 \\
\hline \hline

\end{tabular}\captcont{Continued.}


\end{scriptsize}
\end{center}
\end{table}

\clearpage
\begin{table}
\begin{center}
\begin{scriptsize}\renewcommand\arraystretch{1.1}

\begin{tabular}[H]{ c l c r r r r r r r r r r r }
\hline \hline
$Q^2$ &  $x_{\rm Bj}$ & $\sigma_{r, \rm NC}^{+ }$ & $\delta_{\rm stat}$ & $\delta_{\rm uncor}$ & $\delta_{\rm cor}$ & $\delta_{\rm rel}$ & $\delta_{\gamma p}$ & $\delta_{\rm had}$ &  $\delta_{1}$ & $\delta_{2}$ & $\delta_{3}$ & $\delta_{4}$ &$\delta_{\rm tot}$ \\
${\rm GeV^2}$ & & & \% & \% & \% & \% & \% & \% & \% & \% & \% & \% & \%     \\
\hline
45 & $0.838 \times 10^{-3}$ & 1.432 & 5.99 & 3.37 & 3.16 & $-$0.04 & $-$0.04 & $-$0.12 & $-$0.01 & $-$1.89 & 0.10 & $-$0.08 & 7.80 \\
45 & $0.934 \times 10^{-3}$ & 1.320 & 4.31 & 3.22 & 1.69 & 0.20 & $-$0.08 & $-$0.17 & $-$0.02 & $-$0.48 & 0.07 & $-$0.08 & 5.67 \\
45 & $0.104 \times 10^{-2}$ & 1.349 & 3.41 & 2.34 & 1.16 & 0.27 & $-$0.09 & $-$0.19 & $-$0.01 & 0.27 & 0.07 & 0.09 & 4.32 \\
45 & $0.118 \times 10^{-2}$ & 1.298 & 2.65 & 1.95 & 1.16 & 0.58 & $-$0.12 & $-$0.29 & $-$0.02 & 0.21 & 0.06 & 0.22 & 3.57 \\
45 & $0.127 \times 10^{-2}$ & 1.286 & 2.87 & 2.18 & 1.20 & 0.28 & $-$0.09 & $-$0.12 & $-$0.01 & 0.32 & 0.07 & 0.09 & 3.83 \\
45 & $0.137 \times 10^{-2}$ & 1.292 & 3.08 & 2.20 & 1.32 & 0.67 & $-$0.12 & $-$0.26 & $-$0.02 & 0.28 & 0.06 & 0.36 & 4.10 \\
45 & $0.159 \times 10^{-2}$ & 1.192 & 2.62 & 2.44 & 1.20 & 0.41 & $-$0.09 & $-$0.10 & $-$0.01 & 0.20 & 0.06 & 0.16 & 3.81 \\
45 & $0.181 \times 10^{-2}$ & 1.178 & 1.39 & 1.26 & 1.16 & 0.66 & $-$0.11 & $-$0.08 & $-$0.01 & 0.36 & 0.06 & 0.28 & 2.35 \\
45 & $0.229 \times 10^{-2}$ & 1.114 & 1.91 & 1.47 & 1.23 & 0.58 & $-$0.10 & $-$0.08 & 0.00 & $-$0.12 & 0.07 & 0.31 & 2.79 \\
45 & $0.286 \times 10^{-2}$ & 1.103 & 1.71 & 1.90 & 1.29 & 0.74 & $-$0.10 & 0.05 & 0.00 & 0.55 & 0.06 & 0.36 & 3.04 \\
45 & $0.330 \times 10^{-2}$ & 1.006 & 1.68 & 1.74 & 1.24 & 0.67 & $-$0.11 & $-$0.07 & 0.00 & $-$0.34 & 0.07 & 0.35 & 2.85 \\
45 & $0.444 \times 10^{-2}$ & 0.940 & 1.87 & 1.92 & 1.32 & 0.91 & $-$0.11 & 0.05 & 0.00 & 0.41 & 0.06 & 0.38 & 3.18 \\
45 & $0.550 \times 10^{-2}$ & 0.890 & 1.50 & 1.63 & 1.27 & 0.72 & $-$0.12 & $-$0.09 & $-$0.01 & $-$0.74 & 0.07 & 0.48 & 2.80 \\
45 & $0.740 \times 10^{-2}$ & 0.828 & 3.14 & 2.23 & 1.63 & 1.43 & $-$0.19 & 0.08 & $-$0.02 & $-$0.38 & 0.06 & 0.51 & 4.47 \\
60 & $0.112 \times 10^{-2}$ & 1.276 & 7.05 & 3.22 & 3.06 & $-$0.03 & $-$0.04 & $-$0.11 & $-$0.01 & $-$1.53 & 0.09 & $-$0.07 & 8.47 \\
60 & $0.125 \times 10^{-2}$ & 1.419 & 4.10 & 2.43 & 1.88 & $-$0.02 & $-$0.06 & $-$0.13 & $-$0.01 & $-$0.92 & 0.07 & $-$0.02 & 5.21 \\
60 & $0.139 \times 10^{-2}$ & 1.358 & 4.17 & 2.38 & 1.59 & 0.09 & $-$0.07 & $-$0.13 & $-$0.01 & $-$0.44 & 0.07 & $-$0.09 & 5.08 \\
60 & $0.158 \times 10^{-2}$ & 1.324 & 3.90 & 2.73 & 1.17 & 0.26 & $-$0.09 & $-$0.08 & $-$0.02 & $-$0.20 & 0.07 & 0.08 & 4.92 \\
60 & $0.171 \times 10^{-2}$ & 1.297 & 3.42 & 2.53 & 1.22 & 0.12 & $-$0.07 & $-$0.10 & $-$0.01 & 0.66 & 0.06 & $-$0.11 & 4.48 \\
60 & $0.182 \times 10^{-2}$ & 1.260 & 4.23 & 2.54 & 1.30 & 0.77 & $-$0.11 & $-$0.10 & $-$0.02 & 0.37 & 0.06 & 0.35 & 5.19 \\
60 & $0.211 \times 10^{-2}$ & 1.115 & 3.19 & 2.20 & 1.21 & 0.22 & $-$0.06 & $-$0.08 & 0.01 & 0.40 & 0.07 & $-$0.04 & 4.08 \\
60 & $0.242 \times 10^{-2}$ & 1.169 & 1.59 & 1.38 & 1.16 & 0.48 & $-$0.09 & $-$0.07 & $-$0.01 & 0.38 & 0.06 & 0.24 & 2.49 \\
60 & $0.305 \times 10^{-2}$ & 1.098 & 2.24 & 1.68 & 1.20 & 0.30 & $-$0.08 & $-$0.08 & 0.00 & $-$0.16 & 0.06 & 0.13 & 3.07 \\
60 & $0.382 \times 10^{-2}$ & 1.015 & 1.98 & 1.94 & 1.28 & 0.75 & $-$0.11 & 0.05 & 0.00 & 0.53 & 0.07 & 0.37 & 3.22 \\
60 & $0.460 \times 10^{-2}$ & 0.949 & 1.96 & 1.93 & 1.21 & 0.39 & $-$0.09 & $-$0.08 & 0.00 & $-$0.41 & 0.07 & 0.22 & 3.07 \\
60 & $0.592 \times 10^{-2}$ & 0.994 & 1.97 & 1.96 & 1.30 & 0.91 & $-$0.13 & 0.06 & $-$0.01 & 0.25 & 0.06 & 0.41 & 3.24 \\
60 & $0.730 \times 10^{-2}$ & 0.817 & 1.74 & 1.46 & 1.26 & 0.41 & $-$0.09 & $-$0.08 & $-$0.01 & $-$1.11 & 0.07 & 0.33 & 2.88 \\
60 & $0.986 \times 10^{-2}$ & 0.773 & 3.39 & 2.30 & 1.38 & 0.83 & $-$0.09 & 0.05 & 0.01 & 0.63 & 0.06 & 0.34 & 4.46 \\
90 & $0.167 \times 10^{-2}$ & 1.327 & 7.81 & 2.91 & 3.03 & $-$0.04 & $-$0.05 & $-$0.11 & $-$0.02 & $-$1.49 & 0.06 & $-$0.08 & 9.00 \\
90 & $0.187 \times 10^{-2}$ & 1.392 & 4.56 & 2.39 & 1.89 & 0.08 & $-$0.06 & $-$0.13 & $-$0.01 & $-$0.50 & 0.06 & $-$0.06 & 5.51 \\
90 & $0.209 \times 10^{-2}$ & 1.299 & 4.15 & 2.10 & 1.83 & 0.00 & $-$0.06 & $-$0.13 & $-$0.02 & $-$0.31 & 0.06 & $-$0.05 & 5.01 \\
90 & $0.237 \times 10^{-2}$ & 1.305 & 4.12 & 2.53 & 1.81 & 0.05 & $-$0.06 & $-$0.11 & $-$0.01 & 0.05 & 0.06 & $-$0.09 & 5.16 \\
90 & $0.250 \times 10^{-2}$ & 1.153 & 4.13 & 3.26 & 1.22 & $-$0.04 & $-$0.06 & $-$0.11 & 0.00 & $-$0.17 & 0.07 & $-$0.14 & 5.41 \\
90 & $0.300 \times 10^{-2}$ & 1.159 & 3.56 & 2.80 & 1.23 & $-$0.05 & $-$0.05 & $-$0.09 & 0.00 & 0.52 & 0.07 & $-$0.21 & 4.73 \\
90 & $0.362 \times 10^{-2}$ & 1.124 & 2.13 & 1.63 & 1.15 & 0.34 & $-$0.07 & $-$0.07 & 0.00 & 0.22 & 0.06 & 0.15 & 2.96 \\
90 & $0.460 \times 10^{-2}$ & 1.018 & 2.62 & 1.96 & 1.20 & 0.10 & $-$0.06 & $-$0.08 & 0.00 & $-$0.33 & 0.07 & $-$0.05 & 3.50 \\
90 & $0.573 \times 10^{-2}$ & 0.922 & 2.37 & 2.01 & 1.29 & 0.72 & $-$0.12 & 0.05 & $-$0.02 & 0.41 & 0.06 & 0.39 & 3.49 \\
90 & $0.640 \times 10^{-2}$ & 0.948 & 2.20 & 2.10 & 1.20 & 0.15 & $-$0.06 & $-$0.07 & 0.00 & $-$0.64 & 0.07 & 0.04 & 3.34 \\
90 & $0.888 \times 10^{-2}$ & 0.871 & 2.31 & 2.02 & 1.30 & 0.91 & $-$0.13 & 0.06 & $-$0.01 & 0.23 & 0.06 & 0.42 & 3.49 \\
90 & $0.109 \times 10^{-1}$ & 0.758 & 2.02 & 1.65 & 1.22 & 0.17 & $-$0.07 & $-$0.08 & 0.00 & $-$0.99 & 0.07 & 0.15 & 3.06 \\
90 & $0.148 \times 10^{-1}$ & 0.756 & 3.84 & 2.47 & 1.29 & 0.82 & $-$0.13 & 0.06 & $-$0.01 & 0.31 & 0.06 & 0.41 & 4.84 \\
120 & $0.220 \times 10^{-2}$ & 1.390 & 8.37 & 2.74 & 2.88 & $-$0.04 & $-$0.04 & $-$0.10 & $-$0.01 & $-$1.22 & 0.09 & $-$0.07 & 9.35 \\
120 & $0.250 \times 10^{-2}$ & 1.184 & 7.69 & 2.05 & 2.76 & $-$0.06 & $-$0.03 & $-$0.08 & $-$0.01 & $-$0.50 & 0.06 & $-$0.06 & 8.44 \\
120 & $0.270 \times 10^{-2}$ & 1.081 & 7.61 & 5.88 & 1.23 & $-$0.08 & $-$0.06 & $-$0.15 & $-$0.01 & $-$0.12 & 0.07 & $-$0.05 & 9.70 \\
120 & $0.280 \times 10^{-2}$ & 1.176 & 6.90 & 1.92 & 2.75 & 0.07 & $-$0.04 & $-$0.10 & 0.00 & $-$0.21 & 0.03 & $-$0.03 & 7.68 \\
120 & $0.300 \times 10^{-2}$ & 1.181 & 6.42 & 4.62 & 1.30 & 0.05 & $-$0.06 & $-$0.11 & 0.00 & 0.91 & 0.07 & $-$0.25 & 8.07 \\
120 & $0.320 \times 10^{-2}$ & 1.118 & 6.47 & 1.83 & 2.77 & 0.05 & $-$0.04 & $-$0.08 & $-$0.01 & 0.09 & 0.05 & $-$0.06 & 7.27 \\
120 & $0.340 \times 10^{-2}$ & 1.134 & 4.87 & 3.53 & 1.23 & $-$0.07 & $-$0.06 & $-$0.12 & $-$0.01 & 0.39 & 0.07 & $-$0.10 & 6.15 \\
120 & $0.360 \times 10^{-2}$ & 1.192 & 5.65 & 1.83 & 2.78 & 0.13 & $-$0.06 & $-$0.08 & $-$0.01 & 0.20 & 0.07 & $-$0.07 & 6.56 \\
120 & $0.420 \times 10^{-2}$ & 1.139 & 4.34 & 2.96 & 1.21 & $-$0.13 & $-$0.04 & $-$0.09 & 0.00 & 0.04 & 0.07 & $-$0.15 & 5.39 \\
120 & $0.480 \times 10^{-2}$ & 1.071 & 2.99 & 1.65 & 1.50 & 0.05 & $-$0.05 & $-$0.07 & 0.00 & $-$0.19 & 0.07 & $-$0.13 & 3.74 \\
120 & $0.590 \times 10^{-2}$ & 0.985 & 3.06 & 2.06 & 1.20 & $-$0.14 & $-$0.04 & $-$0.08 & 0.00 & $-$0.38 & 0.07 & $-$0.14 & 3.90 \\
120 & $0.800 \times 10^{-2}$ & 0.869 & 2.66 & 1.90 & 1.21 & $-$0.22 & $-$0.03 & $-$0.08 & 0.00 & $-$0.76 & 0.07 & $-$0.16 & 3.58 \\
120 & $0.130 \times 10^{-1}$ & 0.722 & 2.37 & 2.04 & 1.23 & $-$0.06 & $-$0.04 & $-$0.08 & 0.00 & $-$1.17 & 0.07 & 0.02 & 3.56 \\

\hline \hline

\end{tabular}\captcont{Continued.}


\end{scriptsize}
\end{center}
\end{table}

\clearpage
\begin{table}
\begin{center}
\begin{scriptsize}\renewcommand\arraystretch{1.1}

\begin{tabular}[H]{ c l c r r r r r r r r r r r }
\hline \hline
$Q^2$ &  $x_{\rm Bj}$ & $\sigma_{r, \rm NC}^{+ }$ & $\delta_{\rm stat}$ & $\delta_{\rm uncor}$ & $\delta_{\rm cor}$ & $\delta_{\rm rel}$ & $\delta_{\gamma p}$ & $\delta_{\rm had}$ &  $\delta_{1}$ & $\delta_{2}$ & $\delta_{3}$ & $\delta_{4}$ &$\delta_{\rm tot}$ \\
${\rm GeV^2}$ & & & \% & \% & \% & \% & \% & \% & \% & \% & \% & \% & \%     \\
\hline

150 & $0.280 \times 10^{-2}$ & 1.313 & 10.23 & 2.60 & 3.06 & $-$0.05 & $-$0.03 & $-$0.11 & $-$0.01 & $-$1.86 & 0.08 & $-$0.08 & 11.14 \\
150 & $0.310 \times 10^{-2}$ & 1.184 & 9.42 & 2.04 & 2.79 & 0.02 & $-$0.04 & $-$0.09 & $-$0.01 & $-$0.69 & 0.08 & $-$0.07 & 10.06 \\
150 & $0.350 \times 10^{-2}$ & 1.335 & 7.52 & 2.29 & 2.75 & 0.04 & $-$0.04 & $-$0.09 & 0.00 & $-$0.09 & 0.06 & $-$0.04 & 8.33 \\
150 & $0.390 \times 10^{-2}$ & 1.257 & 6.92 & 1.79 & 2.76 & 0.03 & $-$0.04 & $-$0.08 & $-$0.01 & 0.12 & 0.05 & $-$0.05 & 7.66 \\
150 & $0.450 \times 10^{-2}$ & 1.050 & 6.83 & 1.79 & 2.76 & 0.07 & $-$0.05 & $-$0.09 & $-$0.01 & 0.13 & 0.06 & $-$0.06 & 7.58 \\
150 & $0.600 \times 10^{-2}$ & 1.025 & 3.61 & 1.41 & 2.77 & 0.05 & $-$0.05 & $-$0.08 & 0.00 & 0.26 & 0.06 & $-$0.06 & 4.77 \\
150 & $0.800 \times 10^{-2}$ & 0.973 & 3.93 & 1.15 & 2.78 & 0.14 & $-$0.06 & $-$0.08 & $-$0.01 & 0.28 & 0.07 & $-$0.07 & 4.97 \\
150 & $0.130 \times 10^{-1}$ & 0.863 & 5.20 & 1.84 & 2.79 & 0.04 & $-$0.05 & $-$0.02 & $-$0.01 & 0.28 & $-$0.10 & $-$0.08 & 6.19 \\
150 & $0.200 \times 10^{-1}$ & 0.794 & 6.97 & 2.81 & 2.86 & 0.12 & $-$0.07 & 0.01 & $-$0.01 & 0.28 & $-$0.15 & $-$0.13 & 8.05 \\
200 & $0.370 \times 10^{-2}$ & 1.318 & 12.81 & 2.59 & 3.15 & $-$0.06 & $-$0.04 & $-$0.11 & $-$0.01 & $-$2.14 & 0.06 & $-$0.08 & 13.62 \\
200 & $0.410 \times 10^{-2}$ & 1.311 & 11.44 & 2.11 & 2.85 & 0.08 & $-$0.05 & $-$0.11 & $-$0.01 & $-$0.86 & 0.05 & $-$0.05 & 12.01 \\
200 & $0.460 \times 10^{-2}$ & 1.068 & 10.99 & 2.20 & 2.75 & 0.03 & $-$0.03 & $-$0.08 & 0.00 & $-$0.08 & 0.03 & $-$0.03 & 11.54 \\
200 & $0.520 \times 10^{-2}$ & 1.180 & 8.89 & 1.82 & 2.76 & 0.00 & $-$0.04 & $-$0.08 & $-$0.01 & 0.24 & 0.06 & $-$0.04 & 9.49 \\
200 & $0.610 \times 10^{-2}$ & 1.116 & 8.08 & 1.80 & 2.76 & 0.01 & $-$0.04 & $-$0.08 & $-$0.01 & 0.21 & 0.05 & $-$0.04 & 8.73 \\
200 & $0.800 \times 10^{-2}$ & 0.969 & 4.60 & 1.26 & 2.77 & 0.10 & $-$0.05 & $-$0.08 & $-$0.01 & 0.24 & 0.05 & $-$0.06 & 5.52 \\
200 & $0.130 \times 10^{-1}$ & 0.876 & 4.42 & 1.26 & 2.78 & 0.10 & $-$0.05 & $-$0.07 & $-$0.01 & 0.28 & 0.08 & $-$0.08 & 5.38 \\
200 & $0.200 \times 10^{-1}$ & 0.765 & 4.91 & 0.90 & 2.78 & 0.12 & $-$0.05 & $-$0.11 & $-$0.01 & 0.28 & $-$0.09 & $-$0.03 & 5.72 \\
200 & $0.320 \times 10^{-1}$ & 0.619 & 5.46 & 0.98 & 2.77 & $-$0.01 & $-$0.04 & $-$0.07 & $-$0.01 & 0.27 & $-$0.06 & $-$0.04 & 6.21 \\
200 & $0.500 \times 10^{-1}$ & 0.507 & 6.34 & 1.17 & 2.78 & 0.07 & $-$0.05 & $-$0.06 & $-$0.01 & 0.28 & $-$0.11 & $-$0.06 & 7.03 \\
200 & $0.800 \times 10^{-1}$ & 0.419 & 7.37 & 1.70 & 2.80 & $-$0.09 & $-$0.05 & $-$0.07 & $-$0.01 & 0.27 & $-$0.09 & $-$0.05 & 8.07 \\
200 & 0.130 & 0.384 & 7.16 & 1.89 & 2.82 & 0.11 & $-$0.07 & $-$0.09 & $-$0.01 & 0.28 & $-$0.21 & $-$0.07 & 7.94 \\
200 & 0.180 & 0.311 & 9.17 & 2.63 & 2.94 & 0.44 & $-$0.05 & 0.04 & 0.00 & 0.29 & $-$0.03 & $-$0.13 & 10.00 \\
200 & 0.400 & 0.198 & 10.78 & 3.72 & 3.24 & 0.81 & $-$0.07 & 0.11 & 0.01 & 0.29 & 0.06 & $-$0.20 & 11.89 \\
250 & $0.460 \times 10^{-2}$ & 0.875 & 19.23 & 2.51 & 3.41 & $-$0.01 & $-$0.04 & $-$0.13 & $-$0.01 & $-$2.81 & 0.04 & $-$0.09 & 19.89 \\
250 & $0.520 \times 10^{-2}$ & 1.098 & 13.92 & 2.15 & 2.97 & 0.04 & $-$0.05 & $-$0.11 & $-$0.01 & $-$1.19 & 0.05 & $-$0.06 & 14.45 \\
250 & $0.580 \times 10^{-2}$ & 0.956 & 13.65 & 2.26 & 2.76 & $-$0.03 & $-$0.04 & $-$0.07 & $-$0.01 & 0.09 & 0.04 & $-$0.04 & 14.11 \\
250 & $0.660 \times 10^{-2}$ & 0.954 & 11.44 & 1.84 & 2.76 & 0.02 & $-$0.04 & $-$0.08 & $-$0.01 & 0.22 & 0.06 & $-$0.04 & 11.91 \\
250 & $0.760 \times 10^{-2}$ & 1.025 & 9.60 & 1.81 & 2.76 & 0.04 & $-$0.05 & $-$0.08 & $-$0.01 & 0.19 & 0.05 & $-$0.05 & 10.15 \\
250 & $0.100 \times 10^{-1}$ & 0.960 & 5.09 & 1.24 & 2.77 & 0.04 & $-$0.05 & $-$0.07 & $-$0.01 & 0.28 & 0.06 & $-$0.05 & 5.94 \\
250 & $0.130 \times 10^{-1}$ & 0.854 & 5.09 & 1.07 & 2.77 & $-$0.01 & $-$0.04 & $-$0.06 & $-$0.01 & 0.28 & 0.07 & $-$0.06 & 5.90 \\
250 & $0.200 \times 10^{-1}$ & 0.770 & 5.22 & 1.47 & 2.78 & $-$0.01 & $-$0.05 & $-$0.05 & $-$0.01 & 0.28 & 0.10 & $-$0.08 & 6.10 \\
250 & $0.320 \times 10^{-1}$ & 0.630 & 5.65 & 1.36 & 2.79 & 0.17 & $-$0.05 & $-$0.13 & $-$0.01 & 0.28 & $-$0.08 & $-$0.02 & 6.45 \\
250 & $0.500 \times 10^{-1}$ & 0.549 & 5.84 & 1.35 & 2.78 & 0.17 & $-$0.05 & $-$0.12 & $-$0.01 & 0.28 & $-$0.10 & $-$0.02 & 6.62 \\
250 & $0.800 \times 10^{-1}$ & 0.465 & 6.39 & 1.25 & 2.79 & 0.03 & $-$0.05 & $-$0.13 & $-$0.01 & 0.28 & $-$0.15 & $-$0.01 & 7.09 \\
250 & 0.130 & 0.393 & 6.18 & 1.36 & 2.81 & $-$0.01 & $-$0.05 & $-$0.14 & $-$0.01 & 0.28 & $-$0.18 & 0.00 & 6.93 \\
250 & 0.180 & 0.365 & 6.77 & 2.50 & 2.86 & 0.46 & $-$0.05 & $-$0.04 & 0.00 & 0.29 & $-$0.31 & $-$0.05 & 7.79 \\
250 & 0.400 & 0.167 & 9.74 & 4.15 & 3.40 & 1.35 & $-$0.06 & 0.05 & 0.01 & 0.30 & $-$0.11 & $-$0.12 & 11.21 \\
300 & $0.560 \times 10^{-2}$ & 1.226 & 15.58 & 2.52 & 3.24 & $-$0.08 & $-$0.03 & $-$0.10 & $-$0.01 & $-$2.01 & 0.07 & $-$0.09 & 16.24 \\
300 & $0.620 \times 10^{-2}$ & 0.882 & 17.86 & 2.13 & 2.87 & $-$0.02 & $-$0.04 & $-$0.10 & $-$0.01 & $-$1.13 & 0.09 & $-$0.07 & 18.25 \\
300 & $0.690 \times 10^{-2}$ & 0.979 & 14.75 & 2.08 & 2.77 & 0.06 & $-$0.04 & $-$0.09 & $-$0.01 & 0.28 & 0.06 & $-$0.04 & 15.16 \\
300 & $0.790 \times 10^{-2}$ & 1.046 & 12.67 & 1.86 & 2.77 & 0.09 & $-$0.05 & $-$0.10 & 0.00 & 0.28 & 0.05 & $-$0.03 & 13.10 \\
300 & $0.910 \times 10^{-2}$ & 0.868 & 11.94 & 1.84 & 2.76 & 0.03 & $-$0.04 & $-$0.07 & $-$0.01 & 0.28 & 0.06 & $-$0.05 & 12.40 \\
300 & $0.121 \times 10^{-1}$ & 0.912 & 5.97 & 1.25 & 2.76 & 0.01 & $-$0.04 & $-$0.06 & $-$0.01 & 0.28 & 0.05 & $-$0.05 & 6.71 \\
300 & $0.200 \times 10^{-1}$ & 0.667 & 6.31 & 1.01 & 2.78 & 0.07 & $-$0.05 & $-$0.06 & $-$0.01 & 0.28 & 0.07 & $-$0.07 & 6.97 \\
300 & $0.320 \times 10^{-1}$ & 0.625 & 6.55 & 0.97 & 2.77 & 0.08 & $-$0.04 & $-$0.11 & $-$0.01 & 0.28 & $-$0.07 & $-$0.02 & 7.18 \\
300 & $0.500 \times 10^{-1}$ & 0.545 & 6.74 & 1.17 & 2.78 & 0.17 & $-$0.05 & $-$0.12 & $-$0.01 & 0.28 & $-$0.14 & $-$0.02 & 7.40 \\
300 & $0.800 \times 10^{-1}$ & 0.487 & 7.19 & 1.26 & 2.79 & 0.15 & $-$0.06 & $-$0.13 & $-$0.01 & 0.28 & $-$0.15 & $-$0.01 & 7.83 \\
300 & 0.130 & 0.379 & 7.19 & 1.52 & 2.82 & 0.07 & $-$0.06 & $-$0.15 & $-$0.01 & 0.28 & $-$0.23 & 0.00 & 7.88 \\
300 & 0.180 & 0.330 & 7.91 & 2.50 & 2.87 & 0.58 & $-$0.07 & $-$0.10 & 0.00 & 0.29 & $-$0.26 & $-$0.04 & 8.81 \\
300 & 0.400 & 0.160 & 11.50 & 4.80 & 3.65 & 1.68 & $-$0.06 & 0.08 & 0.01 & 0.31 & $-$0.14 & $-$0.14 & 13.10 \\

\hline \hline

\end{tabular}\captcont{Continued.}


\end{scriptsize}
\end{center}
\end{table}

\clearpage
\begin{table}
\begin{center}
\begin{scriptsize}\renewcommand\arraystretch{1.1}

\begin{tabular}[H]{ c l c r r r r r r r r r r r }
\hline \hline
$Q^2$ &  $x_{\rm Bj}$ & $\sigma_{r, \rm NC}^{+ }$ & $\delta_{\rm stat}$ & $\delta_{\rm uncor}$ & $\delta_{\rm cor}$ & $\delta_{\rm rel}$ & $\delta_{\gamma p}$ & $\delta_{\rm had}$ &  $\delta_{1}$ & $\delta_{2}$ & $\delta_{3}$ & $\delta_{4}$ &$\delta_{\rm tot}$ \\
${\rm GeV^2}$ & & & \% & \% & \% & \% & \% & \% & \% & \% & \% & \% & \%     \\
\hline
400 & $0.740 \times 10^{-2}$ & 0.827 & 22.66 & 2.56 & 3.33 & $-$0.08 & $-$0.04 & $-$0.11 & $-$0.01 & $-$2.42 & 0.09 & $-$0.10 & 23.17 \\
400 & $0.830 \times 10^{-2}$ & 0.602 & 22.83 & 2.26 & 2.79 & $-$0.06 & $-$0.03 & $-$0.08 & $-$0.01 & $-$0.49 & 0.04 & $-$0.05 & 23.11 \\
400 & $0.930 \times 10^{-2}$ & 1.026 & 15.89 & 1.95 & 2.76 & $-$0.02 & $-$0.04 & $-$0.09 & 0.00 & 0.15 & 0.03 & $-$0.02 & 16.24 \\
400 & $0.105 \times 10^{-1}$ & 0.887 & 15.40 & 1.91 & 2.76 & $-$0.01 & $-$0.04 & $-$0.07 & $-$0.01 & 0.28 & 0.06 & $-$0.04 & 15.76 \\
400 & $0.121 \times 10^{-1}$ & 1.013 & 12.81 & 1.85 & 2.76 & 0.06 & $-$0.04 & $-$0.09 & $-$0.01 & 0.28 & 0.02 & $-$0.03 & 13.24 \\
400 & $0.161 \times 10^{-1}$ & 0.884 & 6.97 & 1.22 & 2.77 & 0.04 & $-$0.05 & $-$0.07 & $-$0.01 & 0.28 & 0.06 & $-$0.05 & 7.60 \\
400 & $0.320 \times 10^{-1}$ & 0.650 & 7.15 & 1.15 & 2.78 & 0.04 & $-$0.05 & $-$0.05 & $-$0.01 & 0.28 & 0.09 & $-$0.08 & 7.76 \\
400 & $0.500 \times 10^{-1}$ & 0.519 & 7.83 & 1.07 & 2.79 & 0.19 & $-$0.05 & $-$0.13 & $-$0.01 & 0.28 & $-$0.11 & $-$0.01 & 8.39 \\
400 & $0.800 \times 10^{-1}$ & 0.453 & 8.42 & 1.10 & 2.79 & 0.18 & $-$0.06 & $-$0.13 & $-$0.01 & 0.28 & $-$0.18 & $-$0.01 & 8.95 \\
400 & 0.130 & 0.380 & 8.17 & 1.26 & 2.80 & 0.00 & $-$0.05 & $-$0.15 & $-$0.01 & 0.28 & $-$0.11 & 0.00 & 8.73 \\
400 & 0.180 & 0.353 & 8.44 & 1.98 & 2.82 & 0.40 & $-$0.05 & $-$0.08 & 0.00 & 0.29 & $-$0.34 & $-$0.03 & 9.14 \\
400 & 0.400 & 0.163 & 12.84 & 4.55 & 3.65 & 1.70 & $-$0.06 & 0.09 & 0.01 & 0.31 & $-$0.29 & $-$0.13 & 14.21 \\
500 & $0.930 \times 10^{-2}$ & 0.744 & 27.27 & 2.71 & 3.25 & $-$0.10 & $-$0.03 & $-$0.11 & $-$0.01 & $-$2.16 & 0.07 & $-$0.07 & 27.68 \\
500 & $0.104 \times 10^{-1}$ & 0.739 & 22.59 & 2.27 & 2.76 & $-$0.02 & $-$0.04 & $-$0.08 & $-$0.01 & 0.20 & 0.08 & $-$0.04 & 22.87 \\
500 & $0.116 \times 10^{-1}$ & 1.188 & 16.00 & 1.95 & 2.76 & $-$0.01 & $-$0.04 & $-$0.08 & $-$0.01 & 0.28 & 0.05 & $-$0.03 & 16.36 \\
500 & $0.131 \times 10^{-1}$ & 0.862 & 17.53 & 1.94 & 2.76 & 0.00 & $-$0.04 & $-$0.08 & $-$0.01 & 0.13 & 0.05 & $-$0.04 & 17.86 \\
500 & $0.152 \times 10^{-1}$ & 1.047 & 14.03 & 1.88 & 2.76 & $-$0.05 & $-$0.03 & $-$0.06 & $-$0.01 & 0.27 & 0.06 & $-$0.05 & 14.43 \\
500 & $0.201 \times 10^{-1}$ & 0.738 & 8.97 & 1.28 & 2.76 & $-$0.02 & $-$0.04 & $-$0.06 & $-$0.01 & 0.28 & 0.05 & $-$0.05 & 9.48 \\
500 & $0.320 \times 10^{-1}$ & 0.691 & 8.29 & 1.09 & 2.77 & $-$0.04 & $-$0.04 & $-$0.05 & $-$0.01 & 0.28 & 0.08 & $-$0.06 & 8.81 \\
500 & $0.500 \times 10^{-1}$ & 0.567 & 8.78 & 0.98 & 2.77 & $-$0.04 & $-$0.04 & $-$0.11 & $-$0.01 & 0.28 & 0.01 & $-$0.01 & 9.27 \\
500 & $0.800 \times 10^{-1}$ & 0.454 & 9.61 & 1.05 & 2.77 & 0.03 & $-$0.04 & $-$0.10 & $-$0.01 & 0.28 & $-$0.14 & 0.00 & 10.06 \\
500 & 0.130 & 0.387 & 11.29 & 1.30 & 2.79 & 0.02 & $-$0.04 & $-$0.14 & $-$0.01 & 0.28 & 0.00 & 0.00 & 11.70 \\
500 & 0.180 & 0.351 & 11.27 & 1.46 & 2.81 & 0.07 & $-$0.05 & $-$0.14 & $-$0.01 & 0.28 & $-$0.29 & 0.01 & 11.71 \\
500 & 0.250 & 0.237 & 13.29 & 2.04 & 2.82 & 0.31 & $-$0.04 & $-$0.05 & 0.00 & 0.28 & $-$0.32 & $-$0.03 & 13.75 \\
500 & 0.400 & 0.182 & 15.67 & 4.22 & 3.33 & 1.26 & $-$0.04 & 0.06 & 0.01 & 0.30 & $-$0.52 & $-$0.08 & 16.62 \\
500 & 0.650 & 0.023 & 27.31 & 5.50 & 4.11 & 2.21 & $-$0.04 & 0.15 & 0.02 & 0.31 & $-$0.10 & $-$0.16 & 28.25 \\
650 & $0.121 \times 10^{-1}$ & 0.501 & 38.13 & 2.96 & 2.82 & $-$0.11 & $-$0.02 & $-$0.09 & $-$0.01 & $-$0.92 & 0.02 & $-$0.03 & 38.36 \\
650 & $0.135 \times 10^{-1}$ & 0.705 & 28.06 & 2.22 & 2.77 & $-$0.09 & $-$0.03 & $-$0.07 & $-$0.01 & $-$0.01 & 0.09 & $-$0.04 & 28.29 \\
650 & $0.151 \times 10^{-1}$ & 0.688 & 24.91 & 2.08 & 2.77 & $-$0.02 & $-$0.03 & $-$0.08 & $-$0.01 & 0.04 & 0.03 & $-$0.03 & 25.15 \\
650 & $0.171 \times 10^{-1}$ & 0.701 & 21.18 & 2.04 & 2.76 & $-$0.01 & $-$0.04 & $-$0.07 & $-$0.01 & 0.28 & 0.07 & $-$0.04 & 21.46 \\
650 & $0.197 \times 10^{-1}$ & 0.484 & 22.49 & 1.98 & 2.76 & $-$0.04 & $-$0.04 & $-$0.06 & $-$0.01 & 0.27 & 0.06 & $-$0.05 & 22.75 \\
650 & $0.261 \times 10^{-1}$ & 0.640 & 10.68 & 1.31 & 2.76 & 0.03 & $-$0.04 & $-$0.07 & $-$0.01 & 0.28 & 0.04 & $-$0.05 & 11.11 \\
650 & $0.500 \times 10^{-1}$ & 0.469 & 11.58 & 1.27 & 2.78 & 0.00 & $-$0.04 & $-$0.05 & $-$0.01 & 0.28 & 0.09 & $-$0.07 & 11.98 \\
650 & $0.800 \times 10^{-1}$ & 0.461 & 11.19 & 1.08 & 2.78 & 0.13 & $-$0.05 & $-$0.12 & $-$0.01 & 0.28 & $-$0.15 & $-$0.01 & 11.58 \\
650 & 0.130 & 0.404 & 13.15 & 1.40 & 2.79 & 0.12 & $-$0.05 & $-$0.14 & $-$0.01 & 0.28 & $-$0.05 & 0.00 & 13.51 \\
650 & 0.180 & 0.332 & 13.80 & 1.43 & 2.79 & $-$0.10 & $-$0.04 & $-$0.12 & $-$0.01 & 0.27 & $-$0.17 & 0.01 & 14.16 \\
650 & 0.250 & 0.248 & 15.04 & 2.03 & 2.81 & 0.35 & $-$0.05 & $-$0.11 & 0.00 & 0.28 & $-$0.31 & $-$0.01 & 15.44 \\
650 & 0.400 & 0.169 & 18.17 & 4.47 & 3.33 & 1.42 & $-$0.07 & 0.01 & 0.01 & 0.30 & $-$0.74 & $-$0.07 & 19.07 \\
650 & 0.650 & 0.025 & 31.51 & 6.03 & 4.15 & 2.25 & $-$0.05 & 0.17 & 0.02 & 0.31 & $-$0.18 & $-$0.17 & 32.43 \\
800 & $0.149 \times 10^{-1}$ & 0.672 & 31.65 & 3.07 & 2.76 & $-$0.16 & $-$0.03 & $-$0.06 & $-$0.01 & 0.14 & 0.10 & $-$0.04 & 31.91 \\
800 & $0.166 \times 10^{-1}$ & 0.493 & 38.09 & 2.69 & 2.77 & 0.09 & $-$0.04 & $-$0.12 & $-$0.01 & 0.28 & 0.03 & $-$0.01 & 38.28 \\
800 & $0.185 \times 10^{-1}$ & 0.683 & 26.86 & 2.22 & 2.76 & $-$0.04 & $-$0.04 & $-$0.07 & $-$0.01 & 0.28 & 0.03 & $-$0.04 & 27.09 \\
800 & $0.210 \times 10^{-1}$ & 0.666 & 24.26 & 2.06 & 2.77 & 0.07 & $-$0.05 & $-$0.09 & $-$0.01 & 0.28 & 0.07 & $-$0.04 & 24.51 \\
800 & $0.242 \times 10^{-1}$ & 0.651 & 21.70 & 1.94 & 2.77 & 0.12 & $-$0.05 & $-$0.09 & $-$0.01 & 0.28 & 0.02 & $-$0.04 & 21.96 \\
800 & $0.322 \times 10^{-1}$ & 0.496 & 13.46 & 1.39 & 2.76 & $-$0.07 & $-$0.03 & $-$0.06 & $-$0.01 & 0.27 & 0.04 & $-$0.04 & 13.81 \\
800 & $0.500 \times 10^{-1}$ & 0.585 & 11.87 & 1.22 & 2.77 & 0.03 & $-$0.05 & $-$0.06 & $-$0.01 & 0.28 & 0.08 & $-$0.06 & 12.26 \\
800 & $0.800 \times 10^{-1}$ & 0.562 & 11.86 & 1.08 & 2.78 & 0.13 & $-$0.04 & $-$0.12 & $-$0.01 & 0.28 & 0.08 & $-$0.01 & 12.23 \\
800 & 0.130 & 0.297 & 18.11 & 1.34 & 2.77 & $-$0.01 & $-$0.03 & $-$0.09 & $-$0.01 & 0.28 & $-$0.20 & 0.00 & 18.37 \\
800 & 0.180 & 0.267 & 18.44 & 1.55 & 2.80 & 0.12 & $-$0.05 & $-$0.15 & $-$0.01 & 0.28 & $-$0.10 & 0.00 & 18.72 \\
800 & 0.250 & 0.187 & 20.10 & 1.92 & 2.81 & 0.34 & $-$0.05 & $-$0.11 & $-$0.01 & 0.28 & $-$0.29 & $-$0.01 & 20.39 \\
800 & 0.400 & 0.172 & 20.66 & 3.77 & 3.09 & 0.87 & $-$0.04 & 0.02 & 0.00 & 0.29 & $-$0.38 & $-$0.05 & 21.25 \\
800 & 0.650 & 0.024 & 38.28 & 6.65 & 4.28 & 2.44 & $-$0.02 & 0.22 & 0.02 & 0.31 & $-$0.62 & $-$0.14 & 39.17 \\

\hline \hline

\end{tabular}\captcont{Continued.}


\end{scriptsize}
\end{center}
\end{table}

\clearpage
\begin{table}
\begin{center}
\begin{scriptsize}\renewcommand\arraystretch{1.1}

\begin{tabular}[H]{ c l c r r r r r r r r r r r }
\hline \hline
$Q^2$ &  $x_{\rm Bj}$ & $\sigma_{r, \rm NC}^{+ }$ & $\delta_{\rm stat}$ & $\delta_{\rm uncor}$ & $\delta_{\rm cor}$ & $\delta_{\rm rel}$ & $\delta_{\gamma p}$ & $\delta_{\rm had}$ &  $\delta_{1}$ & $\delta_{2}$ & $\delta_{3}$ & $\delta_{4}$ &$\delta_{\rm tot}$ \\
${\rm GeV^2}$ & & & \% & \% & \% & \% & \% & \% & \% & \% & \% & \% & \%     \\
\hline

1.5 & $0.348 \times 10^{-4}$ & 0.542 & 7.92 & 4.96 & 3.65 & 1.81 & $-$0.24 & $-$0.58 & $-$0.07 & $-$8.62 & 0.06 & 0.58 & 13.38 \\
2 & $0.464 \times 10^{-4}$ & 0.733 & 4.48 & 4.31 & 2.83 & 1.48 & $-$0.26 & $-$0.66 & $-$0.09 & $-$6.25 & 0.07 & 0.58 & 9.42 \\
2 & $0.526 \times 10^{-4}$ & 0.737 & 4.53 & 3.93 & 1.90 & 1.32 & $-$0.25 & $-$0.40 & $-$0.08 & $-$2.96 & 0.07 & 0.57 & 7.11 \\
2.5 & $0.580 \times 10^{-4}$ & 0.815 & 4.08 & 4.15 & 2.93 & 2.07 & $-$0.33 & $-$0.82 & $-$0.10 & $-$7.28 & 0.07 & 0.69 & 10.05 \\
2.5 & $0.658 \times 10^{-4}$ & 0.780 & 2.62 & 3.03 & 1.54 & 1.27 & $-$0.20 & $-$0.57 & $-$0.05 & $-$2.19 & 0.06 & 0.47 & 5.04 \\
2.5 & $0.759 \times 10^{-4}$ & 0.725 & 4.28 & 3.45 & 1.84 & 0.89 & $-$0.21 & $-$0.28 & $-$0.07 & $-$1.48 & 0.07 & 0.51 & 6.08 \\
3.5 & $0.812 \times 10^{-4}$ & 0.819 & 4.14 & 4.05 & 2.65 & 1.70 & $-$0.23 & $-$0.47 & $-$0.05 & $-$5.90 & 0.06 & 0.54 & 8.87 \\
3.5 & $0.921 \times 10^{-4}$ & 0.838 & 2.19 & 2.86 & 1.69 & 1.53 & $-$0.25 & $-$0.38 & $-$0.06 & $-$2.47 & 0.07 & 0.56 & 4.98 \\
3.5 & $0.106 \times 10^{-3}$ & 0.875 & 2.01 & 2.47 & 1.69 & 1.24 & $-$0.24 & $-$0.41 & $-$0.07 & $-$1.58 & 0.07 & 0.55 & 4.19 \\
3.5 & $0.141 \times 10^{-3}$ & 0.810 & 2.53 & 2.45 & 2.27 & 0.62 & $-$0.22 & $-$0.27 & $-$0.07 & $-$0.38 & 0.07 & 0.52 & 4.30 \\
5 & $0.116 \times 10^{-3}$ & 0.980 & 4.10 & 4.01 & 2.85 & 2.15 & $-$0.30 & $-$0.57 & $-$0.08 & $-$6.92 & 0.07 & 0.67 & 9.71 \\
5 & $0.132 \times 10^{-3}$ & 0.940 & 2.05 & 2.80 & 1.53 & 1.41 & $-$0.22 & $-$0.44 & $-$0.05 & $-$1.98 & 0.06 & 0.50 & 4.56 \\
5 & $0.143 \times 10^{-3}$ & 1.048 & 10.24 & 9.23 & 1.66 & 1.38 & $-$0.22 & $-$0.18 & $-$0.05 & 0.98 & 0.05 & 0.55 & 14.01 \\
5 & $0.152 \times 10^{-3}$ & 0.984 & 1.65 & 2.20 & 1.41 & 0.98 & $-$0.19 & $-$0.30 & $-$0.05 & $-$0.78 & 0.07 & 0.46 & 3.39 \\
5 & $0.201 \times 10^{-3}$ & 0.920 & 1.16 & 1.87 & 1.58 & 0.63 & $-$0.16 & $-$0.22 & $-$0.04 & 0.05 & 0.06 & 0.42 & 2.83 \\
6.5 & $0.151 \times 10^{-3}$ & 1.070 & 4.19 & 4.02 & 2.88 & 1.87 & $-$0.23 & $-$0.52 & $-$0.06 & $-$6.52 & 0.06 & 0.56 & 9.42 \\
6.5 & $0.171 \times 10^{-3}$ & 1.012 & 1.96 & 2.54 & 1.46 & 1.98 & $-$0.23 & $-$0.49 & $-$0.05 & $-$1.43 & 0.06 & 0.56 & 4.36 \\
6.5 & $0.183 \times 10^{-3}$ & 1.043 & 6.82 & 7.36 & 2.23 & 2.44 & $-$0.29 & $-$0.34 & $-$0.09 & $-$0.63 & 0.08 & 1.26 & 10.67 \\
6.5 & $0.197 \times 10^{-3}$ & 1.026 & 1.57 & 2.19 & 1.30 & 1.06 & $-$0.17 & $-$0.28 & $-$0.03 & $-$0.45 & 0.06 & 0.43 & 3.25 \\
6.5 & $0.228 \times 10^{-3}$ & 1.067 & 6.81 & 7.39 & 1.62 & 1.03 & $-$0.18 & $-$0.15 & $-$0.04 & $-$0.33 & 0.07 & 0.62 & 10.26 \\
6.5 & $0.262 \times 10^{-3}$ & 1.009 & 0.91 & 1.76 & 1.35 & 0.91 & $-$0.17 & $-$0.22 & $-$0.03 & $-$0.15 & 0.06 & 0.44 & 2.62 \\
6.5 & $0.330 \times 10^{-3}$ & 0.990 & 9.32 & 6.57 & 1.50 & 0.80 & $-$0.12 & $-$0.03 & $-$0.01 & 0.04 & 0.06 & 0.23 & 11.53 \\
6.5 & $0.414 \times 10^{-3}$ & 0.969 & 1.14 & 1.85 & 1.44 & 0.74 & $-$0.15 & 0.06 & $-$0.03 & 0.13 & 0.07 & 0.44 & 2.75 \\
8.5 & $0.197 \times 10^{-3}$ & 1.011 & 4.58 & 4.01 & 3.30 & 2.08 & $-$0.27 & $-$0.49 & $-$0.07 & $-$7.94 & 0.06 & 0.63 & 10.77 \\
8.5 & $0.224 \times 10^{-3}$ & 1.064 & 1.92 & 2.40 & 1.47 & 2.14 & $-$0.22 & $-$0.26 & $-$0.05 & $-$0.96 & 0.06 & 0.64 & 4.20 \\
8.5 & $0.240 \times 10^{-3}$ & 1.195 & 3.52 & 3.88 & 1.65 & 2.39 & $-$0.28 & $-$0.25 & $-$0.07 & 0.47 & 0.05 & 0.91 & 6.09 \\
8.5 & $0.258 \times 10^{-3}$ & 1.119 & 1.58 & 1.90 & 1.42 & 1.54 & $-$0.21 & $-$0.23 & $-$0.04 & $-$0.55 & 0.05 & 0.56 & 3.35 \\
8.5 & $0.299 \times 10^{-3}$ & 1.188 & 4.20 & 3.72 & 1.39 & 1.10 & $-$0.18 & $-$0.08 & $-$0.03 & 0.55 & 0.05 & 0.53 & 5.94 \\
8.5 & $0.342 \times 10^{-3}$ & 1.057 & 0.88 & 1.70 & 1.33 & 0.97 & $-$0.17 & $-$0.20 & $-$0.04 & $-$0.20 & 0.06 & 0.44 & 2.59 \\
8.5 & $0.433 \times 10^{-3}$ & 1.151 & 4.88 & 5.71 & 1.51 & 0.90 & $-$0.14 & $-$0.06 & $-$0.02 & 0.10 & 0.06 & 0.42 & 7.73 \\
8.5 & $0.541 \times 10^{-3}$ & 1.020 & 0.92 & 1.73 & 1.27 & 0.73 & $-$0.12 & 0.03 & $-$0.01 & 0.40 & 0.06 & 0.37 & 2.51 \\
8.5 & $0.838 \times 10^{-3}$ & 0.943 & 1.01 & 1.76 & 1.31 & 0.85 & $-$0.14 & 0.05 & $-$0.02 & 0.12 & 0.06 & 0.43 & 2.61 \\
8.5 & $0.140 \times 10^{-2}$ & 0.864 & 1.20 & 1.83 & 1.47 & 0.75 & $-$0.15 & 0.06 & $-$0.03 & $-$0.01 & 0.07 & 0.45 & 2.78 \\
12 & $0.278 \times 10^{-3}$ & 1.162 & 3.84 & 3.98 & 2.47 & 1.55 & $-$0.22 & $-$0.42 & $-$0.06 & $-$5.29 & 0.06 & 0.54 & 8.22 \\
12 & $0.316 \times 10^{-3}$ & 1.178 & 1.80 & 2.20 & 1.48 & 1.88 & $-$0.23 & $-$0.39 & $-$0.06 & $-$1.04 & 0.07 & 0.66 & 3.94 \\
12 & $0.343 \times 10^{-3}$ & 1.166 & 2.70 & 3.32 & 1.54 & 1.94 & $-$0.26 & $-$0.23 & $-$0.05 & 0.59 & 0.05 & 0.75 & 5.05 \\
12 & $0.364 \times 10^{-3}$ & 1.127 & 1.52 & 1.86 & 1.30 & 1.41 & $-$0.18 & $-$0.23 & $-$0.03 & $-$0.04 & 0.05 & 0.52 & 3.14 \\
12 & $0.423 \times 10^{-3}$ & 1.235 & 2.44 & 2.23 & 1.36 & 1.56 & $-$0.19 & $-$0.10 & $-$0.02 & 0.85 & 0.06 & 0.66 & 4.05 \\
12 & $0.483 \times 10^{-3}$ & 1.120 & 0.95 & 1.52 & 1.29 & 1.10 & $-$0.15 & $-$0.15 & $-$0.02 & 0.14 & 0.05 & 0.45 & 2.52 \\
12 & $0.592 \times 10^{-3}$ & 1.142 & 3.46 & 2.55 & 1.35 & 1.16 & $-$0.16 & $-$0.04 & $-$0.01 & 0.16 & 0.06 & 0.55 & 4.69 \\
12 & $0.764 \times 10^{-3}$ & 1.033 & 0.92 & 1.61 & 1.28 & 0.95 & $-$0.14 & 0.04 & $-$0.01 & 0.14 & 0.07 & 0.44 & 2.50 \\
12 & $0.118 \times 10^{-2}$ & 0.979 & 0.93 & 1.69 & 1.30 & 0.83 & $-$0.14 & 0.05 & $-$0.02 & 0.13 & 0.06 & 0.42 & 2.51 \\
12 & $0.197 \times 10^{-2}$ & 0.880 & 1.04 & 1.71 & 1.39 & 1.04 & $-$0.16 & 0.06 & $-$0.03 & $-$0.27 & 0.06 & 0.48 & 2.72 \\
15 & $0.348 \times 10^{-3}$ & 1.243 & 3.86 & 4.06 & 2.02 & 1.26 & $-$0.18 & $-$0.36 & $-$0.03 & $-$3.58 & 0.06 & 0.47 & 7.09 \\
15 & $0.394 \times 10^{-3}$ & 1.138 & 1.83 & 2.09 & 1.38 & 1.32 & $-$0.19 & $-$0.35 & $-$0.05 & $-$0.61 & 0.05 & 0.47 & 3.48 \\
15 & $0.422 \times 10^{-3}$ & 1.191 & 2.51 & 2.30 & 1.43 & 1.41 & $-$0.21 & $-$0.19 & $-$0.04 & 0.40 & 0.06 & 0.61 & 4.03 \\
15 & $0.455 \times 10^{-3}$ & 1.174 & 1.54 & 1.62 & 1.32 & 1.37 & $-$0.20 & $-$0.23 & $-$0.03 & $-$0.23 & 0.06 & 0.52 & 3.01 \\
15 & $0.529 \times 10^{-3}$ & 1.254 & 1.89 & 1.35 & 1.32 & 1.32 & $-$0.18 & $-$0.10 & $-$0.02 & 0.64 & 0.06 & 0.59 & 3.11 \\
15 & $0.604 \times 10^{-3}$ & 1.170 & 1.03 & 1.18 & 1.28 & 1.15 & $-$0.17 & $-$0.17 & $-$0.03 & 0.14 & 0.06 & 0.51 & 2.40 \\
15 & $0.763 \times 10^{-3}$ & 1.168 & 2.14 & 1.52 & 1.33 & 1.29 & $-$0.17 & $-$0.06 & $-$0.02 & 0.45 & 0.05 & 0.62 & 3.31 \\
15 & $0.955 \times 10^{-3}$ & 1.083 & 0.97 & 1.38 & 1.26 & 1.08 & $-$0.14 & 0.02 & $-$0.01 & 0.30 & 0.06 & 0.46 & 2.44 \\
15 & $0.148 \times 10^{-2}$ & 0.943 & 0.97 & 1.52 & 1.28 & 0.99 & $-$0.14 & 0.04 & $-$0.01 & 0.12 & 0.06 & 0.45 & 2.48 \\

\hline \hline

\end{tabular}

\caption{\label{tab615-225a1}
HERA combined reduced cross sections  $\sigma^{+ }_{r,{\rm NC}}$ for NC $e^{+}p$ scattering at $\sqrt{s} = 225 $~GeV.
The uncertainties are quoted in percent relative to $\sigma^{+ }_{r,{\rm NC}}$.
Other details as for Table~\ref{tab615-318a1}.}

\end{scriptsize}
\end{center}
\end{table}

\clearpage
\begin{table}
\begin{center}
\begin{scriptsize}\renewcommand\arraystretch{1.1}

\begin{tabular}[H]{ c l c r r r r r r r r r r r }
\hline \hline
$Q^2$ &  $x_{\rm Bj}$ & $\sigma_{r, \rm NC}^{+ }$ & $\delta_{\rm stat}$ & $\delta_{\rm uncor}$ & $\delta_{\rm cor}$ & $\delta_{\rm rel}$ & $\delta_{\gamma p}$ & $\delta_{\rm had}$ &  $\delta_{1}$ & $\delta_{2}$ & $\delta_{3}$ & $\delta_{4}$ &$\delta_{\rm tot}$ \\
${\rm GeV^2}$ & & & \% & \% & \% & \% & \% & \% & \% & \% & \% & \% & \%     \\
\hline
20 & $0.464 \times 10^{-3}$ & 1.030 & 5.27 & 4.21 & 2.03 & 1.64 & $-$0.24 & $-$0.62 & $-$0.06 & $-$4.18 & 0.06 & 0.54 & 8.39 \\
20 & $0.526 \times 10^{-3}$ & 1.214 & 2.26 & 2.87 & 1.39 & 1.01 & $-$0.18 & $-$0.36 & $-$0.05 & $-$1.10 & 0.06 & 0.44 & 4.22 \\
20 & $0.607 \times 10^{-3}$ & 1.180 & 1.97 & 2.39 & 1.31 & 0.77 & $-$0.13 & $-$0.31 & $-$0.03 & $-$0.17 & 0.06 & 0.36 & 3.49 \\
20 & $0.805 \times 10^{-3}$ & 1.171 & 1.06 & 1.85 & 1.35 & 1.08 & $-$0.17 & $-$0.13 & $-$0.03 & $-$0.13 & 0.05 & 0.44 & 2.79 \\
20 & $0.127 \times 10^{-2}$ & 1.089 & 1.00 & 1.84 & 1.30 & 0.93 & $-$0.13 & 0.06 & $-$0.02 & 0.27 & 0.06 & 0.40 & 2.68 \\
20 & $0.197 \times 10^{-2}$ & 0.986 & 1.02 & 1.85 & 1.28 & 0.84 & $-$0.12 & 0.05 & $-$0.01 & 0.38 & 0.06 & 0.39 & 2.67 \\
20 & $0.329 \times 10^{-2}$ & 0.877 & 1.14 & 1.87 & 1.34 & 1.03 & $-$0.14 & 0.06 & $-$0.01 & 0.15 & 0.06 & 0.43 & 2.81 \\
25 & $0.616 \times 10^{-3}$ & 1.211 & 2.42 & 2.94 & 1.41 & 1.00 & $-$0.14 & $-$0.46 & $-$0.02 & $-$0.69 & 0.05 & 0.37 & 4.28 \\
25 & $0.657 \times 10^{-3}$ & 1.210 & 2.81 & 2.80 & 1.61 & 1.03 & $-$0.19 & $-$0.27 & $-$0.06 & $-$0.42 & 0.07 & 0.48 & 4.47 \\
25 & $0.700 \times 10^{-3}$ & 1.251 & 2.42 & 2.14 & 1.34 & 0.96 & $-$0.18 & $-$0.17 & $-$0.04 & 0.35 & 0.06 & 0.35 & 3.67 \\
25 & $0.759 \times 10^{-3}$ & 1.223 & 1.50 & 1.44 & 1.26 & 0.83 & $-$0.14 & $-$0.19 & $-$0.02 & 0.33 & 0.06 & 0.39 & 2.63 \\
25 & $0.880 \times 10^{-3}$ & 1.237 & 1.68 & 1.52 & 1.27 & 1.00 & $-$0.15 & $-$0.09 & $-$0.02 & 0.48 & 0.06 & 0.44 & 2.87 \\
25 & $0.101 \times 10^{-2}$ & 1.188 & 1.00 & 1.09 & 1.26 & 1.04 & $-$0.16 & $-$0.14 & $-$0.02 & 0.20 & 0.06 & 0.47 & 2.27 \\
25 & $0.127 \times 10^{-2}$ & 1.146 & 1.52 & 1.02 & 1.30 & 1.13 & $-$0.16 & $-$0.06 & $-$0.01 & 0.48 & 0.06 & 0.53 & 2.61 \\
25 & $0.159 \times 10^{-2}$ & 1.078 & 0.97 & 1.04 & 1.26 & 1.06 & $-$0.14 & 0.00 & $-$0.02 & 0.18 & 0.06 & 0.50 & 2.25 \\
25 & $0.247 \times 10^{-2}$ & 0.989 & 0.94 & 1.24 & 1.28 & 1.08 & $-$0.14 & 0.01 & $-$0.01 & 0.16 & 0.06 & 0.50 & 2.35 \\
25 & $0.411 \times 10^{-2}$ & 0.852 & 1.03 & 1.28 & 1.29 & 1.07 & $-$0.14 & 0.01 & $-$0.01 & $-$0.28 & 0.06 & 0.55 & 2.43 \\
35 & $0.812 \times 10^{-3}$ & 1.336 & 4.48 & 3.79 & 2.90 & 0.52 & $-$0.10 & $-$0.04 & $-$0.01 & $-$2.01 & 0.28 & 0.25 & 6.88 \\
35 & $0.921 \times 10^{-3}$ & 1.210 & 2.44 & 2.07 & 1.41 & 0.98 & $-$0.18 & $-$0.41 & $-$0.06 & $-$0.92 & 0.07 & 0.38 & 3.79 \\
35 & $0.100 \times 10^{-2}$ & 1.363 & 2.63 & 1.87 & 1.27 & 0.41 & $-$0.12 & $-$0.14 & $-$0.03 & 0.16 & 0.07 & 0.21 & 3.51 \\
35 & $0.106 \times 10^{-2}$ & 1.237 & 1.70 & 1.46 & 1.24 & 0.83 & $-$0.15 & $-$0.17 & $-$0.02 & $-$0.09 & 0.06 & 0.32 & 2.73 \\
35 & $0.123 \times 10^{-2}$ & 1.222 & 1.93 & 1.40 & 1.23 & 0.65 & $-$0.12 & $-$0.08 & $-$0.02 & 0.24 & 0.06 & 0.27 & 2.79 \\
35 & $0.141 \times 10^{-2}$ & 1.219 & 1.06 & 1.05 & 1.23 & 0.88 & $-$0.14 & $-$0.15 & $-$0.02 & 0.23 & 0.06 & 0.36 & 2.18 \\
35 & $0.180 \times 10^{-2}$ & 1.143 & 1.53 & 1.18 & 1.25 & 0.85 & $-$0.14 & $-$0.06 & $-$0.01 & 0.35 & 0.06 & 0.39 & 2.51 \\
35 & $0.223 \times 10^{-2}$ & 1.066 & 0.99 & 0.83 & 1.24 & 0.84 & $-$0.12 & $-$0.02 & 0.00 & 0.38 & 0.07 & 0.41 & 2.06 \\
35 & $0.310 \times 10^{-2}$ & 0.984 & 1.27 & 1.59 & 1.28 & 0.97 & $-$0.14 & $-$0.04 & $-$0.01 & $-$0.17 & 0.06 & 0.50 & 2.65 \\
35 & $0.345 \times 10^{-2}$ & 0.977 & 1.27 & 1.91 & 1.36 & 0.69 & $-$0.08 & 0.04 & 0.01 & 0.77 & 0.06 & 0.32 & 2.88 \\
35 & $0.575 \times 10^{-2}$ & 0.833 & 0.94 & 1.19 & 1.31 & 1.07 & $-$0.15 & 0.00 & $-$0.02 & $-$0.46 & 0.06 & 0.55 & 2.39 \\
45 & $0.104 \times 10^{-2}$ & 1.169 & 4.71 & 3.41 & 2.64 & 0.60 & $-$0.11 & $-$0.03 & $-$0.01 & $-$1.65 & 0.19 & 0.26 & 6.63 \\
45 & $0.118 \times 10^{-2}$ & 1.205 & 2.98 & 2.17 & 1.37 & 0.43 & $-$0.11 & $-$0.11 & $-$0.02 & $-$0.37 & 0.07 & 0.16 & 3.98 \\
45 & $0.127 \times 10^{-2}$ & 1.281 & 3.08 & 2.27 & 1.30 & 0.08 & $-$0.09 & $-$0.15 & $-$0.02 & $-$0.09 & 0.07 & $-$0.03 & 4.05 \\
45 & $0.137 \times 10^{-2}$ & 1.243 & 2.09 & 1.47 & 1.21 & 0.37 & $-$0.10 & $-$0.16 & $-$0.02 & 0.20 & 0.06 & 0.16 & 2.87 \\
45 & $0.159 \times 10^{-2}$ & 1.257 & 2.12 & 1.34 & 1.22 & 0.32 & $-$0.10 & $-$0.09 & $-$0.02 & 0.29 & 0.07 & 0.14 & 2.83 \\
45 & $0.181 \times 10^{-2}$ & 1.158 & 1.17 & 1.14 & 1.21 & 0.67 & $-$0.13 & $-$0.12 & $-$0.02 & 0.08 & 0.07 & 0.29 & 2.17 \\
45 & $0.229 \times 10^{-2}$ & 1.094 & 1.69 & 1.00 & 1.21 & 0.56 & $-$0.11 & $-$0.05 & $-$0.01 & 0.22 & 0.06 & 0.23 & 2.40 \\
45 & $0.286 \times 10^{-2}$ & 1.085 & 1.01 & 0.89 & 1.21 & 0.66 & $-$0.11 & $-$0.01 & $-$0.01 & 0.30 & 0.06 & 0.31 & 1.98 \\
45 & $0.444 \times 10^{-2}$ & 0.929 & 0.92 & 1.23 & 1.22 & 0.77 & $-$0.12 & 0.00 & $-$0.01 & 0.05 & 0.06 & 0.38 & 2.15 \\
45 & $0.740 \times 10^{-2}$ & 0.802 & 0.92 & 1.07 & 1.26 & 0.86 & $-$0.14 & $-$0.01 & $-$0.02 & $-$0.57 & 0.07 & 0.49 & 2.21 \\
60 & $0.139 \times 10^{-2}$ & 1.185 & 4.98 & 3.05 & 2.26 & 0.55 & $-$0.11 & $-$0.03 & $-$0.01 & $-$1.14 & 0.14 & 0.27 & 6.39 \\
60 & $0.158 \times 10^{-2}$ & 1.142 & 2.97 & 1.78 & 1.47 & 0.53 & $-$0.11 & $-$0.08 & $-$0.02 & $-$0.44 & 0.10 & 0.13 & 3.83 \\
60 & $0.171 \times 10^{-2}$ & 1.255 & 3.64 & 2.63 & 1.28 & 0.14 & $-$0.10 & $-$0.14 & $-$0.02 & 0.19 & 0.06 & $-$0.07 & 4.68 \\
60 & $0.182 \times 10^{-2}$ & 1.190 & 3.42 & 3.10 & 1.26 & 0.02 & $-$0.09 & $-$0.13 & $-$0.02 & 0.02 & 0.07 & $-$0.10 & 4.79 \\
60 & $0.211 \times 10^{-2}$ & 1.209 & 2.57 & 1.64 & 1.20 & 0.12 & $-$0.07 & $-$0.07 & $-$0.01 & 0.22 & 0.07 & $-$0.05 & 3.29 \\
60 & $0.242 \times 10^{-2}$ & 1.184 & 1.38 & 1.28 & 1.21 & 0.57 & $-$0.13 & $-$0.16 & $-$0.03 & 0.01 & 0.06 & 0.23 & 2.33 \\
60 & $0.305 \times 10^{-2}$ & 1.097 & 1.98 & 1.28 & 1.20 & 0.29 & $-$0.08 & $-$0.05 & $-$0.01 & 0.21 & 0.06 & 0.07 & 2.68 \\
60 & $0.382 \times 10^{-2}$ & 1.021 & 1.16 & 1.00 & 1.20 & 0.48 & $-$0.09 & $-$0.01 & 0.00 & 0.35 & 0.07 & 0.22 & 2.04 \\
60 & $0.592 \times 10^{-2}$ & 0.894 & 1.05 & 1.23 & 1.20 & 0.53 & $-$0.09 & $-$0.01 & 0.00 & 0.17 & 0.06 & 0.27 & 2.11 \\
60 & $0.986 \times 10^{-2}$ & 0.758 & 1.04 & 1.02 & 1.23 & 0.61 & $-$0.11 & $-$0.01 & $-$0.01 & $-$0.68 & 0.07 & 0.37 & 2.15 \\
90 & $0.209 \times 10^{-2}$ & 1.262 & 5.29 & 2.88 & 2.09 & 0.58 & $-$0.11 & $-$0.02 & $-$0.01 & $-$1.01 & 0.16 & 0.27 & 6.49 \\
90 & $0.237 \times 10^{-2}$ & 1.206 & 3.13 & 1.75 & 1.41 & 0.48 & $-$0.11 & $-$0.06 & $-$0.02 & $-$0.08 & 0.10 & 0.17 & 3.89 \\
90 & $0.250 \times 10^{-2}$ & 1.078 & 4.52 & 3.22 & 1.28 & 0.09 & $-$0.08 & $-$0.10 & $-$0.01 & 0.58 & 0.06 & $-$0.16 & 5.73 \\
90 & $0.273 \times 10^{-2}$ & 1.149 & 2.72 & 1.53 & 1.28 & 0.43 & $-$0.10 & $-$0.03 & $-$0.01 & 0.38 & 0.09 & 0.07 & 3.42 \\
90 & $0.317 \times 10^{-2}$ & 1.094 & 3.02 & 2.07 & 1.22 & 0.02 & $-$0.06 & $-$0.05 & 0.00 & 0.26 & 0.06 & $-$0.17 & 3.87 \\
90 & $0.362 \times 10^{-2}$ & 1.106 & 2.10 & 1.26 & 1.19 & 0.30 & $-$0.08 & $-$0.12 & $-$0.01 & 0.01 & 0.07 & 0.05 & 2.74 \\
90 & $0.460 \times 10^{-2}$ & 1.021 & 2.31 & 1.26 & 1.20 & 0.08 & $-$0.06 & $-$0.05 & 0.00 & $-$0.04 & 0.07 & $-$0.07 & 2.90 \\
90 & $0.573 \times 10^{-2}$ & 0.957 & 1.37 & 1.05 & 1.18 & 0.34 & $-$0.08 & $-$0.02 & 0.00 & 0.17 & 0.07 & 0.15 & 2.13 \\
90 & $0.800 \times 10^{-2}$ & 0.906 & 1.64 & 1.48 & 1.19 & 0.21 & $-$0.07 & $-$0.05 & 0.00 & $-$0.41 & 0.07 & 0.07 & 2.56 \\
90 & $0.888 \times 10^{-2}$ & 0.836 & 1.71 & 2.03 & 1.29 & 0.87 & $-$0.13 & 0.06 & $-$0.01 & 0.27 & 0.06 & 0.41 & 3.12 \\
90 & $0.136 \times 10^{-1}$ & 0.712 & 1.51 & 1.26 & 1.22 & 0.23 & $-$0.08 & $-$0.05 & $-$0.01 & $-$0.89 & 0.07 & 0.18 & 2.50 \\
90 & $0.148 \times 10^{-1}$ & 0.717 & 1.90 & 2.10 & 1.29 & 0.88 & $-$0.13 & 0.06 & $-$0.01 & 0.30 & 0.06 & 0.41 & 3.28 \\

\hline \hline

\end{tabular}\captcont{Continued.}


\end{scriptsize}
\end{center}
\end{table}

\clearpage
\begin{table}
\begin{center}
\begin{scriptsize}\renewcommand\arraystretch{1.1}

\begin{tabular}[H]{ c l c r r r r r r r r r r r }
\hline \hline
$Q^2$ &  $x_{\rm Bj}$ & $\sigma_{r, \rm NC}^{+ }$ & $\delta_{\rm stat}$ & $\delta_{\rm uncor}$ & $\delta_{\rm cor}$ & $\delta_{\rm rel}$ & $\delta_{\gamma p}$ & $\delta_{\rm had}$ &  $\delta_{1}$ & $\delta_{2}$ & $\delta_{3}$ & $\delta_{4}$ &$\delta_{\rm tot}$ \\
${\rm GeV^2}$ & & & \% & \% & \% & \% & \% & \% & \% & \% & \% & \% & \%     \\
\hline

120 & $0.280 \times 10^{-2}$ & 1.246 & 6.08 & 2.66 & 2.01 & 0.59 & $-$0.11 & $-$0.02 & $-$0.01 & $-$1.08 & 0.12 & 0.27 & 7.05 \\
120 & $0.320 \times 10^{-2}$ & 1.054 & 4.29 & 1.94 & 1.44 & 0.47 & $-$0.10 & $-$0.03 & $-$0.01 & 0.08 & 0.11 & 0.20 & 4.95 \\
120 & $0.340 \times 10^{-2}$ & 1.185 & 5.29 & 3.23 & 1.26 & 0.01 & $-$0.08 & $-$0.11 & $-$0.02 & 0.40 & 0.07 & $-$0.06 & 6.34 \\
120 & $0.360 \times 10^{-2}$ & 1.096 & 3.22 & 1.82 & 1.27 & 0.41 & $-$0.09 & $-$0.02 & $-$0.01 & 0.36 & 0.10 & 0.15 & 3.96 \\
120 & $0.420 \times 10^{-2}$ & 1.044 & 3.68 & 1.84 & 1.22 & $-$0.06 & $-$0.06 & $-$0.06 & $-$0.01 & 0.27 & 0.07 & $-$0.15 & 4.30 \\
120 & $0.480 \times 10^{-2}$ & 1.029 & 2.67 & 1.80 & 1.47 & 0.72 & $-$0.12 & 0.01 & $-$0.01 & 0.70 & 0.12 & 0.28 & 3.69 \\
120 & $0.500 \times 10^{-2}$ & 1.098 & 3.11 & 1.67 & 1.21 & $-$0.02 & $-$0.05 & $-$0.05 & 0.00 & 0.19 & 0.07 & $-$0.12 & 3.74 \\
120 & $0.590 \times 10^{-2}$ & 0.993 & 2.70 & 1.47 & 1.20 & $-$0.04 & $-$0.05 & $-$0.05 & 0.00 & $-$0.13 & 0.07 & $-$0.13 & 3.31 \\
120 & $0.760 \times 10^{-2}$ & 0.894 & 2.33 & 1.15 & 1.20 & $-$0.05 & $-$0.04 & $-$0.04 & 0.00 & $-$0.30 & 0.07 & $-$0.17 & 2.89 \\
120 & $0.110 \times 10^{-1}$ & 0.818 & 1.97 & 1.44 & 1.21 & $-$0.13 & $-$0.04 & $-$0.04 & 0.00 & $-$0.66 & 0.07 & $-$0.17 & 2.81 \\
120 & $0.180 \times 10^{-1}$ & 0.671 & 1.76 & 1.54 & 1.23 & 0.00 & $-$0.06 & $-$0.05 & $-$0.01 & $-$1.12 & 0.07 & 0.06 & 2.87 \\
150 & $0.350 \times 10^{-2}$ & 1.227 & 7.11 & 2.64 & 2.19 & 0.52 & $-$0.11 & $-$0.02 & $-$0.01 & $-$1.38 & 0.14 & 0.27 & 8.04 \\
150 & $0.390 \times 10^{-2}$ & 1.017 & 5.78 & 1.88 & 1.51 & 0.62 & $-$0.11 & 0.00 & $-$0.01 & 0.11 & 0.12 & 0.29 & 6.30 \\
150 & $0.450 \times 10^{-2}$ & 1.001 & 4.99 & 2.00 & 1.42 & 0.62 & $-$0.10 & 0.01 & $-$0.01 & 0.62 & 0.11 & 0.30 & 5.64 \\
150 & $0.600 \times 10^{-2}$ & 1.048 & 2.57 & 1.77 & 1.45 & 0.67 & $-$0.11 & 0.01 & $-$0.01 & 0.73 & 0.10 & 0.29 & 3.59 \\
150 & $0.800 \times 10^{-2}$ & 0.964 & 2.58 & 1.36 & 1.50 & 0.76 & $-$0.12 & 0.01 & $-$0.01 & 0.77 & 0.12 & 0.28 & 3.47 \\
150 & $0.130 \times 10^{-1}$ & 0.823 & 3.33 & 1.48 & 1.49 & 0.70 & $-$0.12 & 0.03 & $-$0.01 & 0.78 & 0.14 & 0.26 & 4.09 \\
150 & $0.200 \times 10^{-1}$ & 0.690 & 4.51 & 2.16 & 1.61 & 0.82 & $-$0.13 & 0.07 & $-$0.01 & 0.79 & $-$0.19 & 0.24 & 5.38 \\
200 & $0.460 \times 10^{-2}$ & 1.117 & 9.08 & 2.52 & 2.37 & 0.61 & $-$0.11 & $-$0.03 & $-$0.01 & $-$1.58 & 0.10 & 0.26 & 9.87 \\
200 & $0.520 \times 10^{-2}$ & 1.011 & 7.74 & 1.92 & 1.68 & 0.63 & $-$0.11 & $-$0.02 & $-$0.01 & $-$0.28 & 0.09 & 0.31 & 8.19 \\
200 & $0.610 \times 10^{-2}$ & 0.989 & 6.49 & 2.03 & 1.43 & 0.66 & $-$0.11 & 0.00 & $-$0.01 & 0.67 & 0.12 & 0.30 & 7.02 \\
200 & $0.800 \times 10^{-2}$ & 0.947 & 3.40 & 1.73 & 1.44 & 0.68 & $-$0.11 & 0.01 & $-$0.01 & 0.73 & 0.10 & 0.29 & 4.21 \\
200 & $0.130 \times 10^{-1}$ & 0.831 & 3.13 & 1.05 & 1.48 & 0.72 & $-$0.12 & 0.02 & $-$0.01 & 0.78 & 0.11 & 0.28 & 3.78 \\
200 & $0.200 \times 10^{-1}$ & 0.638 & 3.52 & 1.76 & 1.50 & 0.67 & $-$0.12 & 0.04 & $-$0.01 & 0.78 & 0.16 & 0.26 & 4.35 \\
200 & $0.320 \times 10^{-1}$ & 0.523 & 4.04 & 0.85 & 1.44 & 0.68 & $-$0.11 & 0.01 & $-$0.01 & 0.79 & $-$0.01 & 0.31 & 4.50 \\
200 & $0.500 \times 10^{-1}$ & 0.523 & 4.13 & 1.23 & 1.42 & 0.55 & $-$0.11 & 0.04 & $-$0.01 & 0.78 & $-$0.05 & 0.29 & 4.65 \\
200 & $0.800 \times 10^{-1}$ & 0.452 & 4.76 & 1.34 & 1.50 & 0.71 & $-$0.12 & 0.02 & $-$0.01 & 0.78 & $-$0.09 & 0.28 & 5.28 \\
200 & 0.130 & 0.352 & 5.13 & 1.89 & 1.54 & 0.63 & $-$0.12 & 0.01 & $-$0.01 & 0.77 & $-$0.01 & 0.27 & 5.77 \\
200 & 0.180 & 0.319 & 5.82 & 2.28 & 1.65 & 0.91 & $-$0.13 & 0.11 & 0.00 & 0.79 & $-$0.10 & 0.23 & 6.59 \\
200 & 0.400 & 0.178 & 7.99 & 4.50 & 2.84 & 1.82 & $-$0.12 & 0.26 & 0.01 & 0.84 & 0.12 & 0.15 & 9.81 \\
250 & $0.580 \times 10^{-2}$ & 1.050 & 10.44 & 2.45 & 2.42 & 0.50 & $-$0.10 & $-$0.02 & $-$0.02 & $-$1.69 & 0.08 & 0.27 & 11.13 \\
250 & $0.660 \times 10^{-2}$ & 1.029 & 8.85 & 1.93 & 1.78 & 0.58 & $-$0.11 & $-$0.01 & $-$0.01 & $-$0.52 & 0.13 & 0.27 & 9.27 \\
250 & $0.760 \times 10^{-2}$ & 0.938 & 7.74 & 2.03 & 1.42 & 0.59 & $-$0.10 & 0.01 & $-$0.01 & 0.65 & 0.10 & 0.30 & 8.18 \\
250 & $0.100 \times 10^{-1}$ & 0.874 & 3.95 & 1.73 & 1.43 & 0.65 & $-$0.11 & 0.01 & $-$0.01 & 0.74 & 0.10 & 0.29 & 4.67 \\
250 & $0.130 \times 10^{-1}$ & 0.818 & 3.72 & 1.29 & 1.43 & 0.64 & $-$0.11 & 0.02 & $-$0.01 & 0.76 & 0.11 & 0.29 & 4.32 \\
250 & $0.200 \times 10^{-1}$ & 0.670 & 3.84 & 1.04 & 1.47 & 0.70 & $-$0.12 & 0.02 & $-$0.01 & 0.78 & 0.12 & 0.28 & 4.38 \\
250 & $0.320 \times 10^{-1}$ & 0.579 & 4.07 & 1.30 & 1.48 & 0.76 & $-$0.12 & $-$0.04 & $-$0.01 & 0.79 & 0.06 & 0.32 & 4.67 \\
250 & $0.500 \times 10^{-1}$ & 0.499 & 4.15 & 1.26 & 1.47 & 0.72 & $-$0.12 & $-$0.03 & $-$0.01 & 0.79 & 0.01 & 0.33 & 4.72 \\
250 & $0.800 \times 10^{-1}$ & 0.441 & 4.41 & 1.33 & 1.48 & 0.75 & $-$0.12 & $-$0.03 & $-$0.01 & 0.79 & $-$0.14 & 0.33 & 4.97 \\
250 & 0.130 & 0.371 & 4.40 & 1.39 & 1.49 & 0.62 & $-$0.11 & $-$0.04 & $-$0.01 & 0.78 & $-$0.12 & 0.34 & 4.97 \\
250 & 0.180 & 0.361 & 4.47 & 2.06 & 1.53 & 0.93 & $-$0.12 & 0.01 & $-$0.01 & 0.80 & $-$0.20 & 0.32 & 5.31 \\
250 & 0.400 & 0.180 & 6.37 & 4.91 & 2.95 & 2.18 & $-$0.12 & 0.23 & 0.01 & 0.85 & $-$0.07 & 0.22 & 8.88 \\
300 & $0.690 \times 10^{-2}$ & 0.872 & 13.29 & 2.39 & 2.69 & 0.48 & $-$0.10 & $-$0.02 & $-$0.01 & $-$2.12 & 0.08 & 0.26 & 13.94 \\
300 & $0.790 \times 10^{-2}$ & 0.823 & 10.93 & 1.96 & 1.65 & 0.53 & $-$0.10 & 0.01 & $-$0.01 & $-$0.32 & 0.11 & 0.29 & 11.24 \\
300 & $0.910 \times 10^{-2}$ & 0.835 & 9.58 & 1.89 & 1.42 & 0.65 & $-$0.11 & 0.00 & $-$0.01 & 0.67 & 0.09 & 0.31 & 9.92 \\
300 & $0.121 \times 10^{-1}$ & 0.895 & 4.50 & 1.73 & 1.41 & 0.60 & $-$0.10 & 0.02 & $-$0.01 & 0.76 & 0.10 & 0.30 & 5.12 \\
300 & $0.200 \times 10^{-1}$ & 0.719 & 4.23 & 1.04 & 1.44 & 0.63 & $-$0.11 & 0.03 & $-$0.01 & 0.78 & 0.12 & 0.28 & 4.71 \\
300 & $0.320 \times 10^{-1}$ & 0.620 & 4.53 & 0.94 & 1.42 & 0.62 & $-$0.10 & $-$0.02 & $-$0.01 & 0.79 & 0.10 & 0.33 & 4.96 \\
300 & $0.500 \times 10^{-1}$ & 0.519 & 4.74 & 1.08 & 1.46 & 0.72 & $-$0.11 & $-$0.03 & $-$0.01 & 0.79 & 0.05 & 0.33 & 5.20 \\
300 & $0.800 \times 10^{-1}$ & 0.458 & 4.93 & 1.38 & 1.50 & 0.81 & $-$0.12 & $-$0.03 & $-$0.01 & 0.79 & $-$0.13 & 0.33 & 5.47 \\
300 & 0.130 & 0.367 & 5.09 & 1.54 & 1.54 & 0.74 & $-$0.13 & $-$0.06 & $-$0.01 & 0.79 & $-$0.11 & 0.34 & 5.65 \\
300 & 0.180 & 0.335 & 5.22 & 2.22 & 1.57 & 0.97 & $-$0.13 & $-$0.02 & $-$0.01 & 0.80 & $-$0.28 & 0.32 & 6.04 \\
300 & 0.400 & 0.168 & 7.39 & 5.16 & 2.96 & 2.37 & $-$0.14 & 0.16 & 0.01 & 0.86 & $-$0.11 & 0.23 & 9.82 \\
\hline \hline

\end{tabular}\captcont{Continued.}


\end{scriptsize}
\end{center}
\end{table}

\clearpage
\begin{table}
\begin{center}
\begin{scriptsize}\renewcommand\arraystretch{1.1}

\begin{tabular}[H]{ c l c r r r r r r r r r r r }
\hline \hline
$Q^2$ &  $x_{\rm Bj}$ & $\sigma_{r, \rm NC}^{+ }$ & $\delta_{\rm stat}$ & $\delta_{\rm uncor}$ & $\delta_{\rm cor}$ & $\delta_{\rm rel}$ & $\delta_{\gamma p}$ & $\delta_{\rm had}$ &  $\delta_{1}$ & $\delta_{2}$ & $\delta_{3}$ & $\delta_{4}$ &$\delta_{\rm tot}$ \\
${\rm GeV^2}$ & & & \% & \% & \% & \% & \% & \% & \% & \% & \% & \% & \%     \\
\hline
400 & $0.930 \times 10^{-2}$ & 1.026 & 13.30 & 2.43 & 2.68 & 0.56 & $-$0.11 & $-$0.02 & $-$0.01 & $-$1.99 & 0.06 & 0.25 & 13.94 \\
400 & $0.105 \times 10^{-1}$ & 1.071 & 10.14 & 2.19 & 1.49 & 0.61 & $-$0.10 & $-$0.01 & $-$0.01 & 0.18 & 0.11 & 0.31 & 10.51 \\
400 & $0.121 \times 10^{-1}$ & 0.918 & 9.82 & 1.88 & 1.40 & 0.55 & $-$0.10 & 0.02 & $-$0.01 & 0.69 & 0.10 & 0.31 & 10.14 \\
400 & $0.161 \times 10^{-1}$ & 0.806 & 5.45 & 1.72 & 1.42 & 0.64 & $-$0.11 & 0.01 & $-$0.01 & 0.79 & 0.09 & 0.30 & 5.98 \\
400 & $0.320 \times 10^{-1}$ & 0.621 & 5.08 & 1.02 & 1.45 & 0.64 & $-$0.11 & 0.03 & $-$0.01 & 0.78 & 0.12 & 0.28 & 5.48 \\
400 & $0.500 \times 10^{-1}$ & 0.568 & 5.21 & 1.03 & 1.46 & 0.74 & $-$0.11 & $-$0.03 & $-$0.01 & 0.79 & $-$0.07 & 0.33 & 5.63 \\
400 & $0.800 \times 10^{-1}$ & 0.448 & 5.66 & 0.99 & 1.43 & 0.66 & $-$0.11 & $-$0.02 & $-$0.01 & 0.79 & 0.09 & 0.33 & 6.02 \\
400 & 0.130 & 0.427 & 5.36 & 1.24 & 1.45 & 0.61 & $-$0.11 & $-$0.03 & $-$0.01 & 0.79 & $-$0.13 & 0.34 & 5.79 \\
400 & 0.180 & 0.339 & 5.84 & 1.69 & 1.49 & 0.83 & $-$0.12 & $-$0.01 & $-$0.01 & 0.80 & $-$0.21 & 0.33 & 6.37 \\
400 & 0.400 & 0.160 & 8.44 & 4.64 & 2.75 & 2.10 & $-$0.13 & 0.19 & 0.01 & 0.85 & $-$0.26 & 0.25 & 10.28 \\
500 & $0.116 \times 10^{-1}$ & 1.006 & 14.56 & 2.44 & 2.50 & 0.55 & $-$0.09 & $-$0.02 & $-$0.01 & $-$1.44 & 0.05 & 0.28 & 15.06 \\
500 & $0.131 \times 10^{-1}$ & 0.750 & 13.71 & 2.06 & 1.40 & 0.55 & $-$0.10 & 0.02 & $-$0.01 & 0.66 & 0.11 & 0.31 & 13.97 \\
500 & $0.152 \times 10^{-1}$ & 0.686 & 12.27 & 1.93 & 1.40 & 0.59 & $-$0.10 & 0.01 & $-$0.01 & 0.78 & 0.13 & 0.31 & 12.54 \\
500 & $0.201 \times 10^{-1}$ & 0.733 & 6.52 & 1.71 & 1.41 & 0.61 & $-$0.10 & 0.02 & $-$0.01 & 0.79 & 0.08 & 0.31 & 6.96 \\
500 & $0.320 \times 10^{-1}$ & 0.649 & 6.05 & 0.97 & 1.43 & 0.64 & $-$0.11 & 0.02 & $-$0.01 & 0.78 & 0.10 & 0.29 & 6.38 \\
500 & $0.500 \times 10^{-1}$ & 0.599 & 5.92 & 1.29 & 1.46 & 0.60 & $-$0.11 & 0.05 & $-$0.01 & 0.78 & 0.15 & 0.27 & 6.32 \\
500 & $0.800 \times 10^{-1}$ & 0.471 & 6.34 & 1.03 & 1.44 & 0.70 & $-$0.11 & $-$0.02 & $-$0.01 & 0.79 & $-$0.04 & 0.34 & 6.68 \\
500 & 0.130 & 0.410 & 7.51 & 1.35 & 1.49 & 0.72 & $-$0.12 & $-$0.05 & $-$0.01 & 0.79 & $-$0.09 & 0.34 & 7.86 \\
500 & 0.180 & 0.304 & 8.44 & 1.49 & 1.48 & 0.68 & $-$0.11 & $-$0.05 & $-$0.01 & 0.79 & $-$0.13 & 0.35 & 8.77 \\
500 & 0.250 & 0.278 & 8.21 & 1.88 & 1.52 & 0.90 & $-$0.12 & $-$0.02 & $-$0.01 & 0.80 & $-$0.21 & 0.33 & 8.66 \\
500 & 0.400 & 0.138 & 11.44 & 4.14 & 2.18 & 1.66 & $-$0.12 & 0.12 & 0.00 & 0.84 & $-$0.52 & 0.30 & 12.51 \\
500 & 0.650 & 0.019 & 21.66 & 6.96 & 4.15 & 3.47 & $-$0.14 & 0.32 & 0.02 & 0.90 & $-$0.28 & 0.18 & 23.40 \\
650 & $0.151 \times 10^{-1}$ & 0.804 & 19.43 & 2.67 & 1.66 & 0.32 & $-$0.07 & 0.02 & $-$0.01 & $-$0.38 & 0.13 & 0.32 & 19.69 \\
650 & $0.171 \times 10^{-1}$ & 0.914 & 13.91 & 2.02 & 1.40 & 0.58 & $-$0.10 & 0.01 & $-$0.01 & 0.68 & 0.09 & 0.32 & 14.15 \\
650 & $0.197 \times 10^{-1}$ & 0.905 & 12.02 & 1.92 & 1.41 & 0.60 & $-$0.10 & 0.01 & $-$0.01 & 0.70 & 0.09 & 0.31 & 12.29 \\
650 & $0.261 \times 10^{-1}$ & 0.602 & 7.81 & 1.78 & 1.41 & 0.60 & $-$0.10 & 0.02 & $-$0.01 & 0.78 & 0.10 & 0.30 & 8.20 \\
650 & $0.500 \times 10^{-1}$ & 0.488 & 7.87 & 1.12 & 1.45 & 0.66 & $-$0.11 & 0.03 & $-$0.01 & 0.78 & 0.12 & 0.28 & 8.15 \\
650 & $0.800 \times 10^{-1}$ & 0.451 & 7.78 & 1.04 & 1.44 & 0.69 & $-$0.11 & $-$0.03 & $-$0.01 & 0.79 & 0.06 & 0.34 & 8.06 \\
650 & 0.130 & 0.374 & 9.33 & 1.41 & 1.45 & 0.72 & $-$0.11 & $-$0.02 & $-$0.01 & 0.79 & $-$0.16 & 0.34 & 9.62 \\
650 & 0.180 & 0.339 & 9.52 & 1.48 & 1.47 & 0.69 & $-$0.12 & $-$0.04 & $-$0.01 & 0.79 & $-$0.07 & 0.34 & 9.81 \\
650 & 0.250 & 0.252 & 10.12 & 1.70 & 1.44 & 0.69 & $-$0.10 & $-$0.04 & $-$0.01 & 0.79 & 0.06 & 0.35 & 10.42 \\
650 & 0.400 & 0.201 & 10.92 & 3.94 & 2.05 & 1.58 & $-$0.13 & 0.07 & 0.00 & 0.83 & $-$0.61 & 0.32 & 11.94 \\
650 & 0.650 & 0.027 & 21.00 & 7.78 & 4.11 & 3.66 & $-$0.13 & 0.30 & 0.02 & 0.90 & $-$0.62 & 0.24 & 23.09 \\
800 & $0.185 \times 10^{-1}$ & 0.286 & 36.97 & 3.94 & 1.43 & 0.62 & $-$0.10 & 0.00 & $-$0.01 & 0.51 & 0.08 & 0.32 & 37.22 \\
800 & $0.210 \times 10^{-1}$ & 0.658 & 19.13 & 2.27 & 1.38 & 0.51 & $-$0.09 & 0.02 & $-$0.01 & 0.78 & 0.12 & 0.32 & 19.34 \\
800 & $0.242 \times 10^{-1}$ & 0.656 & 15.94 & 2.08 & 1.38 & 0.49 & $-$0.09 & 0.03 & $-$0.01 & 0.78 & 0.10 & 0.31 & 16.17 \\
800 & $0.322 \times 10^{-1}$ & 0.610 & 8.63 & 1.82 & 1.39 & 0.54 & $-$0.10 & 0.03 & $-$0.01 & 0.78 & 0.10 & 0.31 & 8.98 \\
800 & $0.500 \times 10^{-1}$ & 0.479 & 9.02 & 1.23 & 1.40 & 0.55 & $-$0.10 & 0.04 & $-$0.01 & 0.78 & 0.11 & 0.30 & 9.26 \\
800 & $0.800 \times 10^{-1}$ & 0.447 & 9.22 & 1.31 & 1.44 & 0.65 & $-$0.11 & 0.03 & $-$0.01 & 0.78 & 0.11 & 0.28 & 9.48 \\
800 & 0.130 & 0.388 & 10.60 & 1.43 & 1.46 & 0.74 & $-$0.11 & $-$0.02 & $-$0.01 & 0.79 & $-$0.18 & 0.35 & 10.85 \\
800 & 0.180 & 0.363 & 10.93 & 1.46 & 1.42 & 0.62 & $-$0.10 & $-$0.02 & $-$0.01 & 0.79 & 0.04 & 0.34 & 11.17 \\
800 & 0.250 & 0.276 & 11.53 & 1.74 & 1.46 & 0.72 & $-$0.11 & $-$0.03 & $-$0.01 & 0.79 & $-$0.22 & 0.35 & 11.81 \\
800 & 0.400 & 0.131 & 15.53 & 3.53 & 1.81 & 1.23 & $-$0.11 & 0.09 & $-$0.01 & 0.82 & $-$0.53 & 0.33 & 16.10 \\
800 & 0.650 & 0.022 & 27.39 & 7.13 & 3.40 & 2.79 & $-$0.11 & 0.26 & 0.02 & 0.88 & $-$0.50 & 0.27 & 28.66 \\

\hline \hline

\end{tabular}\captcont{Continued.}


\end{scriptsize}
\end{center}
\end{table}

\clearpage
\begin{table}
\begin{center}
\begin{scriptsize}\renewcommand\arraystretch{1.1}

\begin{tabular}[H]{ c l c r r r r r r r r r r r }
\hline \hline
$Q^2$ &  $x_{\rm Bj}$ & $\sigma_{r, \rm NC}^{- }$ & $\delta_{\rm stat}$ & $\delta_{\rm uncor}$ & $\delta_{\rm cor}$ & $\delta_{\rm rel}$ & $\delta_{\gamma p}$ & $\delta_{\rm had}$ &  $\delta_{1}$ & $\delta_{2}$ & $\delta_{3}$ & $\delta_{4}$ &$\delta_{\rm tot}$ \\
${\rm GeV^2}$ & & & \% & \% & \% & \% & \% & \% & \% & \% & \% & \% & \%     \\
\hline
60 & $0.800 \times 10^{-3}$ & 1.483 & 1.01 & 1.84 & 1.19 & 0.16 & $-$0.04 & $-$0.03 & 0.00 & $-$0.04 & 0.01 & $-$0.07 & 2.42 \\
90 & $0.130 \times 10^{-2}$ & 1.466 & 0.82 & 1.63 & 1.12 & 0.03 & $-$0.02 & $-$0.02 & 0.00 & $-$0.04 & 0.03 & $-$0.08 & 2.14 \\
90 & $0.150 \times 10^{-2}$ & 1.422 & 1.20 & 1.10 & 1.43 & 0.98 & $-$1.09 & $-$0.13 & 0.00 & $-$0.04 & $-$0.11 & $-$0.21 & 2.63 \\
90 & $0.200 \times 10^{-2}$ & 1.270 & 3.41 & 3.30 & 1.46 & 0.95 & $-$0.82 & 0.24 & 0.01 & $-$0.03 & 0.06 & $-$0.17 & 5.13 \\
120 & $0.160 \times 10^{-2}$ & 1.439 & 0.92 & 1.71 & 1.10 & 0.05 & $-$0.02 & $-$0.02 & 0.00 & $-$0.04 & $-$0.01 & $-$0.07 & 2.23 \\
120 & $0.200 \times 10^{-2}$ & 1.356 & 0.70 & 0.91 & 1.20 & 0.61 & $-$0.71 & $-$0.10 & 0.00 & $-$0.04 & $-$0.08 & $-$0.18 & 1.92 \\
120 & $0.320 \times 10^{-2}$ & 1.218 & 0.96 & 1.08 & 1.03 & 0.25 & $-$0.17 & $-$0.02 & 0.00 & $-$0.04 & $-$0.02 & $-$0.11 & 1.81 \\
150 & $0.200 \times 10^{-2}$ & 1.355 & 1.03 & 1.77 & 1.07 & 0.04 & $-$0.02 & 0.00 & 0.00 & $-$0.04 & 0.01 & $-$0.07 & 2.31 \\
150 & $0.320 \times 10^{-2}$ & 1.229 & 0.59 & 0.85 & 1.07 & 0.37 & $-$0.38 & $-$0.02 & 0.00 & $-$0.04 & $-$0.02 & $-$0.14 & 1.59 \\
150 & $0.500 \times 10^{-2}$ & 1.109 & 0.71 & 1.13 & 1.03 & 0.15 & $-$0.07 & 0.01 & 0.00 & $-$0.05 & $-$0.01 & $-$0.12 & 1.70 \\
150 & $0.800 \times 10^{-2}$ & 0.952 & 0.97 & 1.72 & 1.09 & 0.07 & $-$0.10 & $-$0.01 & 0.00 & $-$0.04 & $-$0.27 & $-$0.13 & 2.28 \\
150 & $0.130 \times 10^{-1}$ & 0.806 & 1.38 & 2.98 & 1.21 & 0.10 & $-$0.12 & 0.07 & 0.00 & $-$0.04 & $-$0.21 & $-$0.17 & 3.51 \\
200 & $0.260 \times 10^{-2}$ & 1.284 & 1.30 & 1.85 & 1.06 & 0.06 & $-$0.02 & $-$0.01 & 0.00 & $-$0.04 & 0.01 & $-$0.06 & 2.50 \\
200 & $0.320 \times 10^{-2}$ & 1.250 & 1.08 & 0.84 & 1.09 & 0.39 & $-$0.44 & $-$0.06 & 0.01 & $-$0.04 & $-$0.06 & $-$0.14 & 1.85 \\
200 & $0.500 \times 10^{-2}$ & 1.111 & 0.59 & 0.53 & 1.00 & 0.14 & $-$0.10 & 0.00 & 0.00 & $-$0.04 & $-$0.09 & $-$0.09 & 1.30 \\
200 & $0.800 \times 10^{-2}$ & 0.944 & 0.60 & 0.66 & 1.01 & 0.11 & $-$0.06 & 0.00 & 0.00 & $-$0.05 & $-$0.12 & $-$0.07 & 1.36 \\
200 & $0.130 \times 10^{-1}$ & 0.799 & 0.59 & 0.54 & 1.00 & 0.05 & $-$0.05 & $-$0.04 & 0.00 & $-$0.05 & $-$0.20 & $-$0.06 & 1.30 \\
200 & $0.200 \times 10^{-1}$ & 0.696 & 0.67 & 0.62 & 1.00 & 0.03 & $-$0.04 & $-$0.04 & 0.00 & $-$0.05 & $-$0.17 & $-$0.10 & 1.37 \\
200 & $0.320 \times 10^{-1}$ & 0.580 & 0.74 & 0.68 & 1.07 & 0.09 & $-$0.04 & $-$0.09 & 0.00 & $-$0.04 & $-$0.29 & $-$0.16 & 1.51 \\
200 & $0.500 \times 10^{-1}$ & 0.513 & 0.81 & 0.74 & 1.02 & 0.08 & $-$0.04 & $-$0.07 & 0.00 & $-$0.04 & $-$0.20 & $-$0.12 & 1.52 \\
200 & $0.800 \times 10^{-1}$ & 0.438 & 0.77 & 0.74 & 1.17 & 0.18 & $-$0.04 & $-$0.05 & 0.00 & $-$0.04 & $-$0.21 & $-$0.11 & 1.61 \\
200 & 0.130 & 0.364 & 1.67 & 2.15 & 1.17 & 0.29 & $-$0.05 & 0.03 & 0.00 & $-$0.04 & $-$0.21 & $-$0.15 & 2.99 \\
200 & 0.180 & 0.329 & 1.08 & 0.77 & 1.22 & 0.26 & $-$0.05 & $-$0.02 & 0.00 & $-$0.05 & $-$0.15 & $-$0.08 & 1.83 \\
250 & $0.330 \times 10^{-2}$ & 1.270 & 1.49 & 1.84 & 1.06 & 0.05 & $-$0.01 & 0.00 & 0.00 & $-$0.05 & 0.02 & $-$0.06 & 2.60 \\
250 & $0.500 \times 10^{-2}$ & 1.124 & 0.88 & 0.78 & 1.03 & 0.20 & $-$0.23 & 0.01 & 0.00 & $-$0.04 & $-$0.02 & $-$0.11 & 1.60 \\
250 & $0.800 \times 10^{-2}$ & 0.961 & 0.66 & 0.56 & 1.00 & 0.09 & $-$0.05 & $-$0.01 & 0.00 & $-$0.05 & $-$0.06 & $-$0.08 & 1.33 \\
250 & $0.130 \times 10^{-1}$ & 0.817 & 0.69 & 0.54 & 1.02 & 0.17 & $-$0.06 & $-$0.03 & 0.00 & $-$0.05 & $-$0.18 & $-$0.06 & 1.37 \\
250 & $0.200 \times 10^{-1}$ & 0.690 & 0.75 & 0.63 & 1.00 & 0.16 & $-$0.05 & $-$0.08 & 0.00 & $-$0.04 & $-$0.17 & $-$0.09 & 1.43 \\
250 & $0.320 \times 10^{-1}$ & 0.589 & 0.78 & 0.67 & 1.04 & 0.19 & $-$0.04 & $-$0.11 & 0.00 & $-$0.04 & $-$0.26 & $-$0.13 & 1.51 \\
250 & $0.500 \times 10^{-1}$ & 0.508 & 0.83 & 0.60 & 1.04 & 0.07 & $-$0.04 & $-$0.10 & 0.00 & $-$0.05 & $-$0.14 & $-$0.10 & 1.48 \\
250 & $0.800 \times 10^{-1}$ & 0.431 & 0.78 & 0.57 & 1.11 & 0.15 & $-$0.04 & $-$0.08 & 0.00 & $-$0.04 & $-$0.20 & $-$0.09 & 1.50 \\
250 & 0.130 & 0.371 & 1.23 & 1.66 & 1.05 & 0.34 & $-$0.04 & $-$0.09 & 0.00 & $-$0.05 & $-$0.34 & $-$0.07 & 2.37 \\
250 & 0.180 & 0.328 & 1.07 & 0.81 & 1.21 & 0.28 & $-$0.04 & $-$0.06 & 0.00 & $-$0.05 & $-$0.02 & $-$0.09 & 1.83 \\
250 & 0.250 & 0.235 & 7.33 & 8.60 & 1.50 & 0.96 & $-$0.05 & $-$0.78 & 0.00 & $-$0.05 & $-$0.63 & $-$0.01 & 11.48 \\
250 & 0.400 & 0.141 & 9.20 & 6.80 & 1.27 & 0.18 & 0.04 & $-$0.10 & 0.00 & $-$0.04 & $-$0.19 & $-$0.12 & 11.52 \\
300 & $0.390 \times 10^{-2}$ & 1.216 & 1.71 & 1.83 & 1.05 & $-$0.02 & $-$0.01 & $-$0.03 & 0.01 & $-$0.04 & $-$0.01 & $-$0.07 & 2.72 \\
300 & $0.500 \times 10^{-2}$ & 1.157 & 1.49 & 0.89 & 1.05 & 0.31 & $-$0.32 & $-$0.05 & 0.00 & $-$0.04 & $-$0.04 & $-$0.12 & 2.08 \\
300 & $0.800 \times 10^{-2}$ & 0.987 & 0.79 & 0.54 & 1.04 & 0.17 & $-$0.10 & 0.00 & 0.00 & $-$0.04 & $-$0.07 & $-$0.09 & 1.43 \\
300 & $0.130 \times 10^{-1}$ & 0.827 & 0.75 & 0.55 & 1.03 & 0.08 & $-$0.05 & 0.00 & 0.00 & $-$0.05 & $-$0.01 & $-$0.07 & 1.40 \\
300 & $0.200 \times 10^{-1}$ & 0.712 & 0.84 & 0.50 & 0.99 & 0.16 & $-$0.06 & $-$0.06 & 0.00 & $-$0.05 & $-$0.13 & $-$0.07 & 1.41 \\
300 & $0.320 \times 10^{-1}$ & 0.608 & 0.87 & 0.58 & 1.01 & 0.12 & $-$0.05 & $-$0.06 & 0.00 & $-$0.05 & $-$0.11 & $-$0.07 & 1.46 \\
300 & $0.500 \times 10^{-1}$ & 0.517 & 0.90 & 0.57 & 1.03 & 0.11 & $-$0.04 & $-$0.08 & 0.00 & $-$0.05 & $-$0.08 & $-$0.09 & 1.49 \\
300 & $0.800 \times 10^{-1}$ & 0.436 & 0.88 & 0.59 & 1.01 & 0.11 & $-$0.04 & $-$0.09 & 0.00 & $-$0.05 & $-$0.12 & $-$0.08 & 1.48 \\
300 & 0.130 & 0.366 & 1.36 & 1.78 & 1.06 & 0.30 & $-$0.04 & $-$0.10 & 0.00 & $-$0.04 & $-$0.35 & $-$0.06 & 2.53 \\
300 & 0.180 & 0.314 & 1.12 & 0.68 & 1.34 & 0.30 & $-$0.04 & $-$0.07 & 0.00 & $-$0.04 & $-$0.19 & $-$0.12 & 1.92 \\
300 & 0.250 & 0.293 & 6.71 & 9.39 & 1.78 & 1.39 & $-$0.05 & $-$0.93 & 0.01 & $-$0.05 & $-$0.74 & 0.00 & 11.82 \\
300 & 0.400 & 0.157 & 2.17 & 3.28 & 1.68 & 1.10 & $-$0.04 & 0.02 & 0.01 & $-$0.04 & $-$0.25 & $-$0.12 & 4.43 \\

\hline \hline

\end{tabular}

\caption{\label{tab515a1}
HERA combined reduced  cross sections $\sigma^{- }_{r,{\rm NC}}$ for NC $e^{-}p$ scattering at $\sqrt{s} = 318 $~GeV.
 The uncertainties are quoted in percent relative to $\sigma^{- }_{r,{\rm NC}}$.
Other details as for Table~\ref{tab615-318a1}.}

\end{scriptsize}
\end{center}
\end{table}

\clearpage
\begin{table}
\begin{center}
\begin{scriptsize}\renewcommand\arraystretch{1.1}

\begin{tabular}[H]{ c l c r r r r r r r r r r r }
\hline \hline
$Q^2$ &  $x_{\rm Bj}$ & $\sigma_{r, \rm NC}^{- }$ & $\delta_{\rm stat}$ & $\delta_{\rm uncor}$ & $\delta_{\rm cor}$ & $\delta_{\rm rel}$ & $\delta_{\gamma p}$ & $\delta_{\rm had}$ &  $\delta_{1}$ & $\delta_{2}$ & $\delta_{3}$ & $\delta_{4}$ &$\delta_{\rm tot}$ \\
${\rm GeV^2}$ & & & \% & \% & \% & \% & \% & \% & \% & \% & \% & \% & \%     \\
\hline
400 & $0.530 \times 10^{-2}$ & 1.153 & 1.85 & 1.83 & 1.03 & 0.01 & $-$0.01 & $-$0.03 & 0.00 & $-$0.04 & 0.01 & $-$0.08 & 2.81 \\
400 & $0.800 \times 10^{-2}$ & 1.032 & 0.93 & 0.47 & 1.14 & 0.29 & $-$0.16 & 0.00 & 0.00 & $-$0.04 & $-$0.56 & $-$0.15 & 1.69 \\
400 & $0.130 \times 10^{-1}$ & 0.866 & 0.93 & 0.54 & 1.02 & 0.10 & $-$0.06 & $-$0.01 & 0.00 & $-$0.04 & $-$0.14 & $-$0.10 & 1.50 \\
400 & $0.200 \times 10^{-1}$ & 0.712 & 1.01 & 0.72 & 1.03 & 0.05 & $-$0.03 & $-$0.04 & 0.00 & $-$0.04 & $-$0.12 & $-$0.12 & 1.62 \\
400 & $0.320 \times 10^{-1}$ & 0.608 & 0.99 & 0.63 & 1.01 & 0.16 & $-$0.04 & $-$0.05 & 0.00 & $-$0.04 & $-$0.15 & $-$0.08 & 1.57 \\
400 & $0.500 \times 10^{-1}$ & 0.519 & 0.99 & 0.57 & 1.03 & 0.05 & $-$0.04 & $-$0.08 & 0.00 & $-$0.05 & 0.03 & $-$0.06 & 1.54 \\
400 & $0.800 \times 10^{-1}$ & 0.431 & 1.08 & 0.54 & 1.02 & $-$0.01 & $-$0.03 & $-$0.08 & 0.00 & $-$0.05 & 0.05 & $-$0.06 & 1.59 \\
400 & 0.130 & 0.367 & 1.14 & 0.61 & 1.09 & 0.17 & $-$0.04 & $-$0.08 & 0.00 & $-$0.05 & $-$0.07 & $-$0.07 & 1.71 \\
400 & 0.180 & 0.321 & 1.94 & 2.44 & 1.32 & 0.59 & $-$0.03 & 0.02 & 0.01 & $-$0.05 & $-$0.46 & $-$0.08 & 3.47 \\
400 & 0.250 & 0.264 & 1.81 & 0.82 & 1.43 & 0.19 & $-$0.05 & $-$0.07 & 0.00 & $-$0.05 & 0.19 & $-$0.04 & 2.47 \\
400 & 0.400 & 0.157 & 2.51 & 3.00 & 1.78 & 1.05 & $-$0.03 & 0.03 & 0.01 & $-$0.04 & $-$0.17 & $-$0.13 & 4.43 \\
500 & $0.660 \times 10^{-2}$ & 1.059 & 2.05 & 1.79 & 1.03 & $-$0.08 & $-$0.01 & $-$0.01 & 0.00 & $-$0.05 & 0.03 & $-$0.07 & 2.91 \\
500 & $0.800 \times 10^{-2}$ & 1.007 & 2.06 & 0.93 & 1.02 & 0.14 & $-$0.18 & $-$0.01 & 0.00 & $-$0.05 & $-$0.03 & $-$0.11 & 2.49 \\
500 & $0.130 \times 10^{-1}$ & 0.899 & 1.46 & 0.88 & 1.02 & 0.13 & $-$0.07 & 0.01 & 0.00 & $-$0.04 & $-$0.03 & $-$0.10 & 2.00 \\
500 & $0.200 \times 10^{-1}$ & 0.731 & 1.46 & 1.07 & 1.04 & 0.02 & $-$0.02 & 0.03 & 0.00 & $-$0.05 & 0.01 & $-$0.10 & 2.09 \\
500 & $0.320 \times 10^{-1}$ & 0.630 & 1.49 & 0.99 & 1.03 & 0.14 & $-$0.03 & $-$0.07 & 0.00 & $-$0.05 & $-$0.08 & $-$0.05 & 2.07 \\
500 & $0.500 \times 10^{-1}$ & 0.551 & 1.57 & 1.06 & 1.03 & 0.15 & $-$0.03 & $-$0.08 & 0.00 & $-$0.05 & $-$0.27 & $-$0.05 & 2.18 \\
500 & $0.800 \times 10^{-1}$ & 0.432 & 1.78 & 1.06 & 1.06 & 0.15 & $-$0.04 & $-$0.12 & 0.00 & $-$0.05 & $-$0.21 & $-$0.04 & 2.34 \\
500 & 0.130 & 0.377 & 2.06 & 1.27 & 1.05 & 0.08 & $-$0.03 & $-$0.14 & 0.00 & $-$0.05 & $-$0.30 & $-$0.04 & 2.66 \\
500 & 0.180 & 0.338 & 2.21 & 1.76 & 1.09 & 0.44 & $-$0.04 & $-$0.10 & 0.00 & $-$0.05 & $-$0.48 & $-$0.05 & 3.10 \\
500 & 0.250 & 0.267 & 2.61 & 2.41 & 1.39 & 0.74 & $-$0.03 & $-$0.02 & 0.01 & $-$0.04 & $-$0.41 & $-$0.09 & 3.91 \\
500 & 0.400 & 0.149 & 14.88 & 11.98 & 2.70 & 3.17 & $-$0.14 & $-$0.47 & 0.01 & $-$0.07 & $-$0.64 & 0.13 & 19.57 \\
500 & 0.650 & 0.027 & 19.18 & 10.50 & 1.74 & $-$0.73 & 0.02 & $-$1.52 & 0.00 & $-$0.03 & $-$0.92 & $-$0.13 & 22.02 \\
650 & $0.850 \times 10^{-2}$ & 0.987 & 2.19 & 1.54 & 1.02 & $-$0.02 & $-$0.03 & $-$0.05 & 0.00 & $-$0.04 & $-$0.04 & $-$0.07 & 2.87 \\
650 & $0.130 \times 10^{-1}$ & 0.905 & 1.05 & 0.40 & 1.04 & 0.14 & $-$0.12 & $-$0.01 & 0.00 & $-$0.04 & $-$0.10 & $-$0.10 & 1.55 \\
650 & $0.200 \times 10^{-1}$ & 0.778 & 1.20 & 0.43 & 1.00 & 0.07 & $-$0.05 & 0.00 & 0.00 & $-$0.05 & $-$0.12 & $-$0.08 & 1.63 \\
650 & $0.320 \times 10^{-1}$ & 0.625 & 1.32 & 0.55 & 1.03 & 0.09 & $-$0.04 & $-$0.02 & 0.00 & $-$0.04 & $-$0.12 & $-$0.11 & 1.77 \\
650 & $0.500 \times 10^{-1}$ & 0.534 & 1.31 & 0.52 & 0.99 & 0.13 & $-$0.03 & $-$0.05 & 0.00 & $-$0.05 & $-$0.19 & $-$0.07 & 1.74 \\
650 & $0.800 \times 10^{-1}$ & 0.435 & 1.45 & 0.71 & 1.01 & 0.08 & $-$0.04 & $-$0.09 & 0.00 & $-$0.05 & $-$0.13 & $-$0.06 & 1.92 \\
650 & 0.130 & 0.362 & 1.59 & 0.77 & 1.03 & 0.14 & $-$0.04 & $-$0.02 & 0.00 & $-$0.05 & $-$0.08 & $-$0.02 & 2.06 \\
650 & 0.180 & 0.333 & 2.49 & 1.60 & 1.05 & 0.23 & $-$0.03 & $-$0.12 & 0.00 & $-$0.05 & $-$0.34 & $-$0.04 & 3.16 \\
650 & 0.250 & 0.256 & 1.86 & 1.00 & 1.29 & 0.36 & $-$0.05 & $-$0.04 & 0.00 & $-$0.04 & $-$0.23 & $-$0.08 & 2.51 \\
650 & 0.400 & 0.126 & 4.75 & 3.33 & 1.56 & 0.83 & $-$0.04 & 0.10 & 0.02 & $-$0.04 & $-$0.45 & $-$0.09 & 6.09 \\
800 & $0.105 \times 10^{-1}$ & 0.963 & 2.70 & 1.87 & 1.02 & $-$0.08 & $-$0.01 & $-$0.02 & 0.00 & $-$0.05 & 0.00 & $-$0.08 & 3.44 \\
800 & $0.130 \times 10^{-1}$ & 0.888 & 1.45 & 0.36 & 1.11 & 0.23 & $-$0.13 & $-$0.02 & 0.00 & $-$0.04 & $-$0.37 & $-$0.10 & 1.92 \\
800 & $0.200 \times 10^{-1}$ & 0.752 & 1.42 & 0.52 & 1.00 & 0.10 & $-$0.07 & $-$0.01 & 0.00 & $-$0.05 & $-$0.14 & $-$0.08 & 1.83 \\
800 & $0.320 \times 10^{-1}$ & 0.641 & 1.45 & 0.54 & 1.00 & 0.06 & $-$0.05 & 0.01 & 0.00 & $-$0.05 & $-$0.08 & $-$0.07 & 1.85 \\
800 & $0.500 \times 10^{-1}$ & 0.547 & 1.42 & 0.42 & 1.01 & 0.12 & $-$0.04 & $-$0.08 & 0.00 & $-$0.05 & $-$0.22 & $-$0.11 & 1.81 \\
800 & $0.800 \times 10^{-1}$ & 0.452 & 1.56 & 0.48 & 1.05 & 0.05 & $-$0.04 & $-$0.07 & 0.00 & $-$0.05 & $-$0.12 & $-$0.06 & 1.95 \\
800 & 0.130 & 0.373 & 1.80 & 0.49 & 1.00 & 0.12 & $-$0.04 & $-$0.07 & 0.00 & $-$0.05 & $-$0.16 & $-$0.07 & 2.13 \\
800 & 0.180 & 0.326 & 3.07 & 1.68 & 1.05 & 0.25 & $-$0.03 & $-$0.10 & 0.00 & $-$0.05 & $-$0.40 & $-$0.04 & 3.69 \\
800 & 0.250 & 0.254 & 2.16 & 0.64 & 1.14 & 0.28 & $-$0.04 & $-$0.04 & 0.00 & $-$0.05 & $-$0.22 & $-$0.08 & 2.56 \\
800 & 0.400 & 0.141 & 4.71 & 3.85 & 1.82 & 1.19 & $-$0.04 & 0.01 & 0.01 & $-$0.05 & $-$0.75 & $-$0.10 & 6.51 \\
800 & 0.650 & 0.016 & 21.20 & 12.49 & 2.51 & 2.80 & $-$0.10 & $-$0.25 & 0.01 & $-$0.07 & $-$0.46 & 0.09 & 24.89 \\
1000 & $0.130 \times 10^{-1}$ & 0.862 & 2.75 & 1.50 & 1.38 & 0.73 & $-$1.03 & $-$0.17 & 0.00 & $-$0.04 & $-$0.13 & $-$0.18 & 3.66 \\
1000 & $0.200 \times 10^{-1}$ & 0.770 & 2.30 & 0.96 & 1.01 & 0.14 & $-$0.11 & 0.03 & 0.00 & $-$0.05 & $-$0.02 & $-$0.09 & 2.70 \\
1000 & $0.320 \times 10^{-1}$ & 0.658 & 2.32 & 1.05 & 1.01 & 0.06 & $-$0.03 & 0.02 & 0.00 & $-$0.05 & $-$0.02 & $-$0.09 & 2.74 \\
1000 & $0.500 \times 10^{-1}$ & 0.517 & 2.50 & 1.38 & 1.03 & 0.09 & $-$0.03 & 0.00 & 0.00 & $-$0.04 & 0.00 & $-$0.11 & 3.04 \\
1000 & $0.800 \times 10^{-1}$ & 0.440 & 2.72 & 1.01 & 1.02 & 0.06 & $-$0.02 & $-$0.09 & 0.00 & $-$0.05 & $-$0.09 & $-$0.05 & 3.08 \\
1000 & 0.130 & 0.369 & 3.21 & 1.30 & 1.04 & $-$0.04 & $-$0.02 & $-$0.07 & 0.00 & $-$0.05 & $-$0.31 & $-$0.04 & 3.63 \\
1000 & 0.180 & 0.345 & 3.23 & 1.59 & 1.03 & 0.19 & $-$0.03 & $-$0.06 & 0.00 & $-$0.05 & $-$0.37 & $-$0.05 & 3.78 \\
1000 & 0.250 & 0.281 & 3.57 & 1.97 & 1.12 & 0.37 & $-$0.03 & $-$0.11 & 0.00 & $-$0.05 & $-$0.46 & $-$0.05 & 4.27 \\
1000 & 0.400 & 0.137 & 5.13 & 4.02 & 1.76 & 1.04 & $-$0.03 & 0.04 & 0.01 & $-$0.05 & $-$0.54 & $-$0.10 & 6.85 \\

\hline \hline

\end{tabular}\captcont{Continued.}


\end{scriptsize}
\end{center}
\end{table}

\clearpage
\begin{table}
\begin{center}
\begin{scriptsize}\renewcommand\arraystretch{1.1}

\begin{tabular}[H]{ c l c r r r r r r r r r r r }
\hline \hline
$Q^2$ &  $x_{\rm Bj}$ & $\sigma_{r, \rm NC}^{- }$ & $\delta_{\rm stat}$ & $\delta_{\rm uncor}$ & $\delta_{\rm cor}$ & $\delta_{\rm rel}$ & $\delta_{\gamma p}$ & $\delta_{\rm had}$ &  $\delta_{1}$ & $\delta_{2}$ & $\delta_{3}$ & $\delta_{4}$ &$\delta_{\rm tot}$ \\
${\rm GeV^2}$ & & & \% & \% & \% & \% & \% & \% & \% & \% & \% & \% & \%     \\
\hline
1200 & $0.140 \times 10^{-1}$ & 0.916 & 1.93 & 0.46 & 1.51 & 0.40 & $-$0.44 & $-$0.01 & 0.00 & $-$0.04 & $-$0.12 & $-$0.12 & 2.57 \\
1200 & $0.200 \times 10^{-1}$ & 0.801 & 1.62 & 0.40 & 1.04 & 0.17 & $-$0.15 & $-$0.01 & 0.00 & $-$0.04 & $-$0.15 & $-$0.10 & 1.99 \\
1200 & $0.320 \times 10^{-1}$ & 0.639 & 1.58 & 0.48 & 1.00 & 0.09 & $-$0.05 & $-$0.01 & 0.00 & $-$0.05 & $-$0.12 & $-$0.08 & 1.94 \\
1200 & $0.500 \times 10^{-1}$ & 0.557 & 1.47 & 0.39 & 1.00 & 0.06 & $-$0.03 & $-$0.04 & 0.00 & $-$0.05 & $-$0.13 & $-$0.11 & 1.83 \\
1200 & $0.800 \times 10^{-1}$ & 0.462 & 1.53 & 0.39 & 1.00 & 0.08 & $-$0.04 & $-$0.06 & 0.00 & $-$0.05 & $-$0.18 & $-$0.09 & 1.88 \\
1200 & 0.130 & 0.376 & 1.77 & 0.38 & 1.01 & 0.02 & $-$0.03 & $-$0.05 & 0.00 & $-$0.05 & $-$0.06 & $-$0.07 & 2.08 \\
1200 & 0.180 & 0.319 & 3.76 & 1.60 & 1.03 & 0.14 & $-$0.03 & $-$0.09 & 0.00 & $-$0.05 & $-$0.32 & $-$0.03 & 4.23 \\
1200 & 0.250 & 0.247 & 2.01 & 0.52 & 1.02 & 0.18 & $-$0.04 & $-$0.05 & 0.00 & $-$0.05 & $-$0.19 & $-$0.09 & 2.33 \\
1200 & 0.400 & 0.125 & 3.25 & 1.15 & 2.69 & 0.55 & $-$0.03 & $-$0.03 & 0.01 & $-$0.05 & $-$0.58 & $-$0.10 & 4.45 \\
1500 & $0.200 \times 10^{-1}$ & 0.805 & 2.18 & 0.64 & 1.42 & 0.31 & $-$0.45 & $-$0.04 & 0.00 & $-$0.04 & 0.13 & $-$0.14 & 2.74 \\
1500 & $0.320 \times 10^{-1}$ & 0.661 & 2.01 & 0.41 & 0.99 & 0.06 & $-$0.06 & 0.00 & 0.00 & $-$0.05 & $-$0.05 & $-$0.08 & 2.28 \\
1500 & $0.500 \times 10^{-1}$ & 0.545 & 1.86 & 0.42 & 1.00 & 0.10 & $-$0.04 & $-$0.03 & 0.00 & $-$0.05 & $-$0.10 & $-$0.09 & 2.16 \\
1500 & $0.800 \times 10^{-1}$ & 0.490 & 1.83 & 0.41 & 0.99 & 0.05 & $-$0.04 & $-$0.03 & 0.00 & $-$0.05 & $-$0.08 & $-$0.06 & 2.12 \\
1500 & 0.130 & 0.378 & 2.36 & 0.46 & 1.05 & 0.04 & $-$0.04 & $-$0.02 & 0.00 & $-$0.05 & $-$0.02 & $-$0.02 & 2.62 \\
1500 & 0.180 & 0.313 & 2.48 & 0.57 & 1.01 & 0.04 & $-$0.04 & $-$0.02 & 0.00 & $-$0.05 & $-$0.04 & $-$0.02 & 2.74 \\
1500 & 0.250 & 0.260 & 2.85 & 0.74 & 1.01 & 0.17 & $-$0.04 & $-$0.06 & 0.00 & $-$0.05 & $-$0.21 & $-$0.07 & 3.13 \\
1500 & 0.400 & 0.130 & 4.34 & 1.39 & 3.03 & 0.55 & $-$0.04 & $-$0.02 & 0.01 & $-$0.05 & $-$0.49 & $-$0.05 & 5.52 \\
1500 & 0.650 & 0.016 & 11.10 & 6.33 & 2.89 & 2.36 & $-$0.03 & 0.09 & 0.02 & $-$0.06 & $-$0.71 & $-$0.11 & 13.33 \\
2000 & $0.219 \times 10^{-1}$ & 0.878 & 5.51 & 2.19 & 1.60 & 1.01 & $-$1.35 & $-$0.14 & 0.00 & $-$0.03 & $-$0.14 & $-$0.21 & 6.38 \\
2000 & $0.320 \times 10^{-1}$ & 0.658 & 2.41 & 0.56 & 1.16 & 0.16 & $-$0.17 & $-$0.01 & 0.00 & $-$0.04 & 0.10 & $-$0.10 & 2.74 \\
2000 & $0.500 \times 10^{-1}$ & 0.570 & 2.24 & 0.39 & 0.99 & 0.07 & $-$0.04 & 0.00 & 0.00 & $-$0.05 & $-$0.08 & $-$0.06 & 2.49 \\
2000 & $0.800 \times 10^{-1}$ & 0.464 & 2.24 & 0.44 & 1.00 & 0.08 & $-$0.03 & $-$0.05 & 0.00 & $-$0.05 & $-$0.15 & $-$0.12 & 2.50 \\
2000 & 0.130 & 0.370 & 2.69 & 0.54 & 1.00 & 0.06 & $-$0.04 & $-$0.06 & 0.00 & $-$0.05 & $-$0.08 & $-$0.07 & 2.93 \\
2000 & 0.180 & 0.310 & 2.94 & 0.69 & 1.08 & 0.11 & $-$0.05 & $-$0.03 & 0.00 & $-$0.05 & $-$0.04 & $-$0.05 & 3.21 \\
2000 & 0.250 & 0.251 & 3.35 & 0.80 & 1.02 & 0.15 & $-$0.04 & $-$0.04 & 0.00 & $-$0.05 & $-$0.19 & $-$0.06 & 3.60 \\
2000 & 0.400 & 0.123 & 4.60 & 1.58 & 1.17 & 0.24 & $-$0.03 & $-$0.04 & 0.00 & $-$0.05 & $-$0.44 & $-$0.09 & 5.03 \\
2000 & 0.650 & 0.011 & 15.90 & 7.83 & 3.03 & 2.20 & $-$0.02 & 0.22 & 0.03 & $-$0.05 & $-$0.82 & $-$0.14 & 18.13 \\
3000 & $0.320 \times 10^{-1}$ & 0.731 & 3.76 & 1.74 & 1.14 & 0.41 & $-$0.56 & $-$0.06 & 0.00 & $-$0.04 & $-$0.07 & $-$0.14 & 4.36 \\
3000 & $0.500 \times 10^{-1}$ & 0.599 & 2.34 & 0.69 & 1.05 & 0.20 & $-$0.15 & 0.02 & 0.00 & $-$0.04 & $-$0.01 & $-$0.09 & 2.67 \\
3000 & $0.800 \times 10^{-1}$ & 0.508 & 2.34 & 0.56 & 0.99 & 0.05 & $-$0.03 & $-$0.02 & 0.00 & $-$0.05 & $-$0.06 & $-$0.09 & 2.61 \\
3000 & 0.130 & 0.394 & 2.82 & 0.71 & 0.98 & 0.00 & $-$0.03 & $-$0.04 & 0.00 & $-$0.05 & $-$0.05 & $-$0.07 & 3.07 \\
3000 & 0.180 & 0.317 & 3.13 & 0.75 & 1.00 & 0.03 & $-$0.03 & $-$0.03 & 0.00 & $-$0.05 & $-$0.06 & $-$0.03 & 3.37 \\
3000 & 0.250 & 0.265 & 3.40 & 1.02 & 1.19 & 0.35 & $-$0.03 & $-$0.04 & 0.00 & $-$0.05 & $-$0.26 & $-$0.04 & 3.77 \\
3000 & 0.400 & 0.131 & 4.60 & 1.90 & 1.12 & 0.35 & $-$0.04 & $-$0.08 & 0.00 & $-$0.05 & $-$0.34 & $-$0.05 & 5.12 \\
3000 & 0.650 & 0.015 & 8.82 & 3.78 & 1.89 & 1.39 & $-$0.03 & 0.02 & 0.01 & $-$0.05 & $-$1.01 & $-$0.18 & 9.93 \\
5000 & $0.547 \times 10^{-1}$ & 0.642 & 5.01 & 1.90 & 1.11 & 0.42 & $-$0.49 & $-$0.07 & 0.00 & $-$0.04 & $-$0.07 & $-$0.13 & 5.51 \\
5000 & $0.800 \times 10^{-1}$ & 0.546 & 2.34 & 0.58 & 1.06 & 0.15 & $-$0.10 & 0.01 & 0.00 & $-$0.05 & 0.08 & $-$0.07 & 2.64 \\
5000 & 0.130 & 0.471 & 2.85 & 0.72 & 0.99 & 0.11 & $-$0.05 & $-$0.02 & 0.00 & $-$0.05 & $-$0.02 & $-$0.07 & 3.10 \\
5000 & 0.180 & 0.365 & 3.18 & 0.70 & 0.99 & 0.05 & $-$0.03 & $-$0.03 & 0.00 & $-$0.05 & $-$0.05 & $-$0.07 & 3.40 \\
5000 & 0.250 & 0.249 & 4.24 & 0.79 & 1.08 & 0.12 & $-$0.04 & $-$0.05 & 0.00 & $-$0.05 & $-$0.09 & $-$0.03 & 4.45 \\
5000 & 0.400 & 0.130 & 5.18 & 1.65 & 1.20 & 0.43 & $-$0.05 & 0.03 & 0.00 & $-$0.05 & $-$0.37 & $-$0.02 & 5.59 \\
5000 & 0.650 & 0.015 & 12.57 & 6.37 & 1.93 & 1.00 & $-$0.03 & $-$0.01 & 0.01 & $-$0.06 & $-$1.25 & 0.01 & 14.31 \\
8000 & $0.875 \times 10^{-1}$ & 0.644 & 7.39 & 2.49 & 1.27 & 0.71 & $-$0.84 & $-$0.10 & 0.00 & $-$0.04 & $-$0.12 & $-$0.16 & 7.98 \\
8000 & 0.130 & 0.559 & 3.24 & 0.76 & 1.13 & 0.07 & $-$0.12 & 0.00 & 0.00 & $-$0.05 & $-$0.02 & $-$0.09 & 3.52 \\
8000 & 0.180 & 0.417 & 3.95 & 0.98 & 1.01 & $-$0.04 & $-$0.02 & $-$0.02 & 0.00 & $-$0.05 & 0.00 & $-$0.08 & 4.20 \\
8000 & 0.250 & 0.297 & 4.66 & 1.07 & 1.08 & $-$0.01 & $-$0.03 & $-$0.04 & 0.00 & $-$0.05 & $-$0.01 & $-$0.05 & 4.90 \\
8000 & 0.400 & 0.121 & 7.20 & 2.22 & 1.22 & 0.24 & $-$0.02 & $-$0.01 & 0.00 & $-$0.06 & $-$0.24 & $-$0.01 & 7.64 \\
8000 & 0.650 & 0.016 & 11.75 & 3.83 & 1.49 & 0.84 & $-$0.03 & $-$0.01 & 0.00 & $-$0.05 & $-$0.92 & $-$0.08 & 12.51 \\
12000 & 0.130 & 0.733 & 12.72 & 3.66 & 1.44 & 0.92 & $-$1.12 & $-$0.13 & 0.00 & $-$0.04 & $-$0.13 & $-$0.18 & 13.39 \\
12000 & 0.180 & 0.482 & 4.64 & 0.76 & 1.48 & 0.31 & $-$0.15 & $-$0.05 & 0.00 & $-$0.04 & 0.06 & $-$0.06 & 4.94 \\
12000 & 0.250 & 0.334 & 5.85 & 1.24 & 1.21 & 0.10 & $-$0.04 & 0.01 & 0.00 & $-$0.04 & $-$0.01 & $-$0.06 & 6.11 \\
12000 & 0.400 & 0.171 & 7.91 & 2.39 & 1.28 & $-$0.05 & $-$0.05 & $-$0.01 & 0.01 & $-$0.05 & $-$0.11 & 0.01 & 8.36 \\
12000 & 0.650 & 0.015 & 22.68 & 6.33 & 1.67 & 0.58 & 0.01 & 0.02 & 0.00 & $-$0.07 & $-$1.21 & 0.04 & 23.64 \\
20000 & 0.250 & 0.479 & 7.00 & 0.98 & 2.03 & 0.53 & $-$0.46 & $-$0.02 & 0.00 & $-$0.03 & 0.09 & $-$0.13 & 7.38 \\
20000 & 0.400 & 0.206 & 9.67 & 2.24 & 1.26 & 0.28 & $-$0.08 & $-$0.04 & 0.00 & $-$0.04 & $-$0.13 & $-$0.09 & 10.01 \\
20000 & 0.650 & 0.017 & 30.51 & 14.48 & 2.63 & 0.05 & 0.01 & 0.21 & 0.01 & $-$0.09 & $-$0.40 & 0.31 & 33.88 \\
30000 & 0.400 & 0.231 & 13.51 & 1.85 & 2.63 & 0.54 & $-$0.40 & $-$0.16 & 0.00 & $-$0.03 & $-$0.27 & $-$0.07 & 13.91 \\
30000 & 0.650 & 0.044 & 30.34 & 11.33 & 2.21 & 0.04 & 0.09 & $-$0.06 & 0.00 & $-$0.08 & $-$0.43 & 0.17 & 32.46 \\
50000 & 0.650 & 0.082 & 55.75 & 10.55 & 1.78 & 0.11 & 0.05 & $-$0.07 & 0.00 & $-$0.07 & $-$0.35 & 0.11 & 56.77 \\

\hline \hline

\end{tabular}\captcont{Continued.}


\end{scriptsize}
\end{center}
\end{table}

\clearpage
\begin{table}
\begin{center}
\begin{scriptsize}\renewcommand\arraystretch{1.1}

\begin{tabular}[H]{ c l c r r r r r r r r r r r }
\hline \hline
$Q^2$ &  $x_{\rm Bj}$ & $\sigma_{r, \rm CC}^{+ }$ & $\delta_{\rm stat}$ & $\delta_{\rm uncor}$ & $\delta_{\rm cor}$ & $\delta_{\rm rel}$ & $\delta_{\gamma p}$ & $\delta_{\rm had}$ &  $\delta_{1}$ & $\delta_{2}$ & $\delta_{3}$ & $\delta_{4}$ &$\delta_{\rm tot}$ \\
${\rm GeV^2}$ & & & \% & \% & \% & \% & \% & \% & \% & \% & \% & \% & \%     \\
\hline
300 & $0.800 \times 10^{-2}$ & 1.187 & 11.56 & 3.56 & 2.57 & 1.19 & $-$0.49 & $-$0.11 & 0.01 & $-$0.04 & 1.60 & 0.23 & 12.54 \\
300 & $0.130 \times 10^{-1}$ & 1.225 & 5.92 & 1.97 & 2.20 & 1.62 & $-$0.71 & $-$0.28 & 0.00 & $-$0.04 & 1.11 & 0.20 & 6.94 \\
300 & $0.320 \times 10^{-1}$ & 0.859 & 5.40 & 1.49 & 1.54 & 1.00 & $-$0.72 & $-$0.10 & 0.00 & $-$0.04 & 0.68 & 0.20 & 5.98 \\
300 & $0.800 \times 10^{-1}$ & 0.486 & 7.07 & 1.91 & 1.35 & 0.95 & $-$0.73 & $-$0.36 & 0.00 & $-$0.05 & $-$0.46 & 0.24 & 7.57 \\
300 & 0.130 & 0.569 & 27.42 & 1.50 & 1.81 & 0.68 & $-$0.11 & $-$0.01 & 0.01 & $-$0.06 & 0.47 & 0.38 & 27.54 \\
500 & $0.800 \times 10^{-2}$ & 0.733 & 24.79 & 5.57 & 4.13 & 2.09 & $-$0.19 & 0.14 & 0.01 & $-$0.04 & 3.27 & 0.17 & 26.04 \\
500 & $0.130 \times 10^{-1}$ & 0.938 & 5.27 & 1.96 & 1.90 & 1.10 & $-$0.27 & $-$0.28 & 0.00 & $-$0.05 & 1.05 & 0.28 & 6.15 \\
500 & $0.320 \times 10^{-1}$ & 0.863 & 4.15 & 1.23 & 1.13 & 0.86 & $-$0.22 & $-$0.30 & 0.00 & $-$0.05 & 0.53 & 0.30 & 4.61 \\
500 & $0.800 \times 10^{-1}$ & 0.582 & 4.64 & 1.34 & 0.94 & 0.15 & $-$0.15 & $-$0.07 & 0.01 & $-$0.05 & 0.32 & 0.28 & 4.94 \\
500 & 0.130 & 0.428 & 8.45 & 1.72 & 1.74 & 0.75 & $-$0.13 & $-$0.17 & 0.01 & $-$0.05 & $-$1.63 & 0.29 & 8.98 \\
1000 & $0.130 \times 10^{-1}$ & 0.652 & 6.17 & 2.37 & 2.07 & 0.47 & 0.04 & $-$0.38 & $-$0.01 & $-$0.06 & 1.37 & 0.32 & 7.09 \\
1000 & $0.320 \times 10^{-1}$ & 0.730 & 3.44 & 0.97 & 1.01 & 0.20 & $-$0.04 & $-$0.35 & 0.00 & $-$0.06 & 0.62 & 0.30 & 3.79 \\
1000 & $0.800 \times 10^{-1}$ & 0.515 & 3.85 & 1.10 & 0.82 & $-$0.02 & $-$0.03 & $-$0.22 & 0.00 & $-$0.05 & 0.43 & 0.31 & 4.12 \\
1000 & 0.130 & 0.431 & 5.52 & 1.45 & 0.91 & 0.21 & $-$0.02 & $-$0.20 & 0.00 & $-$0.05 & $-$0.28 & 0.32 & 5.80 \\
1000 & 0.250 & 0.232 & 11.66 & 2.58 & 1.13 & 0.24 & $-$0.01 & 0.42 & 0.01 & $-$0.04 & 0.58 & 0.25 & 12.02 \\
1500 & $0.320 \times 10^{-1}$ & 0.566 & 5.49 & 0.77 & 1.18 & 0.08 & $-$0.04 & 0.09 & 0.01 & $-$0.05 & 1.44 & 0.31 & 5.86 \\
1500 & $0.800 \times 10^{-1}$ & 0.475 & 5.17 & 0.47 & 0.92 & 0.05 & $-$0.02 & $-$0.23 & 0.00 & $-$0.06 & 0.85 & 0.36 & 5.35 \\
1500 & 0.130 & 0.346 & 7.37 & 0.49 & 0.89 & $-$0.16 & $-$0.03 & 0.04 & 0.01 & $-$0.05 & 0.57 & 0.33 & 7.47 \\
1500 & 0.250 & 0.227 & 9.22 & 0.94 & 0.89 & 0.10 & $-$0.02 & $-$0.15 & 0.00 & $-$0.05 & 0.40 & 0.33 & 9.33 \\
1500 & 0.400 & 0.072 & 41.90 & 3.35 & 1.35 & 0.42 & $-$0.09 & $-$0.05 & 0.00 & $-$0.06 & 0.16 & 0.43 & 42.06 \\
2000 & $0.320 \times 10^{-1}$ & 0.517 & 4.91 & 1.98 & 1.30 & 0.43 & 0.07 & $-$0.18 & 0.00 & $-$0.06 & 0.45 & 0.29 & 5.50 \\
2000 & $0.800 \times 10^{-1}$ & 0.427 & 4.69 & 1.68 & 0.89 & 0.10 & $-$0.05 & $-$0.20 & 0.00 & $-$0.05 & 0.42 & 0.27 & 5.09 \\
2000 & 0.130 & 0.344 & 6.47 & 2.39 & 0.87 & $-$0.30 & $-$0.05 & $-$0.38 & 0.01 & $-$0.05 & 0.42 & 0.28 & 6.99 \\
2000 & 0.250 & 0.216 & 10.29 & 3.28 & 2.48 & 1.27 & $-$0.01 & $-$0.51 & 0.00 & $-$0.05 & $-$2.99 & 0.26 & 11.56 \\
3000 & $0.320 \times 10^{-1}$ & 0.384 & 12.67 & 1.25 & 2.38 & 1.03 & $-$0.13 & 0.00 & 0.00 & $-$0.05 & 2.25 & 0.32 & 13.19 \\
3000 & $0.800 \times 10^{-1}$ & 0.377 & 3.52 & 1.09 & 0.99 & 0.14 & $-$0.03 & 0.25 & 0.01 & $-$0.05 & 1.03 & 0.25 & 3.97 \\
3000 & 0.130 & 0.312 & 4.41 & 1.00 & 0.87 & 0.06 & $-$0.02 & 0.03 & 0.00 & $-$0.05 & 0.66 & 0.28 & 4.66 \\
3000 & 0.250 & 0.166 & 6.33 & 1.50 & 0.95 & 0.45 & 0.00 & $-$0.02 & 0.00 & $-$0.05 & 0.18 & 0.28 & 6.60 \\
3000 & 0.400 & 0.063 & 15.90 & 2.86 & 1.94 & 3.20 & 0.04 & 0.11 & 0.00 & $-$0.04 & $-$1.23 & 0.19 & 16.63 \\
5000 & $0.800 \times 10^{-1}$ & 0.225 & 5.76 & 1.64 & 1.49 & 0.68 & $-$0.05 & 0.38 & 0.01 & $-$0.05 & 1.54 & 0.23 & 6.41 \\
5000 & 0.130 & 0.219 & 5.08 & 1.35 & 1.21 & 0.40 & $-$0.04 & 0.44 & 0.01 & $-$0.04 & 0.99 & 0.21 & 5.52 \\
5000 & 0.250 & 0.143 & 6.19 & 1.42 & 1.23 & 0.42 & $-$0.01 & $-$0.01 & 0.00 & $-$0.05 & 0.56 & 0.24 & 6.51 \\
5000 & 0.400 & 0.084 & 10.70 & 2.60 & 2.03 & $-$0.42 & 0.05 & 0.20 & 0.00 & $-$0.04 & $-$0.45 & 0.20 & 11.22 \\
8000 & 0.130 & 0.116 & 8.17 & 2.52 & 2.39 & 1.03 & $-$0.07 & 1.03 & 0.03 & $-$0.03 & 1.52 & 0.10 & 9.13 \\
8000 & 0.250 & 0.110 & 7.37 & 2.15 & 1.72 & 0.76 & $-$0.01 & 0.60 & 0.01 & $-$0.04 & 1.23 & 0.17 & 8.02 \\
8000 & 0.400 & 0.049 & 13.87 & 3.88 & 2.46 & 0.95 & 0.06 & 0.33 & 0.00 & $-$0.04 & 0.83 & 0.15 & 14.67 \\
15000 & 0.250 & 0.045 & 13.71 & 3.82 & 3.44 & 1.65 & $-$0.10 & 1.66 & 0.04 & $-$0.03 & 1.88 & $-$0.03 & 14.95 \\
15000 & 0.400 & 0.031 & 15.37 & 4.56 & 2.50 & 2.26 & 0.03 & 0.66 & 0.00 & $-$0.04 & 1.74 & 0.12 & 16.49 \\
30000 & 0.400 & 0.008 & 69.33 & 2.49 & 18.37 & 9.46 & $-$0.50 & 1.29 & 0.02 & 0.01 & $-$2.94 & $-$0.69 & 72.46 \\

\hline \hline

\end{tabular}

\caption{\label{tab3615a1}
HERA combined reduced  cross sections $\sigma^{+ }_{r,{\rm CC}}$  for CC $e^{+}p$ scattering at $\sqrt{s} = 318 $~GeV.
The uncertainties are quoted in percent relative to  $\sigma^{+ }_{r,{\rm CC}}$.
Other details as for Table~\ref{tab615-318a1}.
}

\end{scriptsize}
\end{center}
\end{table}

\clearpage
\begin{table}
\begin{center}
\begin{scriptsize}\renewcommand\arraystretch{1.1}

\begin{tabular}[H]{ c l c r r r r r r r r r r r }
\hline \hline
$Q^2$ &  $x_{\rm Bj}$ & $\sigma_{r, \rm CC}^{- }$ & $\delta_{\rm stat}$ & $\delta_{\rm uncor}$ & $\delta_{\rm cor}$ & $\delta_{\rm rel}$ & $\delta_{\gamma p}$ & $\delta_{\rm had}$ &  $\delta_{1}$ & $\delta_{2}$ & $\delta_{3}$ & $\delta_{4}$ &$\delta_{\rm tot}$ \\
${\rm GeV^2}$ & & & \% & \% & \% & \% & \% & \% & \% & \% & \% & \% & \%     \\
\hline
300 & $0.800 \times 10^{-2}$ & 1.934 & 31.94 & 23.82 & 8.56 & 5.28 & $-$2.92 & $-$0.69 & 0.00 & $-$0.01 & 0.49 & $-$0.35 & 41.21 \\
300 & $0.130 \times 10^{-1}$ & 1.188 & 7.18 & 4.71 & 2.41 & 0.93 & $-$0.30 & $-$0.17 & 0.00 & $-$0.05 & $-$0.35 & $-$0.05 & 8.98 \\
300 & $0.320 \times 10^{-1}$ & 1.091 & 6.58 & 4.18 & 1.89 & 1.51 & $-$0.66 & $-$0.19 & 0.00 & $-$0.05 & $-$0.08 & $-$0.10 & 8.19 \\
300 & $0.800 \times 10^{-1}$ & 0.946 & 7.26 & 2.94 & 2.31 & 2.59 & $-$1.00 & $-$0.26 & 0.00 & $-$0.05 & $-$1.95 & $-$0.14 & 8.84 \\
300 & 0.130 & 0.643 & 26.28 & 3.48 & 2.15 & 0.40 & $-$0.11 & $-$0.12 & 0.00 & $-$0.06 & $-$0.36 & $-$0.03 & 26.60 \\
500 & $0.130 \times 10^{-1}$ & 1.093 & 5.71 & 2.70 & 2.34 & 0.70 & $-$0.02 & $-$0.16 & 0.00 & $-$0.05 & $-$0.30 & $-$0.02 & 6.78 \\
500 & $0.320 \times 10^{-1}$ & 0.968 & 4.93 & 2.92 & 1.59 & 0.67 & $-$0.24 & $-$0.14 & 0.00 & $-$0.05 & $-$0.36 & $-$0.06 & 6.00 \\
500 & $0.800 \times 10^{-1}$ & 0.699 & 5.57 & 2.34 & 1.32 & 0.51 & $-$0.15 & $-$0.12 & 0.00 & $-$0.05 & $-$0.58 & $-$0.05 & 6.23 \\
500 & 0.130 & 0.828 & 8.30 & 2.06 & 2.69 & 3.30 & $-$0.10 & $-$0.29 & 0.00 & $-$0.06 & $-$2.72 & $-$0.03 & 9.94 \\
1000 & $0.130 \times 10^{-1}$ & 0.889 & 5.65 & 3.26 & 2.31 & 0.64 & 0.06 & $-$0.13 & 0.00 & $-$0.06 & $-$0.26 & $-$0.02 & 6.96 \\
1000 & $0.320 \times 10^{-1}$ & 0.918 & 3.96 & 2.60 & 1.31 & 0.18 & $-$0.02 & $-$0.10 & 0.00 & $-$0.06 & $-$0.19 & $-$0.04 & 4.92 \\
1000 & $0.800 \times 10^{-1}$ & 0.743 & 4.13 & 1.92 & 1.18 & 0.08 & $-$0.07 & $-$0.08 & 0.00 & $-$0.05 & 0.00 & $-$0.04 & 4.71 \\
1000 & 0.130 & 0.681 & 5.32 & 1.42 & 1.43 & 0.64 & $-$0.07 & $-$0.14 & 0.00 & $-$0.05 & $-$1.02 & $-$0.04 & 5.82 \\
1000 & 0.250 & 0.519 & 10.81 & 3.33 & 1.19 & 0.10 & $-$0.05 & $-$0.06 & 0.00 & $-$0.05 & $-$0.02 & $-$0.04 & 11.37 \\
1500 & $0.320 \times 10^{-1}$ & 0.827 & 4.99 & 5.42 & 1.30 & 0.16 & $-$0.07 & $-$0.08 & 0.01 & $-$0.05 & $-$0.09 & $-$0.03 & 7.49 \\
1500 & $0.800 \times 10^{-1}$ & 0.751 & 4.95 & 3.15 & 1.18 & 0.11 & $-$0.06 & $-$0.07 & 0.00 & $-$0.05 & $-$0.04 & $-$0.04 & 5.99 \\
1500 & 0.130 & 0.652 & 5.91 & 1.80 & 1.46 & 0.20 & $-$0.08 & $-$0.08 & 0.00 & $-$0.05 & $-$0.13 & $-$0.04 & 6.36 \\
1500 & 0.250 & 0.518 & 7.16 & 1.59 & 1.05 & $-$0.11 & $-$0.03 & $-$0.02 & 0.00 & $-$0.05 & 0.25 & $-$0.03 & 7.41 \\
1500 & 0.400 & 0.348 & 16.43 & 4.08 & 1.59 & $-$0.21 & 0.01 & 0.02 & 0.00 & $-$0.05 & 0.55 & $-$0.04 & 17.01 \\
2000 & $0.320 \times 10^{-1}$ & 0.805 & 4.93 & 2.36 & 1.40 & 0.29 & 0.08 & $-$0.11 & 0.00 & $-$0.06 & $-$0.52 & $-$0.04 & 5.67 \\
2000 & $0.800 \times 10^{-1}$ & 0.754 & 4.48 & 1.92 & 1.11 & $-$0.14 & 0.00 & $-$0.03 & 0.00 & $-$0.05 & 0.24 & $-$0.05 & 5.00 \\
2000 & 0.130 & 0.613 & 6.39 & 2.69 & 1.09 & $-$0.09 & $-$0.03 & $-$0.07 & 0.00 & $-$0.05 & $-$0.71 & $-$0.05 & 7.06 \\
2000 & 0.250 & 0.514 & 10.66 & 3.37 & 4.26 & 2.17 & 0.03 & $-$0.38 & 0.00 & $-$0.05 & $-$5.67 & $-$0.05 & 13.42 \\
3000 & $0.320 \times 10^{-1}$ & 0.700 & 8.57 & 8.30 & 1.48 & 0.21 & $-$0.07 & $-$0.08 & 0.00 & $-$0.06 & $-$0.13 & $-$0.03 & 12.02 \\
3000 & $0.800 \times 10^{-1}$ & 0.661 & 3.20 & 1.90 & 1.04 & $-$0.07 & $-$0.02 & $-$0.06 & 0.00 & $-$0.05 & 0.37 & $-$0.05 & 3.88 \\
3000 & 0.130 & 0.610 & 4.06 & 1.96 & 1.03 & $-$0.04 & $-$0.02 & 0.00 & 0.01 & $-$0.05 & 0.23 & $-$0.05 & 4.62 \\
3000 & 0.250 & 0.478 & 4.55 & 2.27 & 1.08 & 0.10 & $-$0.07 & $-$0.09 & 0.00 & $-$0.05 & $-$0.38 & $-$0.05 & 5.21 \\
3000 & 0.400 & 0.265 & 8.14 & 3.84 & 2.10 & $-$0.24 & 0.05 & 0.18 & 0.00 & $-$0.05 & 0.76 & $-$0.06 & 9.28 \\
5000 & $0.800 \times 10^{-1}$ & 0.629 & 4.13 & 2.09 & 1.13 & 0.05 & 0.01 & 0.01 & 0.00 & $-$0.05 & 0.60 & $-$0.05 & 4.80 \\
5000 & 0.130 & 0.581 & 3.89 & 2.16 & 1.05 & $-$0.01 & $-$0.01 & 0.02 & 0.01 & $-$0.05 & 0.44 & $-$0.05 & 4.59 \\
5000 & 0.250 & 0.433 & 4.46 & 1.99 & 1.13 & $-$0.08 & $-$0.04 & $-$0.03 & 0.00 & $-$0.05 & 0.47 & $-$0.06 & 5.04 \\
5000 & 0.400 & 0.284 & 6.68 & 3.90 & 2.17 & 0.46 & 0.06 & 0.18 & 0.00 & $-$0.05 & $-$0.18 & $-$0.06 & 8.06 \\
5000 & 0.650 & 0.078 & 27.12 & 12.82 & 7.03 & $-$0.38 & 0.24 & 0.31 & 0.01 & $-$0.05 & 2.38 & $-$0.05 & 30.91 \\
8000 & 0.130 & 0.633 & 4.51 & 2.97 & 1.48 & $-$0.09 & 0.02 & 0.11 & 0.00 & $-$0.05 & 1.36 & $-$0.07 & 5.76 \\
8000 & 0.250 & 0.419 & 4.71 & 2.77 & 1.33 & $-$0.20 & 0.02 & 0.08 & 0.00 & $-$0.05 & 1.10 & $-$0.07 & 5.73 \\
8000 & 0.400 & 0.244 & 7.12 & 3.77 & 2.25 & $-$0.14 & 0.07 & 0.22 & 0.00 & $-$0.05 & 1.11 & $-$0.08 & 8.44 \\
8000 & 0.650 & 0.035 & 46.57 & 27.23 & 9.72 & $-$0.05 & 0.36 & 0.45 & 0.01 & $-$0.05 & 3.32 & $-$0.06 & 54.92 \\
15000 & 0.250 & 0.454 & 5.17 & 4.05 & 2.62 & 0.30 & 0.08 & 0.23 & 0.01 & $-$0.04 & 2.03 & $-$0.07 & 7.37 \\
15000 & 0.400 & 0.204 & 7.20 & 4.18 & 2.79 & $-$0.17 & 0.09 & 0.29 & 0.01 & $-$0.04 & 2.23 & $-$0.08 & 9.07 \\
15000 & 0.650 & 0.036 & 46.02 & 24.89 & 11.35 & 0.28 & 0.44 & 0.54 & 0.01 & $-$0.05 & 3.91 & $-$0.07 & 53.68 \\
30000 & 0.400 & 0.231 & 11.22 & 9.16 & 4.69 & $-$1.32 & 0.13 & 0.12 & 0.01 & $-$0.04 & 4.37 & $-$0.10 & 15.89 \\
30000 & 0.650 & 0.040 & 43.25 & 29.71 & 19.04 & 3.06 & 0.83 & 1.02 & 0.01 & $-$0.04 & 6.96 & $-$0.10 & 56.35 \\

\hline \hline

\end{tabular}

\caption{\label{tab3515a1}
HERA combined reduced  cross sections  $\sigma^{- }_{r,{\rm CC}}$ for CC $e^{-}p$ scattering at $\sqrt{s} = 318 $~GeV.
The uncertainties are quoted in percent relative to $\sigma^{- }_{r,{\rm CC}}$.
Other details as for Table~\ref{tab615-318a1}. }

\end{scriptsize}
\end{center}
\end{table}

\end{document}